\documentclass[english,12pt,a4paper]{book}

\usepackage{lmodern}
\usepackage[T1]{fontenc}
\usepackage{fix-cm}
\usepackage{anyfontsize}
\usepackage[utf8]{inputenc}

\usepackage[english]{babel}
\usepackage{caption}
\usepackage{amsmath,amssymb}

\usepackage{graphicx}
\usepackage{xcolor}
\definecolor{Valentia}{RGB}{233,78,82}
\definecolor{Titleblue}{RGB}{114, 146, 162}

\usepackage{geometry}
\usepackage{fancyhdr}
\usepackage{pdfpages}

\usepackage{sectsty}
\usepackage{titlesec}
\usepackage[Sonny]{fncychap}

\usepackage{array, colortbl, multirow, multicol, longtable, booktabs}
\usepackage{mdframed}

\usepackage[absolute]{textpos}
\usepackage{scrextend}
\usepackage{tikz}
\usepackage{setspace}
\usepackage[normalem]{ulem}
\usepackage{wrapfig}

\usepackage[square, numbers, sort&compress]{natbib}
\usepackage[pdftex,colorlinks,linkcolor=black,citecolor=red,urlcolor=blue]{hyperref}

\fancyheadoffset{0.005\textwidth}
\begin{document}
\pagestyle{empty}
\begin{titlepage}

\newgeometry{left=2.5cm, bottom=3cm, top=2cm, right=2.5cm}

\tikz[remember picture,overlay]
  \node[inner sep=0pt, anchor=north west] at (current page.north west) {
    \includegraphics[width=\paperwidth, height=\paperheight, keepaspectratio=false]{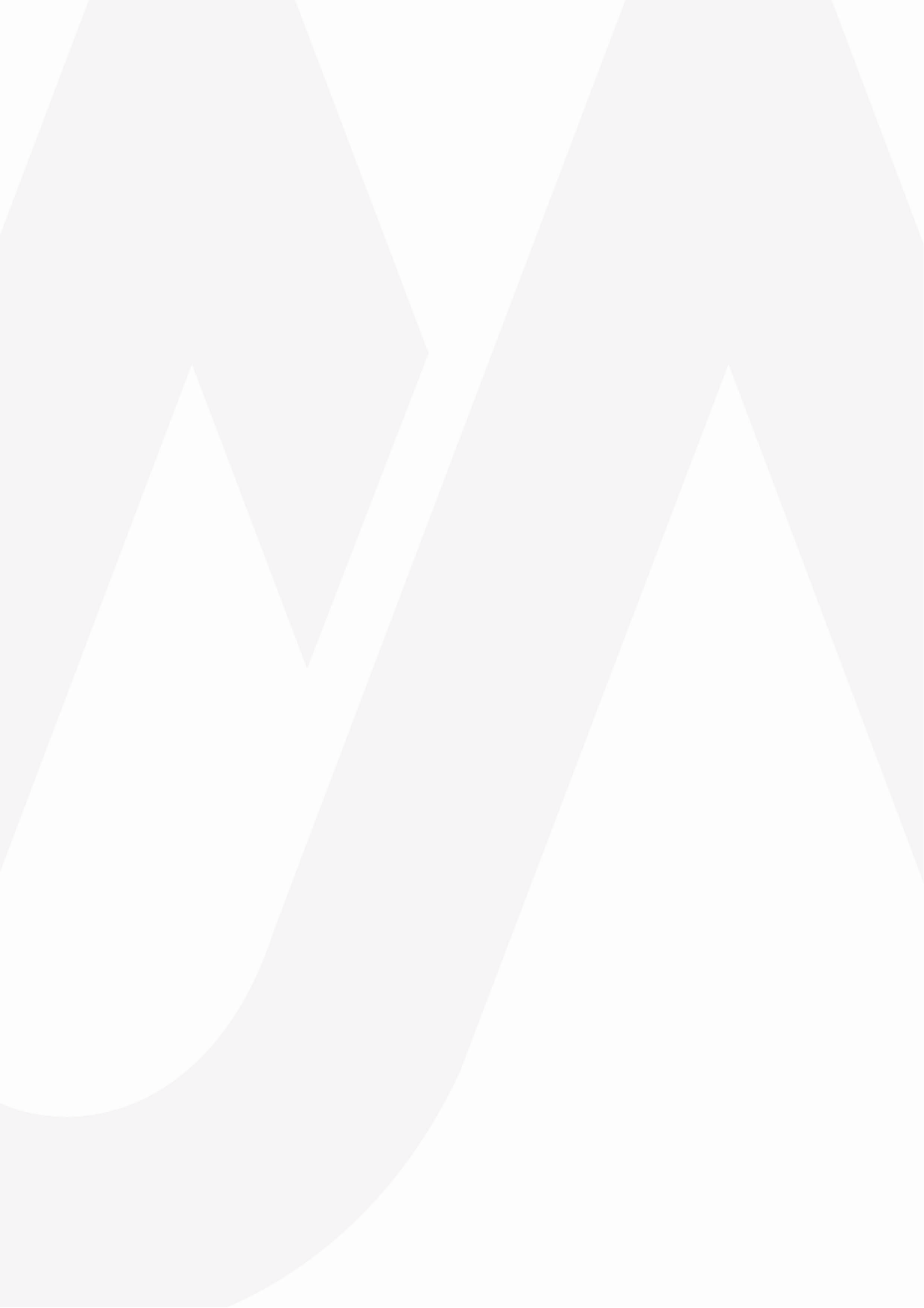}  };

\fontfamily{lmss}\selectfont 
\fontfamily{phv}\fontseries{b}\selectfont
\centering
\color{Valentia}
\fontsize{18}{20}\selectfont
\textbf{L’UNIVERSIT\'{E} DE MONTPELLIER}

\normalsize
\color{black}

\bigskip
\textbf{M\'{e}moire d'Habilitation \`{a} Diriger des Recherches}

\bigskip
\textbf{\'{E}cole doctorale: Physique}

\bigskip
\textbf{Unit\'{e} de recherche: Laboratoire Charles Coulomb (L2C)}
~\\~ ~\\~

\bigskip
\color{Titleblue}
\fontsize{17}{20.4}\selectfont
\vspace{2cm}
\textbf{Pseudo-relativistic fermions and topological states \\ \makebox[\textwidth][c]{in zinc-blende semiconductors and their heterostructures}}

\vspace{4cm}
\fontsize{15}{18}\selectfont
\color{black}
\textbf{Pr\'{e}sent\'{e}e par Sergey KRISHTOPENKO\\
le 25 Mars 2026}

\bigskip
\fontsize{13}{15}\selectfont

\vspace{1.5cm}
\normalsize
\textbf{Devant le jury compos\'{e} de}\\
\bigskip
\fontsize{10}{12}\selectfont
\vspace{1.5mm}
\begin{tabular}{p{14cm}p{3.5cm}}
\textbf{Alexey KAVOKIN, Professeur, Universit\'{e} du Lac de l'Ouest, Hangzhou, Chine} & \textbf{Rapporteur} \\
\textbf{Robson FERREIRA, Directeur de recherche, LPENS CNRS, Paris, France} & \textbf{Rapporteur} \\
\textbf{Michel GOIRAN, Professeur, LNCMI CNRS, Toulouse, France} & \textbf{Pr\'{e}sident de jury\newline et Rapporteur} \\
\textbf{Alexei BARANOV, Directeur de recherche, IES CNRS, Montpellier, France} & \textbf{Membre du jury} \\
\textbf{Benoit JOUAULT, Directeur de recherche, L2C CNRS, Montpellier, France} & \textbf{Membre du jury} \\
\textbf{Fr\'{e}d\'{e}ric TEPPE, Directeur de recherche, L2C CNRS, Montpellier, France} & \textbf{Membre du jury}
\end{tabular}

\vspace{\fill}
\includegraphics[scale=1]{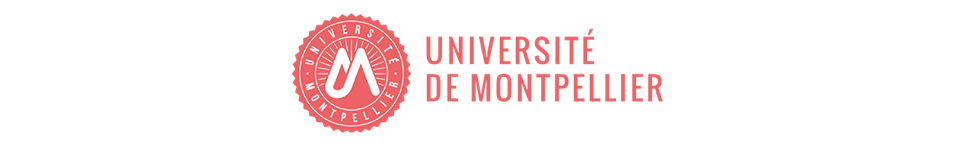}
\vspace{-15mm}
\end{titlepage}

\newgeometry{top=2cm, bottom=2.5cm, left=2cm, right=2cm}
\setstretch{1.3}

\clearpage
\null
\thispagestyle{empty}

\pagestyle{fancy}
\fancyhf{}
\fancyfoot[C]{\thepage}
\pagenumbering{arabic}
\setcounter{page}{1}

\includepdf[
  pages=1,
  pagecommand={%
     \begin{textblock*}{\textwidth}(2cm,2cm)
      \chapter*{Declarations}
    \end{textblock*}%
  }]{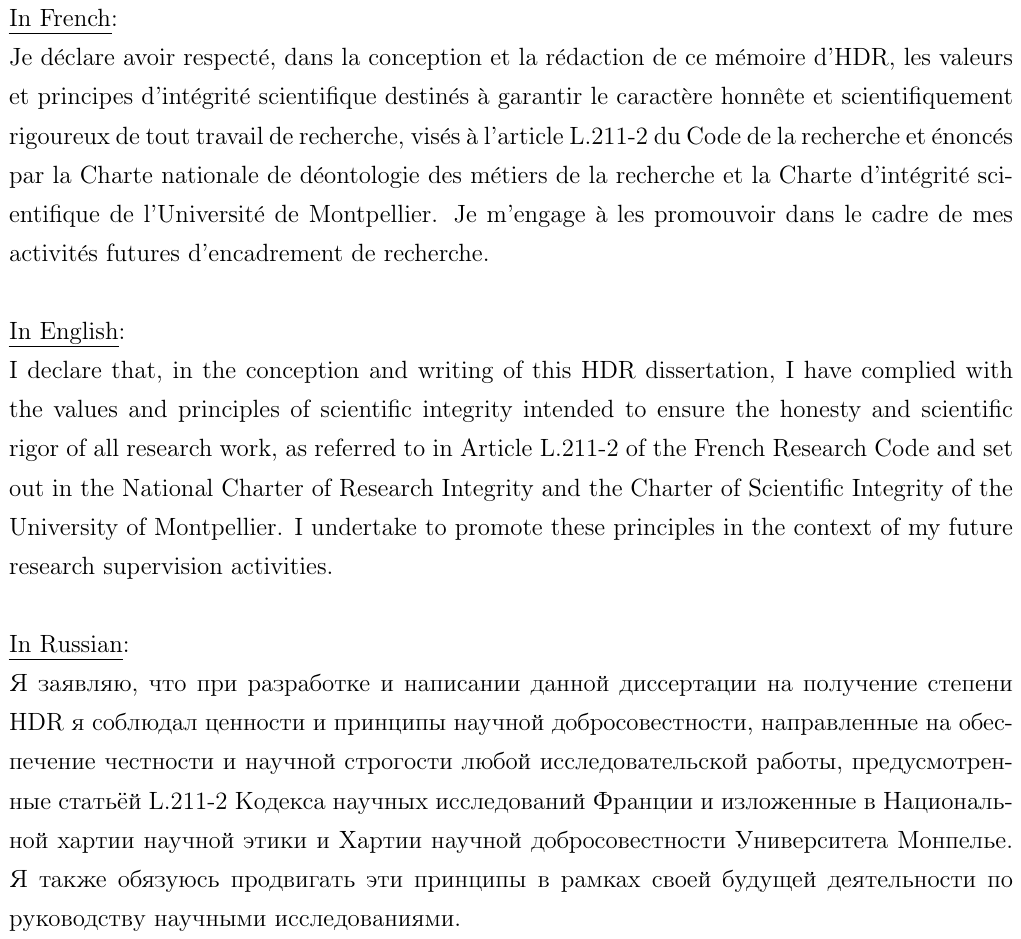}




\titleformat{\section}
{\normalfont\fontsize{14}{14}\selectfont}
{\thesection}{1em}{\textbf}

\tableofcontents


\chapter*{\label{sec:Int}Introduction}
\phantomsection 
\addcontentsline{toc}{chapter}{Introduction}
The writing of this \emph{m\'{e}moire d'Habilitation \`{a} diriger des Recherches} (HdR) is of course dictated by administrative requirements. But beyond these formal motivations, I also see it as a rare opportunity to look back at my scientific path, to organize a wide range of results obtained over the years, and to reflect on possible future directions. What perhaps characterizes my way in science most is the fact that, although I am formally a ``theorist'', I have spent my entire career working hand in hand with experimentalists being always interested in physics arising in realistic low-dimensional semiconductor structures. This experience has shaped a conviction that theory must above all explain experimentally established facts, rather than exist only as a collection of elegant but ultimately sterile models. My own work has shown me how easily theoretical descriptions can lose their relevance when confronted with the complexity of real systems such as quantum wires, quantum wells (QWs), and superlattices (SLs) based on narrow-gap or gapless semiconductors. In these materials, the proximity of conduction and valence bands often produces phenomena that defy intuition, demanding realistic and careful modeling of physical phenomena in their heterostructures. Many of my papers reflect this interplay: some were written in response to questions raised by experiments, while others proposed predictions that later inspired new measurements. In several experimental publications where I appear as first or second author, one can see this fruitful ``dialogue'' between theory and experiment particularly clearly.

This m\'{e}moire aims to familiarize the reader with the particularities of my scientific career path and to provide an overview of how it has shaped the researcher I have become today. Writing this m\'{e}moire has been a true challenge for me, as this largely autobiographical text differs in both style and purpose from all scientific papers, project proposals, and even my two PhD theses written earlier. To be honest, I had never faced anything similar before, since dissertations written for the French HdR are subject to specific academic traditions and expectations, which are different from the ones of other countries. For example, in Russia, Poland, and several other Eastern European countries, an HdR dissertation typically resembles a scientific review summarizing the researcher's own published works. Since I intend to make my HdR m\'{e}moire public after its defense, I would like to emphasize its especial style, comes from the requirements of the University of Montpellier. I am aware that, for many of my colleagues from outside France, this narrative and reflective format may appear somewhat unusual, yet it provides a rare opportunity to combine personal reflection with a scientific overview.

At the same time, I sought to reveal a unifying framework that connects a large number of studies performed in the THz~team of Laboratoire Charles Coulomb in the University of Montpellier. In my opinion, the m\'{e}moire's title ``\emph{Pseudo-relativistic fermions and topological states in zinc-blende semiconductors and their heterostructures}'' represents perfectly the concept of investigations made there over the past ten years I have spent in Montpellier. Naturally, a significant portion of my published works could not be included in this m\'{e}moire simply because they did not fit neatly within the chosen structure. I also made a deliberate effort not to overburden the text with mathematics in the parts devoted to theoretical results, or with excessive experimental details in the experimental reviews. All relevant technical details and quantitative discussions can be found in the original publications cited throughout the text. Finally, I would like to extend my heartfelt thanks to my colleagues Fr\'{e}d\'{e}ric Teppe and Beno\^{\i}t Jouault for their invaluable assistance in navigating the intricacies of French academic bureaucracy during the preparation of this m\'{e}moire. Their support has been instrumental in allowing me to focus on the writing process itself.

\includepdf[
  pages=1, 
  pagecommand={%
    \thispagestyle{fancy}%
    \phantomsection 
    \addcontentsline{toc}{chapter}{Curriculum Vitae}%
    \begin{textblock*}{\textwidth}(2cm,2cm) 
          \chapter*{\label{sec:CV} Curriculum Vitae}
    \end{textblock*}%
  }]{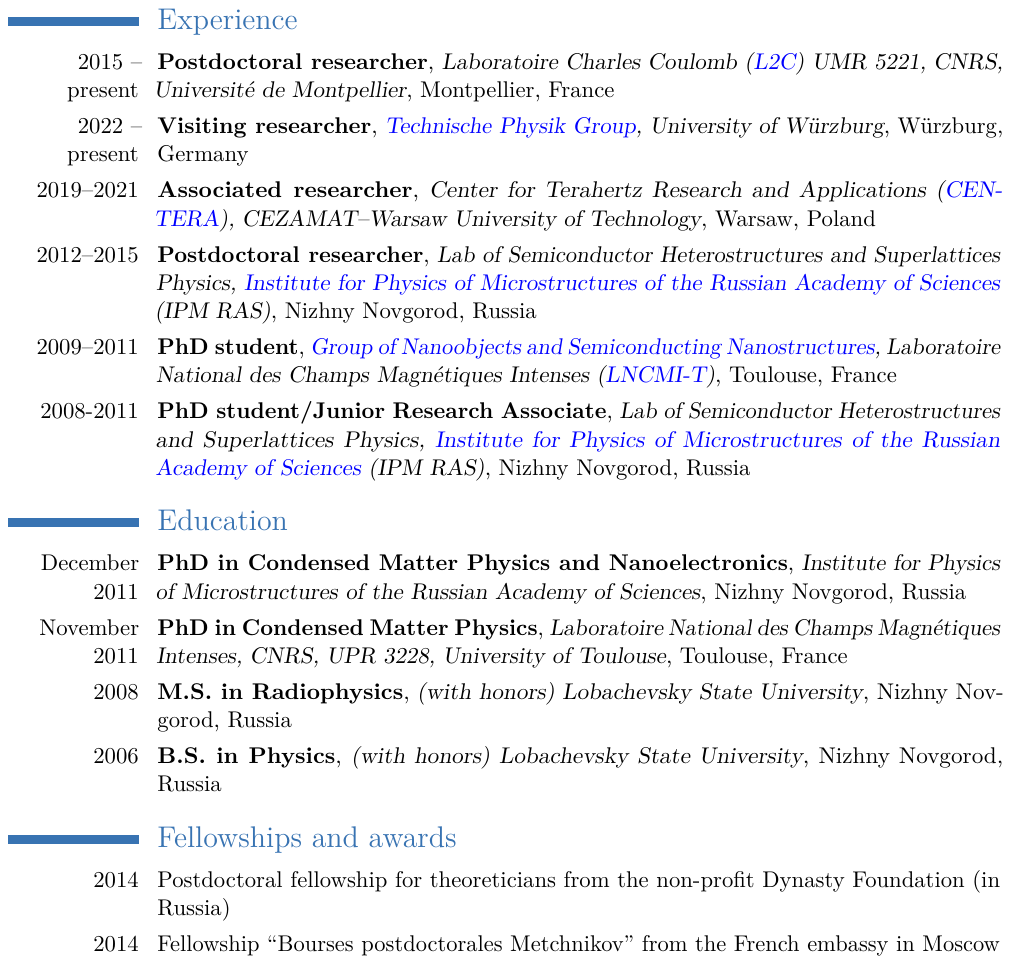}
\includepdf[pages=2-, pagecommand=\thispagestyle{fancy}]{CVthesis2.pdf}

\chapter{\label{Chap:Career} Scientific career path}
My scientific path began in Nizhny Novgorod, when, as a third-year student at the \href{http://eng.unn.ru/about/academic-units/faculty-of-radiophysics}{Faculty of Radiophysics} of Lobachevsky State University, I joined the laboratory of Vladimir Gavrilenko at the \href{http://www.ipmras.ru/en/}{Institute for Physics of Microstructures} of the Russian Academy of Sciences (IPM RAS). Vladimir Gavrilenko, himself an experimentalist in the field of semiconductor heterostructures, naturally encouraged me to take part in experimental studies of magnetotransport and magneto-optics in narrow-gap QWs. During those years, I made a genuine effort to become a capable experimentalist, and I can say that I partially succeeded: I was able to carry out measurements of magnetotransport, cyclotron resonance, and persistent photoconductivity in InAs/AlSb QWs. Yet even then, I realized that what excited me most was not the measurement itself, but rather the attempt to understand and explain the underlying physics -- for instance, why the cyclotron mass depended on electron concentration in such a non-trivial way. This curiosity gradually drew me toward theoretical modeling of the processes studied at IPM RAS, and marked the beginning of my scientific identity: a theorist deeply rooted in experimental reality.

After graduating from Lobachevsky State University with honors, I entered the PhD program at IPM RAS and joined Vladimir Gavrilenko's group as a doctoral student. At the same time, having won a French government scholarship (the ``Vernadsky'' doctoral grant) through the Embassy of France in Moscow, I also became a PhD student at the Universit\'{e} Paul Sabatier in Toulouse, under the supervision of Michel Goiran, who led a research group at the National Laboratory for High Magnetic Fields (\href{http://lncmi.cnrs.fr/informations-pratiques-toulouse/}{LNCMI-T}). This made me a ``double'' PhD student, spending roughly six months each year in Toulouse and six months in Nizhny Novgorod -- a truly formative experience, both scientifically and personally. Much like my Russian advisor, my French supervisor did everything possible as well to turn me into an experimentalist. Each time I came to Toulouse, I brought with me InAs/AlSb QW samples prepared at IPM RAS for experimental studies in strong magnetic fields. However, since my access to high-field experiments at LNCMI-T was quite limited, I had a lot of time to analyze the results of magnetotransport measurements I had performed myself, as well as the magneto-absorption data obtained on similar samples by my senior colleagues Anton Ikonnikov and Milan Orlita in Grenoble. Through a detailed analysis of these numerous experimental datasets, I succeed to identify reproducible effects that could not be explained within the usual single-particle picture of Landau levels in strong magnetic fields. This marked the beginning of my deep interest in electron-electron (\emph{e-e}) interaction effects in narrow-gap semiconductor QWs in the quantum Hall effect regimes. The small band-gap in these materials not only leads to strong non-parabolic (Dirac- or Kane-like) energy dispersion in the QW subbands, but also gives rise to the spinor character of their wave-functions -- a consequence of the coupling between the $|S\rangle$-like conduction and $|P\rangle$-like valence states. This, in turn, profoundly modifies the nature of \emph{e-e} interactions compared to those in a conventional two-dimensional (2D) electron gas with parabolic band dispersion.

Fascinated by these questions, I spent all of my free time developing a theoretical framework capable of describing the manifestations of many-body effects in narrow-gap semiconductor QWs. I am deeply grateful to both of my supervisors, Vladimir Gavrilenko and Michel Goiran, for recognizing my growing passion for theory and for encouraging my independent research, even though it diverged from their experimental orientation. Thanks to their support, I succeed to solve several nontrivial many-body problems, including the generalization of the theory of exchange-enhanced g-factor -- originally proposed by (\href{http://journals.jps.jp/doi/10.1143/JPSJ.37.1044}{Ando and Uemura} -- to the case of 2D electron gases in narrow-gap QWs. These theoretical developments, together with the results of my own experimental studies, formed the foundation of my PhD thesis ``Spin splitting and collective effects in InAs/AlSb QW heterostructures''. Because of differing administrative requirements between the two institutions, I had to defend my thesis twice: first in Toulouse, in \href{http://utheme.univ-tlse3.fr/s/fr/item/1459}{English}, with reviewers Wlodek Zawadzki and Guy Fishman, and later in Nizhny Novgorod, in Russian (\href{http://www.ipmras.ru/UserFiles/Diss/Krishtopenko.pdf}{autoreferat}), with reviewers \href{http://scholar.google.com/citations?user=boexuIkAAAAJ&hl=en}{Leonid Golub} and \href{http://scholar.google.com/citations?hl=en&user=mEk2E3IAAAAJ}{Denis Vodolazov}. Although the defenses took place in two different academic cultures, they both reflected the same guiding idea of my work -- the constant dialogue between experiment and theory in understanding the physics of low-dimensional semiconductor systems.

After completing the defenses of my PhD theses, I continued my research career in the same environment that had shaped me as a scientist -- joining Vladimir Gavrilenko's group at the Institute for Physics of Microstructures of the Russian Academy of Sciences (IPM RAS) as a postdoctoral researcher. The main focus of my postdoctoral studies was the investigation of various collective excitations in QWs based on narrow-gap and gapless semiconductors in the integer and fractional quantum Hall effect regimes. In many respects, this work was a natural continuation of the research that had formed the basis of my doctoral theses, but it also allowed me to explore new conceptual aspects of many-body physics in low-dimensional systems. During this period, I succeed to obtain several results that had not been known before. These findings reinforced my conviction that narrow-gap semiconductor heterostructures offer a remarkably rich platform for studying fundamental many-body effects in strong magnetic fields. For my theoretical contributions in this area, I was honored in 2014 with a fellowship from the \href{http://web.archive.org/web/20190829233545/https://www.dynastyfdn.com/english/}{Dynasty Foundation} (see Table~\ref{tab:Fellows}), which supported young theoretical physicists holding a PhD degree. This recognition was a meaningful milestone for me, as it confirmed the scientific relevance and originality of my independent theoretical work. In parallel with research, I was actively involved in teaching and mentoring at IPM RAS. I conducted practical courses for university students and supervised their Bachelor and Master theses -- an experience I found both challenging and rewarding (further details about this activity are presented in Section~\ref{sec:StudentRU} of this m\'{e}moire). I discovered that supervising students often provides a unique perspective on scientific problems revealing solutions, unexpected previously. Based on my own experience, I can confidently say that mentoring students contributes not only to their development but also to the professional growth of their supervisor. An interesting moment in my career came just four years after obtaining my PhD, when I was invited to serve as an external reviewer for the PhD defense of \href{http://www.ioffe.ru/theory/sector/kokurin.html}{Ivan Kokurin}, a student of \href{http://www.ioffe.ru/theory/sector/averkiev.html}{Nikita Averkiev} from the Ioffe Institute. His dissertation, entitled ``\emph{Spin-orbit interaction effects in ultrathin semiconductor nanostructures}'' (\href{http://www.ipmras.ru/UserFiles/Diss/Kokurin/Kokurin_autoreferat.pdf}{autoreferat in Russian}), dealt with topics very close to my own research interests. Being asked to act as an official reviewer at such an early stage in my career was both an honor and an affirmation of the recognition my work had gained within the Russian condensed matter community.

\begin{table}[h]
\caption{\label{tab:Fellows} Post-doctoral fellowships}
\begin{tabular}{|p{0.515\textwidth}|p{0.35\textwidth}|p{0.05\textwidth}|}
\hline
\centering{Title} & \centering{Who awarded} & Year  \\
\hline
\centering { Postdoctoral fellowship for theoreticians in Physics } & \centering{\href{http://web.archive.org/web/20190829233545/https://www.dynastyfdn.com/english/}{Dynasty Foundation}} & 2014 \\
\hline
\centering { ``Bourse postdoctorale Metchnikov'' } & \centering{French Embassy in Moscow} & 2014 \\
\hline
\end{tabular}
\end{table}

The years spent at IPM RAS also significantly broadened the scope of my scientific interests. The long-standing collaboration between Vladimir Gavrilenko's group and the {\href{http://coulomb.umontpellier.fr/}{Laboratoire Charles Coulomb} (L2C) in Montpellier played a particularly important role in this evolution. These two groups had a rich history of joint research projects, and at that time the most active collaboration was focused on the investigation of HgTe/CdHgTe QWs by means of terahertz spectroscopy methods. It was precisely the experimental results obtained by {\href{http://scholar.google.com/citations?user=-HRLT2wAAAAJ&hl=en}{Fr\'{e}d\'{e}ric Teppe} and his PhD student Maxim Zholudev -- under joint supervision with Vladimir Gavrilenko -- that first drew my attention to the fascinating physics of gapless semiconductor heterostructures and, more specifically, to quantum spin Hall insulators (QSHIs). The field of topological phases in condensed matter was at that moment of great interest, and HgTe-based QWs provided a unique platform to probe experimentally new ideas of topological physics. My own path into this emerging field began with the theoretical study of tunnel-coupled QSHI systems, which can be realized experimentally in double HgTe QWs. These structures, consisting of two HgTe layers separated by a thin CdHgTe barrier, allow one to probe the interplay between topological states, tunneling, and symmetry breaking -- topics that have remained at the heart of my research ever since.

In 2014, I received a postdoctoral fellowship ``Bourse postdoctorale Metchnikov'' from the French Embassy in Moscow (see Table~\ref{tab:Fellows}), which offered the opportunity to carry out a short-term research stay in any laboratory in France. I chose to spend two months in ``Terahertz spectroscopy team'' (THz team) of Fr\'{e}d\'{e}ric Teppe in Montpellier. Our discussions of my preliminary theoretical results on double HgTe QWs, combined with our shared enthusiasm for the rapidly developing field of topological materials, defined the direction of my subsequent research. From that moment, the study of HgTe-based topological systems and their unique electronic properties became one of the central themes of my scientific career. In 2015, I made the decision to leave IPM RAS and join  Fr\'{e}d\'{e}ric Teppe's group in Montpellier. The main motivation for this step was not only attributed to the personal reasons but also the opportunity to see many of my theoretical ideas come to life through direct experimental realizations -- a possibility that had always been central to my way of research in physics. Another important factor was the greater scientific freedom offered by Fr\'{e}d\'{e}ric, which allowed me to explore new directions more independently than before. At that time, the field of HgTe-based topological insulators was developing extremely rapidly, and THz team provided a uniquely stimulating environment, in which theoretical insights could be quickly tested experimentally. Several ideas that had first emerged from my theoretical studies soon found tangible confirmation in experiments carried out within L2C. Among these were investigations of temperature-driven topological phase transitions in single HgTe QWs, as well as the first experimental studies of double HgTe QWs -- a system whose concept had originally arisen from my theoretical research. Interestingly, the samples with double HgTe QWs initially grown in Novosibirsk for THz team also became a central object of research in several other groups worldwide.

Being free to choose my research directions, my interest in topological states was by no means limited to those realized in HgTe-based QW heterostructures. For a long time, I sought to find ways to ``realize'' the physics of single and double HgTe QWs but in III-V semiconductor QW heterostructures. This motivation stemmed from a very practical consideration: at that time, high-quality HgTe-based QWs necessary for the experimental investigation of topological states were grown only in three research groups worldwide -- the \href{http://scholar.google.com/citations?user=ptmy2MsAAAAJ&hl=en}{
 Molenkamp}'s group at the University of W\"{u}rzburg, the group of \href{http://www.researchgate.net/profile/Nikolay-Mikhailov}{Mikhailov} and \href{http://www.researchgate.net/profile/S-Dvoretskii}{Dvoretskii} in Novosibirsk, and (to a very limited extent) at \href{http://www.cea.fr/cea-tech/leti}{CEA
 Leti} in Grenoble. By contrast, high-quality III-V semiconductor heterostructures are routinely grown in dozens of laboratories around the world, offering a much broader experimental platform. Another critical advantage lies in the relative ease of device fabrication: processing III-V materials is considerably simpler than working with HgCdTe-based heterostructures, which are mechanically fragile and chemically less stable. Finally, III-V semiconductor heterostructures can be grown on silicon substrates -- as demonstrated, for instance, by the \href{http://www.ies.umontpellier.fr/la-recherche-et-linnovation/les-equipes-de-recherche/nanomir/}{NanoMIR} group in Montpellier -- opening perspectives for their large-scale integration with existing semiconductor technologies. Taken together, these factors make III-V semiconductor heterostructures an extremely attractive playground for the realization of novel devices operating on the principles of topological physics.

Among III-V semiconductor heterostructures, the only system in which a QSHI state had been both theoretically predicted and experimentally observed at that time was the InAs/GaSb QW bilayer. The main limitation of these structures stemmed from the relatively small topological band-gap, on the order of 3--4~meV, which opened for bulk states away from the $\Gamma$ point of the Brillouin zone. As a result, the observation of quantized edge conductance in such systems was restricted to extremely low temperatures, typically below 100~mK. I remember quite clearly that around 2015--2016, while preparing for publication my paper on topological states in double HgTe QWs, an idea occurred to me that could not only overcome the limitations associated with the small band-gap in InAs/GaSb QW bilayers, but also allow the full reproduction of the physics of single and double HgTe QWs -- including the emergence of massless Dirac fermions, relatively large-gap QSHI state, and the semi-metallic ``bilayer graphene''-like state -- but this time within III-V semiconductor heterostructures. Subsequent theoretical development of this concept demonstrated the possibility of realizing a QSHI state with a topological band-gap as large as 60~meV, which would in principle permit the observation of the quantum spin Hall effect at temperatures well above 77~K (the liquid nitrogen temperature). Inspired by these revolutionary theoretical predictions, Fr\'{e}d\'{e}ric and I shared a common dream -- to obtain experimental confirmation of these results within the THz team in Montpellier. This strong desire to achieve a complete theoretical-experimental verification of the proposed concept also explains the deliberate delay in publishing my theoretical findings: our paper was finally released only in 2018, in \href{http://www.science.org/doi/10.1126/sciadv.aap7529}{Science Advances}.
Interestingly, in the same year 2018, the possibility of observing the quantum spin Hall effect up to temperatures of about 100~K was experimentally demonstrated in \href{https://www.science.org/doi/10.1126/science.aan6003}{monolayer WTe$_2$}, which possesses a band-gap of approximately 45~meV.

\begin{table}[h]
\caption{\label{tab:Rej} The projects on ``topological'' III-V semiconductor structures from the THz team rejected for funding by ``\href{http://anr.fr/en/}{l'Agence nationale de la recherche}''.}
\begin{tabular}{|p{0.68\textwidth}|p{0.18\textwidth}|p{0.05\textwidth}|}
\hline
\centering{Title} & \centering{Acronim} & Year  \\
\hline
\centering{Dirac fermions in III-V semiconductor structures: towards relativistic terahertz electronics} & \centering{DIRACTRONICS} & 2016 \\
\hline
\centering{Towards topological electronics in III-V-based semiconductor structures} & \centering{TOPTRONICS} & 2017 \\
\hline
\centering{Quantum spin Hall insulator: a bridge to topological superconductivity} & \centering{HIBRID} & 2018 \\
\hline
\centering{Towards Quantum Anomalous Hall effect with III-V heterostructures} & \centering{QUIVER} & 2019 \\
\hline
\end{tabular}
\end{table}

Unfortunately, none of the experimental projects proposed by the THz team -- which aimed at realizing and studying the suggested QW heterostructures to observe a high-temperature quantum spin Hall effect -- received the necessary funding (see Table~\ref{tab:Rej}). The lack of financial support severely limited the development of the technological expertise required for the fabrication of gated micron-sized Hall bar devices in the cleanroom facilities of the University of Montpellier. Such devices are essential for detecting quantized edge conductance in the quantum spin Hall effect regime. Nevertheless, even in the absence of device processing capabilities, we succeeded in performing the first experimental investigations of the proposed III-V semiconductor heterostructures by means THz spectroscopy techniques. The samples were grown for us in the \href{http://www.ies.umontpellier.fr/la-recherche-et-linnovation/les-equipes-de-recherche/nanomir/}{NanoMIR} group of \'{E}ric Tourni\'{e}, with whom we established a very fruitful collaboration. These studies provided the first experimental evidence of \href{http://www.ies.umontpellier.fr/la-recherche-et-linnovation/les-equipes-de-recherche/nanomir/}{massless Dirac fermions} in the designed QW heterostructures and revealed a remarkable \href{http://journals.aps.org/prb/abstract/10.1103/PhysRevB.97.245419}{temperature independence of the inverted band-gap} -- a crucial prerequisite for the realization of the high-temperature quantum spin Hall effect. My leading role in coordinating these experimental studies is reflected in the fact that I appear as the first author of both publication, released in 2019 and 2018, respectively.

The situation regarding the observation of the quantum spin Hall effect changed significantly after the beginning of our collaboration with the \href{http://www.physik.uni-wuerzburg.de/tep/}{Technische Physik} group at the University of W\"{u}rzburg, led by \href{https://scholar.google.com/citations?user=ZVggr60AAAAJ&hl=en}{Sven H\"{o}fling} (further details are presented in Section~\ref{sec:Wurz}). The development of advanced fabrication techniques for gated micron-sized Hall bar devices finally made it possible to observe quantized conductance values arising from edge-state transport in the QW heterostructures up to temperatures of about 60~K (\href{http://www.science.org/doi/10.1126/sciadv.adz2408}{Science Advances 2025}). Importantly, the observed temperature limitation of about 60~K was found to originate from the instability of the gate operation, a purely technical issue that can be resolved by improving the device processing. These results provide clear evidence that the realization of the quantum spin Hall effect in III-V semiconductor heterostructures at temperatures above 77~K is now only a matter of time. This experimental achievement not only confirms the validity of the theoretical predictions made in Montpellier (\href{http://www.science.org/doi/10.1126/sciadv.aap7529}{Science Advances 2018}) but also opens the door to the practical implementation of topological devices operating under technologically relevant conditions.

Obviously, the QSHI state (except integer and fractional quantum Hall insulators) is not the only topological phase that can, in principle, be realized in QW heterostructures, including both HgCdTe-based and III-V-semiconductor-based QWs.
A wide variety of topological states, including those of higher-order, arises from the interplay between band inversions, crystalline symmetries, and external factors such as different types of interactions or applied fields. For quite some time now, one of the central goals of my theoretical research in Fr\'{e}d\'{e}ric Teppe's group has been to demonstrate that nearly all known topological states in condensed matter can be realized on the basis of III-V-semiconductor heterostructures. The key idea guiding this work is that both the required band inversion and the necessary crystalline symmetry can be engineered in purposefully designed 2D and three-dimensional (3D) systems. As clear, the band inversion can be tuned through quantum confinement in QWs or within the unit cell of superlattices, as well as by applying external fields. In turn, ``symmetry engineering" can be performed by growing the heterostructure along specific crystallographic orientations, which can be varied in the wide range for the case of zinc-blende semiconductors. The second way is the symmetry engineering of isolate heterointerface. Particularly, the molecular beam epitaxy (MBE) of III-V semiconductors allows creating specific chemical bounds at the interfaces, which does not share common atoms. Quantum wells and superlattices with common-atom and non-common-atom interfaces exhibit different crystalline symmetries, which in turn may lead to different topological properties.

To date, I have theoretically demonstrated the existence of a 2D higher-order topological insulator (HOTI) phase in both HgCdTe- and III-V-semiconductor-based QWs with double band inversion. In superlattices, I have predicted the emergence of higher-order three-dimensional (3D) Dirac semimetal (SM) and 3D HOTI states that appear for specific layer thicknesses within the unit cell. I hope that, in the near future, the THz team, Sven H\"{o}fling's group, or another research team will be able to obtain experimental confirmation of these theoretical predictions. At present, I am developing several original ideas aimed at realizing a higher-order 3D Weyl SM state in III-V-semiconductor superlattices and at investigating the conditions for observing the \href{http://journals.aps.org/prl/abstract/10.1103/5yxp-djy9}{quantum electric Hall effect} in III-V-semiconductor QWs. In parallel, I am also exploring the possibility of controlling phase transitions between different topological states in semiconductor QWs and superlattices by means of external electromagnetic fields. These include both TE and TM modes of microwave cavity, as well as the intrinsic modes of photonic crystals, which may provide entirely new routes for dynamic manipulation of topological electronic states.

Apart from my work on topological states, my research activity in the THz team also expanded into an entirely new field for me -- the physics of non-equilibrium (hot) charge carriers in crossed electric and magnetic fields. The starting point for this new direction was the experimental observation of \href{http://www.nature.com/articles/s41566-019-0496-1}{spontaneous cyclotron (Landau) emission from bulk HgCdTe films}, induced by an in-plane electric field applied to the structure. Historically, the studies of cyclotron emission of hot charge carriers had been performed since the second half of the 1970s in several leading research centers around the world. Among them were the group of \href{http://www.researchgate.net/profile/Erich-Gornik}{Erich Gornik} in Europe, the group of \href{http://scholar.google.com.au/citations?user=UoFf61gAAAAJ&hl=en}{Susumu Komiyama} in Japan, and several laboratories in the USSR that conducted extensive experimental and theoretical research on hot-carrier physics in bulk Si, Ge, GaAs (and other semiconductor materials). In particular, key contributions to the understanding of hot-carrier behavior in crossed electric and magnetic fields were made by research groups in Moscow, Gorky (now Nizhny Novgorod), Leningrad (now Saint-Petersburg), and Vilnius. Unfortunately, many of the most important papers published in this field remain accessible only in printed form, preserved in the libraries of scientific institutes. For this reason, the reviews issued between 1983 and 1990 by the \href{http://www.ipfran.ru/}{Institute of Applied Physics} of the Soviet Academy of Sciences represent an especially valuable source of knowledge. These books (written in Russian), effectively serve as detailed reviews summarizing both experimental and theoretical progress achieved in the Soviet research community during that period:
\begin{itemize}
    \item{ \href{http://old.ipfran.ru/biblio/img/he.pdf}{Hot Electrons in Semiconductors: Streaming and Anisotropic Distributions in Crossed Fields} (1983)};
    \item{ \href{http://old.ipfran.ru/biblio/img/hep.pdf}{Inverted distributions of hot electrons in semiconductors} (1983)};
    \item{ \href{http://old.ipfran.ru/biblio/img/las.pdf}{Submillimeter lasers on hot holes in semiconductors} (1986)};
    \item{ \href{http://old.ipfran.ru/biblio/img/pm.pdf}{Cyclotron resonance semiconductor masers} (1990)}.
\end{itemize}
These studies formed the foundation for the creation of Landau lasers based on bulk p-type germanium (\emph{p}-Ge) of several types, covering a frequency range between 65 GHz and 375 GHz in magnetic fields up to 5--8 T. Although Landau emission has also been studied intensively in various materials such as \emph{n}-type InSb, GaAs, and Si inversion layers, the \emph{p}-Ge lasers remain the only semiconductor Landau lasers ever successfully implemented to date. The uniqueness of the experimental results obtained by the THz team in Montpellier lies in the range of energies and magnetic fields corresponding to the observed Landau emission. The detected emission covered 2--7~meV (0.5--1.6~THz) and was observed in magnetic fields up to 0.25~T. Taking into account that modern MBE growth technologies, which were unavailable forty years ago, now make it possible to produce not only bulk films but also QW heterostructures of extremely high quality, the idea of developing a powerful terahertz Landau laser based on new materials and operating in the magnetic fields of permanent magnets appears highly realistic. It is worth noting that the last experimental studies of Landau emission had been carried out in Montpellier by \href{http://scholar.google.com/citations?user=S7v731YAAAAJ&hl=en}{Wojciech Knap} more than fifteen years ago. Therefore, there have been no specialists in L2C capable of providing theoretical support for the ongoing experiments. Since Russian is my native language, I naturally had access to a wide range of Soviet-era publications on the physics of hot carriers in crossed electric and magnetic fields. This background allowed me to deepen my own expertise in this entirely new field, primarily in order to assist with the interpretation of experimental data obtained by the THz team.

The main difficulty in describing non-equilibrium charge carriers in crossed electric and magnetic fields, whether in bulk materials, quantum wells, or superlattices based on narrow-gap and gapless semiconductors, arises from the fact that a strong electric field not only transfers energy to charged particles -- substantially modifying their distribution function -- but also alters the Landau-level spectrum itself. The latter effect stems from the nonparabolic band structure of such materials, which contrasts sharply with the simple parabolic dispersion typical of wide-gap semiconductors. A striking example of this phenomenon is the pseudo-relativistic collapse of Landau quantization of Dirac fermions, which occurs when the carrier drift velocity, $V_d = c\mathcal{E}/\mathcal{B}$ (where $c$ is the speed of light, and $\mathcal{B}$ and $\mathcal{E}$ are the magnetic and electric field strengths, respectively) approaches the Fermi velocity $V_F$. Physically, this collapse of Landau levels means that charged particles no longer move along closed cyclotron orbits -- their trajectories become effectively unclosed. This remarkable effect was first described in semiconductor physics by Aronov and Pikus in 1966 (\href{http://jetp.ras.ru/cgi-bin/e/index/e/24/1/p188?a=list}{here} and \href{http://jetp.ras.ru/cgi-bin/e/index/e/24/2/p339?a=list}{here}), and four decades later, similar results were independently obtained by \href{http://journals.aps.org/prl/abstract/10.1103/PhysRevLett.98.116802}{Lukose~et~al.} for graphene. In the case of bulk HgCdTe films, the band structure in the small wave-vector limit can be represented by Kane fermions, whose energy spectrum near the Brillouin-zone center ($\Gamma$-point) consists of two Dirac-like bands and a ``flat'' $|\Gamma_8,{\pm}3/2\rangle$ band. The latter is attached either to the valence or conduction band, depending on whether the material exhibits a direct or inverted band structure (see Chapter~\ref{Chap:Kane}). Consequently, in such system, one can expect a pronounced dependence of the Landau-level energies not only on the magnetic field but also on the applied electric field. It is most likely this effect that underlies the pronounced shifts of cyclotron emission lines observed experimentally from Landau-quantized massless Kane electrons at high current densities -- well above those used in (\href{http://www.nature.com/articles/s41566-019-0496-1}{Nat. Photonics 2019}). These experimental results remain unpublished to date and still require further systematic investigation. Nevertheless, they became the primary motivation to develop a rigorous theoretical description of the \href{http://journals.aps.org/prb/abstract/10.1103/PhysRevB.105.125203}{relativistic collapse of Landau levels of Kane fermions in crossed electric and magnetic fields} within the fully quantum-mechanical framework (see Section~\ref{sec:LLKane}). I should particularly emphasize that the last researcher who addressed this problem, in the context of InSb, was Wlodek Zawadzki in 1985 (\href{http://journals.aps.org/prl/abstract/10.1103/PhysRevLett.55.983}{here} and \href{http://journals.aps.org/prb/abstract/10.1103/PhysRevB.33.6916}{here}). As demonstrated in our paper, the approximate solution obtained by Zawadzki corresponds to the quasiclassical treatment of the problem. After my PhD thesis defense in Toulouse, I had a warm and friendly relationship with Wlodek, and I regret that I was unable to discuss these new theoretical results with him in person. Wlodek passed away just a few months before the problem of the Landau levels in crossed fields was finally solved.

In addition to bulk HgCdTe films, the THz team also reported for the first time the observation of cyclotron emission from Dirac fermions in 2D systems, such as HgTe QWs and graphene. From a theoretical point of view, the interpretation of experimental data in these systems is, in a certain sense, simpler than in the case of Kane fermions. As experimental studies on HgTe QWs have shown, when the applied electric field remains moderate -- so that the positions of spontaneous cyclotron emission lines coincide with those of absorption resonances -- the behavior of these systems can be well described within the quasiclassical approximation (\href{http://www.nature.com/articles/s41566-022-01129-1}{Nat.~Photonics~2023}, \href{http://pubs.aip.org/aip/app/article/8/11/116106/2920162/Gate-tunable-terahertz-cyclotron-emission-from-two}{APL~Photonics~2023} and \href{http://arxiv.org/abs/2502.07897}{Arxiv~2025}). This approach, compared to the fully quantum treatment, greatly simplifies the theoretical modeling of carrier dynamics under conditions of strong applied currents. At present, my ongoing theoretical work is focused precisely on the interpretation of recent (yet unpublished) experimental results obtained by the THz team in HgTe QWs and graphene under strong electric fields. Particularly, I am developing a theoretical framework that allows numerical modeling of the dynamic conductivity tensor in conditions of strongly non-equilibrium carrier distributions in crossed electric and magnetic fields. This task is far from trivial, yet I must admit that I have found invaluable guidance in the Soviet scientific heritage summarized in the four books mentioned earlier. The expected outcome of this work will be the identification of conditions under which the real part of the conductivity tensor becomes negative, indicating the amplification of incident radiation. Such a result would represent a great step toward the realization of powerful terahertz Landau lasers operating with permanent magnets.

In conclusion of this part, which summarizes the main stages of my scientific path, I would like to note that at present I have a large number of ongoing research projects, whose development could be significantly accelerated through the involvement of young scientists, particularly PhD students. Unfortunately, in most European countries, I cannot officially supervise either Master or PhD students without holding the \emph{Habilitation \`{a} diriger des Recherches} (HdR) degree. Beyond these formal aspects, I believe that, sooner or later, every scientist feels an inner need to transmit accumulated knowledge and experience to the next generation. For me, this deeply personal motivation -- the desire to become a scientific superviser to young researchers at the beginning of their careers -- constitutes the principal reason for pursuing the HdR degree. A more detailed description of my previous experience with students is presented in Sections~\ref{sec:StudentRU} and \ref{sec:StudentFR} of this m\'{e}moire.

\section{\label{sec:Visites}Visits and collaborations}
My stay in the THz team has been marked by a steady expansion of my research collaborations beyond Montpellier, which opened new possibilities for both theoretical and experimental studies. Two of these collaborations have had a particularly strong impact on the evolution of my scientific work. The first one is with the \href{https://centera.eu/}{Center for Terahertz Research and Applications} (CENTERA), an interdisciplinary research center established in Poland with the mission of conducting fundamental studies using THz spectroscopy and developing new THz devices based on Dirac materials. The second collaboration that has become central to my research on III-V semiconductor heterostructures is with the \href{http://www.physik.uni-wuerzburg.de/tep/}{Technische Physik} group at the University of W\"{u}rzburg, headed by \href{https://scholar.google.com/citations?user=ZVggr60AAAAJ&hl=en}{Sven H\"{o}fling}. This cooperation has created a natural link between my long-standing theoretical work in this field and the outstanding experimental expertise of the W\"{u}rzburg team in quantum device growth, fabrication, and characterization. Below, I will briefly highlight the main points of both collaborations.

\subsection{\label{sec:Poland}Center for Terahertz Research and Applications in Warsaw (2019--2021)}
My collaboration with \href{http://centera.eu/}{Center for Terahertz Research and Applications (CENTERA)} was by no means accidental. This center was founded by \href{http://scholar.google.com/citations?user=S7v731YAAAAJ&hl=en}{Wojciech Knap}, the former head of the THz team in Montpellier, and built upon several research groups in Warsaw that had previously participated in the international cooperation ``Laboratory of Terahertz and Mid-Infrared Collective Phenomena in Semiconductor Nanostructures'' (TERAMIR). This international laboratory brought together researchers from France, Poland, and Russia under the patronage of the CNRS within the framework of the CNRS programs ``LIA TERAMIR'' (2015--2018) and ``IRP TERAMIR'' (2019--2023). On the French side, both projects were coordinated by Fr\'{e}d\'{e}ric Teppe. Therefore, when I joined his group in Montpellier in 2015, I automatically became a member of the TERAMIR laboratory. As in Montpellier, one of the main research areas in Warsaw was the study of HgTe-based QWs by means of terahertz spectroscopy methods. Being familiar with my theoretical results on \href{http://journals.aps.org/prb/abstract/10.1103/PhysRevB.94.245402}{topological phase transitions driven by hydrostatic pressure and temperature in single and double HgTe QWs}, Wojciech proposed me to provide theoretical support for similar experimental studies carried out in Poland. Moreover, the theoretical results obtained in Montpellier laid the conceptual foundation for the \href{http://www.unipress.waw.pl/index.php?option=com_content&view=article&id=1428:project-fnp-team&catid=82:teragan&Itemid=146&lang=en}{FNP TEAM project ``Terahertz Sensor Based on Topological Materials''}, funded by the \emph{Foundation for Polish Science}. This project made it possible for Wojciech Knap to establish his own research group at the Institute of High Pressure Physics in Warsaw. Thus, it is not surprisinh that after being founded in September 2018, I joined CENTERA as a \emph{visiting scientist}, regularly spending several months each year in Warsaw between 2019 and 2021.

During this period, my main activity was providing theoretical support to the experimental work of \href{http://www.researchgate.net/profile/Ivan-Yahniuk}{Ivan Yahniuk} on the study of topological phase transitions in HgTe quantum wells under hydrostatic pressure. I would like to emphasize that I was particularly impressed by the exceptional talent of this young experimentalist, who independently mastered the fabrication of fully operational gated Hall bars from HgTe QWs in the cleanroom facilities. These devices enabled high-quality magnetotransport measurements and were, at that time, produced in only three laboratories worldwide -- in W\"{u}rzburg, Novosibirsk, and Grenoble. The results obtained by Ivan, with my theoretical guidance and support, formed the basis of his PhD thesis \href{http://bip.unipress.waw.pl/images/stories/doktoraty/yahniukivan.pdf}{``Hydrostatic pressure- and temperature-driven topological properties of HgTe/HgCdTe quantum wells''}, successfully defended in 2021.

\subsection{\label{sec:Wurz}Technische Physik group at the University of W\"{u}rzburg  (2022--2025)}
The history of the THz team's collaboration with the \href{http://www.physik.uni-wuerzburg.de/tep/}{Technische Physik} group at the University of W\"{u}rzburg began in 2020 with the internship of PhD student \href{http://www.researchgate.net/profile/Sebastian-Gebert}{Sebastian Gebert} from Montpellier in \href{https://scholar.google.com/citations?user=ZVggr60AAAAJ&hl=en}{Sven H\"{o}fling}'s group. In many respects, Sebastian became the first to build a bridge connecting the two research teams and to merge their complementary expertise originating from distinct but related scientific domains. Our initial exchanges focused on the possibility of applying the technological experience of the W\"{u}rzburg group in the fabrication of gated Hall bar devices to the III-V semiconductor QWs hosting the large-gap QSHI state, emerged from my theoretical proposals. This collaboration led to the first experimental demonstration of an inverted energy gap of about 45~meV in III-V semiconductor QWs (\href{http://journals.aps.org/prresearch/abstract/10.1103/PhysRevResearch.4.L042042}{Phys. Rev. Research 2022}), a key condition for the observation of the high-temperature quantum spin Hall effect. Since 2022, after becoming an official visiting scientist within the Technische Physik group, I have spent several months each year -- typically from July to September -- working in W\"{u}rzburg. During this period, I collaborated closely with \href{http://scholar.google.com/citations?user=s8lYHXcAAAAJ&hl=en}{Fabian Hartmann} and his PhD student \href{http://scholar.google.com/citations?hl=en&user=SCZ1D5AAAAAJ}{Manuel Meyer}, who successfully defended his thesis at the end of 2024. I can say without hesitation that this has been one of the most fruitful and intellectually stimulating collaborations with experimentalists in my career. The synergy that developed between theory and experiment allowed us to obtain a large number of important results, some of which have already been published, while others are still in preparation.

At present, this collaboration between Montpellier and W\"{u}rzburg has grown into a broad scientific partnership that encompasses several major directions. These include the experimental realization of quantum spin Hall states in III-V heterostructures, the investigation of various higher-order topological phases in such systems, and the development of a field-effect transistor operating on the phase transition between topologically trivial and nontrivial states. All these topics form the core of the joint French--German research project \emph{``Construction and manipulation of topological insulators from III-V heterostructures'' (Cantor)} (2024--2026), supported by the \emph{l'Agence nationale de la recherche} and the \emph{German Research Foundation}. The joint efforts of the two teams have already led to significant achievements, among them the first observation of quantized conductance in the proposed III-V heterostructures at temperatures up to 60~K (\href{http://www.science.org/doi/10.1126/sciadv.adz2408}{Science Advances 2025}). My personal collaboration with the \emph{Technische Physik} group, however, will not end in 2025. Next year, I will join the group as a full member, which I believe will play a decisive role in advancing my future research program -- one that combines realistic theoretical modeling and prediction of complex effects in semiconductor heterostructures with world-class experimental verification.

\section{\label{sec:StudentRU}Management of Bachelor and Master students in IPM RAS (2011--2015)}
As already mentioned in the previous sections, in parallel with my scientific research at the Institute for Physics of Microstructures, I was also actively involved in teaching and supervising students. In particular, I regularly guided Bachelor and Master students who joined Vladimir Gavrilenko's group for their course internships and diploma projects. Such internships and the subsequent public defense of diploma theses form an essential part of the Russian higher education system, which consists of a four-year Bachelor program followed by a two-year Master stage. This educational structure naturally integrates students into real research activities at scientific institutes, allowing them to gain first-hand experience of experimental and theoretical work. My participation in this process provided me with an early and valuable opportunity to develop my skills as a scientific advisor of young researchers at the very beginning of their scientific paths.

\href{http://scholar.google.com/citations?hl=en&user=K8FYj4wAAAAJ}{\textbf{\uline{Alexander Malyzhenkov}}}. My first student was Alexander Malyzhenkov, who joined me for his internship and Bachelor thesis project during the spring semester of 2012. Since Alexander expressed a strong interest in conducting experimental work, his research project was devoted to the detection of the spin resonance of 2D electrons in InAs/AlSb QWs. At that time, I was actively developing a theoretical model describing the influence of electron-electron interactions on the spin-resonance energy as a result of the violation of the so-called \emph{Larmor theorem} in narrow-gap QWs (\href{http://journals.aps.org/prb/abstract/10.1103/PhysRevB.87.155113}{Phys. Rev. B 2013}). In this context, possible experimental observations could serve as a valuable test for the approximations used in the theoretical calculations. The main experimental approach to detect spin resonance in a 2D electron (or hole) gas is based on measuring the photoconductivity in the integer quantum Hall effect regime. In this method, the resonance manifests itself as a narrow peak corresponding to spin-flip transitions between Zeeman-split Landau levels. Using a backward-wave oscillator as a radiation source and a small electromagnet of his own construction -- capable of generating magnetic fields up to 3~T at liquid helium temperature -- Alexander devoted his entire internship to the photoconductivity measurement in the submillimeter range. Although the limited magnetic field strength did not allow us to confirm unambiguously that the observed peaks were indeed associated with the spin resonance, this work provided Alexander with his first hands-on experience in low-temperature experimental studies of semiconductor structures in magnetic fields. I would also like to acknowledge the valuable assistance of \href{http://www.ipmras.ru/en/institute/persons/staff/maremyanin-kirill/}{Kirill Maremyanin}, who helped in supervising this purely experimental project and provided important technical guidance throughout the measurements.

\href{http://scholar.google.com/citations?user=tFihXQQAAAAJ&hl=en}{\textbf{\uline{Kirill P. Kalinin}}}. Another student whose work I would particularly like to highlight is Kirill Kalinin, who proved to be the most extraordinary among all the students I have supervised, displaying equally strong abilities as both a theoretician and an experimentalist. He stayed with me for two semesters during the 2012/2013 academic year, participating in the investigation of various effects observed in narrow-gap QWs. I organized his research project in such a way that he combined experimental measurements with their theoretical interpretation, developing his own computer programs to model the quantities obtained experimentally. As a result, Kirill independently solved the problem of calculating the energy spectrum of 2D electrons in the subbands of narrow-gap asymmetric QWs. The main difficulty of this problem was due to the multi-band Hamiltonian (linear in the wave-vector) that accounts for the coupling between the $\Gamma_6$, $\Gamma_8$ and $\Gamma_7$ bands in bulk materials. This makes the problem highly nontrivial for a fourth-year student. Nevertheless, Kirill successfully solved it, obtaining theoretical dependencies for the Rashba spin splitting and the cyclotron mass at the Fermi level as functions of the electron concentration in such 2D systems as InAs/AlSb and InGaAs/InP QWs. The experimental part of Kirill's research project focused mainly on magnetotransport and magnetoabsorption studies of InGaAs/InP QWs, while he also took part in transport and submillimeter photoconductivity measurements on other 2D systems. The exceptional character of Kirill's work is reflected in the fact that the results of his Bachelor thesis were published as a scientific paper ``\href{http://link.springer.com/article/10.1134/S1063782613110092}{Rashba spin splitting and cyclotron resonance in strained InGaAs/InP QW heterostructures with a 2D electron gas}'' in \emph{Semiconductors}. Although Kirill later devoted himself entirely to theoretical physics, his early experience (much like my own scientific path) convinced me that even a theorist should acquire at least a basic understanding of experimental work. This integration of theory and experiment forms the most solid foundation for a comprehensive and realistic view of physics.

\href{http://scholar.google.com/citations?user=wGsBYWMAAAAJ&hl=en}{\textbf{\uline{Aleksandr Kadykov}}}. Aleksandr Kadykov joined Vladimir Gavrilenko's group as a Master student of the \href{http://eng.unn.ru/about/academic-units/faculty-of-radiophysics}{Faculty of Radiophysics} of Lobachevsky State University. The central part of his research project, carried out during 2013--2014 under the supervision of \href{http://www.ipmras.ru/en/institute/persons/staff/morozov-sergey/}{Sergey Morozov}, was devoted to time-resolved studies of long-wavelength photoluminescence from HgCdTe-based QWs. One of the characteristic features of research organization within Vladimir Gavrilenko's group in IPM RAS was the active involvement of students in experiments performed using different experimental techniques. In Aleksandr's case, in addition to photoluminescence, his work also included magnetotransport studies of HgTe-based QWs. Particularly, he analyzed the temperature dependence of Shubnikov--de Haas oscillations, which allows to determine the effective mass and the $g$-factor of charge carriers in 2D structures. Both of these quantities, being intrinsic characteristics of quasiparticles, may significantly deviate from their single-particle values due to electron-electron interaction effects. These two parameters are fundamental quasiparticle characteristics and may significantly differ from their single-particle values due to electron-electron interaction effects. At that time, Aleksandr was performing such measurements within the framework of my research project ``Collective excitations in single and double QWs based on narrow-band and gapless semiconductors in the integer and fractional quantum Hall effect regime'' (see Section~\ref{sec:Projects}). Consequently, I became the scientific supervisor of Aleksandr's work, which was aimed at studying the influence of electron-electron interactions on the magnetotransport properties of HgTe/CdHgTe QWs in the quantum Hall effect regime. This experience later influenced his research in Fr\'{e}d\'{e}ric Teppe's group in Montpellier (see Section~\ref{sec:StudentFR}), where he continued the studies of HgTe/CdHgTe QWs, as a PhD student.

As often happens in research, students working on their projects may also contribute to the broader scientific activities of their supervisors. This was also the case in my experience: through their dedication and curiosity, both Alexander Malyzhenkov and Kirill~P.~Kalinin became co-authors of several research papers published in peer-reviewed journals:
\begin{enumerate}
    \item{S. S. Krishtopenko,  \uline{K. P. Kalinin}, V. I. Gavrilenko, Yu. G. Sadofyev, M. Goiran.
\href{http://link.springer.com/article/10.1134/S1063782612090138}{\emph{Rashba spin splitting and exchange enhancement of the g-factor in InAs/AlSb heterostructures with a two-dimensional electron gas.}}
\emph{Semiconductors} \textbf{46}, 1163 (2012)}

  \item{K. E. Spirin, \uline{K. P. Kalinin}, S. S. Krishtopenko, K. V. Maremyanin, V. I. Gavrilenko, Yu. G. Sadofyev.
\href{http://link.springer.com/article/10.1134/S1063782612110206}{\emph{Features of the persistent photoconductivity in InAs/AlSb heterostructures with double quantum wells and tunneling-transparent barrier.}} \emph{Semiconductors} \textbf{46}, 1396 (2012)}

\item{S. S. Krishtopenko, K. V. Maremyanin, \uline{A. V. Malyzhenkov}, \uline{K. P. Kalinin}, A. V. Ikonnikov, V. I. Gavrilenko, M. Goiran and Yu. G. Sadofyev.
\href{http://iopscience.iop.org/article/10.1088/1742-6596/456/1/012021}{\emph{Effects of Rashba spin splitting and exchange interaction in electron spin resonance in narrow-gap quantum well heterostructures.}} \emph{J. Phys.: Conf. Ser.} \textbf{456}, 012021 (2013)}

\item{S. S. Krishtopenko, \uline{K. P. Kalinin}, \uline{A. V. Malyzhenkov}, K. V. Maremyanin, V. I. Gavrilenko, M. Goiran and Yu. G. Sadofyev
\href{http://iopscience.iop.org/article/10.1088/1742-6596/461/1/012037}{\emph{Exchange enhancement of quasiparticle and ESR spin-gap in symmetric and asymmetric narrow-gap quantum wells.}} \emph{J. Phys.: Conf. Ser.} \textbf{461}, 012037 (2013)}

\item{\uline{K. P. Kalinin}, S. S. Krishtopenko, K. V. Maremyanin, K. E. Spirin, V. I.~Gavrilenko, A. A.~Biryukov, N. V.~Baidus, B. N.~Zvonkov. \href{http://link.springer.com/article/10.1134/S1063782613110092}{\emph{Rashba spin splitting and cyclotron resonance in strained InGaAs/InP quantum well heterostructures.}}
\emph{Semiconductors} \textbf{47}, 1485 (2013)}

\item{S. S. Krishtopenko, K. V. Maremyanin, \uline{K. P. Kalinin}, K. E. Spirin, N. V. Baidus, B. N. Zvonkov, V. I. Gavrilenko.
\href{http://link.springer.com/article/10.1134/S1063782615020141}{\emph{Exchange enhancement of electron g-factor in strained InGaAs/InP heterostructures.}} \emph{Semiconductors} \textbf{49}, 191 (2015)}
\end{enumerate}

\section{\label{sec:StudentFR}Unofficial supervising of PhD students in L2C}
After moving to Montpellier, I continued to work with students. Unfortunately, the structure of the French academic system is such that a researcher holding only a PhD degree cannot serve as an official supervisor unless he/she occupies a permanent position as a \emph{charg\'{e} de recherche} in CNRS or as a \emph{ma\^{i}tre de conf\'{e}rences} at a university. In Russia, on the contrary, the decision to admit researchers to supervise postgraduate students is made by the academic council of the research institute. In France, however, this flexibility does not exist. As a result, my role in the supervision of PhD students within the THz~team remained informal, yet it was nonetheless active and scientifically meaningful. As mentioned in Section~\ref{sec:Poland}, Fr\'{e}d\'{e}ric Teppe's group in Montpellier had long-standing research connections with Vladimir Gavrilenko's laboratory at IPM~RAS through the framework of international cooperation projects under the CNRS program TERAMIR. This collaboration naturally led to the emergence of several PhD students enrolled in joint doctoral programs between the University of Montpellier and IPM~RAS.

The first of these joint PhD students was \href{http://scholar.google.com/citations?user=wGsBYWMAAAAJ&hl=en}{\textbf{\uline{Aleksandr Kadykov}}} (2015--2018), whom I had already known since the time he was completing his Master thesis. Much like my own earlier experience, Aleksandr spent roughly half of each academic year in Montpellier. His main research topic focused on the experimental observation of temperature-driven topological phase transitions in single HgTe QWs -- phenomena that had been theoretically predicted in Montpellier. The second PhD student to come from Nizhny Novgorod was \href{http://www.researchgate.net/profile/Mikhail-Fadeev}{\textbf{\uline{Mikhail Fadeev}}} (2017--2020). His experimental work in the THz~team was devoted to spectroscopic investigations of inter-Landau-level transitions in double HgTe QWs and in multilayer QWs based on III-V semiconductors. Both systems are characterized by a rich phase diagram of topological states that had also been studied theoretically in Montpellier. The central objective of Mikhail's research was to identify specific absorption features that could be directly linked to a given topological phase of the sample. My role in supervising both Aleksandr and Mikhail consisted primarily in guiding them through the physical understanding of newly predicted effects and in helping them interpret the experimental data within a consistent theoretical framework. I should emphasize that their remarkably rapid progress in new research area was greatly facilitated by the fact that I could train and advise them in their native language -- a factor that proved crucial for clear conceptual understanding the essential points of their researches.

Finally, the third student in Montpellier, for whose PhD project I effectively became an informal co-superviser, was \href{http://www.researchgate.net/profile/Sebastian-Gebert}{\textbf{\uline{Sebastian Gebert}}} (2019-2022). Interestingly, it was Sebastian's internship at the University of W\"{u}rzburg that initiated the collaboration between the THz~team and the \href{http://www.physik.uni-wuerzburg.de/tep/}{Technische Physik} group already mentioned in Section~\ref{sec:Wurz}. His doctoral research consisted of two main experimental parts: magnetoabsorption spectroscopy of inter-Landau-level transitions and the study of spontaneous Landau emission from HgTe/HgCdTe QWs and graphene. The analysis of spontaneous cyclotron emission, in particular, required an understanding of the physics of hot charge carriers in crossed electric and magnetic fields -- an area in which I had developed significant expertise, as described earlier. My role in supervising his PhD project was to provide continuous theoretical consulting and conceptual guidance, ensuring a coherent interpretation of the experimental results.

\section{\label{sec:Projects}Research projects}
The submission of research proposals and the management of funded projects have always been an integral part of a scientist's professional life. My first research grant, entitled ``Collective excitations in single and double quantum wells based on narrow-band and gapless semiconductors in the integer and fractional quantum Hall effect regime'', was awarded by the Russian Foundation for Basic Research in 2013, during my postdoctoral stay in Vladimir~Gavrilenko's group. It was a particularly important step in my career, as it marked my transition toward independent project leadership. It is worth noting that in the Russian scientific system, a substantial part of researchers' income is directly generated through project. Thus it is especially crucial for early-career scientists holding a PhD degree, as well as for postgraduate, Master, and Bachelor students involved in scientific researches. This structure naturally encourages young scientists to engage in grant applications and project coordination from the very beginning of their independent careers. To support this process, a number of dedicated funding programs exist in Russia, including those of the Russian Foundation for Basic Research (RFBR), the Russian Science Foundation (RSF), and the presidential grants for young scientists (projects labeled ``MK''). Over the years, I have had the opportunity to lead several research projects supported by these foundations, each of which has contributed to the development of new directions in my scientific activity and to the training of students and young researchers. The complete list of projects for which I served as the principal investigator is presented in Table~\ref{tab:Projects}.
\begin{table}[h!]
\caption{\label{tab:Projects} Principal investigator in projects}
\begin{tabular}{|p{0.13\textwidth}|p{0.06\textwidth}|p{0.48\textwidth}|p{0.23\textwidth}|}
\hline
\centering{Project} & \centering{Years} & \centering{Title} & {\centering Student participation} \\
\hline
{\centering \href{http://www.rfbr.ru/project_search/280794/}{RFBR-13-02-00894} } & {\centering 2013--2015} & ``Collective excitations in single and double QWs based on narrow-band and gapless semiconductors in the integer and fractional quantum Hall effect regime'' &
{\centering - Kirill P. Kalinin \newline  - Aleksandr Kadykov } \\
\hline
{\centering \href{http://www.elibrary.ru/ifebqg}{MK-4758.2014.2} } & {\centering 2014--2015} & ``Collective excitations in two-dimensional semiconductor systems with spin-orbit interaction'' &
 \\
 \hline
{\centering \href{http://www.elibrary.ru/item.asp?id=53992620}{MK-1136.2017.2} } & {\centering 2017--2018} & ``Phase state investigations in two-dimensional topological insulators based on multi-layer semiconductor heterostructures'' &
{\centering - Aleksandr Kadykov \newline  - Mikhail Fadeev } \\
 \hline
{\centering \href{http://rscf.ru/project/qZDFUyWvkvpBgN8iPQVWoS3g_N8Xs_LcrmAJKbxXI_cxUYVc81SkCZEjjdvAqHDcX2AxlSYN/}{RSF-17-72-10158} } & {\centering 2017--2019} & ``2D topological insulators with large bulk energy gap based on strained InAs/GaInSb structures'' &
{\centering - Aleksandr Kadykov \newline  - Mikhail Fadeev } \\
 \hline
{\centering TARFEP } & {\centering 2023} & Post-doctoral grant from the \href{http://iqoc.fr/}{Institute for Quantum Technologies in Occitanie}:
``Topological states And Relativistic-like Fermions of III-V semiconductors in quantum
Electronics and Photonics'' &
\\
\hline
\end{tabular}
\end{table}

After moving to France, I discovered that without holding a permanent research position, it was not possible for me to serve as a principal Investigator on projects funded by the \emph{Agence nationale de la recherche}. As a result, I had no formal opportunity to gain direct experience in leading ANR-funded projects. This limitation existed despite the fact that a significant part of the research programs on HgTe-based QWs and III-V semiconductor QW heterostructures within the ANR projects of the THz~team was actually based on my theoretical ideas and proposals. The most illustrative example of this situation is the French part of the joint French-German research project \emph{``Construction and manipulation of topological insulators from III-V heterostructures'' (Cantor)}, supported by both the ANR and the \emph{German Research Foundation}. Nevertheless, thanks to the long-standing cooperation between the THz~team and Vladimir~Gavrilenko's group within the framework of the international collaboration \emph{TERAMIR}, I was able to continue submitting project proposals to the \emph{Russian Foundation for Basic Research} and the \emph{Russian Science Foundation}. This explains my participation in several RFBR and RSF projects both as principal Investigator and as a collaborating researcher during my stay in Montpellier. It is important to emphasize that these projects also played an important practical role in providing financial support to our joint PhD students. In particular, a significant part of the salaries of our doctoral students \emph{Aleksandr~Kadykov} and \emph{Mikhail~Fadeev} during their stays in Nizhny~Novgorod came from these Russian-funded projects, while their research periods in Montpellier were covered by the French government programs through \href{http://www.campusfrance.org/fr}{Campus~France}. The involvement of both Aleksandr and Mikhail in my RFBR and RSF projects (see Tables~\ref{tab:Projects} and~\ref{tab:TheorProject}) provides an additional confirmation of my effective contribution in scientific supervision of their PhD research (see Section~\ref{sec:StudentFR}).

\begin{table}[h!]
\caption{\label{tab:TheorProject} Projects with my leading role in theoretical support and investigations}
\begin{tabular}{|p{0.13\textwidth}|p{0.06\textwidth}|p{0.48\textwidth}|p{0.23\textwidth}|}
\hline
\centering{Project} & \centering{Years} & \centering{Title} & {\centering Student participation} \\
\hline
{\centering \href{http://www.rfbr.ru/project_search/322046/}{RFBR-15-02-08274} } & {\centering 2015--2017} & ``Studies of the optical properties of tunnel-coupled narrow-gap HgTe/CdHgTe quantum wells with a graphene-like dispersion law'' &
{\centering - Aleksandr Kadykov } \\
\hline
{\centering \href{https://www.rfbr.ru/project_search/313456/}{RFBR-15-52-16011} } & {\centering 2015--2017} & ``Terahertz spectroscopy of three-layer QW heterostructures based on InAs/GaSb (topological insulators)'' &
{\centering - Aleksandr Kadykov } \\
\hline
{\centering \href{http://anr.fr/Project-ANR-19-CE30-0032}{ANR-19-CE30-0032} } & {\centering 2019--2023} & ``Collective excitations in topological matter (Colector)'' &
\\
\hline
{\centering \href{http://rscf.ru/project/w_83wiIx3vty05Ep-j02_xBixBRUayJwCooVhN8rW8L94LQpKdNRpkiRwpxPpRCFsP7vUGfQ/}{RSF-22-22-00382} } & {\centering 2022--2023} & ``Energy band structure and topological states of various orders in double HgTe quantum wells'' &
 \\
 \hline
 {\centering \href{http://anr.fr/Project-ANR-23-CE24-0022}{ANR-23-CE24-0022} } & {\centering 2024--2026} & ``Construction and manipulation of topological insulators from III-V heterostructures (Cantor)'' &
\\
\hline
\end{tabular}
\end{table}

\section{\label{sec:AuthPub}List of papers in peer-reviewed journals after PhD defense}
\begin{enumerate}
\item[ ]
\item{B. Benhamou-Bui, C. Consejo, \uline{S.S. Krishtopenko}, S. Ruffenach, C. Bray, J. Torres, J. Dzian, F. Le Mardel\'{e}, A. Pagot, X. Baudry, S.V. Morozov, N.N. Mikhailov, S.A. Dvoretskii, B. Jouault, P. Ballet, M. Orlita, C. Ciuti, F.Teppe. \href{https://doi.org/10.1103/7rcf-fd3g}{\emph{Nonlinear Terahertz Electroluminescence from Dirac-Landau Polaritons.}} \emph{\textbf{Phys. Rev. Lett.}} \textbf{136}, 096902 (2026).}

\item{M. Meyer, J. Baumbach, \uline{S.S. Krishtopenko}, A. Wolf, M. Emmerling, S. Schmid, M. Kamp, B. Jouault, J.-B. Rodriguez, E. Tourni\'{e}, T. M\"{u}ller,  R. Thomale, G. Bastard, F. Teppe, F. Hartmann and S. H\"{o}fling. \href{https://www.science.org/doi/10.1126/sciadv.adz2408}{\emph{Quantum spin Hall effect in III-V semiconductors at elevated temperatures: advancing topological electronics.}} \emph{\textbf{Science Adv.}} \textbf{11}, eadz2408 (2025).}    

\item{\uline{S.S. Krishtopenko}, A.V. Ikonnikov, B. Jouault, F. Hartmann, S. H\"{o}fling, F. Teppe. \href{http://doi.org/10.1103/PhysRevResearch.7.033116}{\emph{Quasiparticle picture of topological phase transitions induced by interactions.}} \emph{\textbf{Phys. Rev. Research}} \textbf{7}, 033116 (2025).}
    
\item{M.L. Savchenko, D.A. Kozlov, \uline{S.S. Krishtopenko}, N.N. Mikhailov, S.A. Dvoretsky, Z. D. Kvon, A. Pimenov, D. Weiss \href{http://doi.org/10.1103/fc7w-cq8f}{\emph{Quantum Hall effect and zero-resistance plateau in bulk HgTe.}} \emph{\textbf{Phys. Rev. Research}} \textbf{7}, L042047 (2025).}
    
\item{I.~Yahniuk, D.A. Kozlov, M.D. Moldavskaya, L.E. Golub, V.V. Bel'kov, I.A. Dmitriev, \uline{S.S. Krishtopenko}, F. Teppe, Y. Ivonyak, A. Bercha, G. Cywi\'{n}ski, W. Knap, S.D.  Ganichev. \href{http://doi.org/10.1103/hfn6-v2y6}{\emph{Positive Terahertz Photoconductivity in CdHgTe Under Hydrostatic Pressure.}} \emph{\textbf{Phys. Rev. B}}
    112, 075417 (2025).}

\item{\uline{S.S. Krishtopenko}, A.V. Ikonnikov, B. Jouault, F. Teppe. \href{https://doi.org/10.1103/PhysRevB.111.045431}{\emph{Disorder-induced phase transitions in double HgTe quantum wells.}} \emph{\textbf{Phys. Rev. B}} \textbf{111}, 045431 (2025).}

\item{K. Dinar, J. Delgado-Notario, C. Bray, K. Maussang, E. Perez-Martin, B. Benhamou-Bui, C.~Consejo, S. Ruffenach, \uline{S.S. Krishtopenko}, L. Bonnet, M. Paillet, J. Torres, Y.M. Meziani, I. Rozhansky, B. Jouault, S. Nanot, F. Teppe. \href{http://doi.org/10.1103/PhysRevApplied.23.044043}{\emph{Spin-dependent photovoltage in graphene/MoS$_2$-based field-effect transistors.}} \emph{\textbf{Phys. Rev. Appl.}} \textbf{23}, 044043 (2025).}
\item{S. Benlenqwanssa, \uline{S.S. Krishtopenko}, M. Meyer, B. Benhamou-Bui, L. Bonnet, A. Wolf, C.~Bray, C.~Consejo, S. Ruffenach, S. Nanot, J.-B. Rodriguez, E. Tourni\'{e}, F. Hartmann, S. H\"{o}fling, F. Teppe, B. Jouault. \href{http://doi.org/10.1103/PhysRevApplied.22.064059}{\emph{Multi-probe analysis to separate edge currents from bulk currents in quantum spin Hall insulators and to analyze their temperature dependence.}} \emph{\textbf{Phys. Rev. Applied}} \textbf{22}, 064059 (2024).}

\item{M.D. Moldavskaya, L.E. Golub, V.V. Bel'kov, S.N. Danilov, D.A. Kozlov, J. Wunderlich, D.~Weiss, N.N. Mikhailov, S.A. Dvoretsky, \uline{S.S. Krishtopenko}, B. Benhamou-Bui, F. Teppe, S.D. Ganichev. \href{http://doi.org/10.1103/PhysRevB.110.205204}{\emph{Magnetophotogalvanic Effects Driven by Terahertz Radiation in CdHgTe Crystals with Kane Fermions.}} \emph{\textbf{Phys. Rev. B}} \textbf{110}, 205204 (2024).}

\item{D. Soranzio, E. Abreu, S.~Houver, J. D\"{o}ssegger, M. Savoini, F. Teppe, \uline{S.S. Krishtopenko}, N.N. Mikhailov, S.A. Dvoretsky, S.L. Johnson. \href{http://doi.org/10.1103/PhysRevB.110.094303}{\emph{Roles of band gap and Kane electronic dispersion in the terahertz-frequency nonlinear optical response in HgCdTe.}} \emph{\textbf{Phys. Rev. B}} \textbf{110}, 094303 (2024).}

\item{M.~Meyer, T. F\"{a}hndrich, S. Schmid, A. Wolf, \uline{S.S. Krishtopenko}, B. Jouault, G. Bastard, F. Teppe, F. Hartmann, S. H\"{o}fling. \href{http://doi.org/10.1103/PhysRevB.109.L121303}{\emph{Coexistence of topological and normal insulating phases in electro-optically tuned InAs/GaSb bilayer quantum wells.}} \emph{\textbf{Phys. Rev. B}} \textbf{109}, L121303 (2024).}
\item{S. Gebert, C. Consejo, \uline{S.S. Krishtopenko}, S. Ruffenach, M. Szo{\l}a, J. Torres, C. Bray, B. Jouault, M. Orlita, X. Baudry, P. Ballet, S.V. Morozov, V.I. Gavrilenko, N.N. Mikhailov, S.A. Dvoretskii, F. Teppe. \href{http://doi.org/10.1038/s41566-022-01129-1}{\emph{Terahertz cyclotron emission from two-dimensional Dirac fermions.}} \emph{\textbf{Nat. Photonics}} \textbf{17}, 244 (2023).}

\item{B. Benhamou-Bui, C. Consejo, \uline{S.S. Krishtopenko}, M. Szo{\l}a, K. Maussang, S. Ruffenach, E. Chauveau, S. Benlemqwanssa, C. Bray, X. Baudry, P. Ballet, S.V. Morozov, V.I. Gavrilenko, N.N. Mikhailov, S.A. Dvoretskii, B. Jouault, J. Torres, F. Teppe. \href{http://doi.org/10.1063/5.0168578}{\emph{Gate tunable terahertz cyclotron emission from two-dimensional Dirac fermions.}} \emph{\textbf{APL Photonics}} \textbf{8}, 116106 (2023).}

\item{L.S.~Bovkun, \uline{S.S. Krishtopenko}, V.Ya. Aleshkin, N.N. Mikhailov, S.A. Dvoretsky, F. Teppe, M. Orlita, V.I. Gavrilenko, A.V. Ikonnikov.
 \href{http://doi.org/10.1134/S0021364023603536}{\emph{Simultaneous observation of the cyclotron resonances of electrons and holes in a HgTe/CdHgTe double quantum well under ``Optical Gate'' effect.}} \emph{\textbf{JETP Letters}} \textbf{118}, 867 (2023).}
\item{\uline{S.S. Krishtopenko}, M. Antezza, F. Teppe. \href{http://doi.org/10.1103/PhysRevB.106.115203}{\emph{Disorder-induced topological phase transition in HgCdTe crystals.}} \emph{\textbf{Phys. Rev. B}} \textbf{106}, 115203 (2022).}

\item{\uline{S.S. Krishtopenko}, F. Teppe. \href{http://doi.org/10.1103/PhysRevB.105.125203}{\emph{Relativistic collapse of Landau levels of Kane fermions in crossed electric and magnetic fields.}} \emph{\textbf{Phys. Rev. B}} \textbf{105}, 125203 (2022).}

\item{A.V.~Ikonnikov, \uline{S.S.~Krishtopenko}, L.S. Bovkun, N.N. Mikhailov, S.A. Dvoretskii, B.A. Piot, M. Potemski, M. Orlita, F. Teppe, V.I. Gavrilenko.
 \href{http://doi.org/10.1134/S0021364022601889}{\emph{Origin of structure inversion asymmetry in double HgTe quantum wells.}} \emph{\textbf{JETP Letters}} \textbf{116}, 547 (2022).}

\item{C. Avogadri, S. Gebert, \uline{S.S.~Krishtopenko}, I.~Castillo, C.~Consejo, S.~Ruffenach, C.~Roblin, C.~Bray, Y.~Krupko, S.~Juillaguet, S.~Contreras, A.~Wolf, F.~Hartmann, S.~H\"{o}fling, G.~Boissier, J.-B.~Rodriguez, S.~Nanot, E.~Tourni\'{e}, F.~Teppe, B.~Jouault. \href{http://doi.org/10.1103/PhysRevResearch.4.L042042}{\emph{Large inverted band gap in strained three-layer InAs/GaInSb quantum wells.}} \emph{\textbf{Phys. Rev. Research}} \textbf{4}, L042042 (2022).}

\item{C.~Bray, K.~Maussang, C.~Consejo, J.A.~Delgado-Notario,\uline{S.S.~Krishtopenko}, I.~Yahniuk, S.~Gebert, S.~Ruffenach, K.~Dinar, E.~Moench, J.~Eroms, K.~Indykiewicz, B.~Jouault, J.~Torres, Y.M.~Meziani, W.~Knap, A.~Yurgens, S.D.~Ganichev, F.~Teppe. \href{http://doi.org/10.1103/PhysRevB.106.245141}{\emph{Temperature-dependent zero-field splittings in graphene.}} \emph{\textbf{Phys. Rev. B}} \textbf{106}, 245141 (2022).}
\item{\uline{S.S. Krishtopenko}. \href{http://doi.org/10.1038/s41598-021-00577-z}{\emph{Higher-order topological insulator in cubic semiconductor quantum wells.}} \emph{\textbf{Sci. Rep.}} \textbf{11}, 21060 (2021).}

\item{\uline{S.S. Krishtopenko}, M. Antezza, F. Teppe. \href{http://doi.org/10.1103/PhysRevB.101.205424}{\emph{Disorder-induced phase transition in Dirac systems beyond the linear approximation.}} \emph{\textbf{Phys. Rev. B}} \textbf{101}, 205424 (2020).}

\item{\uline{S.S. Krishtopenko}, A.M. Kadykov, S. Gebert, S. Ruffenach, C. Consejo, J. Torres, C. Avogadri, B. Jouault, W. Knap, N.N. Mikhailov, S.A. Dvoretskii, F. Teppe. \href{http://doi.org/10.1103/PhysRevB.102.041404}{\emph{Many-particle effects in optical transitions from zero-mode Landau levels in HgTe quantum wells.}} \emph{\textbf{Phys. Rev. B}} \textbf{102}, 041404(R) (2020).}

\item{M.V.~Yakunin, \uline{S.S. Krishtopenko}, W. Desrat, S.M. Podgornykh, M.R. Popov, V.N. Neverov, S.A. Dvoretskii, N.N.~ Mikhailov, F. Teppe, B.~ Jouault. \href{http://doi.org/10.1103/PhysRevB.102.165305}{\emph{Unconventional reentrant quantum Hall effect in a HgTe/CdHgTe double quantum well.}}  \emph{\textbf{Phys. Rev. B}} \textbf{102}, 165305 (2020).}

\item{S. Mantion, C. Avogadri, \uline{S.S. Krishtopenko}, S. Gebert, S. Ruffenach, C. Consejo, S.V. Morozov, N.N. Mikhailov, S.A. Dvoretskii, W. Knap, S. Nanot, F. Teppe, B. Jouault.
    \href{http://doi.org/10.1103/PhysRevB.102.075302}{\emph{Quantum Hall states in inverted HgTe quantum wells probed by transconductance fluctuations.}} \emph{\textbf{Phys. Rev. B}} \textbf{102}, 075302 (2020).}

\item{\uline{S.S. Krishtopenko}, M. Antezza, F. Teppe. \href{http://dx.doi.org/10.1088/1361-648X/ab6741}{\emph{Hybridization of topological surface states with a flat band.}} \emph{\textbf{J. Phys.: Condens. Matter}} \textbf{32}, 165501 (2020).}

\item{L.S.~Bovkun, A.V.~Ikonnikov, \uline{S.S. Krishtopenko}, V.Ya. Aleshkin, M.S. Zholudev, S.~Ruffenach, C.~Consejo, F.~Teppe, S.A.~Dvoretskii, N.N.~Mikhailov, M. Potemski, M. Orlita, V.I.~Gavrilenko \href{http://doi.org/10.1134/S0021364020200059}{
    \emph{Effects of electron-electron interaction in magnetoabsorption spectra of HgTe/CdHgTe quantum wells with inverted band structure.}} \emph{\textbf{JETP Letters}}
    \textbf{112}, 508 (2020).}

\item{V.V. Utochkin, M.A. Fadeev, \uline{S.S. Krishtopenko}, V.V. Rumyantsev, V.Ya. Aleshkin, A.A.~Dubinov, S.V. Morozov, B.R. Semyagin, M.A. Putyato, E.A. Emelyanov, V.V.~Preobrazhenskii, V.I.~Gavrilenko \href{http://doi.org/10.1134/S1063782620090304}{\emph{Photoluminescence spectra of InAs/GaInSb/InAs quantum wells in the mid-infrared region.}} \emph{\textbf{Semiconductors}} \textbf{54}, 1119 (2020).}
\item{\uline{S.S. Krishtopenko}, W. Desrat, K.E. Spirin, C. Consejo, S. Ruffenach, F.~Gonzalez-Posada, B. Jouault, W. Knap, K.V. Maremyanin, V.I. Gavrilenko, G. Boissier, J.~Torres, M. Zaknoune, E. Tourni\'{e}, F. Teppe. \href{http://doi.org/10.1103/PhysRevB.99.121405}{\emph{Massless Dirac fermions in III-V semiconductor quantum wells.}} \emph{\textbf{Phys. Rev. B}} \textbf{99}, 121405(R) (2019).}

\item{I.~Yahniuk, \uline{S.S. Krishtopenko}, G.~Grabecki, B.~Jouault, C.~Consejo, W.~Desrat, M.~Majewicz, A.M.~Kadykov, K.E.~Spirin, V.I.~Gavrilenko, N.N.~Mikhailov, S.A.~Dvoretsky, D.B.~But, F.~Teppe, J.~Wr\'{o}bel, G.~Cywi\'{n}ski, S.~Kret, T.~Dietl, W.~Knap. \href{http://doi.org/10.1038/s41535-019-0154-3}{\emph{Magneto-transport in inverted HgTe quantum wells.}} \emph{\textbf{npj Quantum Mater.}} \textbf{4}, 13 (2019).}

\item{\uline{S.S. Krishtopenko}, S.~Ruffenach, F.~Gonzalez-Posada, C.~Consejo, W. Desrat, B. Jouault, W. Knap, M.A. Fadeev, A.M. Kadykov, V.V. Rumyantsev, S.V. Morozov,  G. Boissier, E. Tourni\'{e}, V.I. Gavrilenko, F. Teppe, \href{http://doi.org/10.1134/S0021364019020085} {\emph{Terahertz spectroscopy of two-dimensional semimetal in three-layer InAs/GaSb/InAs quantum well.}} \emph{\textbf{JETP Letters}} \textbf{109}, 96 (2019).}

\item{L.S. Bovkun, A.V. Ikonnikov, V.Ya. Aleshkin, K.V. Maremyanin, N.N. Mikhailov, S.A. Dvoretskii, \uline{S.S. Krishtopenko}, F. Teppe, B. A.Piot, M. Potemski, M. Orlita, V. I. Gavrilenko. \href{http://doi.org/10.1016/j.opelre.2019.06.001}{\emph{Magnetospectroscopy of double HgTe/CdHgTe quantum wells with inverted band structure in high magnetic fields up to 30 T.}} \emph{\textbf{Opto-Electron. Rev.}} \textbf{27}, 213 (2019).}

\item{S.M. Podgornykh, M.V. Yakunin, \uline{S.S. Krishtopenko}, M.R. Popov, N.N. Mikhailov, S.A. Dvoretskii. \href{http://link.springer.com/article/10.1134/S1063782619070194}{\emph{On the thermal activation of conductivity electrons in a p-type HgTe/CdHgTe double quantum well with HgTe layers of critical width.}} \emph{\textbf{Semiconductors}} \textbf{53}, 919 (2019).}

\item{A.A. Dobretsova, Z.D. Kvon, \uline{S.S. Krishtopenko}, N.N. Mikhailov and S.A. Dvoretsky, \href{http://doi.org/10.1063/1.5086405} {\emph{Spin splitting of surface states in HgTe quantum wells.}} \emph{\textbf{Low Temp. Phys.}} \textbf{45}, 159 (2019).}

\item{\uline{S.S. Krishtopenko}, F. Teppe.    \href{http://advances.sciencemag.org/content/4/4/eaap7529}{\emph{Quantum spin Hall insulator with a large bandgap, Dirac fermions, and bilayer graphene analog.}} \emph{\textbf{Science Adv.}} \textbf{4}, eaap7529 (2018).}

\item{A.M. Kadykov, \uline{S.S. Krishtopenko}, B. Jouault, W. Desrat, W. Knap, S. Ruffenach, C. Consejo, J. Torres, S.V. Morozov, N.N. Mikhailov, S.A. Dvoretskii, F. Teppe.    \href{http://doi.org/10.1103/PhysRevLett.120.086401}{\emph{Temperature-induced topological phase transition in HgTe quantum wells.}} \emph{\textbf{Phys. Rev. Lett.}} \textbf{120}, 086401 (2018).}

\item{\uline{S.S. Krishtopenko}, F. Teppe.    \href{http://doi.org/10.1103/PhysRevB.97.165408}{\emph{Realistic picture of helical edge states in HgTe quantum wells.}} \emph{\textbf{Phys. Rev. B}} \textbf{97}, 165408 (2018).}

\item{\uline{S.S. Krishtopenko}, S.~Ruffenach, F.~Gonzalez-Posada, G. Boissier, M. Marcinkiewicz, M.A. Fadeev, A.M. Kadykov, V.V. Rumyantsev, S.V. Morozov, V.I. Gavrilenko, C. Consejo, W.Desrat, B.Jouault, W. Knap, E.Tourni\'{e}, F. Teppe \href{http://doi.org/10.1103/PhysRevB.97.245419}{\emph{Temperature-dependent terahertz spectroscopy of inverted-band three-layer InAs/GaSb/InAs quantum well.}} \emph{\textbf{Phys. Rev. B}} \textbf{97}, 245419 (2018).}

\item{W. Desrat, \uline{S.S. Krishtopenko}, B.A. Piot, M. Orlita, C. Consejo, S. Ruffenach, W.~Knap, A. Nateprov, E. Arushanov, F. Teppe. \href{http://doi.org/10.1103/PhysRevB.97.245203}{\emph{Band splitting in Cd$_3$As$_2$ measured by magnetotransport.}} \emph{\textbf{Phys. Rev. B}} \textbf{97}, 245203 (2018).}

\item{L.S.~Bovkun, A.V.~Ikonnikov, V.Ya.~Aleshkin, \uline{S.S.~Krishtopenko}, N.N. Mikhailov, S.A. Dvoretsky, M. Potemski, B. Piot, M. Orlita, V.I. Gavrilenko, \href{http://doi.org/10.1134/S0021364018170058} {\emph{Polarization-sensitive fourier-transform spectroscopy of HgTe/CdHgTe quantum wells in the far infrared range in a magnetic field.}} \emph{\textbf{JETP Letters}} \textbf{108}, 329 (2018).}
\item{M. Marcinkiewicz, S. Ruffenach, \uline{S.S. Krishtopenko}, A.M. Kadykov, C. Consejo, D.B.~But, W. Desrat, W. Knap, J. Torres, A.V. Ikonnikov, K.E. Spirin, S.V. Morozov, V.I.~Gavrilenko, N.N.~Mikhailov, S.A.~Dvoretskii, F. Teppe. \href{http://doi.org/10.1103/PhysRevB.96.035405}{\emph{Temperature-driven single-valley Dirac fermions in HgTe quantum wells.}} \emph{\textbf{Phys. Rev. B}} \textbf{96}, 035405 (2017).}

\item{S. Ruffenach, A. Kadykov, V.V. Rumyantsev, J. Torres, D. Coquillat, D. But, \uline{S.S. Krishtopenko}, C. Consejo, W. Knap, S. Winnerl, M. Helm, M.A. Fadeev, N.N. Mikhailov, S.A. Dvoretskii, V.I.~Gavrilenko, S.V.~Morozov, F. Teppe. \href{http://dx.doi.org/10.1063/1.4977781}{\emph{HgCdTe-based heterostructures for terahertz photonics.}} \emph{\textbf{APL Materials}} \textbf{5}, 035503 (2017).}

\item{S. Ruffenach, \uline{S.S. Krishtopenko}, L.S. Bovkun, A.V. Ikonnikov, M. Marcinkiewicz, C. Consejo, M. Potemski, B. Piot, M. Orlita, B.R. Semyagin, M.A. Putyato, E.A. Emelyanov, V.V. Preobrazhenskii, W. Knap, F. Gonzalez-Posada, G. Boissier, E. Tourni\'{e}, F. Teppe, V.I. Gavrilenko. \href{http://doi.org/10.1134/S0021364017230102} {\emph{Magneto-absorption of Dirac fermions in gapless "three-layer" InAs/GaSb/ InAs quantum wells.}} \emph{\textbf{JETP Letters}} \textbf{106}, 727 (2017).}

\item{\uline{S.S. Krishtopenko}, A.V. Ikonnikov, K.V. Maremyanin, L.S. Bovkun, K.E. Spirin, A.M. Kadykov, M. Marcinkiewicz, S. Ruffenach, C. Consejo, F. Teppe, W. Knap, B.R. Semyagin, M.A. Putyato, E.A. Emelyanov, V.V. Preobrazhenskii, V. I. Gavrilenko. \href{http://dx.doi.org/10.1134/S1063782617010109}{\emph{Cyclotron Resonance of Dirac Fermions in InAs/GaSb/InAs Quantum Wells.}} \emph{\textbf{Semiconductors}} \textbf{51}, 38 (2017).}

\item{A.V. Ikonnikov, L.S. Bovkun, V.V. Rumyantsev, \uline{S.S. Krishtopenko}, V.Ya. Aleshkin, A.M. Kadykov, M. Orlita, M. Potemski, V.I. Gavrilenko, S.V. Morozov, S.A. Dvoretsky, N.N. Mikhailov. \href{http://link.springer.com/article/10.1134/S1063782617120090}{\emph{On the band spectrum in p-type HgTe/CdHgTe heterostructures and its transformation under temperature variation.}} \emph{\textbf{Semiconductors}} \textbf{51}, 1531 (2017).}

\item{V.I. Gavrilenko, L.S. Bovkun, A.V. Ikonnikov, V.Ya. Aleshkin, \uline{S.S. Krishtopenko}, A.V. Antonov, K.E. Spirin, N.N. Mikhailov, S.A. Dvoretsky. \href{http://link.springer.com/article/10.1134/S106378261712003X}{\emph{Activation conductivity in HgTe/ CdHgTe quantum wells at integer Landau level filling factors: Role of the random potential.}} \emph{\textbf{Semiconductors}} \textbf{51}, 1562 (2017).}
\item{\uline{S.S. Krishtopenko}, W. Knap, F. Teppe. \href{http://www.nature.com/articles/srep30755}{\emph{Phase transitions in two tunnel-coupled HgTe quantum wells: Bilayer graphene analogy and beyond.}} \emph{\textbf{Sci. Rep.}} \textbf{6}, 30755 (2016).}

\item{F. Teppe, M. Marcinkiewicz, \uline{S.S. Krishtopenko}, S. Ruffenach, C. Consejo, A.M. Kadykov, W. Desrat, D. But, W. Knap, J. Ludwig, S. Moon, D. Smirnov, M. Orlita, Z. Jiang, S.V. Morozov, V.I. Gavrilenko, N.N. Mikhailov, S.A. Dvoretskii. \href{http://www.nature.com/ncomms/2016/160830/ncomms12576/full/ncomms12576.html}{\emph{Temperature-driven massless Kane fermions in HgCdTe crystals.}} \emph{\textbf{Nat. Commun.}} \textbf{7}, 12576 (2016).}

\item{\uline{S.S. Krishtopenko}, I. Yahniuk, D.B. But, V.I. Gavrilenko, W. Knap, F. Teppe. \href{http://doi.org/10.1103/PhysRevB.94.245402}{\emph{Pressure- and temperature-driven phase transitions in HgTe quantum wells.}} \emph{\textbf{Phys. Rev. B}} \textbf{94}, 245402 (2016).}

\item{A.V. Ikonnikov, \uline{S.S. Krishtopenko}, O. Drachenko, M. Goiran, M.S. Zholudev, V.V. Platonov, Yu.B. Kudasov, A.S.~Korshunov, D.A. Maslov, I.V. Makarov, O.M. Surdin, A.V. Philippov, M. Marcinkiewicz, S. Ruffenach, F. Teppe, W. Knap, N.N. Mikhailov, S.A. Dvoretsky, V.I. Gavrilenko. \href{http://doi.org/10.1103/PhysRevB.94.155421}{\emph{Temperature-dependent magnetospectroscopy of HgTe quantum wells.}} \emph{\textbf{Phys. Rev. B}} \textbf{94}, 155421 (2016).}

\item{A.M. Kadykov, J. Torres, \uline{S.S. Krishtopenko}, C.~Consejo, S.~Ruffenach, M.~Marcinkiewicz, D. But, W. Knap, S.V. Morozov, V.I. Gavrilenko, N.N. Mikhailov, S.A. Dvoretsky, F.~Teppe. \href{http://scitation.aip.org/content/aip/journal/apl/108/26/10.1063/1.4955018}{\emph{Terahertz imaging of Landau levels in HgTe-based topological insulators.}} \emph{\textbf{Appl. Phys. Lett.}} \textbf{108}, 262102 (2016).}

\item{M.V. Yakunin, \uline{S.S. Krishtopenko}, S.M. Podgornykh, M.R. Popov, V.N. Neverov, N.N. Mikhailov, S.A. Dvoretsky. \href{http://doi.org/10.1134/S0021364016180132} {\emph{HgTe/CdHgTe Double Quantum Well with a Spectrum of Bilayer Graphene and Peculiarities of Its Magnetotransport.}} \emph{\textbf{JETP Letters}} \textbf{104}, 403 (2016).}

\item{L.S. Bovkun, \uline{S.S. Krishtopenko}, A.V. Ikonnikov, V.Ya. Aleshkin, A.M. Kadykov, S.~Ruffenach, C.~Consejo, F.~Teppe, W. Knap, M. Orlita, B. Piot, M. Potemski, N.N.~Mikhailov, S. A. Dvoretskii, V.I. Gavrilenko. \href{http://dx.doi.org/10.1134/S1063782616110063}{\emph{Magnetospectroscopy of double HgTe/CdHgTe quantum wells.}} \emph{\textbf{Semiconductors}} \textbf{50}, 1532 (2016).}

\item{A.M. Kadykov, C. Consejo, M. Marcinkiewicz, L. Viti, M.S. Vitiello, \uline{S.S. Krishtopenko}, S.~Ruffenach, S.V. Morozov,  W. Desrat, N. Dyakonova, W Knap, V.I. Gavrilenko, N.N. Mikhailov, S.A. Dvoretsky, F. Teppe. \href{http://dx.doi.org/10.1002/pssc.201510264}{\emph{Observation of topological phase transition by terahertz photoconductivity in HgTe-based transistors.}} \emph{\textbf{Phys. Status Solidi C}} \textbf{13}, 534 (2016).}
\item{A.M. Kadykov, F. Teppe, C. Consejo, L. Viti, M.S. Vitiello, \uline{S.S. Krishtopenko}, S. Ruffenach, S.V.~Morozov, M.~Marcinkiewicz, W. Desrat, N. Dyakonova, W Knap, V.I. Gavrilenko, N.N. Mikhailov, S.A. Dvoretsky. \href{http://dx.doi.org/10.1063/1.4932943}{\emph{Terahertz detection of magnetic field-driven topological phase transition in HgTe-based transistors.}} \emph{\textbf{Appl. Phys. Lett.}} \textbf{107}, 152101 (2015).}

\item{\uline{S.S. Krishtopenko}, A.V. Ikonnikov, M. Orlita, Yu.G. Sadofyev, M. Goiran, F.~Teppe, W.~Knap, V.I.~Gavrilenko. \href{http://dx.doi.org/10.1063/1.4913927}{\emph{Effect of e-e interaction on cyclotron resonance in high-mobility InAs/AlSb quantum wells.}} \emph{\textbf{J. Appl. Phys.}} \textbf{117}, 112813 (2015).}

\item{\uline{S.S. Krishtopenko}. \href{http://link.springer.com/article/10.1134/S106378261502013X}{\emph{Effects of electron-electron interaction in the spin resonance in 2D systems with Dresselhaus spin-orbit coupling.}} \emph{\textbf{Semiconductors}} \textbf{49}, 174 (2015).}

\item{\uline{S.S. Krishtopenko}, K.V. Maremyanin, K.P. Kalinin, K.E. Spirin, N.V. Baidus, B.N. Zvonkov, V.I. Gavrilenko.
\href{http://link.springer.com/article/10.1134/S1063782615020141}{\emph{Exchange enhancement of electron g-factor in strained InGaAs/InP heterostructures.}} \emph{\textbf{Semiconductors}} \textbf{49}, 191 (2015).}

\item{L.S. Bovkun, \uline{S.S. Krishtopenko}, M.S. Zholudev, A.V. Ikonnikov, K.E.Spirin, S.A. Dvoretsky, N.N. Mikhailov, F. Teppe, W. Knap, V.I. Gavrilenko. \href{http://dx.doi.org/10.1134/S1063782615120052}{\emph{Exchange enhancement of electron g-factor in two-dimensional semimetal in HgTe quantum wells.}} \emph{\textbf{Semiconductors}} \textbf{49}, 1627 (2015).}

\item{K.E. Spirin, \uline{S.S. Krishtopenko}, Yu.G.Sadofyev, O. Drachenko, M. Helm, F. Teppe, W. Knap, V.I. Gavrilenko. \href{http://dx.doi.org/10.1134/S1063782615120210}{\emph{Cyclotron resonance in InAs/AlSb quantum wells in magnetic field up to 45 T.}} \emph{\textbf{Semiconductors}} \textbf{49}, 1616 (2015).}
\item{\uline{S.S. Krishtopenko}. \href{http://iopscience.iop.org/0268-1242/29/8/085005}{\emph{Many-body effects in electron spin resonance in 2D systems with Rashba spin-orbit interaction.}} \emph{\textbf{Semicond. Sci. Technol.}} \textbf{29}, 085005 (2014).}
\item{\uline{S.S. Krishtopenko}, V.I. Gavrilenko, M. Goiran.
\href{http://prb.aps.org/abstract/PRB/v87/i15/e155113}{\emph{Spin-wave excitations and electron spin resonance in symmetric and asymmetric narrow-gap quantum wells.}}
\emph{\textbf{Phys. Rev. B}} \textbf{87}, 155113 (2013).}

\item{\uline{S.S. Krishtopenko}. \href{http://iopscience.iop.org/0953-8984/25/36/365602}{\emph{Magnetoplasmon excitations from integer-filled Landau levels in narrow-gap quantum wells.}}
\emph{\textbf{J. Phys.: Condens. Matter}} \textbf{25}, 365602 (2013).}

\item{\uline{S.S. Krishtopenko}.
\href{http://iopscience.iop.org/0953-8984/25/10/105601}{\emph{Electron spin resonance and cyclotron resonance for fractional Quantum Hall States in narrow-gap QW heterostructures.}}
\emph{\textbf{J. Phys.: Condens. Matter}} \textbf{25}, 105601 (2013).}

\item{K.P. Kalinin, \uline{S.S. Krishtopenko}, K.V. Maremyanin, K.E. Spirin, V.I. Gavrilenko, A.A. Biryukov, N.V. Baidus, B.N. Zvonkov. \href{http://link.springer.com/article/10.1134/S1063782613110092}{\emph{Rashba spin splitting and cyclotron resonance in strained InGaAs/InP quantum well heterostructures.}}
\emph{\textbf{Semiconductors}} \textbf{47}, 1485 (2013).}
\item{\uline{S.S. Krishtopenko}, V.I. Gavrilenko, M. Goiran.
\href{http://iopscience.iop.org/0953-8984/24/13/135601}{\emph{The effect of exchange interaction on quasiparticle Landau levels in narrow-gap quantum well heterostructures.}}
\emph{\textbf{J. Phys.: Condens. Matter}} \textbf{24}, 135601 (2012).}

\item{\uline{S.S. Krishtopenko}, V.I. Gavrilenko, M. Goiran.
\href{http://iopscience.iop.org/0953-8984/24/25/252201/}{\emph{Exchange interaction effects in electron spin resonance: Larmor theorem violation in narrow-gap quantum well heterostructures.}}
\emph{\textbf{J. Phys.: Condens. Matter}} \textbf{24}, 252201 (2012).}

\item{\uline{S.S. Krishtopenko}, V.I. Gavrilenko, M. Goiran.
\href{http://doi.org/10.4028/www.scientific.net/SSP.190.554}{\emph{Exchange Enhancement of g-Factor in Narrow-Gap InAs/AlSb Quantum Well Heterostructures.}}
\emph{\textbf{Solid State Phenomena}} \textbf{190}, 554 (2012).}

\item{\uline{S.S. Krishtopenko}, A.V. Ikonnikov, K.V. Maremyanin, K.E. Spirin, V.I. Gavrilenko, Yu.G. Sadofyev, M.Goiran, M.Sadowsky, Yu.B. Vasilyev.
\href{http://jap.aip.org/resource/1/japiau/v111/i9/p093711_s1}{\emph{Cyclotron resonance study in InAs/AlSb quantum well heterostructures with two occupied electronic subbands.}}
\emph{\textbf{J.~Appl. Phys.}} \textbf{111}, 093711 (2012).}

\item{\uline{S.S. Krishtopenko}, K.P. Kalinin, V.I. Gavrilenko, Yu.G. Sadofyev, M. Goiran.
\href{http://link.springer.com/article/10.1134/S1063782612090138}{\emph{Rashba spin splitting and exchange enhancement of the g-factor in InAs/AlSb heterostructures with a two-dimensional electron gas.}}
\emph{\textbf{Semiconductors}} \textbf{46}, 1163 (2012).}

\item{K.E. Spirin, K.P. Kalinin, \uline{S.S. Krishtopenko}, K.V. Maremyanin, V.I. Gavrilenko, Yu.G. Sadofyev.
\href{http://link.springer.com/article/10.1134/S1063782612110206}{\emph{Features of the persistent photoconductivity in InAs/AlSb heterostructures with double quantum wells and tunneling-transparent barrier.}} \emph{\textbf{Semiconductors}} \textbf{46}, 1396 (2012).}

\end{enumerate}

\chapter{\label{Chap:Kane}Kane fermions in bulk narrow-gap and zero-gap semiconductors}
According to the solutions of the Dirac equation, essential in high energy physics, relativistic particles are classified into Dirac, Majorana or Weyl fermions.
In solids, the interaction of electrons with the periodic potential of the crystal lattice can give rise to dispersion relations at high symmetry points of the Brillouin zone, which mimic the relativistic particles. This relativistic analogy is usually used for materials, in which conduction and valence bands are well separated from the other bands, considered as the remote bands. In the early 1960s, Keldysh~\cite{Flds1} and Wolff~\cite{Flds2} independently discovered that the two-band description of the bismuth, PbTe, PbSe and PbS-type semiconductors was equivalent to the Dirac equation for free particles. If the band-gap vanishes, the band structure mimics either by massless Dirac fermions or Weyl fermions. Solid state analogues of massless relativistic particles were demonstrated in graphene~\cite{Dirac1}, semiconductor QWs~\cite{Dirac2,Dirac3,Dirac4}, topological insulators~\cite{Dirac3DTI1,Dirac3DTI2,Dirac3DTI3}, 3D Dirac~\cite{3DDiracSM1,3DDiracSM2,3DDiracSM3} and 3D Weyl~\cite{3DWeylSM1,3DWeylSM2,3DWeylSM3} semimetals.

A fundamental difference between electrons in solids and those at high energy is that Lorentz invariance is not always preserved in condensed matter physics. By generalizing the Dirac equation, one can find many other free fermionic excitations that have no high-energy analogs~\cite{Exotic1,Exotic2,Exotic3,Exotic4}. Massless ``pseudo-spin-1 and -3/2'' Dirac and Weyl fermions are a remarkable example of such particles, whose low-energy Hamiltonian has the form of $\mathbf{k}\cdot\mathbf{S}$, where $\mathbf{S}$ is the vector of spin-1 or -3/2 matrices. Massless Kane fermions, observed in gapless HgCdTe crystals~\cite{Kane1}, represent another type of fermionic excitations formed by the crossing of three bands. In the lowest order in quasimomentum, its band structure is characterized by doubly degenerate conical bands intersected at the vertex by an additional flat band, closely resembling a pseudo-spin-1 Dirac semimetal~\cite{Exotic4}. However, unlike the pseudo-spin-1 Dirac fermions, Kane fermions are not protected by symmetry or topology, and their band structure in Hg$_{1-x}$Cd$_{x}$Te can be tuned intrinsically by adjusting the chemical composition $x$, or externally, by changing temperature~\cite{Kane5}. The discussion of the nature of Kane fermions has become more confusing after Malcolm and Nicol~\cite{Kane2}, who noted that the three-band Kane Hamiltonian involving the $\Gamma_{8v}$ and $\Gamma_{6c}$ bands maps onto an intermediate value ($\alpha=1/\sqrt{3}$) of the $\alpha$-$\mathcal{T}_3$ model~\cite{Kane3}. Since the $\alpha$-$\mathcal{T}_3$ model interpolates between the spin-1/2 (graphene, $\alpha=0$) and pseudo-spin-1 (dice or $\mathcal{T}_3$ lattice, $\alpha=1$) Dirac-Weyl systems, it was concluded that the Kane fermions are hybrids of pseudo-spin-1 and -1/2 Dirac fermions.

Up to first-order in $\mathbf{k}\cdot\mathbf{p}$ theory, the three-band Hamiltonian involving the $\Gamma_{6c}$ and $\Gamma_{8v}$ bands can be presented in the form~\cite{Kane4}:
\begin{equation}
\label{Keq:1}
\mathcal{\hat{H}}_{6\times6}(\mathbf{k},\theta)=\begin{pmatrix}
\hat{\mathcal{H}}_{0}(k_x,k_y,\theta) & \hat{\mathcal{H}}_{z}(k_z) \\ \hat{\mathcal{H}}_{z}(k_z)^{\dag} & \hat{\mathcal{H}}_{0}^{*}(-k_x,-k_y,\theta)\end{pmatrix},
\end{equation}
where the asterisk stands for complex conjugation and ``$\dag$'' corresponds to Hermitian conjugation. The $3\times3$ blocks $\hat{H}_{0}({k}_x,{k}_y,{k}_z)$ and $\hat{H}_{z}({k}_z)$ in Eq.~(\ref{Keq:1}) are written as
\begin{equation}
\label{Keq:2}
\hat{H}_{0}=\begin{pmatrix}
C_{0}+M_{0} & V{k}_{+}\sin\theta & V{k}_{-}\cos\theta \\
V{k}_{-}\sin\theta & C_{0}-M_{0} & 0\\
V{k}_{+}\cos\theta & 0 & C_{0}-M_{0}\end{pmatrix}
\end{equation}
and
\begin{equation}
\label{Keq:3}
\hat{H}_{z}=\begin{pmatrix}
0 & 0 & V{k}_{z} \\
0 & 0 & 0\\
V{k}_{z} & 0 & 0\end{pmatrix},
\end{equation}
where $k_{\pm}=k_x\pm i k_y$ with $k_x$, $k_y$, $k_z$ being momentum operators, $C_0$ is an energy reference; $V/\hbar$ is the ``effective speed of the light'' $\tilde{c}$ related with the momentum matrix element $P$ between the $\Gamma_{6c}$ and $\Gamma_{8v}$ bands~\cite{Kane4}; and $M_0$ is the mass parameter describing the ordering of $\Gamma_{6c}$ and $\Gamma_{8v}$ bands. In Eq.~(\ref{Keq:2}), we have additionally introduced a ``hybridization angle'' $\theta$ describing the mixing between light- and heavy-holes at finite quasimomentum~\cite{Kane4}. For the Kane Fermions, $\theta=\pi/3$. Since $\mathcal{\hat{H}}_{6\times6}(\mathbf{k},\theta)$, $\mathcal{\hat{H}}_{6\times6}(\mathbf{k},-\theta)$ and $\mathcal{\hat{H}}_{6\times6}(\mathbf{k},\pi/2\pm\theta)$ are all related by unitary transformation, one can only consider the $\theta$ values within the range $[0,\pi/2)$. As clear from Eq.~(\ref{Keq:1}), $\mathcal{\hat{H}}_{6\times6}(\mathbf{k},\theta)$ at $k_z=0$ and $M_0=0$ indeed interpolates between the spin-1/2 and pseudo-spin-1 Dirac-Weyl systems. In our notation, $\mathcal{\hat{H}}_{6\times6}(\mathbf{k},\theta)$ describes the physics of graphene (dice lattice) at $\theta=0$ ($\theta=\pi/4$), where two diagonal blocks pertain to the distinct chiral centers $K$ and $K'$, respectively~\cite{Kane3}. In the 3D case, non-zero $k_z$ takes the role of connecting two otherwise independent chiral blocks.

\begin{figure}
\centering
\includegraphics [width=0.53\columnwidth, keepaspectratio] {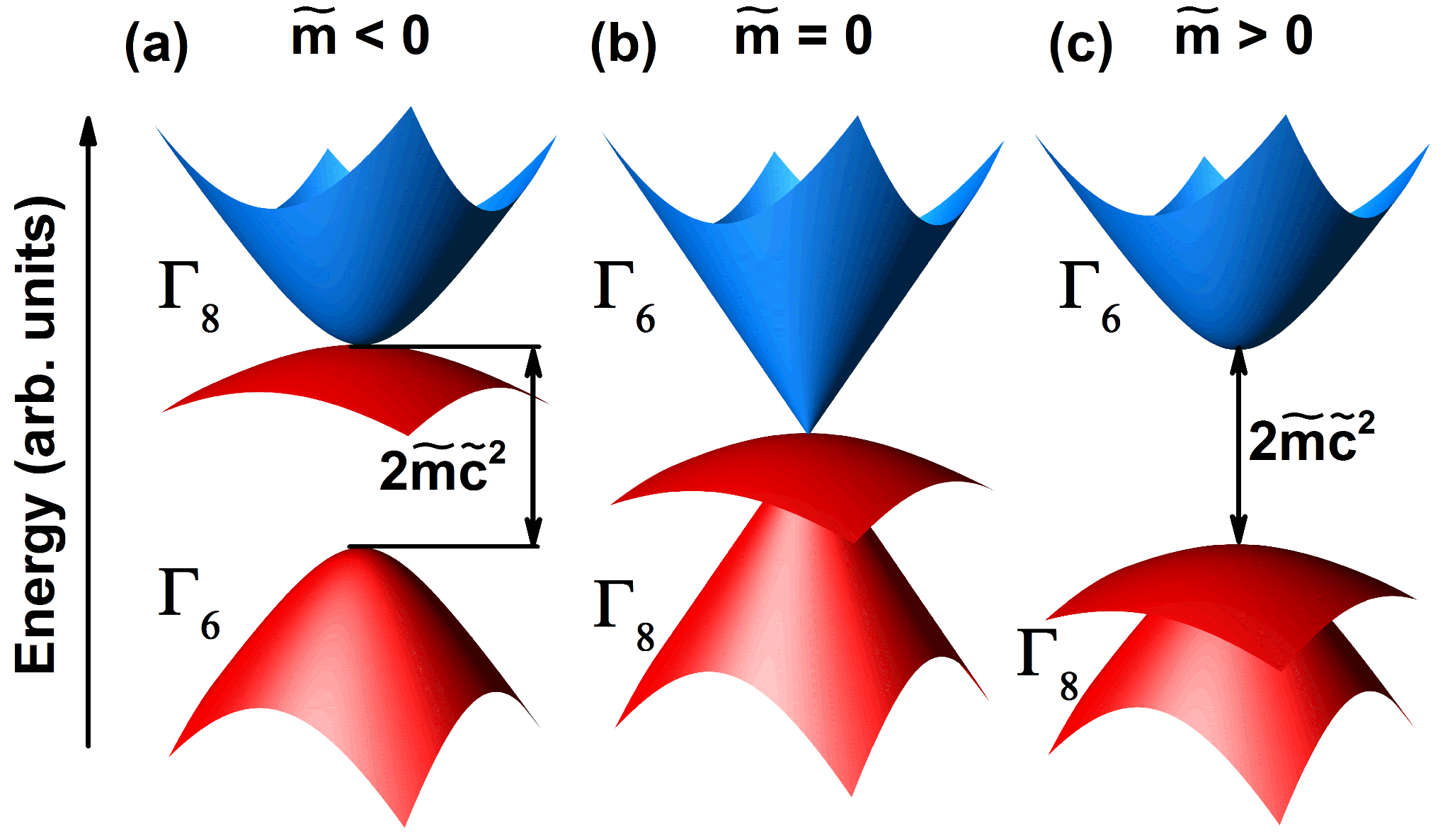} 
\caption{\label{Fig:Chapter2Fig1} Schematic band structure of Kane fermions in 3D Hg$_{1-x}$Cd$_{x}$Te as a function of the rest-mass $\tilde{m}$. For clarity, the ``flat heavy-hole'' band is shown with the curvature characteristic of the real crystals beyond
$\mathcal{\hat{H}}_{6\times6}(\mathbf{k},\theta)$ in Eq.~(\ref{Keq:1}).}
\end{figure}

An inherent feature of $\mathcal{\hat{H}}_{6\times6}(\mathbf{k},\theta)$ is the independence of its eigenvalues from $\theta$, that makes the Kane fermion energy dispersion ($\theta=\pi/3$) the same as the one of Dirac fermion ($\theta=0$) with independent flat band:
\begin{equation}
\label{Keq:4}
E_{\pm}(\mathbf{k})=C_0\pm\sqrt{M_0^2+V^2k^2},~~~~~~~E_0=C_0-M_0,
\end{equation}
where $k^2=k_x^2+k_y^2+k_z^2$.  To emphasize the similarity between Kane and Dirac fermions, Eq.~(\ref{Keq:4}) is often rewritten in terms of the effective rest-mass $\tilde{m}=M_0/\tilde{c}^2$ and Fermi velocity $\tilde{c}=V/\hbar{k}$, representing the asymptotic velocity of Kane fermions~\cite{Kane6,Kane7,Kane8}:
\begin{equation}
\label{eq:x1}
E_{\xi}(p)=\xi^2\tilde{m}\tilde{c}^2+\left(-1\right)^{1-\tilde{\theta}(\tilde{m})}\xi
\sqrt{\tilde{m}^2\tilde{c}^4+{p}^2\tilde{c}^2},
\end{equation}
where $p=\hbar{k}$ is an absolute momentum value and $\tilde{\theta}(x)$ is a Heaviside step function, which equals to $1$ for $x{\geq}0$ and to $0$ if $x$ is negative. Here, the values of $C_0$ are chosen so that the flat band energy is zero.

The first eigenvalue with $\xi=0$ in Eq.~(\ref{Keq:4}) corresponds to the heavy-hole $\Gamma_8$ band, which is dispersionless in the linear approximation. The two other eigenvalues describe the electron $\Gamma_6$ ($\xi=+1$) and light-hole $\Gamma_8$ ($\xi=-1$) conical bands separated by $E_g=2\tilde{m}\tilde{c}^2$. The evolution of Kane fermions in Hg$_x$Cd$_x$Te is illustrated in Fig.~\ref{Fig:Chapter2Fig1}. If the rest-mass $\tilde{m}$ is positive, the crystal is a typical narrow-gap semiconductor with the $s$-type $\Gamma_6$ band lying above the $p$-type $\Gamma_8$ bands. On the other hand, if $\tilde{m}<0$, the band ordering is inverted: the $\Gamma_6$ band lies below the $\Gamma_8$ bands. As the two $\Gamma_8$ bands always touch each other at the $\Gamma$ point of the Brillouin zone (in the absence of strain), the band structure is gapless and the crystal is a semimetal. The two distinct phases with
different sign of the rest-mass are not topologically equivalent, as characterized by different sets of the $\mathbb{Z}_2$ topological invariants~\cite{3DTI1}. In the THz team in Montpellier, we actively investigated the properties of Kane fermions that emerge not only in bulk Hg CdTe films but also in Cd\textsubscript{3}As\textsubscript{2}~\cite{Kane9,Kane10}. In the following sections, I will briefly present a selection of our experimental and theoretical results, which, I hope, will help the reader gain an overall picture of these pseudo-relativistic particles.

\section{\label{sec:TempKane}Temperature-driven Kane fermions in bulk HgCdTe films}
Part of our experimental research in Montpellier included a systematic optical investigation of the dispersion of Kane fermions in HgCdTe crystal as a function of temperature and magnetic field in the vicinity of the topological phase transition~\cite{Kane6}. Despite the long-standing experimental interest in bulk HgCdTe films -- largely motivated by their use in infrared detectors -- no comparable optical studies had previously been performed in the vicinity of the topological transition point. The absence of such results was primarily due to the difficulty of obtaining sufficiently high-quality samples. This limitation was overcome through the collaboration between the THz team in Montpellier and Vladimir Gavrilenko's group in Nizhny Novgorod, which provided us with high-quality bulk Hg\textsubscript{1–x}Cd\textsubscript{x}Te samples grown by molecular beam epitaxy in Novosibirsk. Two representative Cd compositions were selected for detailed experimental studies: sample A with $x\simeq 0.175$ and sample B with $x\simeq0.155$. According to the phenomenological band-gap dependence $E_g(x,T)$, derived from a comprehensive set of earlier experimental data~\cite{Kane5}, one can expect that sample A behaves as a trivial semiconductor at any temperature, while sample B represents a topological semimetal at low temperatures with a ``negative'' band-gap corresponding to inverted band ordering. Upon increasing temperature, the inverted gap in sample B gradually closes, and the system undergoes a semimetal-to-semiconductor phase transition at the critical temperature $T_c$, followed by the reopening of a normal (trivial) gap.

The central goal of our study was to determine experimentally the temperature dependence of the rest-mass $\tilde{m}$ and the Fermi velocity $\tilde{c}$ of Kane fermions in the range of $T_c$, by analyzing the positions of inter–Landau-level transition lines in the magnetoabsorption spectra. This analysis can be performed by fitting the experimental data using the analytical expressions for the energy differences between Landau levels derived within the Kane model, thus providing direct access to the evolution of the parameters across the topological phase transition. Assuming the orientation of magnetic field $\mathcal{B}$ in the $z$ direction, the Landau level energies at arbitrary $\theta$ values in Eq.~(\ref{Keq:5}) is written as~\cite{Flds8}:
\begin{eqnarray}
\label{Keq:5}
E^{(\sigma)}_{\pm}=C_0\pm\sqrt{M_0^2+V^2k_z^2+\dfrac{2V^{2}}{a_B^2}(n+1+\sigma\sin^2\theta)},\nonumber\\
E_0=C_0-M_0,~~~~~~~~~~~~~~~~~~~~~~
\end{eqnarray}
where $n\geq{-1}$ is the Landau level index, $\sigma=-1$ for $n\geq{0}$, while $\sigma=+1$ for $n\geq{-1}$, and $a_B$ is the magnetic length ($a_B^2=c\hbar/e\mathcal{B}$, $c$ is the speed of light in vacuum). As seen from Eq.~(\ref{Keq:5}), Landau levels of Dirac fermion are doubly degenerate at $n\geq{0}$, while the degeneracy is lifted for the Kane fermion case.

Alternatively, Eq.~(\ref{Keq:5}) with $\theta=\pi/3$ can be represented in terms of $\tilde{m}$ and $\tilde{c}$ in the form of Eq.~(\ref{eq:x1}):
\begin{equation}
\label{eq:x2}
E_{\xi,n,\sigma}(p_z)=\xi^2\tilde{m}\tilde{c}^2+\left(-1\right)^{1-\tilde{\theta}(\tilde{m})}\xi\sqrt{\tilde{m}^2\tilde{c}^4+\dfrac{e\hbar\tilde{c}^2\mathcal{B}}{2}\left(4n-2+\sigma\right)+\tilde{c}^2p_{z}^2},
\end{equation}
where $p_z$ is a momentum projection onto direction of magnetic field. Here, the \uline{redefined} LL index $n$ runs over positive integers $n=1,2,\ldots$ for the states in the electron and light-hole bands ($\xi={\pm}1$). For the zero energy, flat heavy-hole band ($\xi=0$), $n$ now runs over all non-negative integers, except~$1$: $n=0,2,3\ldots$. The \uline{redefined} quantum number $\sigma$ accounts for the Kramer's degeneracy lifted by the magnetic field. The pairs of Landau levels, whose energy difference contributes to magnetoabsorption, obey by the following electric-dipole selection rules: $\Delta{n}=n{\pm}1$ with ``$\pm$'' corresponding to the two circular polarizations, ${\Delta}p_z=0$, ${\Delta}\sigma=0$, and no restriction on changing of $\xi$.

\begin{figure}
\includegraphics [width=1.0\columnwidth, keepaspectratio] {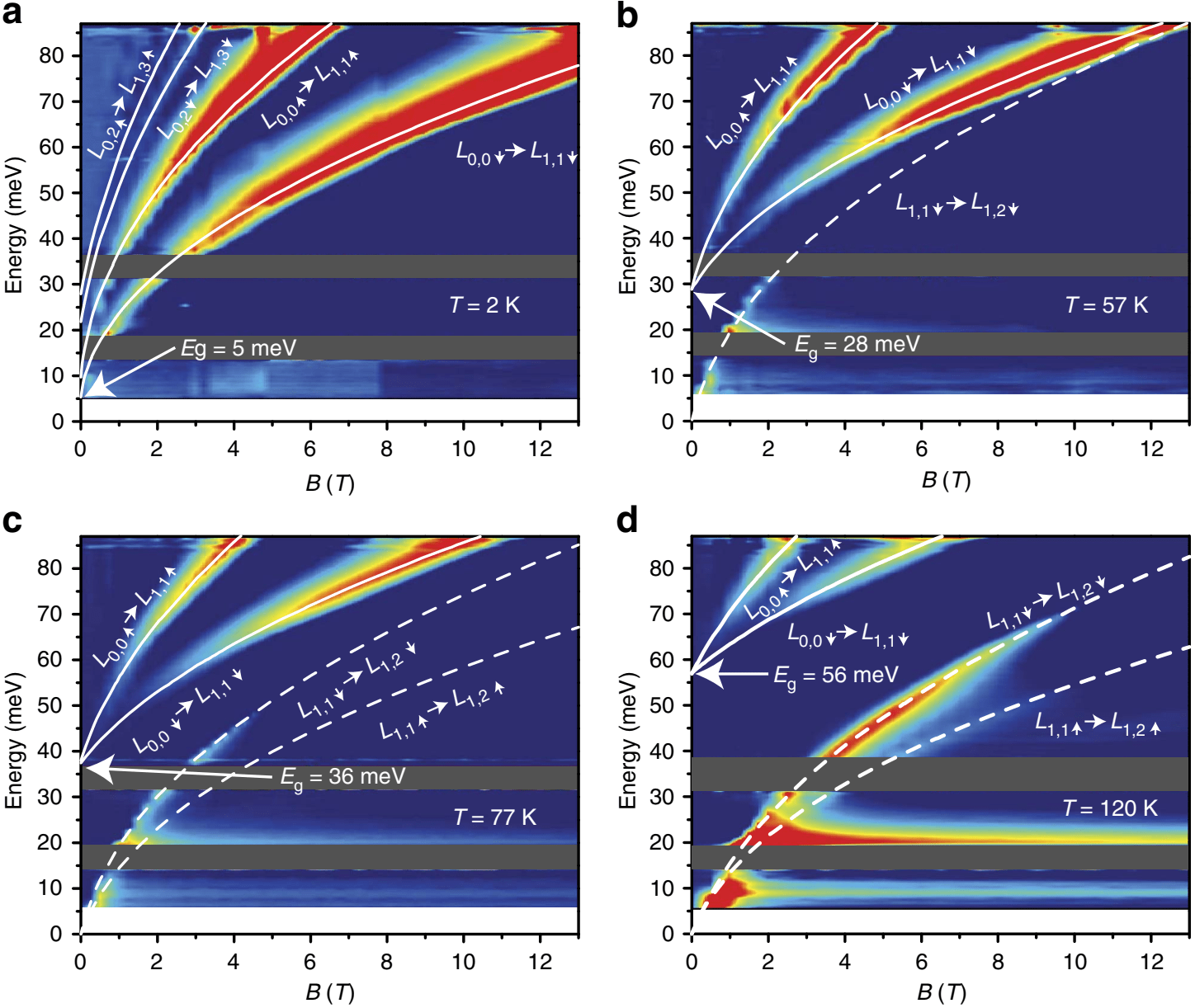} 
\caption{\label{Fig:Chapter2Fig2} \textbf{False colour maps present inter-Landau-level transitions as a function of magnetic field in the sample A}. White lines represent the fits using Eq.~(\ref{eq:x2}), allowing for determination of $\tilde{m}$ and $\tilde{c}$. Solid lines correspond to inter-band transitions, whereas dashed lines account for intra-band transitions. The band-gap values at different temperatures, determined by the inter-band transition energies at zero magnetic field, are depicted with arrows. Shaded areas represent the Reststrahlen bands (between 15--20 meV for Hg$_{1-x}$Cd$_{x}$Te and 32--37 meV for GaAs substrate band). \textbf{(a)} At 2 K, only inter-band transitions are seen and a $\sqrt{\mathcal{B}}$-like behaviour of inter-Landau-level resonances. \textbf{(b)} At 57 K, intra-band transitions become visible in addition to previous lines. \textbf{(c,d)} At higher temperatures the energy difference between inter-band and intra-band lines at zero field clearly increases, corresponding to a gap opening as a function of temperature. The horizontal line observed at 21 meV in (c,d) corresponds to the energy of TO CdTe-like phonons, arising in magnetoabsorption due to the electron-phonon interaction. In addition, the horizontal line observed at 9~meV in (c,d) is attributed to optical transitions from the impurities, represented by mercury vacancies~\cite{Kane11}.}
\end{figure}

\begin{figure}
\includegraphics [width=1.0\columnwidth, keepaspectratio] {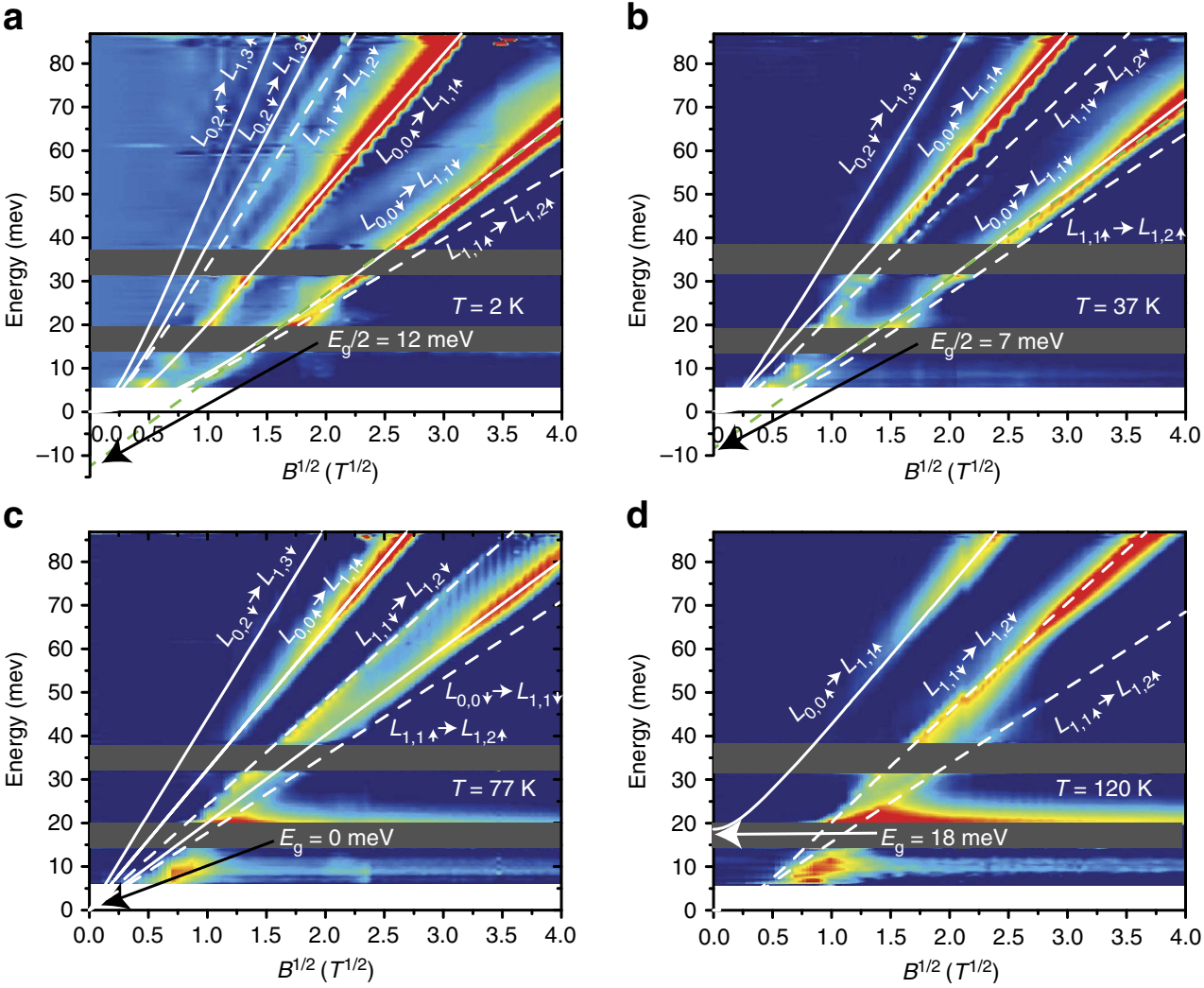} 
\caption{\label{Fig:Chapter2Fig3} \textbf{Colour maps of the inter-Landau-level transitions as a function of magnetic field in the sample B at temperatures from 2 to 120 K in a square-root scale}. White lines represent the fits using using Eq.~(\ref{eq:x2}). Solid lines represent
inter-band optical transitions, whereas dashed lines correspond to intra-band transitions. Green dotted lines are guides for the eyes accounting for half the band-gap energy at each temperature (with arrows indicating half the band-gap values). \textbf{(a,b)} At 2 and 37 K, some discrepancies to the $\sqrt{\mathcal{B}}$ behaviour of inter-Landau-level optical transitions are observed, corresponding to the existence of a negative band-gap. \textbf{(c)}~At~77~K, a pure $\sqrt{\mathcal{B}}$ behaviour corresponds to the gapless state and the presence of genuine massless Kane fermions. \textbf{(d)} At 120 K, a positive gap opens as seen with the presence of an inter-band optical transition. Shaded areas denote the Reststrahlen bands (between 15--20 meV for Hg$_{1-x}$Cd$_{x}$Te and 32--37 meV for GaAs substrate band). The horizontal line at 21 meV in (c,d) is related to the energy of TO CdTe-like phonons, arising in magnetoabsorbance due to the electron-phonon interaction. The line at 9 meV is attributed to optical transitions from impurities~\cite{Kane11}.}
\end{figure}

The measurements of magnetoabsorption at 2~K revealed the presence of 3D fermions with $\tilde{m}\simeq0$ in the sample A by $\sqrt{\mathcal{B}}$-like dependence of optical transitions and a spin splitting of Landau levels (see Fig.~\ref{Fig:Chapter2Fig2}(a)). The fitting of the two main lines based on Eq.~(\ref{eq:x2}) shows that they are related to the inter-band transitions between the heavy-hole band remaining at zero energy ($\xi=0$) and the $n=1$ spin-up and spin-down LLs with $\xi=1$). The extracted band-gap value at 2 K equals $2\tilde{m}\tilde{c}^2=(5\pm2)$~meV. Magneto-optical results obtained at temperatures from 57 to 120 K are shown in Fig.~\ref{Fig:Chapter2Fig2}(b-d). The band-gap, visualized by intersect of the inter-band transitions with the energy axis (shown by white arrows), increases with temperature. In addition to inter-band transitions (${\Delta}\xi=1$), intra-band Landau-level transitions (${\Delta}\xi=0$) are also observed and fitted with Eq.~(\ref{eq:x2}).
Magnetoabsorption of the sample B at different temperatures is presented in Fig~\ref{Fig:Chapter2Fig3}, in a $\sqrt{\mathcal{B}}$-scale for the sake of clarity. It is seen that at low temperatures and low magnetic field values, the Landau-level transitions exhibit some
discrepancies compared with a pure $\sqrt{\mathcal{B}}$ behaviour. Indeed, as shown in Fig.~\ref{Fig:Chapter2Fig1}, the linear energy dispersion in conduction and valence bands arises in gapless samples only. However, even in the case of small negative (or positive) gap, the bands could be considered as parabolic in the vicinity of the $\Gamma$ point, defined by $p/|\tilde{m}\tilde{c}|\ll1$. In this case, the band parabolicity at low magnetic fields results in a linear behaviour of the Landau-level transitions as a function of
$\mathcal{B}$. Therefore, only a precise $\sqrt{\mathcal{B}}$ behaviour down to the lowest applied magnetic fields implies a system with genuine massless particles. At temperatures below $T_c$, the deviation from a pure $\sqrt{\mathcal{B}}$ behaviour is well reproduced by the theory and gives absolute values of the rest-mass $\tilde{m}$ approaching to zero when temperature increases. As seen in Eq.~(\ref{eq:x2}), linear extrapolation of optical transitions in the square-root scale intersects the energy axis at $E=E_g/2=\tilde{m}\tilde{c}^2$, as represented by arrows in Fig.~\ref{Fig:Chapter2Fig3}(a,b). An accurate $\sqrt{\mathcal{B}}$ behaviour for all optical transitions is obtained as an evident proof of gap closing at 77~K. Above the critical temperature, the difference in energies of inter- and intra-band transitions in low magnetic fields becomes visible, as it is for the sample A, meaning that a positive gap between the $\Gamma_6$ and $\Gamma_8$ bands opens. This allows us claiming that at $T_c=77$~K, a temperature-driven topological
phase transition with massless Kane fermions occurs.

\begin{figure}
\centering
\includegraphics [width=0.75\columnwidth, keepaspectratio] {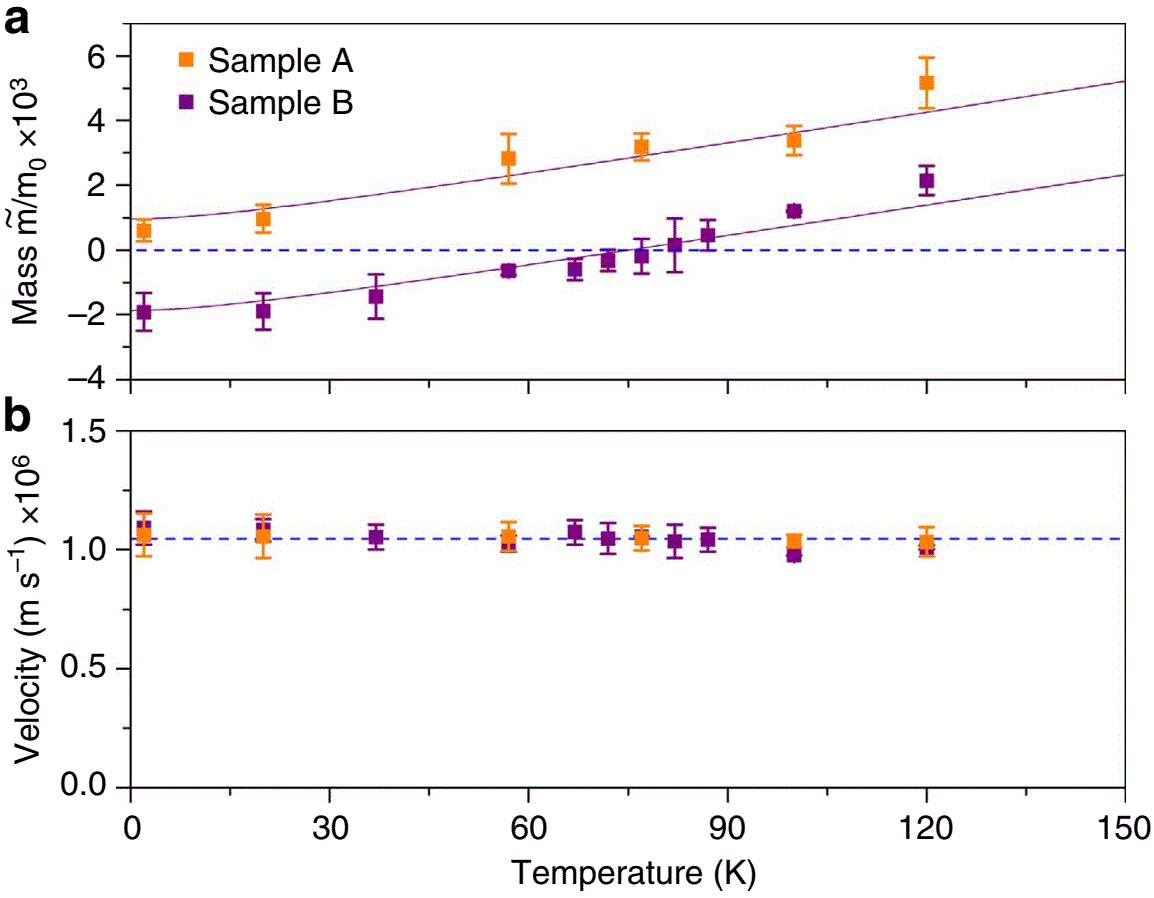} 
\caption{\label{Fig:Chapter2Fig4} \textbf{Parameters $\tilde{m}$ and $\tilde{c}$ determined by fitting magneto-optical data with Eq.~(\ref{eq:x2})}. \textbf{(a)}~The values of  Kane fermion rest-mass for both samples. The rest-mass in the sample~A is close to zero at low temperatures and increases with temperature up to 120~K. In the sample B, the rest-mass is negative at low temperatures, it vanishes at 77~K and then becomes positive above $T_c$. The curves represent the values, calculated on the basis of phenomenological dependence $E_g(x,T)$~\cite{Kane5}. \textbf{ (b)}~The Fermi velocity $\tilde{c}$ is nearly constant in both samples between 2 and 120 K. The blue dashed line corresponds to the theoretical value.}
\end{figure}

Figure~\ref{Fig:Chapter2Fig4} summarises parameters of Kane fermions extracted from experimental data for both samples. In the
sample A, $\tilde{m}$ is positive and increases in the whole  range of temperatures, while in the sample B, the Kane fermion rest-mass experiences a
change of sign corresponding to the temperature-induced semimetal-to-semiconductor topological phase transition at 77~K. Remarkably, the experimentally obtained temperature dependence of the rest-mass in both samples is in very good agreement with the phenomenological relation $E_g(x,T)$ derived earlier from a wide set of experimental data on Hg\textsubscript{1–x}Cd\textsubscript{x}Te far away from the topological phase transition~\cite{Kane5} (see also our original publication~\cite{Kane6}). Another striking result is the near constancy of the Fermi velocity $\tilde{c}$ for both samples, regardless of temperature or Cd composition ($x=0.155$ and $x=0.175$). The extracted value $\tilde{c}=(1.07\pm0.05)\cdot10^6$~m/s shows excellent agreement with the theoretical value defined by $\tilde{c}^2=2P^{2}/(3\hbar^2)$, which equals to $1.05\cdot10^6$~m/s for the well accepted value of $E_P=2m_{0}P^{2}/\hbar^2\approx18.8$~eV~\cite{HgTeQWs1} (where $m_0$ is the rest-mass of free electron). This universality of $\tilde{c}$ underscores its fundamental role as a material constant of HgCdTe and enables a straightforward determination of the rest-mass from the band-gap value near the semimetal-to-semiconductor phase transition induced by temperature, Cd content, or other external parameters (such as hydrostatic pressure \cite{Kane7}).

In the end of this section, it is worth discussing the general applicability of the three-band Kane model with the $6\times6$ Hamiltonian $\mathcal{\hat{H}}_{6\times6}(\mathbf{k},\theta)$ introduced in Eq. (\ref{Keq:1}), which is linear in the quasi-momentum. By its construction, this model accurately describes the electronic states only in a narrow vicinity of the $\Gamma$ point of the Brillouin zone. However, the main magneto-optical transitions observed in our experiments are precisely associated with the Landau levels originating from this region of quasi-momentum space. Therefore, the energies of these transitions, and hence the positions of the absorption lines in Figs. 2 and 3, are well reproduced by fitting the data with the analytical expression~(\ref{eq:x2}). Nevertheless, this good agreement should not be misinterpreted as evidence that such a simplified model is universally valid. Its applicability is confined to conditions where the contribution of other bands can be neglected. In the following sections, I will discuss several phenomena in which the influence of additional remote bands -- beyond the $\Gamma_6$ and $\Gamma_8$ bands included in $\mathcal{\hat{H}}_{6\times6}(\mathbf{k},\theta)$ -- becomes essential for understanding the electronic properties of narrow-gap and gapless semiconductor systems.

\section{\label{sec:SSKane}Domain-wall--localized states of Kane fermions}
Domain-wall-localized states of Dirac fermions represent one of the most striking manifestations of topology in both quantum field theory and condensed matter systems. When the mass term in a Dirac Hamiltonian varies spatially and changes sign across a boundary, the resulting spectrum necessarily hosts fermionic modes confined to the region where the mass term vanishes. These domain-wall states arise from the interplay between the local inversion of the mass gap and the global topology of the underlying field configuration. First predicted in the seminal work of Jackiw and Rebbi~\cite{SS3DKane1}, such bound modes appear as zero-energy solutions of the Dirac equation with a position-dependent mass and are now regarded as the canonical example of topologically protected interface states. The energy dispersion of Kane fermions closely resembles that of Dirac fermions, which is why these two types of quasiparticles are often confused. However, their fundamental differences become apparent in the presence of a magnetic field, as described by Eq.~(\ref{Keq:5}). In this section, we examine what kinds of states emerge within the framework of the $6\times6$ Hamiltonian $\mathcal{\hat{H}}_{6\times6}(\mathbf{k},\theta)$ introduced in Eq. (\ref{Keq:1}) when a domain wall is present. In this model, the spatial variation of the mass parameter creates a boundary region analogous to that in the Jackiw–Rebbi problem, but with an additional degree of complexity introduced by the coupling to the ``flat'' heavy-hole band inherent to Kane systems. By continuously varying the parameter $\theta$, which controls the interaction between the Dirac-like and flat bands, one can trace the smooth transformation of domain-wall–localized states from the pure Dirac limit ($\theta=0$) to the Kane limit ($\theta=\pi/3$). This interpolation provides a framework for understanding how the presence of the flat band modifies the topological character and spatial localization of interface states.

Following the notations used in the original work~\cite{Kane4}, I slightly redefine $\hat{H}_{0}$  used in $\mathcal{\hat{H}}_{6\times6}(\mathbf{k},\theta)$ as follows:
\begin{equation}
\label{eq:DW2}
\hat{H}_{0}=\begin{pmatrix}
C_{0}+M_{0} & V\hat{k}_{+}\sin\theta & V\hat{k}_{-}\cos\theta \\
V\hat{k}_{-}\sin\theta & C_{0}+S_{0} & 0\\
V\hat{k}_{+}\cos\theta & 0 & C_{0}-M_{0}\end{pmatrix}
\end{equation}
where $S_0$ describes the position of the ``flat'' band, which may now differ from $-M_{0}$, and $\hat{k}_{\pm}=\hat{k}_x\pm i\hat{k}_y$ with $\hat{k}_x$, $\hat{k}_y$, $\hat{k}_z$ being momentum operators.
In this case, the three eigenvalues $E$ of  $\mathcal{\hat{H}}_{6\times6}(\mathbf{k},\theta)$, each double degenerate due to the time-reversal symmetry, follow the equation~\cite{Kane4}:
\begin{equation}
\label{eq:DW4}
V^2\left(k_x^2+k_y^2\right)\left(E_{hh}\cos^2\theta+E_{lh}\sin^2\theta\right)+V^2k_z^2E_{hh}=E_cE_{hh}E_{lh},
\end{equation}
where $k_x$, $k_y$, $k_z$ are the quantum numbers of the momentum operators, $E_c(E)=C_{0}+M_{0}-E$, $E_{hh}(E)=C_{0}+S_{0}-E$ and $E_{lh}(E)=C_{0}-M_{0}-E$.

To analyze the emergence of domain-wall--localized states in a realistic heterostructure, let us consider a simplified model consisting of two semi-infinite regions forming a junction along the $z=0$ plane. We assume that region~I corresponds to CdTe, characterized by a positive mass term $M_0^{(I)}>0$, while region~II represents HgTe, where the band ordering is inverted and the corresponding parameter satisfies $M_0^{(II)}<0$. In this configuration, the spatial inversion of the mass term $M_0(z)$ across the interface plays the role of a topological domain wall, giving rise to the appearance of interface-bound fermionic modes:
\begin{eqnarray}
\label{eq:BC100}
M_0(z)=M_0^{(I)}+\left(M_0^{(II)}-M_0^{(I)}\right)\tilde{\theta}(z),\nonumber\\
S_0(z)=S_0^{(I)}+\left(S_0^{(II)}-S_0^{(I)}\right)\tilde{\theta}(z),
\end{eqnarray}
where $\tilde{\theta}(z)$ is a  Heaviside step function defined as $\tilde{\theta}(z)=0$ at $z<0$ and $\tilde{\theta}(z)=1$ at $z\geq0$. For the sake of completeness, we note that the parameters $S_0^{(I)}$ and $S_0^{(II)}$ are assumed to differ only slightly from
$-M_0^{(I)}$ and $-M_0^{(II)}$, respectively. Such deviations may arise due to the presence of biaxial strain in the epitaxially grown layers, which slightly modifies the local band-edge energies. The function $C_0(z)$ in turn describes the electrostatic potential associated with a perpendicular electric field across the heterointerface. Since the goal of this section is to focus on the intrinsic topological doman-wall effects alone, we set $C_0(z)=0$, neglecting any electrostatic contributions. All other parameters, including $V$ and $\theta$, are considered to be independent of coordinates with the same values from both side of the CdTe/HgTe interface, which is justified by the experimental results discussed in the previous section. Under these assumptions, the bulk Hamiltonian $\mathcal{\hat{H}}_{6\times6}(\mathbf{k},\theta)$ remains Hermitian even in the presence of the junction, presuming the operators $\hat{k}_x$, $\hat{k}_y$ and $\hat{k}_z$ do not commute.

The initial Schr\"{o}dinger equation with $\mathcal{\hat{H}}_{6\times6}(\mathbf{k},\theta)$ can be considered as a set of differential equations for the $(\varphi_1,\varphi_2,\varphi_3,\varphi_4,\varphi_5,\varphi_6)^{T}$ envelope functions, resulting in an 6$\times$6 differential matrix. The absence of $\textbf{k}$-dependent terms in the diagonal elements allows one to express four of the six envelope functions by the two other functions for the same band. Such procedure, which is known as Gaussian elimination, is often used for multi-band Hamiltonians~\cite{Kane12,Kane13,Kane14}. Considering the pair of $(\varphi_1,\varphi_4)$ envelope functions we can express the four functions $\varphi_2$, $\varphi_3$, $\varphi_5$, $\varphi_6$ in terms of $\varphi_1$ and $\varphi_4$ by keeping the right order of non-commuting operators. Then, substituting the expressions for $\varphi_2$, $\varphi_3$, $\varphi_5$, $\varphi_6$ into the other two equations, we obtain the $2\times2$ energy-dependent Hamiltonian $\hat{H}_E$, which describes the evolution of the vector $\mathbf{\Phi}=(\Phi_1,\Phi_2)^T=(\varphi_1,\varphi_4)^T$ for two spin states~\cite{Kane4}:
\begin{equation}
\label{eq:DW6}
\begin{pmatrix}
\hat{A}+i\hat{B}-E & \hat{K} \\ \hat{K}^{\dag} & \hat{A}-i\hat{B}-E\end{pmatrix}\begin{pmatrix} \Phi_1(x,y,z) \\ \Phi_2(x,y,z) \end{pmatrix}=0,
\end{equation}
where $E$ is the eigenvalue, and
\begin{equation}
\label{eq:DW7}
\hat{A}=E_c-\hbar^2v^2\hat{k}_z\dfrac{1}{E_{lh}}\hat{k}_z-\hbar^2v^2\hat{k}_x\left(\dfrac{\cos^2\theta}{E_{lh}}+\dfrac{\sin^2\theta}{E_{hh}}\right)\hat{k}_x-\hbar^2v^2\hat{k}_y\left(\dfrac{\cos^2\theta}{E_{lh}}+\dfrac{\sin^2\theta}{E_{hh}}\right)\hat{k}_y,
\end{equation}
\begin{equation}
\label{eq:DW8}
\hat{B}=\hbar^2v^2\hat{k}_y\left(\dfrac{\cos^2\theta}{E_{lh}}-\dfrac{\sin^2\theta}{E_{hh}}\right)\hat{k}_x
-\hbar^2v^2\hat{k}_x\left(\dfrac{\cos^2\theta}{E_{lh}}-\dfrac{\sin^2\theta}{E_{hh}}\right)\hat{k}_y,
\end{equation}
\begin{equation}
\label{eq:DW9}
\hat{K}=\hbar^2v^2\hat{k}_z\dfrac{\cos\theta}{E_{lh}}\hat{k}_{-}-\hbar^2v^2\hat{k}_{-}\dfrac{\cos\theta}{E_{lh}}\hat{k}_z.
\end{equation}
As clear, the eigenvalue problem in Eq.~(\ref{eq:DW6}) is very similar to conventional Schr\"{o}dinger equation with the single-band Hamiltonian including non-diagonal spin-orbit interaction~\cite{Kane12,Kane13,Kane14}. Although the $2\times2$ energy-dependent Hamiltonian $\hat{H}_E$ in Eq.~(\ref{eq:DW6}) is non-Hermitian, its eigenvalues are real and are the same as those of $\mathcal{\hat{H}}_{6\times6}(\mathbf{k},\theta)$. Note that the similar procedure described above can be also performed for the other pairs $(\varphi_2,\varphi_5)$ and $(\varphi_3,\varphi_6)$ of the envelope functions. However, the resulting effective Hamiltonians will have the same eigenvalues. In general, non-Hermiticity in electronic system arises from the presence of its interaction with environment~\cite{Kane15,Kane16}. In our case the presence of heterointerface in the crystal induces additional interaction between all the bands from the opposite sides of the boundary, which yields the non-Hermiticity of $\hat{H}_E$ describing the pair $(\varphi_1,\varphi_4)$. In order to find the domain-wall-localized states, one needs first establish the boundary conditions for the envelope functions. This can be achieved by integrating the Schrödinger equation~(\ref{eq:DW6}) with the energy-dependent Hamiltonian across a small region in the vicinity of $z=0$. Such a procedure ensures  the continuity of the following functions across the heterointerface~\cite{Kane4}:
\begin{equation}
\label{eq:DW11}
\begin{pmatrix} \Phi_1 \\[5pt] \Phi_2 \end{pmatrix},~~~~\begin{pmatrix}
\dfrac{1}{E_{lh}(E)}\dfrac{\partial}{\partial z} & -ik_{-}\dfrac{\cos\theta}{E_{lh}(E)} \\[9pt] ik_{+}\dfrac{\cos\theta}{E_{lh}(E)} & \dfrac{1}{E_{lh}(E)}\dfrac{\partial}{\partial z}\end{pmatrix}\begin{pmatrix} \Phi_1 \\[5pt]  \Phi_2 \end{pmatrix}.
\end{equation}
Then, applying this boundary conditions to $\Phi_{1}$ and $\Phi_{2}$ that must decay exponentially away from the heterointerface, one can find the secular equation for the domain-wall-localized states~~\cite{Kane4}:
\begin{equation}
\label{eq:DW12}
\left(\lambda_{z}^{(I)}E_{lh}^{(II)}+\lambda_{z}^{(II)}E_{lh}^{(I)}\right)^2=
\left(E_{lh}^{(II)}-E_{lh}^{(I)}\right)^2k^2\cos^2\theta,
\end{equation}
where $\lambda_{z}^{(I)}$ and $\lambda_{z}^{(II)}$ are written as:
\begin{equation}
\label{eq:DW10}
\lambda_{z}^{(n)}=\sqrt{k^2\left(1+\dfrac{E_{lh}^{(n)}-E_{hh}^{(n)}}{E_{hh}^{(n)}}\sin^2\theta\right)-\dfrac{E_{c}^{(n)}E_{lh}^{(n)}}{V^2}}.
\end{equation}
with $n=I,II$ and $k^2=k_x^2+k_y^2$. Equations (\ref{eq:DW10}) and (\ref{eq:DW12}) give the energy dispersion of the domain-wall states. It is seen that these states at $k=0$ exist if $E_{lh}^{(I)}$ and $E_{lh}^{(II)}$ are of different sign, which requires different signs of $M_0^{(I)}$ and $M_0^{(II)}$. Substitution of Eq.~(\ref{eq:DW10}) into Eq.~(\ref{eq:DW12}) results in a biquadratic equation, which can be solved analytically.

In the case $S_0^{(n)}=-M_0^{(n)}$, representing the Kane fermions in unstrained HgCdTe crystals, considered in the previous section, Eq.~(\ref{eq:DW12}) for the domain-wall states is reduced to
\begin{equation}
\label{eq:DW13}
A\left(V^2k^2\right)^2-B{\cdot}V^2k^2+E^2\left(E+M_0^{(I)}\right)^2\left(E+M_0^{(II)}\right)^2=0,
\end{equation}
where
\begin{equation*}
A=\dfrac{\left(M_0^{(I)}-M_0^{(II)}\right)^2}{4}\sin^4\theta+\left(E+M_0^{(I)}\right)\left(E+M_0^{(II)}\right)\sin^2\theta,
\end{equation*}
\begin{equation}
\label{eq:DW14}
B=\left(E^2-M_0^{(I)}M_0^{(II)}\right)\left(E+M_0^{(I)}\right)\left(E+M_0^{(II)}\right)\sin^2\theta+\left(E+M_0^{(I)}\right)^2\left(E+M_0^{(II)}\right)^2.
\end{equation}
As clear from these equations, the domain-wall states for the Dirac fermion case at $\theta=0$ have conical dispersion $E={\pm}Vk$, while the solutions with $E=-M_0^{(I)}$ and $E=-M_0^{(II)}$ are spurious as they both lead to the zero values of
$\lambda_{z}^{(n)}$ in Eq.~(\ref{eq:DW10}). Non-zero $\theta$ values split initial Dirac-like domain-wall states $E=\pm\hbar vk$ into several branches below and above the flat bands in the bulk materials at both sides of the heterpinterface (see Fig.~\ref{Fig:Chapter2Fig4}). Particularly, the asymptotic energies of the domain-wall states lying between the flat bands of the bulk materials $E=-M_0^{(I)}$ and $E=-M_0^{(II)}$, at large values of $k$ are written as
\begin{equation}
\label{eq:DW16}
E_{\mathrm{as\pm}}=-\dfrac{M_0^{(I)}+M_0^{(II)}}{2}\pm\dfrac{M_0^{(I)}-M_0^{(II)}}{2}\cos\theta,
\end{equation}
which can be found from Eqs~(\ref{eq:DW13}) and~(\ref{eq:DW14}) with $A=0$.

\begin{figure}
\centering
\includegraphics [width=0.96\columnwidth, keepaspectratio] {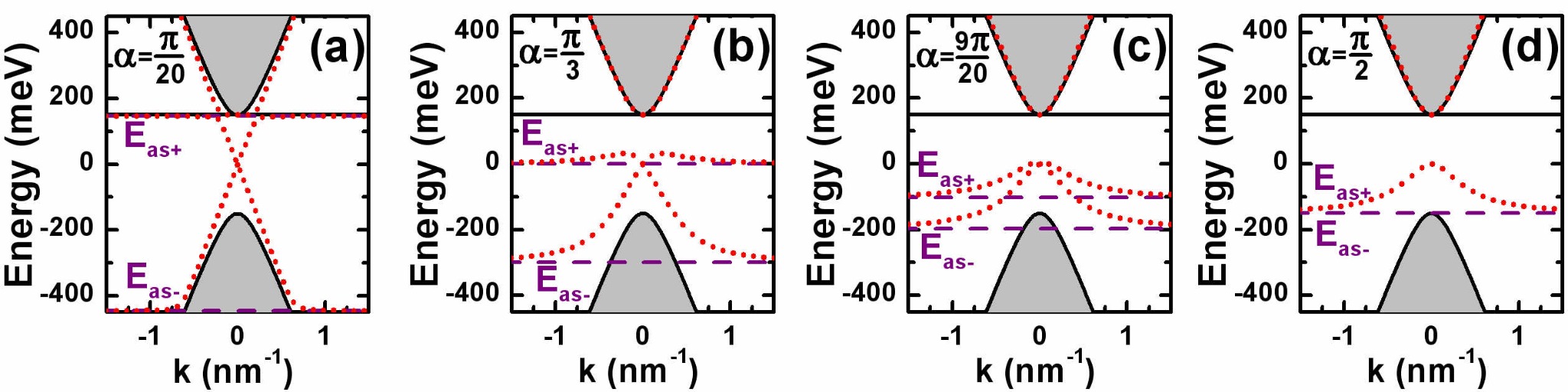}
\caption{\label{Fig:Chapter2Fig5} \textbf{(a-d)} Dispersion of the domain-wall-localised states (dotted red) as a function of $k$ for $S_0^{(II)}=-M_0^{(II)}$ at different $\theta$ values: \textbf{(a)} $\theta=\pi/20$, \textbf{(b)} $\theta=\pi/3$, \textbf{(c)} $\theta=9\pi/20$, \textbf{(d)} $\theta=\pi/2$. The grey regions correspond to the bulk states of HgTe projected onto the heterointerface. The solid black curves represent the edges of the projected bulk bands. The bulk states in CdTe arising at $|E|\geq M_0^{(I)}$ are beyond the scale of the panels. The dashed purple lines represent the asymptotic energies $E_{\mathrm{as\pm}}$ at large $k$, given by Eq.~(\ref{eq:DW16}). The value of $\alpha=\pi/3$ corresponds to the Kane fermions in unstrained HgTe crystals.}
\end{figure}

Figure~\ref{Fig:Chapter2Fig5} shows the dispersion of the bulk and domain-wall states for different values of $\theta$ in the range $(0,\pi/2]$ in bulk HgTe film with the HgTe/CdTe interface. All the parameters used in the calculations ($V=850$~meV$\cdot$nm, $M_0^{(I)}=450$~meV and $M_0^{(II)}=-150$~meV) are consistent with the experimental values of $\tilde{m}$ and $\tilde{c}$ in Section~\ref{sec:TempKane}. Note that the energy range $-M_0^{(I)}<E<M_0^{(I)}$ corresponds to the band-gap region of CdTe. Although the bulk dispersion is independent of $\theta$ (due to $S_0^{(n)}=-M_0^{(n)}$, see Eq.~(\ref{eq:DW4}) at $E_{hh}=E_{lh}$), the dispersion of the domain-wall states strongly depends on the hybridization between conical and flat bands. At small values of $\theta$ (see Fig.~\ref{Fig:Chapter2Fig5}(a)), it consists of four branches $E=E_{\mathrm{as+}}$, $E=\pm\hbar vk$ and $E=E_{\mathrm{as-}}$ anticrossing in the vicinity of the crossing points. In this case, the values of $E_{\mathrm{as+}}$ and $E_{\mathrm{as-}}$ are very close to $-M_0^{(II)}$ and $-M_0^{(I)}$, respectively, since $\cos\theta\approx1$ in Eq.~(\ref{eq:DW16}). With increasing of $\theta$, the domain-wall states for  $-M_0^{(I)}<E<-M_0^{(II)}$ are pushed away from the energies of the flat bulk bands in the materials at both sides of the heterointerface toward the regions where these bulk states are absent. This is clearly represented by the evolution of the asymptotic energies $E_{\mathrm{as+}}$ and $E_{\mathrm{as-}}$, which are getting closer to each other when $\theta$ increases as shown in Fig.~\ref{Fig:Chapter2Fig5}(a-c). Note that dispersion of the domain-wall states remains linear at small quasimomentum values. At the specific value of $\theta=\pi/2$, the asymptotic energies coincide both being equal to $-(M_0^{(I)}+M_0^{(II)})/2$, and the domain-wall states become degenerate as seen in Fig.~\ref{Fig:Chapter2Fig5}(d). The latter means the absence of odd-in-$k$ terms in their dispersion. We note that the domain-wall states, similar to those provided in Fig.~\ref{Fig:Chapter2Fig5}(b) for $\alpha=\pi/3$, were also obtained by more sophisticated numerical calculations based on tight-binding extension of the $6\times6$ Kane Hamiltonian with square terms~\cite{Kane17}. Although, the results of Ref.~\cite{Kane17} depend on the constant of artificial cubic lattice used in the calculations, they qualitatively reproduce the dispersion of the domain-wall states at small quasimomentum obtained from $\mathcal{\hat{H}}_{6\times6}(\mathbf{k},\theta)$ at $\theta=\pi/3$.

In addition to the modification of the domain-wall states in the range of $-M_0^{(I)}<E<-M_0^{(II)}$, the hybridization of Dirac states with the flat bulk bands at non-zero $\theta$ also yields new ``massive'' branches in the regions above and below the flat bands. We refer to the upper ``massive'' domain-wall states above the flat band as the \emph{Dyakonov-Khaetskii} branch. Dyakonov and Khaetskii~\cite{Kane18} were the first, who predicted the massive states at the surface of HgTe crystal. They derived analytically this branch in 1981 by using the $4\times4$ Luttinger Hamiltonian for the $\Gamma_8$ bands~\cite{Kane19} with an open boundary conditions. In 1985, existence of the localized states at the HgTe/CdTe interface was also predicted for the quantum wells~\cite{Kane20} and superlattices~\cite{Kane21}. Although the Luttinger Hamiltonian, used in Refs~\cite{Kane18,Kane20}, does not formally consider the inverted $|\Gamma_6,\pm1/2\rangle$ band, this Hamiltonian can be obtained from the $6\times6$ Kane Hamiltonian with the HgTe/CdTe interface by assuming $M_0^{(II)}\rightarrow-\infty$ and $M_0^{(I)}\rightarrow\infty$. Therefore, the upper ``massive'' domain-wall states in Fig.~\ref{Fig:Chapter2Fig5} and solution of Dyakonov and Khaetskii~\cite{Kane18}, obtained for particular case of $\theta=\pi/3$, have the same origin. As seen from Fig.~\ref{Fig:Chapter2Fig5}, such Dyakonov-Khaetskii branch is caused by the band inversion in the presence of hybridization with the flat bulk band.

When considering a bulk crystal, in which the flat band no longer coincides with the bottom of the conduction band $S_0^{(II)}\neq-M_0^{(II)}$, the physics becomes even richer. The case $S_0^{(II)}<-M_0^{(II)}$ corresponds to tensile biaxial strain, which opens a finite band-gap and drives the system into a 3D topological insulator state~\cite{Kane22,Kane22}. The opposite case, $S_0^{(II)}>-M_0^{(II)}$, describes either compressively strained HgTe films~\cite{Kane24} or unstrained Cd$_3$As$_2$ crystals~\cite{Kane9,Kane10}. The latter case, however, will not be discussed here; readers interested in this regime are referred to our original work~\cite{Kane4}. Figure~\ref{Fig:Chapter2Fig6} presents the calculated spectra of bulk and domain-wall-localized states for various values of the parameter $\theta$ within the interval $(0,\pi/2]$, using $S_0^{(I)}=-450$~meV and $S_0^{(II)}=110$~meV. One can clearly see that both the energy of the bulk states and the band-gap value between the flat and conduction bands now depend strongly on $\theta$, in sharp contrast to the unstrained case discussed earlier. The largest strain-induced gap appears in the absence of hybridization, while increasing $\theta$ progressively closes the band-gap. At $\theta=\pi/2$, the system becomes a semimetal featuring a circular nodal line at $k_z=0$ and $k=k_N$, where $k_N^2=2M_0^{(II)}(M_0^{(II)}+S_0^{(II)})/V^2$. Importantly, as illustrated in Fig.~\ref{Fig:Chapter2Fig6}, the domain-wall states in tensile-strained HgTe films remain qualitatively similar to those in the unstrained case (Fig.~\ref{Fig:Chapter2Fig5}). The main difference lies in the behavior of the Dyakonov-Khaetskii  branch, which now lies in the bulk band-gap for all $\theta$ values. This observation is fully consistent with the general topological argument that tensile-strained HgTe represents a 3D topological insulator with gapless surface states~\cite{Kane22}. However, these interface modes cannot be described as massless Dirac fermions, contrary to what is sometimes claimed in experimental works on bulk strained HgTe by CdTe substrates~\cite{Kane25,Kane26,Kane27}. Instead, Fig.~\ref{Fig:Chapter2Fig6} reveals that the surface states at the HgTe/CdTe interface of the strained 3D TI are ``massive'', arising from hybridization with the heavy-hole $|\Gamma_8,\pm3/2\rangle$ band. These hybridized modes correspond precisely to the Dyakonov-Khaetskii branch originally predicted in Ref.~\cite{Kane18}.

\begin{figure}
\centering
\includegraphics [width=1.0\columnwidth, keepaspectratio] {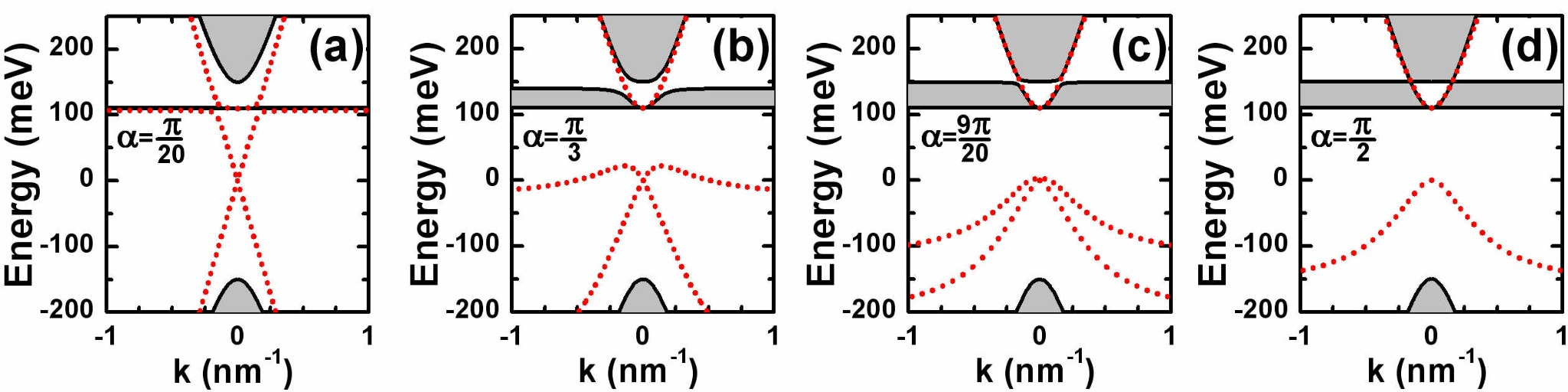}
\caption{\label{Fig:Chapter2Fig6} \textbf{(a-d)} Dispersion of the domain-wall-localised states (dotted red) as a function of $k$ for $S_0^{(II)}=110$~meV ($S_0^{(II)}<-M_0^{(II)}$) at different $\theta$ values: \textbf{(a)} $\theta=\pi/20$, \textbf{(b)} $\theta=\pi/3$, \textbf{(c)} $\theta=9\pi/20$, \textbf{(d)} $\theta=\pi/2$. The grey regions correspond to the bulk states of HgTe projected onto the heterointerface. The solid black curves represent the edges of the projected bulk bands. The bulk states in CdTe arising at $|E|\geq M_0^{(I)}$ are beyond the scale of the panels. The value of $\theta=\pi/3$ corresponds to the Kane fermions in tensile strained HgTe crystals.}
\end{figure}

In conclusion, the existence of domain-wall-localized solutions in $\mathcal{\hat{H}}_{6\times6}(\mathbf{k},\theta)$~(\ref{Keq:1}) provides a compelling argument that Kane fermions can indeed be regarded as pseudo-relativistic fermions. Nevertheless, their description in real materials such as HgCdTe and Cd$_3$As$_2$ is not without limitations. The inclusion of quadratic terms in the $6\times6$ Hamiltonian for $|\Gamma_6\rangle$ and $|\Gamma_8\rangle$ band, which are responsible for introducing curvature to the heavy-hole $|\Gamma_8,\pm3/2\rangle$ band, significantly alters the picture of the localized states. These additional terms couple the Dirac-like domain-wall branches with the bulk heavy-hole continuum, transforming the lower interface branches in Figs~\ref{Fig:Chapter2Fig5} and \ref{Fig:Chapter2Fig6} into the resonant states. In this case, the fermions spend only a finite time localized near the interface, with their lifetime decreasing as $k$ increases. Even so, such resonant states still contribute appreciably to the local density of states~\cite{Kane17} and, consequently, can be experimentally observed by angle-resolved photoemission spectroscopy (ARPES). On the other hand, introducing quadratic corrections formally breaks the pseudo-relativistic symmetry, which by its nature is linear in momentum. This raises an intriguing conceptual question: what exactly is a Kane fermion? Is it a hybrid entity combining the properties of pseudo-spin-1 and pseudo-spin-1/2 Dirac fermions~\cite{Kane2}, or is this interpretation merely a byproduct of an incomplete theoretical framework? Many of these questions will become clear after the next section, which offers a new and perhaps unexpected perspective on Kane fermions.

\section{\label{sec:LLKane}Relativistic collapse of Landau levels of Kane fermions}
Beyond the existence of domain-wall-localized states, there is another remarkable property of relativistic Dirac fermions that can be explored even in ``nonrelativistic'' condensed-matter systems. This phenomenon emerges in the presence of mutually perpendicular electric and magnetic fields and is known as the relativistic collapse of Landau quantization. It occurs when the carrier drift velocity $V_d = c\mathcal{E}/\mathcal{B}$, where $\mathcal{E}$ and $\mathcal{B}$ are the electric field and magnetic strengths, respectively, approaches the speed of light $c$. In semiconductor physics, this effect was first theoretically analyzed in 1966 by Aronov and Pikus~\cite{Flds3,Flds4}, and later -- independently -- by Zawadzki and Lax~\cite{Flds5,Flds6}, who applied a relativistic analogy to describe inter-Landau-level transitions within a two-band Hamiltonian in crossed $\mathcal{E}$ and $\mathcal{B}$ fields. In their pioneering works, Aronov and Pikus~\cite{Flds3,Flds4}  were the first who proposed to apply the Lorentz transformation with ``an effective speed of the light'' to eliminate the magnetic field when $\mathcal{E}>\mathcal{B}$ in the moving coordinate system, and the electric field when $\mathcal{E}<\mathcal{B}$. In the former case, the particle motion is infinite and no quantization takes place, while in the latter case the Landau quantization remains until its collapse, when $V_d$ reaches the ``effective speed of the light'' $\tilde{c}$ in the system. Physically, this ``collapse'' means that charge carriers no longer move along closed cyclotron orbits but instead follow unbounded trajectories, signaling the quantization breakdown. Forty years later, similar results were obtained for graphene~\cite{Flds7}. Note that the ``effective speed of the light'' in solids represents the maximum particle velocity in the system -- for instance, the Fermi velocity in graphene~\cite{Dirac1} or the one of Kane fermions (see Section~\ref{sec:TempKane}).

The fundamental difference between Kane and Dirac fermions lies in the mathematical structure of their description. While Dirac particles are represented within the framework of the SU(2) algebra corresponding to spin-1/2, the three-band model of Kane fermions, described by $\mathcal{\hat{H}}_{6\times6}(\mathbf{k},\theta)$ in Eq.~(\ref{Keq:1}) in Eq.~(\ref{Keq:1}), necessarily invokes the Lie algebra of the SU(3) group, inherent to particles with a spin $S=1$. However, in contrast to the Dirac fermions, the Lorentz transformation for particles with $S>1/2$ does not allow one to eliminate the electric field $\mathcal{E}$ by moving to a reference frame drifting with velocity $V_d$. This can be illustrated by explicitly attempting to perform a rotation of the spin matrices for $S=1$ by an imaginary angle. As a result, an analytic solution describing the relativistic collapse of Landau levels becomes fundamentally inaccessible for Kane fermions, within the linear-in-$k$ $\mathcal{\hat{H}}_{6\times6}(\mathbf{k},\theta)$. From the numerical computational point of view, the problem is no less challenging. A numerical approach that treats the electric-field term in the Hamiltonian (which may include the quadratic-in-$k$ terms) as a perturbation expanded over the Landau-level basis (the harmonic-oscillator wave-functions) faces an inevitable divergence. As the magnetic field $\mathcal{B}$ decreases or the electric field $\mathcal{E}$ increases, higher-order Landau levels must be included to preserve accuracy. This rapidly leads to the mathematical instability of the problem solution, since the number of coupled states grows without bound in the limit $\mathcal{B}\rightarrow\mathcal{E}$. Consequently, both analytical and numerical treatments of Landau quantization for Kane fermions in crossed electric and magnetic fields appear, at first sight, to be an intractable problem.

\subsection{\label{sec:Zawadzki}Approximate solution proposed by Wlodek Zawadzki}
The first attempt to solve the problem of particle motion in crossed fields within the three-band approximation was made by Zawadzki~\emph{et~al.}~\cite{Flds9,Flds10} in the context of InSb semiconductor ($M_0>0$). In his works, \emph{neglecting} the electric-field-induced inter-band terms arising for the squared Hamiltonian $\mathcal{\hat{H}}_{6\times6}(\mathbf{k},\theta=\pi/3)$, the Landau level energies for the $\Gamma_6$ conduction band were written as
\begin{equation}
\label{Keq:6}
E_{\Gamma_6}=C_0+\hbar{V_d}k_y+\sqrt{1-\delta^2}\left[M_0^2+V^2k_z^2+\sqrt{1-\delta^2}\dfrac{2V^{2}}{a_B^2}\left(n'+\frac{1}{2}+\frac{\sigma'}{4}\right)\right]^{1/2},
\end{equation}
where $n'\geq{0}$ is the Landau level index, $\sigma'={\pm}1$ for all values of $n'$. In order to obtain this expression in our notations from those of Ref.~\cite{Flds9}, one has to use the following expressions for the band gap $\epsilon_g$, the effective mass $m_0^{*}$ and the spin-splitting factor $g_0^{*}$ at the band edge~\cite{Flds10}:
\begin{equation}
\label{Keq:7}
\epsilon_g=2M_0,~~~~~m_0^{*}=\dfrac{\hbar^{2}M_0}{V^2},~~~~~g_0^{*}=-\dfrac{m_0}{m_0^{*}},
\end{equation}
where $m_0$ is the free-electron mass. The most important quantity in Eq.~(\ref{Keq:6}) is $\delta$ defined as $\delta={\hbar}V_d/V=(c\hbar\mathcal{E})/(V\mathcal{B})$. In deriving Eq.~(\ref{Keq:6}) on the basis of $\mathcal{\hat{H}}_{6\times6}(\mathbf{k},\theta=\pi/3)$, it was assumed that the electric field $\mathcal{E}$ is oriented along the $x$ axis. As seen, the results of Zawadzki~\emph{et~al.}~\cite{Flds9,Flds10} are very similar to the case of Dirac fermions~\cite{Flds3,Flds4,Flds5,Flds6,Flds7} -- the Landau levels collapses when $V_d$ approaches the ``effective speed of the light'', $\tilde{c}=V/\hbar$. By redefining $n'$ and $\sigma'$, $E_{\Gamma_6}$ in Eq.~(\ref{Keq:6}) in the absence of electric field is reduced to $E^{(\sigma)}_{+}$ in Eq.~(\ref{Keq:5}) at $\theta=\pi/3$:
\begin{eqnarray*}
E^{(\sigma)}_{\pm}=C_0\pm\sqrt{M_0^2+V^2k_z^2+\dfrac{2V^{2}}{a_B^2}(n+1+\sigma\sin^2\theta)},\nonumber\\
E_0=C_0-M_0,~~~~~~~~~~~~~~~~~~~~~~
\end{eqnarray*}

Let us now show that the assumptions made by Zawadzki~\emph{et~al.} in the derivation of Eq.~(\ref{Keq:6}) are equivalent to the semiclassical quantization rule for $E_{+}(\mathbf{k})$ in Eq.~(\ref{Keq:4}). Indeed, according to the generalization of Lifshitz-Onzager quantization rule made by Lifshitz and Kaganov~\cite{Flds11}, semiclassical Landau levels in crossed electric and magnetic fields can be found from the relation:
\begin{equation}
\label{Keq:8}
\mathcal{S}(\varepsilon^{*},k_z)=\dfrac{2\pi}{a_B^2}(n+\gamma),
\end{equation}
where $\gamma$ is constant and $\mathcal{S}(\varepsilon^{*},k_z)$ is the cross-section  of the surface $E(\mathbf{k})-{\hbar}V_{d}k_y=\varepsilon^{*}$ in the momentum space. This generalization is due to the fact that in the crossed fields, instead of the energy, the quantity $E(\mathbf{k})-{\hbar}V_{d}k_y$ is conserved. The latter is nothing but an ellipse in the momentum space
\begin{multline}
\label{Keq:9}
V^2k_x^2+\left(V^2-\hbar^2V_d^2\right)\left[k_y-\dfrac{{\hbar}V_d}{V^2-\hbar^2V_d^2}\left(\varepsilon^{*}-C_0\right)\right]^2=\\
=\dfrac{V^2}{V^2-\hbar^2V_d^2}\left(\varepsilon^{*}-C_0\right)^2-M^2-V^2k_z^2.
\end{multline}

By calculation the ellipse cross-section, the energies of the semiclassical Landau levels are written as
\begin{equation}
\label{Keq:10}
E_{\pm}=C_0+\hbar{V_d}k_y\pm\sqrt{1-\delta^2}\biggl[M_0^2+V^2k_z^2
+\sqrt{1-\delta^2}\dfrac{2V^{2}}{a_B^2}\left(n+\gamma\right)\biggr]^{1/2},
\end{equation}
where $\gamma$ can be defined from the comparison of Eq.~(\ref{Keq:10}) at $\delta=0$ with Eq.~(\ref{Keq:5}). The latter makes $E_{+}$ to be equivalent to $E_{\Gamma_6}$ at $\theta=\pi/3$ in Eq.~(\ref{Keq:6}).

The main problem with the Zawadzki's approximate solution in crossed electric and magnetic fields appears when the inter-band coupling cannot be neglected. For the $\Gamma_{6}$ band, this arises when $M_0^2$ becomes comparable with $\beta{\hbar}^2V^2\cos\theta/a_B^2$ and $\beta{\hbar}^2V^2\sin\theta/a_B^2$, which represent the coupling strength with the light- and heavy-hole bands, respectively. On the contrary, for the $\Gamma_{8}$ band, the inter-band coupling is always relevant due to the proximity of the light- and heavy-hole bands. Thus, the impossibility of neglecting the inter-band coupling in materials such as HgCdTe makes it impossible to clearly understand the behavior of Landau levels at relatively high electric fields.

\subsection{\label{sec:novelKane}A novel look at Kane Fermions}
To accurately solve the problem of Landau levels in crossed electric and magnetic fields, it is necessary to look at Kane fermions in a different way -- beyond the three-band approximation. In our original work~\cite{Flds8}, we propose to consider an additional conduction band, whose $\mathbf{k}\cdot\mathbf{p}$ interaction induces the non-zero curvature of the heavy-hole band. For the zinc-blende materials, this is the second conduction $|\Gamma_{7c},{\pm}1/2\rangle$ band~\cite{Flds12,Flds13,Flds14}. Further, we briefly outline the main consequences of taking into account the $|\Gamma_{7c},{\pm}1/2\rangle$ band, while a more rigorous derivation of their proper (3+1)-band Hamiltonian from the 14$\times$14 \textbf{k$\cdot$p} model~\cite{Flds12,Flds13,Flds14} of cubic semiconductors can be found in Ref.~\cite{Flds8}. The first point is that the  $|\Gamma_{6c},{\pm}1/2\rangle$--$|\Gamma_{8v},{\pm}1/2\rangle$ and $|\Gamma_{7c},{\pm}1/2\rangle$--$|\Gamma_{8v},{\pm}3/2\rangle$ band pairs represents two Dirac subsystems (see Fig.~\ref{Fig:Chapter2Fig7}) each described by \emph{anisotropic} relativistic 3D Dirac Hamiltonian:
\begin{equation}
\label{L1eq:1}
\mathcal{\hat{H}}^{(1,2)}_{D}(\mathbf{k})=
C_{1,2}\mathcal{I}_4+M_{1,2}\alpha_0+V\cos{\theta}\left(k_{x}\alpha_{x}+k_{y}\alpha_{y}\right)+
Vk_z\alpha_z,
\end{equation}
where $\mathcal{I}_4$ is a $4\times4$ identity matrix and the $\alpha$ matrices are related to the Pauli matrices as
\begin{equation}
\label{L1eq:2}
\alpha_0=\begin{pmatrix}
\sigma_0 &0 \\
0 & -\sigma_0
\end{pmatrix},~~~~~
\alpha_{x,y,z}=\begin{pmatrix}
0 & \sigma_{x,y,z} \\
\sigma_{x,y,z} & 0
\end{pmatrix}.
\end{equation}
Here, the forth Pauli matrix $\sigma_0$ is defined as a $2\times2$ identity matrix. In Eq.~(\ref{L1eq:1}), the constants $C_{1,2}$ and mass parameters $M_{1,2}$ are defined by the energies of $|\Gamma_{6c},{\pm}1/2\rangle$, $|\Gamma_{8v},{\pm}1/2\rangle$, $|\Gamma_{7c},{\pm}1/2\rangle$ and $|\Gamma_{8v},{\pm}3/2\rangle$ bands:
\begin{eqnarray}
\label{L1eq:3}
C_1+M_1=E_{6c},~~~~~~~~~~~C_1-M_1=E_{8v,\pm{1/2}},\nonumber\\
C_2+M_2=E_{7c},~~~~~~~~~~~C_2-M_2=E_{8v,\pm{3/2}}.
\end{eqnarray}
The last expression determines the position of the heavy-hole band at $\mathbf{k}=0$. For the case of uniaxially strained HgCdTe or Cd$_3$As$_2$, $C_2-M_2=E_{8v,\pm{3/2}}+\delta_{\epsilon}$, where $\delta_{\epsilon}$ represents the gap between the light- and heavy-hole subbands.

\begin{figure}
\includegraphics [width=1.0\columnwidth, keepaspectratio] {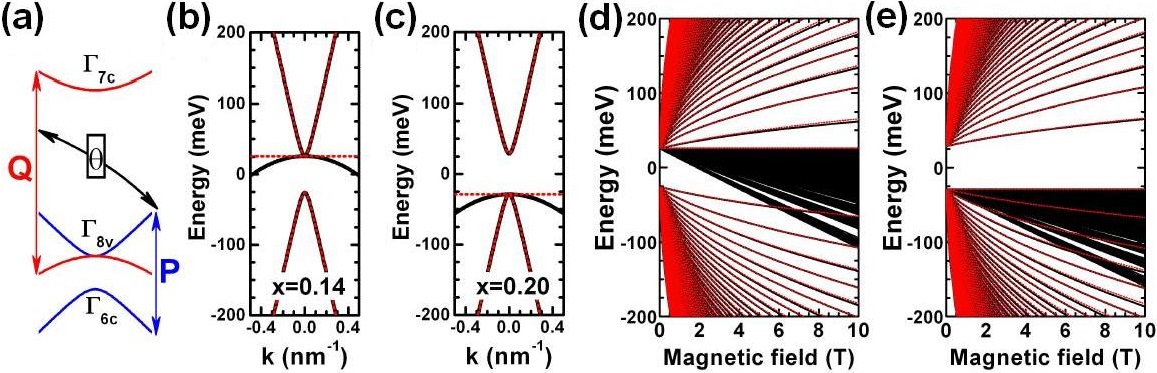}
\caption{\label{Fig:Chapter2Fig7}  \textbf{(a)} Schematic representation of the (3+1)-band model involving $\Gamma_{8v}$, $\Gamma_{6c}$ and $\Gamma_{7c}$ bands as two coupled Dirac systems described by the ``hybridization angle'' (for zinc-blende crystal, $\theta=\pi/3$). The ``effective speed of light'' in each system is determined by the corresponding momentum matrix element $P$ or $Q$. As shown in Ref.~\cite{Flds8}, for narrow-gap HgCdTe crystals, one can assume $V=\sqrt{2/3}P=\sqrt{2/3}Q$. \textbf{(b,c)} Band dispersion of Hg$_{1-x}$Cd$_{x}$Te bulk crystals in the three-band (in dotted red, Eq.~(\ref{Keq:4})) and (3+1)-band (in solid black) models. For the latter case, the second conduction $\Gamma_{7c}$ band is high in energy lying beyond the figures' scale, and $\mathbf{k}$ is oriented along [100] crystallographic direction. \textbf{(d,e)} Landau levels in the absence of electric field in (d) Hg$_{0.14}$Cd$_{0.86}$Te and (e) Hg$_{0.20}$Cd$_{0.80}$Te bulk crystals with inverted and non-inverted band structures, respectively (see (b) and (c) panels). The solid black curves are the calculations within the (3+1)-band model, while the dotted red curves represent the calculations based on Eq.~(\ref{Keq:5}). For both cases, $\mathcal{B}\parallel[001]$. The second conduction $\Gamma_{7c}$ band is high in energy lying beyond the figure scale.}
\end{figure}

The second point is that the two Dirac subsystems are coupled because of the inevitable hybridization between $|\Gamma_{8v},{\pm}1/2\rangle$ and $|\Gamma_{8v},{\pm}3/2\rangle$ bands in cubic semiconductors. As a result, the generalized (3+1)-band Hamiltonian can be represented as
\begin{equation}
\label{L1eq:4}
\mathcal{\hat{H}}^{(\mathbf{3+1})}_{8\times8}(\mathbf{k})
=\begin{pmatrix}
\mathcal{\hat{H}}^{(1)}_{D}(\mathbf{k}) & V\sin{\theta}\left(k_{x}\beta_{y}-k_{y}\beta_{x}\right) \\
V\sin{\theta}\left(k_{y}\beta_{x}-k_{x}\beta_{y}\right) & \mathcal{\hat{H}}^{(2)}_{D}(\mathbf{k})
\end{pmatrix},
\end{equation}
where we have introduced the $\beta$ matrices as
\begin{equation}
\label{L1eq:5}
\beta_0^{(z)}=\begin{pmatrix}
i\sigma_z &0 \\
0 & -i\sigma_z
\end{pmatrix},~~~~~
\beta_{x,y}=\begin{pmatrix}
0 & \sigma_{y,x} \\
-\sigma_{y,x} & 0
\end{pmatrix}.
\end{equation}
Note the indices order in the definition of $\beta_{x}$ and $\beta_{y}$. Importantly, the Cartesian axes $x$, $y$, $z$ in the above equations are oriented along [100], [010] and [001], respectively. For other orientations of the Cartesian axes, the form of $\mathcal{\hat{H}}^{(\mathbf{3+1})}_{8\times8}(\mathbf{k})$ differs from Eq.~(\ref{L1eq:4})~\cite{Flds8}.

The most important property of the (3+1)-band Hamiltonian is that it considers the Kane fermion as a superposition of two Dirac particles with different rest-masses hybridized in a special way. This situation is partially reminiscent of the case of relativistic Dirac fermion itself, which is also a superposition of two Weyl particles mutually hybridized  due to a finite rest-mass. Thus, the structure of the (3+1)-band Hamiltonian proves the origin of Kane fermions as spin-1/2 particles, which was previously impossible to see due to the truncated treatment within the three-band Hamiltonian $\mathcal{\hat{H}}_{6\times6}(\mathbf{k},\theta)$ in Eq.~(\ref{Keq:1}). In its turn, $\mathcal{\hat{H}}_{6\times6}(\mathbf{k},\theta)$ can be directly obtained from Eq.~(\ref{L1eq:4}) by projecting $\mathcal{\hat{H}}^{(\mathbf{3+1})}_{8\times8}(\mathbf{k})$ on the $\Gamma_{6c}+\Gamma_{8v}$ subspace in the limit of $E_{7c}-E_{8v,\pm{3/2}}\rightarrow\infty$.

Comparison of the energy dispersion for Hg$_{0.14}$Cd$_{0.86}$Te and Hg$_{0.20}$Cd$_{0.90}$Te bulk crystals ($C_1=0$, $\theta=\pi/3$) calculated within $\mathcal{\hat{H}}_{6\times6}(\mathbf{k},\theta)$ and $\mathcal{\hat{H}}^{(\mathbf{3+1})}_{8\times8}(\mathbf{k})$ is provided in Fig.~\ref{Fig:Chapter2Fig7}(b,c). The parameters used in the calculations for these two cases can be found in our original work~\cite{Flds8}. Figure~\ref{Fig:Chapter2Fig7}(c,d) provides the energies of Landau levels for HgCdTe bulk crystals with inverted and non-inverted band structures in the absence of electric field when magnetic field $\mathcal{B}$ is oriented along [001] crystallographic axis. It is seen that although the Landau levels for the light-hole and $\Gamma_{6c}$ bands are in good agreement with Eq.~(\ref{Keq:5}), the non-zero curvature of the heavy-hole band in the absence of rotational symmetry in the plane perpendicular to the magnetic field leads to a non-equidistant spectrum of the Landau levels.

The fact that the Kane fermion consists of two Dirac fermions allows us to solve exactly the problem of finding the Landau levels in crossed and magnetic fields using the Lorentz transformation (Lorentz boost) for spin-1/2 particles. The resulting transformation applied to the (3+1)-band Hamiltonian is mathematically involved, requiring careful consideration of the coupling between different blocks. The detailed derivation of this procedure is presented in our original publication~\cite{Flds8}, and therefore only the final results will be outlined here. A crucial feature revealed by this analysis is the strong anisotropy induced by the cubic symmetry of $\mathcal{\hat{H}}^{(\mathbf{3+1})}_{8\times8}(\mathbf{k})$. Unlike the isotropic Dirac case in graphene, where the collapse of Landau levels occurs when the drift velocity  $V_d$ approaches a single critical value corresponding to the Fermi velocity~\cite{Flds7}, in the Kane-fermion-systems this threshold depends sensitively on the relative orientation of the electric and magnetic fields with respect to the principal crystallographic axes. As demonstrated in Ref.~\cite{Flds8}, for the configuration $\mathcal{B}\parallel[001]$ and $\mathcal{E}\parallel[100]$, the collapse occurs when the drift velocity reaches $V_d=V\cos{\theta}/{\hbar}$. In contrast, for the orthogonal configuration $\mathcal{B}\parallel[010]$ and $\mathcal{E}\parallel[100]$, the critical velocity equals $V_d=V/{\hbar}$. The first configuration thus corresponds to a collapse occurring at precisely one-half of the ``effective speed of light'' $\tilde{c}=V/{\hbar}$, since $\theta=\pi/3$ for the Kane fermions. For arbitrary orientations of the magnetic field, while maintaining the condition $\mathcal{E}\perp\mathcal{B}$, the critical drift velocity $V_d^{*}$ was found to take intermediate values between one-half and the full $\tilde{c}$~\cite{Flds8}.

Another fundamental property of Kane fermions, which manifests itself in the presence of crossed electric and magnetic fields, can likewise be derived from the analysis of the Lorentz boost applied to $\mathcal{\hat{H}}^{(\mathbf{3+1})}_{8\times8}(\mathbf{k})$~\cite{Flds8}. In particular, this analysis reveals that the degree of hybridization between the two Dirac particles forming the Kane fermion is governed by a factor $\exp(-\xi_0^2)$, whose magnitude for the case of $\mathcal{B}\parallel[001]$ and $\mathcal{E}\parallel[100]$ is defined as follows~\cite{Flds8}:
\begin{equation}
\label{L3eq:0a}
e^{-\xi_0^2}=
\exp\left[{-\dfrac{\beta^2{a_B^2}(C_2-C_1)^2}{4V^2\cos^2{\theta}(1-\beta^2)^{3/2}}}\right],
\end{equation}
where $\beta={\hbar}V_d/(V\cos{\theta})=(c{\hbar}\mathcal{E})/(\mathcal{B}V\cos{\theta})$. Since $C_1{\neq}C_2$, which is always the case of HgCdTe and Cd$_{3}$As$_{2}$ crystals, Eq.~(\ref{L3eq:0a}) makes it possible to immediately understand the behavior of Landau levels in strong electric fields. Indeed, since the Gaussian function $e^{-\xi_0^2}$ falls off very quickly, the coupling between two Dirac particles becomes negligibly small long before $\beta$ reaches $1$. In this case, the Landau level energies can be found analytically~\cite{Flds8}:
\begin{equation}
\label{L3eq:1}
E_{\mathrm{Dirac}}^{[001]}=C_{1,2}-{\beta}Vk_{y}\cos{\theta}\pm\sqrt{1-\beta^2}\sqrt{M_{1,2}^2+V^{2}k_{z}^{2}+\sqrt{1-\beta^2}\dfrac{2V^{2}(n+1)\cos^2{\theta}}{a_B^2}},
\end{equation}
where $n\geq-1$. Thus, Kane fermion in crossed electric and magnetic fields decays into two independent Dirac fermions by increasing the electric field.

\begin{figure}
\centering
\includegraphics [width=0.6\columnwidth, keepaspectratio] {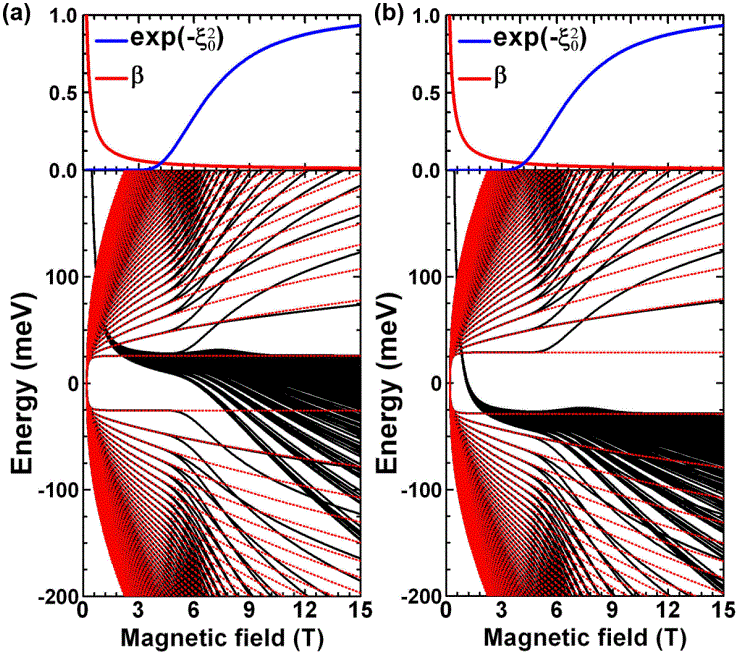}
\caption{\label{Fig:Chapter2Fig8} \textbf{(a,b)} Landau levels at $\mathcal{E}=1000$~V/cm and $k_y=k_z=0$ in \textbf{(a)} Hg$_{0.14}$Cd$_{0.86}$Te and \textbf{(b)} Hg$_{0.20}$Cd$_{0.80}$Te bulk crystals for $\mathcal{B}\parallel[001]$ and $\mathcal{E}\parallel[100]$. The solid black curves are the calculations within the (3+1)-band model, while the dotted red curves represent the calculations based on Eq.~(\ref{L3eq:1}). Here, the critical magnetic field $B_c=(c\hbar\mathcal{E})/(V\cos{\theta})$, at which Landau levels of Kane fermions ($\theta=\pi/3$) collapse, approximately equals to $0.19$~T. The upper panels represent $\beta$ and  $e^{-\xi_0^2}$ (see Eq.~(\ref{L3eq:0a})) as functions of magnetic field.}
\end{figure}

Figure~\ref{Fig:Chapter2Fig8} provides the Landau levels calculations in crossed electric and magnetic fields in HgCdTe bulk crystals with inverted and non-inverted band structure for $\mathcal{B}\parallel[001]$ and $\mathcal{E}\parallel[100]$.  In small magnetic fields exceeding $B_c=(c\hbar\mathcal{E})/(V\cos{\theta})$ at which $\beta=1$, the Landau levels energies are reproduced by the picture involving two independent Dirac fermions in accordance with the conclusion made above. Above 3~T, the Dirac fermion regime, characterized by double-degenerated Landau levels (except the one at $n=-1$ in Eq~(\ref{L3eq:1})) progressively transforms into the Kane fermion picture characterized by strong hybridization between light- and heavy-hole bands. In high magnetic fields, corresponding to the small $\beta$ values, the Landau level energies asymptotically tend to the ones in the absence of electric fields. A qualitatively similar behavior of Landau levels is observed for other orientations of the electric and magnetic fields~\cite{Flds8}. Thus, we can conclude that the Kane fermion in crossed fields always decays into two independent Dirac particles, which is represented by a vanishing of the mixing between the light and heavy hole bands as the magnetic field decreases. The magnetic field range of such Dirac fermion regime is defined by the difference $C_2-C_1$ and the orientation of electric and magnetic fields with respect to the main crystallographic axes of the zinc-blende crystal~\cite{Flds8}. We also note that the conclusions above holds in the presence of a gap between the light- and heavy-hole subbands, which includes the uniaxially strained HgCdTe and Cd$_3$As$_2$ bulk crystals. As seen from above, the decay of Kane fermion is absent if the (3+1)-band Hamiltonian preserves the particle-hole symmetry, i.e. at $C_1=C_2$.

Importantly, since the Kane fermion decay into two Dirac particles in the (3+1)-band model occurs at any values of $M_2$ and $0<\theta<\pi/2$, one can also conclude that the flat band described by $\mathcal{\hat{H}}_{6\times6}(\mathbf{k},\theta)$ in Eq.~(\ref{Keq:1}) in the limiting case of $M_2\rightarrow{\infty}$ becomes fully decoupled from the conical bands when magnetic field approaches its critical value. This fact in a sense justifies the approximations made by Wlodek Zawadzki within the three-band model (see Sec.~\ref{sec:Zawadzki}), which ignores the inter-band coupling. Although in this case the critical magnetic field does not depend on the orientation of $\mathcal{B}$, this approach can nevertheless qualitatively describe the evolution of Landau levels if $\mathcal{E}$ is used as a fitting parameter. This particularly explains a good agreement between the Landau level transition energies calculated on the basis of Eq.~(\ref{Keq:6}) and experimental results on magneto-absorption of InSb observed by Zawadzki~\emph{et~al.}~\cite{Flds9}.

In conclusion, it is worth emphasizing that our interpretation of Kane fermions as a superposition of two Dirac particles provides a qualitatively richer and more versatile framework for describing a broad range of physical phenomena observed in real narrow-gap and gapless semiconductors than the simplified model with a flat band. In particular, by treating the conduction $\Gamma_{7c}$ band perturbatively, the (3+1)-band Hamiltonian can be reduced explicitly to the so-called three-band Kane Hamiltonian for the $\Gamma_{6c}$ and $\Gamma_{8v}$ bands~\cite{Flds8}. This resulting Hamiltonian naturally includes all quadratic-in-momentum terms that account for the cubic symmetry of semiconductors and ensures a realistic description of their electronic properties. It is precisely this form of the three-band Kane Hamiltonian that often serves as the minimal yet comprehensive starting point for modeling the majority of phenomena encountered in the heterostructures based on narrow-gap and gapless semiconductors -- some of which will be discussed in Chapters~\ref{Chap:HgTe} and \ref{Chap:III-V} of this m\'{e}moire.

\section{\label{sec:DisorderKane}Disorder-induced topological phase transition for Kane femions}
As demonstrated in Section~\ref{sec:TempKane}, the band structure of bulk HgCdTe films described by Kane fermions can be efficiently tuned by varying temperature. However, such a mechanism of band ordering is a specific property of HgCdTe, unavailable in other narrow-gap semiconductors such as InAs, InSb, and InAsSb. In the present section, we discuss a more general and universal mechanism that leads to the inversion of the Kane-fermion mass parameter, corresponding to a topological phase transition. This mechanism, originating from the influence of disorder, is inherent to all narrow-gap and gapless semiconductors whose band structures are governed by Kane fermions.

The crucial role of disorder in driving topological phase transitions was first identified in 2D systems hosting the quantum spin Hall insulator (QSHI) state, such as HgTe/CdHgTe QWs~\cite{TAI1}. In their seminal work, by means of tight-binding calculations, Li~\emph{et~al.}~\cite{TAI1} demonstrated that disorder can induce a topological phase transition in QWs that are initially in the trivial insulating phase. Subsequently, Groth~\emph{et~al.}~\cite{TAI2} showed that this disorder-induced topological transition in HgTe/CdHgTe QWs originates from the quadratic terms $\propto k^2\sigma_z$ in the 2D Bernevig–Hughes–Zhang (BHZ) Hamiltonian~\cite{HgTeQWs2}. The latter can be viewed as a 2D Dirac Hamiltonian supplemented by quadratic terms necessary for the correct topological classification of the system~\cite{HgTeQWs3}. Following Li~\emph{et~al.}~\cite{TAI1}, this disorder-driven transition toward a topological insulator state has often been referred to as the ``topological Anderson insulator''. Yet, as emphasized by Groth~\emph{et~al.}~\cite{TAI2}, such a weak-disorder topological transition is not an Anderson transition at all, but rather a continuous modification of the effective band parameters that can be described within the framework of the self-consistent Born approximation (SCBA). Later studies extended these ideas to 3D topological insulators~\cite{TAI3,TAI4,TAI5} and to Dirac and Weyl semimetals~\cite{TAI6,TAI7,TAI8,TAI9}, revealing that disorder may fundamentally alter the topology of the band structure even in materials without an intrinsic gap. This section summarizes the main results of our original work devoted to the disorder-induced topological phase transition in 3D systems with Kane fermions~\cite{TAI10}.

For a general description of the band structure in narrow-gap and gapless semiconductors, we employed the three-band Kane Hamiltonian for the $\Gamma_{6c}$ and $\Gamma_{8v}$ bands, including quadratic terms that account for the mixing between light- and heavy-hole states within the Luttinger block of the Hamiltonian. As mentioned at the end of the previous section, this model can be rigorously derived from the (3+1)-band Hamiltonian by treating the second conduction band within the framework of perturbation theory~\cite{Flds8}. The only approximation made in Ref.~\cite{TAI10} was the neglect of the terms that break spherical symmetry. While the inclusion of such terms describing cubic anisotropy would make the theoretical analysis of disorder significantly more cumbersome, it does not qualitatively affect the main physical conclusions. It is worth emphasizing that the three-band Kane Hamiltonian in the spherical approximation has long been recognized as an effective and reliable model for describing a wide variety of phenomena in narrow-gap and gapless semiconductors, as well as in their heterostructures. Particularly, the spherical approximation allows one to use a series of unitary transformations that bring the three-band Kane Hamiltonian to the form independent of the momentum orientation $\mathbf{k}$:
\begin{multline}
\label{eq:TAI8}
\tilde{H}_{3D}(k)=CI_{6{\times}6}+MI_{M}+\\
+\begin{pmatrix}
-(D+B)k^2 & 0 & 0 & Ak & 0 & 0 \\
0 & -(D+B)k^2 & 0 & 0 & Ak & 0 \\
0 & 0 & -\dfrac{\hbar^2k^2}{2m_{hh}} & 0 & 0 & 0 \\
Ak & 0 & 0 & -(D-B)k^2 & 0 & 0 \\
0 & Ak & 0 & 0 & -(D-B)k^2 & 0 \\
0 & 0 & 0 & 0 & 0 & -\dfrac{\hbar^2k^2}{2m_{hh}}
\end{pmatrix},
\end{multline}
where $k=|\mathbf{k}|$, $I_{6{\times}6}$ is the $6{\times}6$, $I_M=\mathrm{diag}\{1,1,-1,-1,-1,-1\}$ is a diagonal matrix; other parameters are defined as $A=\sqrt{6}P/3$, $C=(E_c+E_v)/2$, $M=(E_c-E_v)/2$,
\begin{eqnarray}
\label{eq:TAI9}
D=\dfrac{\hbar^2}{2m_0}\dfrac{\gamma_{1}+2\gamma_{2}-2F-1}{2},~~\notag\\
B=-\dfrac{\hbar^2}{2m_0}\dfrac{\gamma_{1}+2\gamma_{2}+2F+1}{2},\notag\\
\dfrac{1}{m_{hh}}=\dfrac{1}{m_0}\left(\gamma_{1}-2\gamma_{2}\right).~~~~~~
\end{eqnarray}
The band parameters used in $\tilde{H}_{3D}(k)$ for Hg$_{1-x}$Cd$_{x}$Te crystals at different $x$ can be found in Ref.~\cite{TAI10}. The Hamiltonian $\tilde{H}_{3D}(k)$ has three doubly-degenerate eigenvalues, corresponding to the energy dispersion of the conduction (c), light-hole (lh) and heavy-hole bands:
\begin{eqnarray}
\label{eq:TAI10}
E_{c,lh}=C-Dk^2
\pm\sqrt{\left(M-Bk^2\right)^2+A^2k^2},\notag\\
E_{hh}=C-M-\dfrac{\hbar^2k^2}{2m_{hh}}.~~~~~~~~~~~~~
\end{eqnarray}
The mass parameter $M$ in Eq.~(\ref{eq:TAI8}) describes the band inversion: $M>0$ corresponds to a trivial semiconductor, while $M<0$ represents a topological semimetal (see Fig.~\ref{Fig:Chapter2Fig1}).

To provide a quantitative description of the disorder-induced renormalization of the mass parameter $M$, it is convenient to evaluate the density of states (DOS) of the system. As a source of disorder, we consider two types of random potentials: the impurity potential $V_{imp}(\textbf{r})I_{6{\times}6}$ and the ``band-gap fluctuation'' potential $U_M(\textbf{r})I_{M}$ to $H_{3D}(\mathbf{k})$, both assumed to be Gaussian distributed with zero mean values. Their spatial correlations are defined as follows:
\begin{eqnarray}
\label{eq:TAI11n}
\langle{V_{imp}(\textbf{r})}\rangle=\langle{U_M(\textbf{r})}\rangle=0,~~~~~\notag\\
\langle{V_{imp}(\textbf{r})V_{imp}(\textbf{r}')}\rangle=v^2g\left(\left|\textbf{r}-\textbf{r}'\right|\right),\notag\\
\langle{U_M(\textbf{r})U_M(\textbf{r}')}\rangle=u^2\xi\left(\left|\textbf{r}-\textbf{r}'\right|\right),\notag\\
\langle{V_{imp}(\textbf{r})U_M(\textbf{r}')}\rangle=0,~~~~~~~
\end{eqnarray}
where $\langle...\rangle$ represents the average over all realizations of the random potentials; $g(r)$ are $\xi(r)$ are the normalized correlation functions~\cite{TAI11,TAI12}, while $v$ and $u$ represent the corresponding disorder strength. The last expression in Eq.~(\ref{eq:TAI11n}) means that the random potentials $V_{imp}(\textbf{r})$ and $U_M(\textbf{r})$ are supposed to be independent.

Due to the full rotational symmetry inherent to the spherical approximation, the disorder-averaged Green's function can be presented as~\cite{TAI10}:
\begin{equation}
\label{eq:TAI15}
\hat{G}(k,\varepsilon)=\left[\varepsilon-\tilde{H}_{3D}(k)-\hat{\Sigma}(k,\varepsilon)\right]^{-1},
\end{equation}
which $\varepsilon$ is the energy, $\hat{\Sigma}(k,\varepsilon)$ is the self-energy matrix, which has the following form~\cite{TAI10}:
\begin{equation}
\label{eq:TAI17}
\hat{\Sigma}(k,\varepsilon)=\begin{pmatrix}
\Sigma_c & 0 & 0 & \Sigma_{p} & 0 & 0 \\
0 & \Sigma_c & 0 & 0 & \Sigma_{p} & 0 \\
0 & 0 & \Sigma_{hh} & 0 & 0 & 0 \\
\Sigma_{p} & 0 & 0 & \Sigma_{lh} & 0 & 0 \\
0 & \Sigma_{p} & 0 & 0 & \Sigma_{lh} & 0 \\
0 & 0 & 0 & 0 & 0 & \Sigma_{hh}
\end{pmatrix},
\end{equation}
where $\Sigma_c$, $\Sigma_{lh}$, $\Sigma_{hh}$ and $\Sigma_{p}$ are functions of $k$ and $\varepsilon$.

After some calculations, which can be found in our original work~\cite{TAI10}, the self-energy matrix $\hat{\Sigma}(k,\varepsilon)$ within the SCBA for correlated and \emph{mutually independent} disorder potentials is written as
\begin{equation*}
\hat{\Sigma}(k,\varepsilon)=\int\limits_0^{K_c}\dfrac{{k^\prime}^{2}dk^\prime}{2\pi}
\int\limits_0^{\pi}\dfrac{\sin{\theta}d\theta}{2\pi}
W_{-}(k,k',\theta)G_{p}^\prime\cos{\theta}\begin{pmatrix}
0 & 0 & 0 & 1 & 0 & 0 \\
0 & 0 & 0 & 0 & 1 & 0 \\
0 & 0 & 0 & 0 & 0 & 0 \\
1 & 0 & 0 & 0 & 0 & 0 \\
0 & 1 & 0 & 0 & 0 & 0 \\
0 & 0 & 0 & 0 & 0 & 0
\end{pmatrix}+~~~~~~~~~~~
\end{equation*}
\begin{multline}
\label{eq:TAI18}
~~~~~~~~~~~+\int\limits_0^{K_c}\dfrac{{k^\prime}^{2}dk^\prime}{2\pi}
\int\limits_0^{\pi}\dfrac{\sin{\theta}d\theta}{2\pi}W_{+}(k,k',\theta)
\begin{pmatrix}
G_{c}^\prime & 0 & 0 & 0 & 0 & 0 \\
0 & G_{c}^\prime & 0 & 0 & 0 & 0 \\
0 & 0 & G_{hh}^\prime & 0 & 0 & 0 \\
0 & 0 & 0 & G_{lh}^\prime & 0 & 0 \\
0 & 0 & 0 & 0 & G_{lh}^\prime & 0 \\
0 & 0 & 0 & 0 & 0 & G_{hh}^\prime
\end{pmatrix}-\\
-\dfrac{3}{4}
\int\limits_0^{K_c}\dfrac{{k^\prime}^{2}dk^\prime}{2\pi}
\int\limits_0^{\pi}\dfrac{\sin^{3}{\theta}d\theta}{2\pi}W_{+}(k,k',\theta)
\left(G_{hh}^\prime-G_{lh}^\prime\right)\begin{pmatrix}
0 & 0 & 0 & 0 & 0 & 0 \\
0 & 0 & 0 & 0 & 0 & 0 \\
0 & 0 & 1 & 0 & 0 & 0 \\
0 & 0 & 0 & -1 & 0 & 0 \\
0 & 0 & 0 & 0 & -1 & 0 \\
0 & 0 & 0 & 0 & 0 & 1
\end{pmatrix},
\end{multline}
where $G_{n}^\prime\equiv G_{n}(k^\prime,\varepsilon)$ are the component of the Green's function $\hat{G}(k,\varepsilon)$ having the same form as $\hat{\Sigma}(k,\varepsilon)$ in Eq.~(\ref{eq:TAI17}), while $W_{\pm}(k,k',\theta)$ are defined as
\begin{equation}
\label{eq:TAI18n}
W_{\pm}(k,k',\theta)=v^2\tilde{g}\left(\sqrt{k^2+{k^\prime}^{2}-2{k}{k^\prime}\cos{\theta}}\right)
{\pm}u^2\tilde{\xi}\left(\sqrt{k^2+{k^\prime}^{2}-2{k}{k^\prime}\cos{\theta}}\right),
\end{equation}
where $\tilde{g}(k)$ and $\tilde{\xi}(k)$ are the Fourier transform of the disorder correlation functions defined by Eq.~(\ref{eq:TAI11n}):
\begin{equation*}
g(r)=\int\dfrac{d^3\textbf{q}}{(2\pi)^3}e^{i\textbf{q}\cdot\textbf{r}}\tilde{g}(q),~~~~
\xi(r)=\int\dfrac{d^3\textbf{q}}{(2\pi)^3}e^{i\textbf{q}\cdot\textbf{r}}\tilde{\xi}(q).
\end{equation*}
In Eq.~(\ref{eq:TAI18}), we also introduce a cut-off wave-vector $K_c=\pi/a_{\mathrm{0}}$ (where $a_{\mathrm{0}}$ is the lattice constant of bulk semiconductor), corresponding to the size of the Brillouin zone.

To advance the analysis, we assume $\tilde{g}(q)=\tilde{\xi}(q)=1$, which corresponds to uncorrelated disorder potentials $U_M(\textbf{r})$ and $V_{imp}(\textbf{r})$. The latter, for instance, can be interpreted as the electrostatic potential created by short-range impurities. Under this assumption, the self-energy matrix $\hat{\Sigma}(\varepsilon)$ becomes independent of the wave vector $k$ and takes a diagonal form, with $\Sigma_p(\varepsilon)=0$. Consequently, the integrals in Eq.~(\ref{eq:TAI18}) can be evaluated analytically, transforming Eqs.~(\ref{eq:TAI15})--(\ref{eq:TAI18}) into a set of algebraic self-consistent equations, which can be solved numerically by straightforward iteration~\cite{TAI10}. Once the Green's function is determined, the spectral function $A(k,\varepsilon)$ and the density of states $D(\varepsilon)$ can be obtained as follows:
\begin{eqnarray}
\label{eq:TAI19}
A(k,\varepsilon)=-\dfrac{1}{\pi}\textrm{Im}\left\{\textrm{Tr}\left(\hat{G}(k,\varepsilon+i0)\right)\right\},\notag\\
D(\varepsilon)=\int\limits_0^{K_c}\dfrac{k^2dk}{2{\pi}^2}A(k,\varepsilon).~~~~~~~~~~
\end{eqnarray}
Figure~\ref{Chapter2Fig9} shows the evolution of DOS with the disorder strength $W$ introduced as $W=\sqrt{v^2+u^2}$~\cite{TAI10}  for Hg$_{1-x}$Cd$_{x}$Te crystals with different Cd concentration $x$. If $M>0$, the band-gap in the DOS decreases with $W$ until it vanishes above a critical value $W^{*}$. Conversely, the DOS at $M<0$ remains gapless for all values of $W$.

\begin{figure}
\centering
\includegraphics [width=1.0\columnwidth, keepaspectratio] {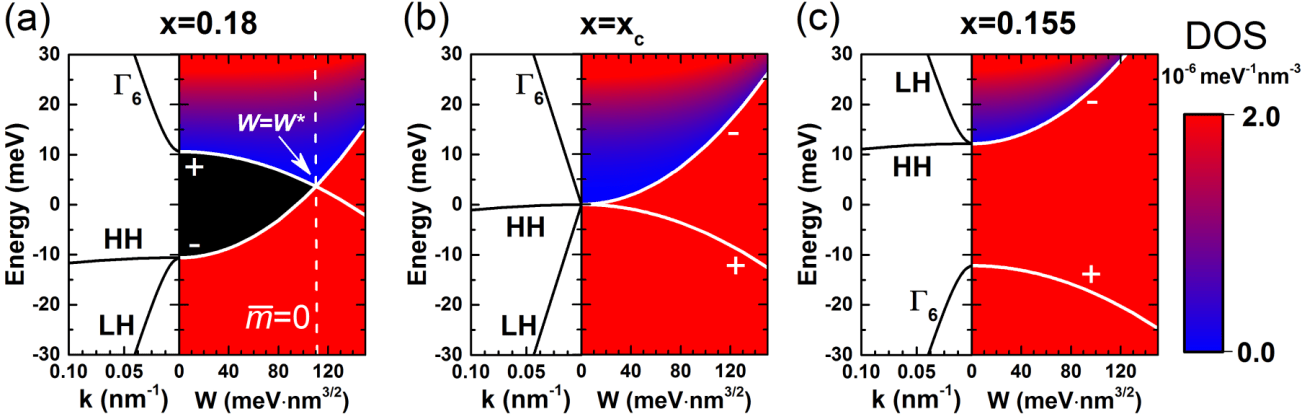}
\caption{\label{Chapter2Fig9} \textbf{(a-c)} Band structure and color map of the DOS as a function of the disorder strength~$W=\sqrt{v^2+u^2}$ calculated on the basis of three-band Kane Hamiltonian for bulk Hg$_{1-x}$Cd$_{x}$Te crystals at different Cd concentration: \textbf{(a)} $x=0.18$ ($M>0$), \textbf{(b)} $x=x_c$ ($M=0$ for $x_c\simeq0.168$ at 2~K~\cite{Kane5}) and \textbf{(c)} $x=0.155$ ($M<0$). The band parameters are provided in Ref.~\cite{TAI10}. The white solid curves show the band edge positions determined by Eq.~(\ref{eq:TAI25}). The sign on the solid curves conforms to the corresponding sign in the formula. The white dotted curve found from $\overline{m}(\varepsilon,W)=0$ represents the topological phase transition.}
\end{figure}

Let us now demonstrate that such behavior of the DOS in HgCdTe crystals is attributed to the disorder-induced topological phase transition. Indeed, according to Groth~\emph{et~al.}~\cite{TAI2}, the presence of disorder leads to the renormalization of both topological mass $\overline{m}(\varepsilon,W)$ and chemical potential $\overline{\mu}(\varepsilon,W)$:
\begin{eqnarray}
\label{eq:TAI24}
\overline{m}(\varepsilon,W)=M+\dfrac{\overline{\Sigma}_{c}(\varepsilon)-\overline{\Sigma}_{lh}(\varepsilon)}{2},\notag\\
\overline{\mu}(\varepsilon,W)=C-\dfrac{\overline{\Sigma}_{c}(\varepsilon)+\overline{\Sigma}_{lh}(\varepsilon)}{2},~
\end{eqnarray}
where the overbar stresses the values found on the set of real numbers. Since the finite DOS is associated with a finite imaginary part of the self-energy matrix $\hat{\Sigma}(\varepsilon)$, the band-gap region is characterized by the purely real quantities $\overline{\Sigma}_{c}(\varepsilon)$, $\overline{\Sigma}_{lh}(\varepsilon)$ and $\overline{\Sigma}_{hh}(\varepsilon)$. Note that $\overline{\Sigma}_{hh}(\varepsilon)=\overline{\Sigma}_{lh}(\varepsilon)$ for the Kane fermions. A topological transition occurs when the renormalized mass parameter changes the sign, i.e. $\overline{m}(\varepsilon,W)=0$~\cite{TAI2}. The band edge positions on the ($\varepsilon$,$W$)-diagram are found from the condition:
\begin{equation}
\label{eq:TAI25}
\overline{\mu}(\varepsilon,W)=C\pm\overline{m}(\varepsilon,W),
\end{equation}
where ``$+$'' and ``$-$'' correspond to the edge of $\Gamma_6$ and $\Gamma_8$ bands, respectively. As seen from Fig.~\ref{Chapter2Fig9}(a), two curves corresponding to the band edges cross at the transition point $W=W^{*}$, where $\overline{m}(\varepsilon,W)=0$. Interestingly, these curves not only describe the boundaries of the region with vanishing DOS, but also allow one to trace the band edge positions in the absence of the band-gap at $\overline{m}(\varepsilon,W)<0$. The latter is also seen in Fig.~\ref{Chapter2Fig9}(b,c).

The description of disorder-induced phase transition in terms of the renormalized topological mass and chemical potential implies the implicit replacement of Kane fermions by \emph{new quasiparticles} with the the \emph{quasiparticle} Hamiltonian
\begin{equation}
\label{eq:TAInew1}
\mathcal{H}_{qp}(k,\varepsilon)=\tilde{H}_{3D}(k)+\hat{\Sigma}(\varepsilon).
\end{equation}
In the most general case, $\mathcal{H}_{qp}(k,\varepsilon)$ is \emph{non-Hermitian} and therefore has complex eigenvalues. Physically, the real part of its eigenvalues characterizes the ``renormalized'' band structure, while the imaginary part describes the quasiparticle decay resulting in the finite quasiparticle lifetime. As was shown in our recent work~\cite{TAI13} and will be discussed in Section~\ref{sec:qpQW}, using this quasiparticle approach significantly simplifies the calculation of topological invariants in the presence of various interactions.
\begin{figure}
\centering
\includegraphics [width=1.0\columnwidth, keepaspectratio] {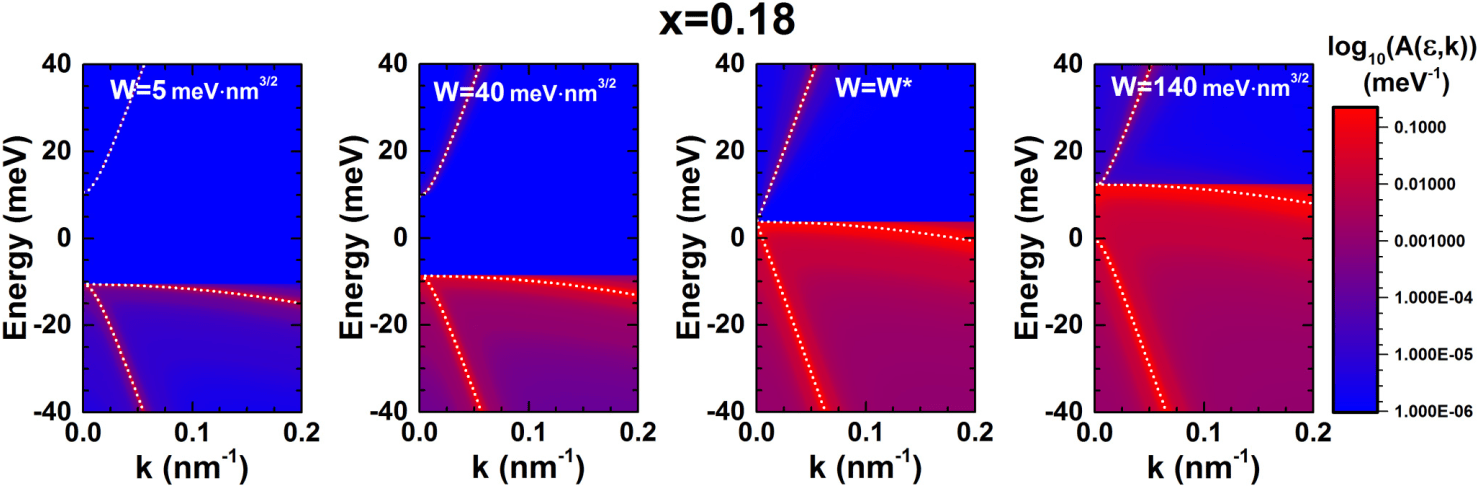}
\caption{\label{Chapter2Fig10}  Color map of the spectral function $A(k,\varepsilon)$ for bulk Hg$_{0.82}$Cd$_{0.18}$Te crystal at different values of the disorder strength~$W=\sqrt{v^2+u^2}$. The white dotted curves represent the quasiparticle energy dispersion defined by Eq.~(\ref{eq:TAI26}).}
\end{figure}

In order to shed further light on the quasiparticle picture, Fig.~\ref{Chapter2Fig10} provides color maps of the spectral function $A(k,\varepsilon)$ for bulk Hg$_{0.82}$Cd$_{0.18}$Te crystal at different values of the disorder strength. As clear, at $W=0$ corresponding to the ``clean'' crystal, $\hat{\Sigma}(\varepsilon)=0$ and the spectral function is represented by sum of $\delta$-functions centered at the energies $E_{c}$, $E_{lh}$ and $E_{hh}$ given by Eq.~(\ref{eq:TAI10}). In the presence of disorder, when $W{\neq}0$, the spectral function becomes broadened; nevertheless, its maxima continue to trace the quasiparticle energy dispersion defined by
\begin{equation}
\label{eq:TAI26}
\mathrm{det}\left|\tilde{H}_{3D}(k)+\hat{\overline{\Sigma}}(\varepsilon)-{\varepsilon}I_{6{\times}6}\right|=0,
\end{equation}
where the self-energy matrix is found on the set of real numbers. The solutions $\varepsilon(k)$ of Eq.~(\ref{eq:TAI26}) are shown by the white dotted curves. As can also be observed from the evolution of the spectral function $A(k,\varepsilon)$, the quasiparticle energy dispersion indeed reproduces the characteristic form expected for Kane fermions. This remarkable correspondence provides direct evidence that the topological phase transition induced by short-range disorder in narrow-gap semiconductors can, in fact, be fully understood within the framework of disorder-driven Kane fermions. Thus, the Kane fermion conception captures not only intrinsic band-structure effects but also the subtle interplay between topology and disorder in real materials.

\section{\label{sec:SummaryKane}Perspectives}
Of course, the studies of narrow-gap and gapless bulk semiconductors carried out in the THz team of Fr\'{e}d\'{e}ric Teppe with my active participation are by no means limited to the results presented in the preceding sections. The main goal of this Chapter was to introduce the reader to the concept of Kane fermions as one of the types of pseudo-relativistic quasiparticles that emerge beyond the Standard Model of particle physics. At the same time, I would like to emphasize that, despite the substantial amount of work already accomplished on various phenomena in these systems, narrow-gap and gapless semiconductors will undoubtedly remain at the core of my future research interests. As I have already mentioned in Chapter~\ref{Chap:Career}, when describing my scientific career path, I am deeply motivated to continue developing a theoretical framework that enables numerical modeling of the dynamic conductivity tensor under strongly non-equilibrium carrier distributions in crossed electric and magnetic fields. Such an approach, applied to materials hosting Kane fermions, will, I believe, shed light on the conditions under which the real part of the dynamic conductivity becomes negative -- thus leading to the amplification of the incoming radiation in the cyclotron resonance regime. This effect is a key prerequisite for realizing the long-sought Landau laser, a concept that inspires the ongoing experimental efforts on spontaneous Landau emission in Fr\'{e}d\'{e}ric Teppe's group.

Should it turn out, however, that an efficient Landau laser based on current pumping in narrow-gap and gapless semiconductors cannot be achieved in practice, the theoretical insights gained from these studies will still have significant value. In that case, they will contribute to improving the performance of cyclotron-resonance detectors -- devices that remain indispensable in the THz frequency range. The next generation of such detectors, as I envision it, will not rely on equilibrium carriers but will instead exploit hot-carrier dynamics, thereby offering enhanced sensitivity and faster response.

\chapter{\label{Chap:HgTe}Topological states and phase transitions in HgTe QWs}
The theoretical prediction of the quantum spin Hall effect (QSHE)~\cite{HgTeQWs4} marked the beginning of a renewed interest in the study of topological states in condensed matter systems. Subsequent works predicted~\cite{HgTeQWs2} and experimentally demonstrated~\cite{HgTeQWs5} the existence of the quantum spin Hall insulator (QSHI) state -- also known as the time-reversal-invariant 2D topological insulator (TI) -- in HgTe/CdHgTe QWs with inverted band ordering. The key feature of QSHIs is that, while being insulating in their 2D ``bulk'', they host a pair of spin-polarized 1D edge states. Charge transport through these helical edge channels gives rise to quantized values of conductance~\cite{HgTeQWs5,HgTeQWs8,HgTeQWs9}. For a long time, it was commonly believed that a direct backscattering from non-magnetic impurities between these edge states is prohibited by time-reversal symmetry. The reader of this m\'{e}moire has certainly encountered such statements in many papers devoted to the study of topological states. However, I would like to point out that this assumption is not entirely correct -- at least not for the QSHI states in QW heterostructures, including those discussed later in Chapter~\ref{Chap:III-V}. Demonstrating the falsity of this statement, and clarifying the mechanisms that enable the backscattering in the presence of time reversal symmetry, will be the subject of one of my future works. For now, let us return to the inverted HgTe/CdHgTe QWs themselves.

As was already predicted in 1990 by Gerchikov and Subashiev~\cite{HgTeQWs6}, the transition from a normal to an inverted band structure in HgTe/CdHgTe QWs occurs when the QW width $d$ exceeds a critical value $d_c$. When $d$ is smaller than $d_c$, the lowest electron-like (E1) subband lies above the highest hole-like (H1) subband in energy. This band ordering is referred to as trivial or direct, since it represents a conventional trivial (band) insulator state. At the critical thickness $d = d_c$, the band-gap closes, and the band dispersion in the vicinity of the $\Gamma$ point of the Brillouin zone is represented by massless Dirac fermions~\cite{Dirac2,Dirac3}. When the QW width exceeds this critical value ($d > d_c$), the band ordering becomes inverted: the electron-like E1 subband moves below the hole-like H1 subband, and a new band-gap opens in the system. It is precisely this inverted-band-gap state that is characterized by a nontrivial value of the $\mathbb{Z}_2$ topological invariant, signaling the emergence of the QSHI phase~\cite{HgTeQWs2}. With a further increase in $d$, the E1 subband descends below the second heavy-hole-like (H2) subband, and the minimum of the conduction band (located at the $\Gamma$ point) and the side maximum of the valence band approach each other. When the indirect band-gap vanishes at $d = d_{sm}$, the HgTe/CdHgTe QW enters a 2D semimetallic (SM) state~\cite{HgTeQWs7}, as illustrated in Fig.~\ref{Chapter3Fig1}. This SM state, independent of the QW orientation, is a universal property of wide HgTe/CdHgTe QWs grown on CdTe buffers~\cite{HgTeQWs10}, whereas it is absent in compressively biaxial-strained HgTe QWs~\cite{HgTeQWs12}.

\begin{figure}
\centering
\includegraphics [width=1.0\columnwidth, keepaspectratio] {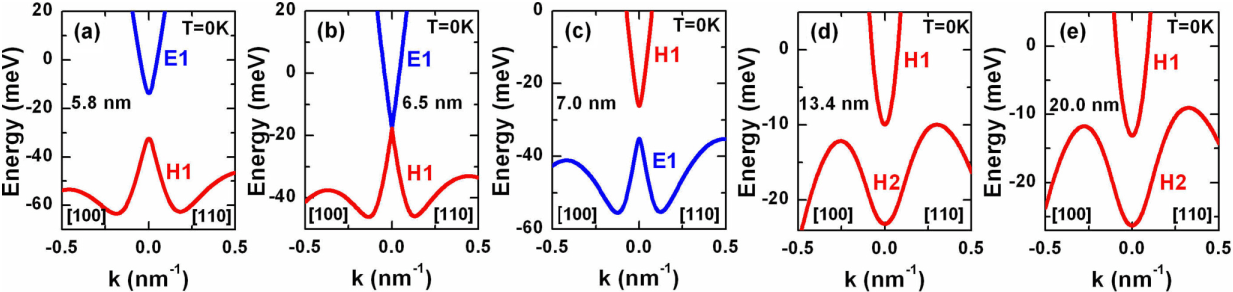}
\caption{\label{Chapter3Fig1}  \textbf{(a-e)} Typical band structure of (001)-oriented HgTe QWs at zero temperature and at different QW width~\cite{HgTeQWs10}: \textbf{(a)} trivial (band) insulator, $d<d_c$, \textbf{(b)} massless Dirac fermions, $d=d_c$, \textbf{(c)} QSHI states, $d>d_c$, \textbf{(d)} gapless state, $d=d_{SM}$, \textbf{(e)} SM state, $d>d_{SM}$. Electron-like E1 subband is shown in blue, while red curves correspond to the heavy-hole- subbands. In the panels (d) and (e), E1 subband lies significantly lower in energy. Here, I do not differ the QSHI state, in which the band-gap is defined by the gap between H1 and E1 subband, and the QSHI, in which the E1 subband lies below several heavy-hole-like subbands~\cite{HgTeQWs11}.}
\end{figure}

\begin{figure}
\centering
\includegraphics [width=0.85\columnwidth, keepaspectratio] {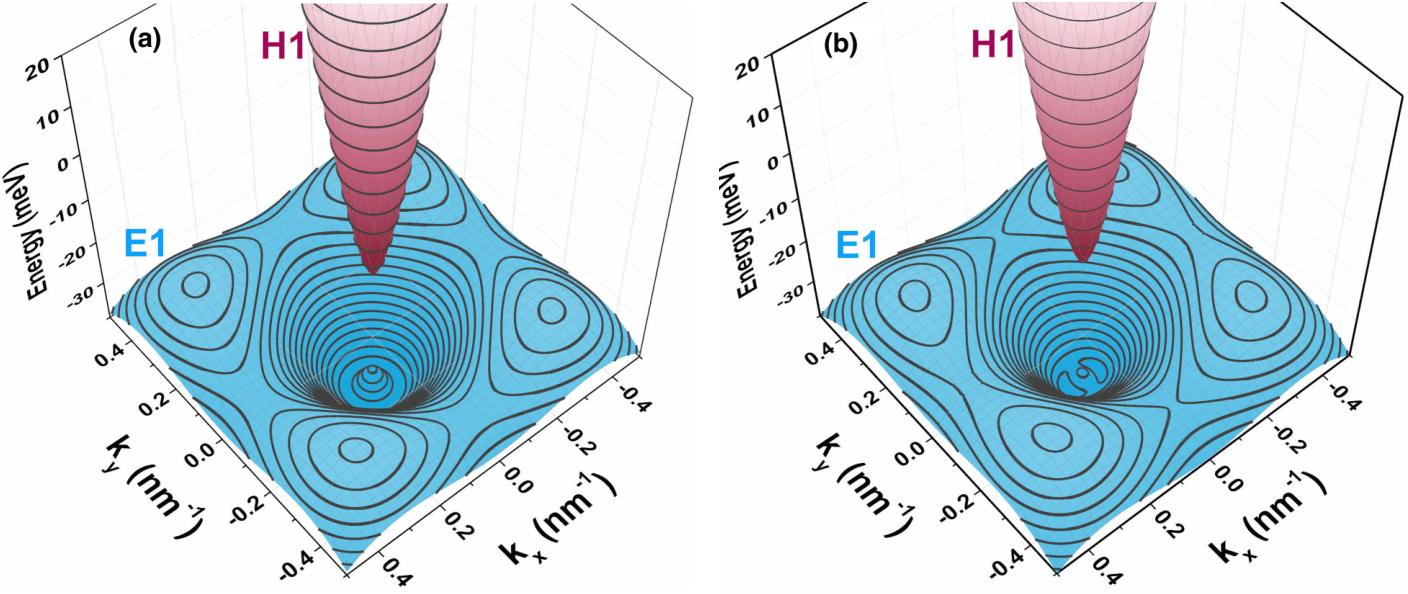}
\caption{\label{Chapter3Fig2} \textbf{(a,b)} Band structure and contour lines for HgTe/Cd$_{0.7}$Hg$_{0.3}$Te QW of $8$~nm width, grown on \textbf{(a)} (001) and \textbf{(b)} (013) planes at zero temperature and hydrostatic pressure. The E1 subband is shown in blue, the red surface corresponds to the H1 subband. The $x$ and $y$ axes for the (001) QW are oriented along [100] and [010] crystallographic directions, while for the (013) QW, the axes correspond to the directions of [100] and [0$3\bar{1}$], respectively~\cite{HgTeQWs10}.}
\end{figure}

At present, high-quality HgTe/CdHgTe QW structures available for scientific research worldwide are grown by molecular beam epitaxy in only a few laboratories: the University of W\"{u}rzburg~\cite{HgTeQWs5,HgTeQWs8,HgTeQWs9}, CEA Leti in Grenoble~\cite{Kane26,Kane27}, and the Rzhanov Institute of Semiconductor Physics in Novosibirsk~\cite{HgTeQWs13}. A distinctive feature of the most Novosibirsk-grown samples lies in their rather unusual [013] growth direction, in contrast to the conventional [001] orientation characteristic of the French and German samples. However, despite certain differences in the band structure between the QWs grown along different crystallographic directions (see Fig.~\ref{Chapter3Fig2}), according to my own experience, I would say that most physical phenomena in these systems manifest themselves qualitatively in the same way, independent of the growth orientation. Because of the long-standing collaboration established through the international TERAMIR cooperation (see Chapter~\ref{Chap:Career}), the most samples studied in the THz team of Fr\'{e}d\'{e}ric Teppe were grown in Novosibirsk.

Finally, another problematic feature of HgCdTe-based structures is the exceptionally delicate process of fabricating gated Hall bars, which must be performed at relatively low temperatures due to the weak bonding energy of mercury atoms. This technological constraint makes device processing significantly more challenging than in III-V semiconductor systems. In fact, the fabrication of micron-scale gated Hall bars -- necessary for observing conductance quantization arising due to the transport via topological edge state -- has been successfully achieved only at the University of W\"{u}rzburg. Therefore, the most experimental studies of HgTe/CdHgTe QWs, performed in Montpellier, have focused on the bulk properties of these QWs, both in the trivial and inverted band regimes by means of various techniques. In this Chapter, I have chosen to present several key works that, in my opinion, most clearly highlight the contribution of the Montpellier group to the field of HgTe QW research.

\section{\label{sec:PandTinQWs}Pressure- and temperature-driven phase transitions in HgTe QWs}
Following our experimental observation of the temperature-driven topological phase transition in bulk HgCdTe films~\cite{Kane6}, a natural question arose: could a similar transition be observed in HgTe/CdHgTe QWs? However, unlike the relatively simple case of bulk HgCdTe -- where the relevant parameters reduce to the rest mass and the Fermi velocity of Kane fermions -- the band structure of HgTe/CdHgTe QWs depends on a considerably larger number of parameters describing both constituent bulk materials. To determine the critical temperature $T_c$ for a given QW width $d$, it is therefore essential not only to include the temperature dependence of all relevant material parameters and the variation of the valence-band offset between HgTe and CdTe, but also to correctly account for the confinement effects responsible for the formation of QW subbands.

For a rigorous description of these confinement effects, we employed the eight-band $\mathbf{k}\cdot\mathbf{p}$ Kane Hamiltonian~\cite{HgTeQWs1}, which explicitly incorporates the interaction between the $\Gamma_6$, $\Gamma_8$, and $\Gamma_7$ bands. This approach provides an accurate description of the experimentally observed band structures and has become the standard theoretical tool for narrow-gap and gapless II-VI heterostructures. It is worth emphasizing that our earlier studies of temperature-dependent band inversion in bulk HgCdTe films provided a crucial insight: the interband momentum matrix element $P$, which defines the Fermi velocity $\tilde{c}$ of Kane fermions (see Section~\ref{sec:TempKane}), remains independent of temperature. Later experimental works~\cite{Kane7,Kane8} confirmed our assumption, extending its validity also to hydrostatic pressure. Based on a comprehensive analysis of numerous experimental and theoretical works on bulk HgCdTe, we were able to establish reliable temperature- and pressure-dependent variations for all band parameters required to calculate the electronic structure of HgTe/CdHgTe QWs. The technical details, as well as the full set of parameters used in these calculations, can be found in our original paper~\cite{HgTeQWs10}. Therefore, in what follows, I will focus only on the main numerical results demonstrating pressure- and temperature-induced phase transitions in HgTe/CdHgTe QWs.

\begin{figure}
\centering
\includegraphics [width=0.6\columnwidth, keepaspectratio] {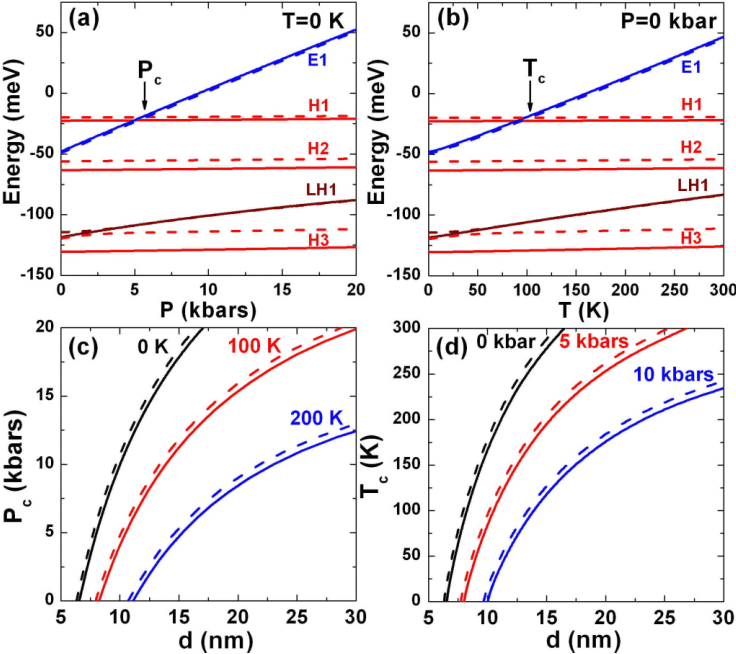}
\caption{\label{Chapter3Fig3} \textbf{(a,b)} Band edge of the electron-like E1, the heavy-hole-like H1, H2, and H3 and the light-hole-like LH1 subbands at the $\Gamma$ point in 8 nm HgTe/Cd$_{0.7}$Hg$_{0.3}$Te QWs as a function of hydrostatic pressure (a) and temperature (b). \textbf{(c,d)} Dependence of critical pressure $P_c$ and temperature $T_c$, at which the Dirac cone in the $\Gamma$ point arises (see Fig.~\ref{Chapter3Fig1}b), as a function of $d$ for HgTe/Cd$_{0.7}$Hg$_{0.3}$Te QWs. In all the panels, solid curves represent the calculations for (001)-oriented QWs, while the dashed curves correspond to the structures grown on (013) plane.}
\end{figure}

The calculated results presented in Fig.~\ref{Chapter3Fig2} provide a clear quantitative picture of how temperature and hydrostatic pressure influence the band ordering of HgTe/CdHgTe QWs. The upper panels of the figure show the calculated band-edge energies of the electron-like E1, heavy-hole-like H1, H2, and H3, and light-hole-like LH1 subbands at the $\Gamma$ point for an 8~nm-wide HgTe/Cd${0.7}$Hg${0.3}$Te QW as functions of both hydrostatic pressure $P$ and temperature $T$. These results demonstrate a remarkable similarity between the effects of pressure and temperature on the band ordering in such structures. At low pressures and temperatures, the band structure remains inverted, corresponding to the topological QSHI regime. However, both $P$ and $T$ strongly affect the position of the E1 subband, leading to a crossing between the E1 and H1 subbands at certain critical values of hydrostatic pressure ($P_c$) and temperature ($T_c$). At these critical points, the energy dispersion near the $\Gamma$ point becomes linear in quasimomentum, forming a Dirac-like cone (see Fig.~\ref{Chapter3Fig1}b). A further increase in either $P$ or $T$ drives the system into a trivial band-insulator (BI) state with direct band ordering. The slight difference between the critical values $P_c$ and $T_c$ for (001)- and (013)-oriented QWs arises primarily from lattice-mismatch strain, which depends on the specific growth direction of the heterostructure~\cite{HgTeQWs10}. The lower panels of Fig.~\ref{Chapter3Fig2} summarize the calculated dependence of these critical parameters on the QW width for different values of $T$ and $P$. The overall behavior clearly indicates that both increasing temperature and applying hydrostatic pressure drive a topological phase transition into a conventional BI state.

\begin{figure}
\centering
\includegraphics [width=0.8\columnwidth, keepaspectratio] {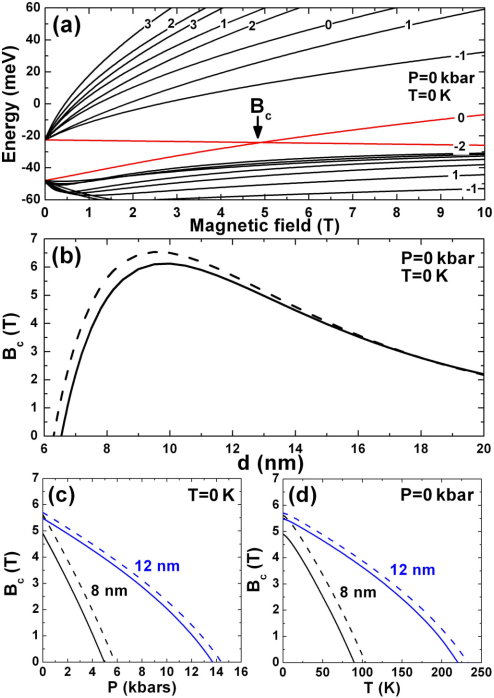}
\caption{\label{Chapter3Fig4} \textbf{(a)} Landau levels for (001)-oriented HgTe/Cd$_{0.7}$Hg$_{0.3}$Te QW of 8 nm thickness at $T=$0 K and $P=$0 kBar. Pair of zero-mode Landau levels is shown by red curves. \textbf{(b)} A critical field $\mathcal{B}_c$ as a function of QW width $d$ at zero pressure and temperature. \textbf{(c,d)} A critical field $\mathcal{B}_c$ as a function of hydrostatic pressure and temperature for 8 nm and 12 nm QWs. The solid curves in the panels (b-d) display the calculations for (001)-oriented QWs, while the dashed curves correspond to the (013)-oriented structures.}
\end{figure}

The inherent property of HgTe QWs, arising under perpendicular magnetic field $\mathcal{B}$ , is the presence of a particular pair of zero-mode Landau levels, which split from the edges of E1 and H1 subbands~\cite{HgTeQWs5}. As these zero-mode Landau levels disperse in opposite directions with increasing $\mathcal{B}$, they cross at the critical magnetic field $\mathcal{B}_c$ under band inversion. If $\mathcal{B}>\mathcal{B}_c$, HgTe QW has a direct band ordering, while below $\mathcal{B}_c$ the band structure remains inverted and the helical edge states still exists~\cite{HgTeQWs14,HgTeQWs15}. The top panel in Fig.~\ref{Chapter3Fig4} shows Landau levels in (001) HgTe/Cd$_{0.7}$Hg$_{0.3}$Te QW calculated for zero $P$ and $T$ values. The numbers over the curves correspond to the Landau level indices~\cite{HgTeQWs10}. Two red curves represent the zero-mode Landau level, which are identified within a simplified approach, based on 2D Dirac-type BHZ Hamiltonian~\cite{HgTeQWs2}. The panel below displays $\mathcal{B}_c$ as a function of QW width $d$ for the (001)- and (013)-oriented QWs for zero pressure and temperature. The bottom panels in Fig.~\ref{Chapter3Fig4} show the dependence of $\mathcal{B}_c$ on $P$ and $T$ for the 8 nm and 12 nm QWs of different orientations. It is seen that critical magnetic field decreases with pressure and temperature. The latter is related with the collapse of the gap between E1 and H1 subbands, if pressure and temperature tends to $P_c$ and $T_c$ respectively (see Fig.~\ref{Chapter3Fig3}). Above $P_c$ and $T_c$, the band structure is direct and the zero-mode Landau levels are not crossed.

I would like to emphasize that the discussion so far has been limited to the local picture of the band structure in HgTe/CdHgTe QWs in the vicinity of the $\Gamma$ point of the Brillouin zone. However, such a local description does not fully capture the whole electronic properties of HgTe QWs with an inverted band structure. As illustrated in Fig.~\ref{Chapter3Fig1}, with increasing well width $d$, the system first becomes indirect-gap, due to the emergence of side maxima in the valence band, and eventually evolves into a SM state. As demonstrated in our work~\cite{HgTeQWs10}, the overlap between the side maxima of the valence band and the conduction-band minimum is also strongly dependent on both temperature and pressure. Figure~\ref{Chapter3Fig5} presents the calculated pressure- and temperature- phase diagrams for (001)-oriented HgTe/Cd${0.7}$Hg${0.3}$Te QWs grown on CdTe buffer layers. The open (white) regions correspond to insulating states, whereas the striped areas mark the semimetallic regime, where the conduction and valence bands overlap. Two important trends can be immediately identified. First, over a wide range of QW widths, variation of hydrostatic pressure and temperature enables transitions between the SM, BI, and QSHI states. Second, the QSHI states exists only within a finite range of pressure and temperature. Indeed, for any given pressure, there exists a critical temperature above which the QSHI state collapses, caused eigther by the nonlocal overlap between the conduction and valence bands, or by the change in the band ordering. Interestingly, the phase diagrams in Fig.~\ref{Chapter3Fig5} reveal that at high pressures and temperatures, variation of the QW width drives the system directly from the BI to the SM state, bypassing the topological QSHI regime altogether.

\begin{figure}
\centering
\includegraphics [width=1.0\columnwidth, keepaspectratio] {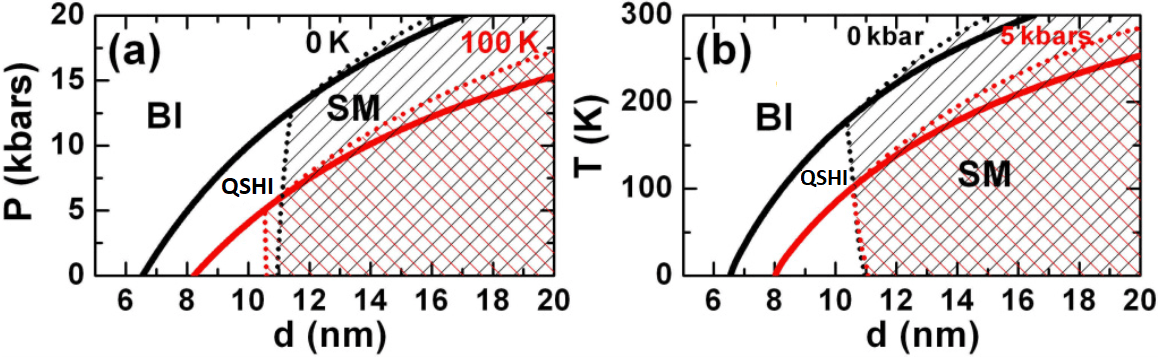}
\caption{\label{Chapter3Fig5} \textbf{(a)} Pressure and \textbf{(b)} temperature phase diagram for (001) HgTe/Cd$_{0.7}$Hg$_{0.3}$Te QWs, grown on CdTe buffer. The solid curves represent the dependence of $T_c$ and $P_c$ on the QW width. The dotted curves correspond to a formation of the gapless states, shown in Fig.~\ref{Chapter3Fig1}(d).}
\end{figure}

In conclusion, the results presented in this section provided the theoretical foundation and motivation for subsequent experimental studies aimed at observing temperature- and pressure-driven topological phase transitions in HgTe QW. These studies have performed in the THz team in Montpellier, in the Vladimir Gavrilenko' group in Nizhny Novgorod and in \href{http://centera.eu/}{CENTERA} in Warsaw. The latter includes the study of topological phase transitions under hydrostatic pressure performed by \href{http://www.researchgate.net/profile/Ivan-Yahniuk}{Ivan Yahniuk}. In the next section, I will focus only on the experimental confirmation of temperature-induced phase transitions in HgTe QWs, while Ivan's results can be found in his \href{http://bip.unipress.waw.pl/images/stories/doktoraty/yahniukivan.pdf}{PhD thesis} defended in 2021.

\section{\label{sec:ExpTinQWs}Experimental observation of the band structure tuned by temperature}
Following the theoretical analysis presented in the previous section, it became essential to verify experimentally how temperature influences the band structure of HgTe/CdHgTe QWs. Over the past years, our group in Montpellier has conducted a series of experimental studies by means of temperature-dependent Landau-level spectroscopy and magnetotransport measurements. In this section, I will present the main results of these investigations. I have also decided to include here the temperature-dependent studies of interband photoconductivity in HgTe QWs, performed by Anton Ikonnikov in the group of Vladimir Gavrilenko in Nizhny Novgorod, for which I provided the theoretical interpretation. The aim of this section is to demonstrate how these complementary experimental approaches offer direct insight into the evolution of the electronic spectrum across the temperature-driven topological phase transition. For clarity, I will focus primarily on the obtained results, while those interested in the detailed description of the experimental procedures are encouraged to refer to the original publications.

\subsection{\label{sec:LLspectHgTeQWs}Temperature-dependent Landau-level spectroscopy}
As demonstrated earlier for bulk HgCdTe films (see Section~\ref{sec:TempKane}), temperature-dependent spectroscopy of inter-Landau-level transitions can serve as a highly effective tool for the experimental observation of temperature-driven topological phase transitions. At present, only two studies devoted to temperature-dependent Landau-level spectroscopy of HgTe/CdHgTe QWs have been reported in the literature, both carried out within the framework of the TERAMIR cooperation by the teams in Montpellier and Nizhny Novgorod. The first study~\cite{HgTeQWs16} employed THz quantum cascade lasers as monochromatic radiation sources in combination with pulsed magnetic fields up to 45~T. Its results primarily revealed a pronounced difference in the temperature evolution of the magnetoabsorption lines between samples with normal and inverted band ordering at low temperatures. The second work~\cite{HgTeQWs17}, based on measurements performed in Montpellier, is more representative in the context of observing the temperature-driven topological phase transition, since it provides a clear picture of how the inter-Landau-level transition energies evolve with magnetic field as the temperature increases. Therefore, further we will only focus on the results of our second work~\cite{HgTeQWs17}.

The two QW samples studied in this work were grown by molecular beam epitaxy on (013)-oriented semi-insulating GaAs substrates with a relaxed CdTe buffer~\cite{HgTeQWs13}. The nominal QW widths were $d=6$~nm (sample~A) and $d=8$~nm (sample~B). Each HgTe QW was embedded in Cd$x$Hg${1-x}$Te barriers about 40~nm thick, with Cd concentrations of $x=0.62$ for sample~A and $x=0.80$ for sample~B. A thin CdTe cap layer was deposited on top of the structures to protect the surface from oxidation. The 8~nm QW (sample~B) was remotely doped on both sides by 15-nm-thick In-doped regions with a doping concentration of $6.5\times10^{16}$~cm$^{-3}$, which resulted in an electron sheet density of approximately $3\times10^{11}$~cm$^{-2}$ at 2~K. The 6~nm QW (sample~A), in contrast, was nominally undoped and contained a residual two-dimensional hole concentration of about $3\times10^{10}$~cm$^{-2}$ at low temperature. The results of the band-structure calculations~\cite{HgTeQWs10}, presented in Fig.~\ref{Chapter3Fig6}, clearly illustrate the contrasting nature of these two QWs. Sample~A is expected to exhibit a normal band ordering corresponding to the BI state with a small band-gap at 2~K that gradually increases with temperature. Sample~B, on the other hand, is predicted to possess an inverted band structure corresponding to the QSHI state with an indirect gap of about 10~meV at low temperatures. The critical temperature $T_c$, marking the transition between the QSHI and BI regimes, is estimated to be around 90~K.

\begin{figure}
\centering
\includegraphics [width=1.0\columnwidth, keepaspectratio] {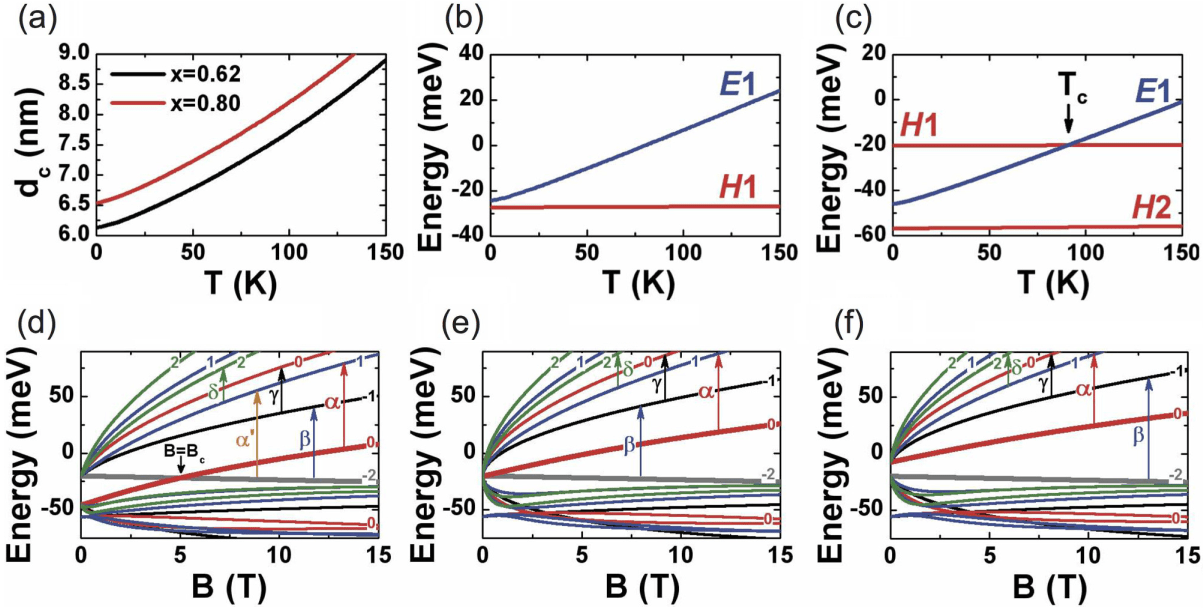}
\caption{\label{Chapter3Fig6} \textbf{(a)} Critical QW width $d_c$, corresponding to the phase transition and arising of massless Dirac fermions, as a function of temperature in (013) HgTe/Cd$_{x}$Hg$_{1-x}$Te QWs for $x=0.62$ and 0.80. \textbf{(b,c)} Temperature dependence of the electron-like \emph{E}1 and the heavy-hole-like \emph{H}1, \emph{H}2 subbands as a function of temperature at zero quasimomentum (b) for sample A and (c) sample B. The \emph{E}1 subband in the top panels is shown in blue, the red surface corresponds to the \emph{H}1 subband. \textbf{(d-f)} Landau levels in sample B at different temperatures: (d) $T=2$~K, (e) $T=T_c$ and (f) $T=130$~K. The numbers over the curves in the bottom panels correspond to the LL indices~\cite{HgTeQWs10}. A pair of zero-mode Landau levels with indices $-2$ and $0$ is shown by bold curves. The arrows and Greek letters denote Landau-level transitions, observed in magneto-absorption spectra of sample B.}
\end{figure}

\begin{figure}
\centering
\includegraphics [width=1.0\columnwidth, keepaspectratio] {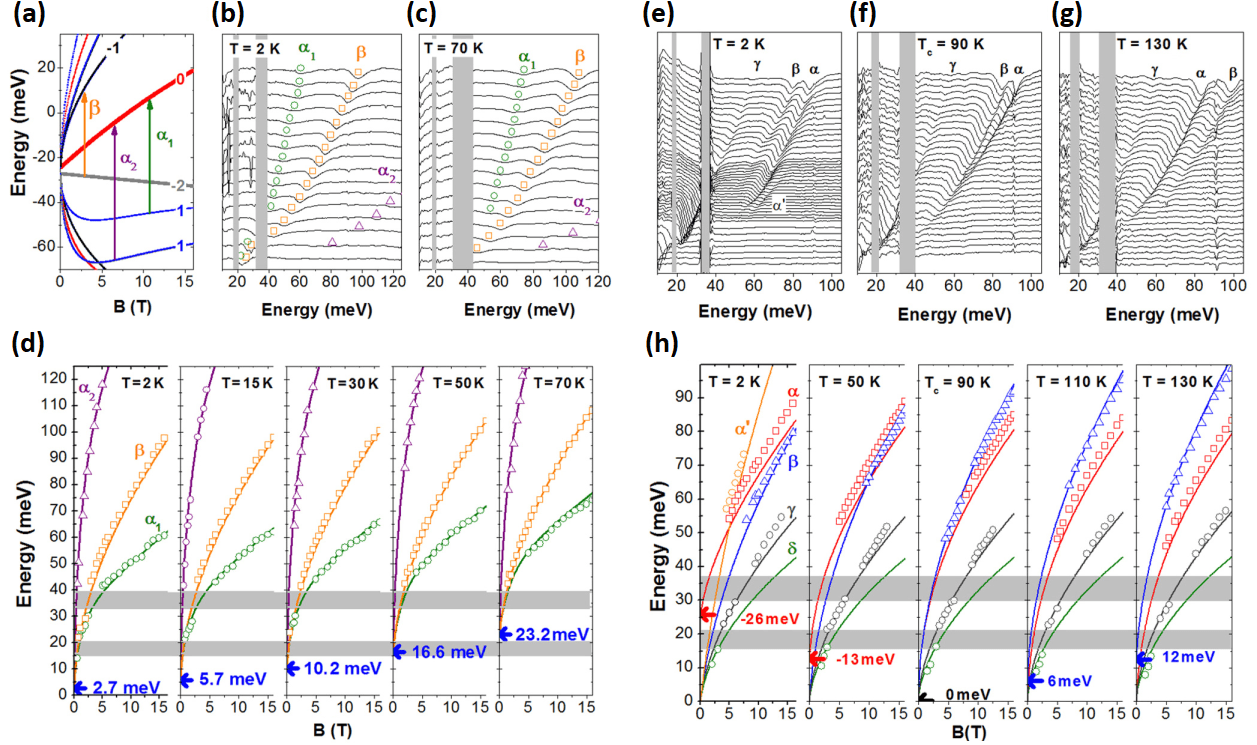}
\caption{\label{Chapter3Fig7} \textbf{(a)} The energies of Landau levels as a function of magnetic field in sample A at 2~K. The numbers conform to the Landau-level indices. The zero-mode Landau levels are shown by bold curves. The arrows and Greek letters denote Landau-level transitions, observed in magneto-absorption spectra. \textbf{(b,c)} Transmission spectra of sample A at 2~K and 70~K from 1~T (the first plot at the bottom) to 16~T (the last plot on the top) with the step for magnetic field of 1.0~T. \textbf{(d)} Energy of $\alpha_1$ (green curve), $\alpha_2$ (purple curve) and $\beta$ (orange curve) transitions as a function of magnetic field. The experimental data are represented by symbols: circles, triangles and squares for the $\alpha_1$, $\alpha_2$ and $\beta$ transitions, respectively. The value of the band-gap is shown by blue arrows. \textbf{(e-g)} Transmission spectra of sample B at 2~K, 90~K and 130~K from 0.5~T (the first plot at the bottom) to 16~T (the last plot on the top). The step for magnetic field is 0.5~T, except the panel for 2~K, in which a 0.25~T step was used between 4~T and 9~T allowing to accurately follow the evolution of $\alpha'$ transition. The Greek letters denote Landau-level transitions, shown in Fig.~\ref{Chapter3Fig6}. \textbf{(h)} Fan chart of inter-Landau-level transitions in sample B. The $\alpha'$, $\alpha$, $\beta$, $\gamma$ and $\delta$ calculated transitions are shown in solid lines in orange, red, dark grey and green respectively. Experimental data are represented by symbols in the same colors as the theoretical curves. The gap at $k=0$ is shown by blue and red arrows respectively for negative and positive values, and black for the gapless state. The samples are completely opaque in the range of the Cd$_x$Hg$_{1-x}$Te- and GaAs-reststrahlen bands, indicated by shaded areas in (d) and (h). For simplicity, the evolution of the reststrahlen bands with temperature is not shown in the figure}
\end{figure}

The magnetotransmission spectra of sample~A, measured at 2~K and 70~K, are shown in Fig.~\ref{Chapter3Fig7}(b,c), while the complete temperature series can be found in our original publication~\cite{HgTeQWs17}. Because of the very low hole concentration in this sample, only a few Landau levels in the valence band are occupied. For example, at a magnetic field of $\mathcal{B}\simeq1.2$~T, the filling factor corresponds to $\nu=1$ for the carrier density determined at 2~K. As a consequence, only a limited number of interband transitions can be observed. In all spectra, three intense absorption lines are clearly visible, which can be identified as interband Landau-level transitions involving the zero-mode Landau levels. These transitions obey the conventional electric-dipole selection rules $\Delta n=\pm1$ in the Faraday configuration and are indicated in Fig.~\ref{Chapter3Fig7}(a) by small Greek letters, in accordance with the previously used notations~\cite{HgTeQWs18,HgTeQWs19,HgTeQWs20}. Comparison between experimental and theoretical values of the transition energies is presented in Fig.~\ref{Chapter3Fig7}(d). The extrapolation of the interband transition energies to $B=0$ provides an estimate of the band-gap at zero quasimomentum, as shown by the blue arrows in Fig.~\ref{Chapter3Fig7}(d).

Figure~\ref{Chapter3Fig7}(e--g) shows the magnetotransmission spectra of sample~B, measured at 2~K, 90~K, and 130~K. The full set of spectra for other temperatures can be found in Ref.~\cite{HgTeQWs17}. As in the previous case, three intense absorption lines dominate the spectra at all temperatures. These lines can be identified as the inter–Landau-level transitions involving the zero-mode levels (the $\alpha$ and $\beta$ transitions) and the cyclotron resonance (CR) absorption of two-dimensional electrons (the $\gamma$ and $\delta$ transitions), as schematically illustrated in Fig.~\ref{Chapter3Fig6}(d--f). By taking into account the electron concentration determined at 2~K, one can see that the CR observed at magnetic fields below 3.4~T (corresponding to $\nu\approx3$) originates from both $\gamma$ and $\delta$ transitions, while for higher fields only the $\gamma$ transition remains possible. The $\gamma$ line disappears from the spectrum at approximately $\mathcal{B}=12$~T, which corresponds to the filling factor $\nu=1$. This observation provides excellent quantitative agreement with the electron concentration independently extracted from magnetotransport measurements. The overall agreement between experimental data and theoretical calculations obtained for both samples (see Fig.~\ref{Chapter3Fig7}(d,h)) unambiguously confirms the validity of the band-structure modeling presented in Section~\ref{sec:PandTinQWs}, predicting both pressure- and temperature-driven topological phase transitions in HgTe QWs.

Finally, I would like to discuss one additional feature observed in the magnetoabsorption spectra of sample~B. At 2~K, an extra transition, labeled $\alpha'$ in Fig.~\ref{Chapter3Fig6}(d), appears in a relatively narrow range of magnetic fields $B=5$–$7.5$~T, corresponding to the crossing of the zero-mode Landau levels. This transition does not satisfy the dipole selection rules $\Delta n=\pm1$ and should therefore be forbidden within the electric-dipole approximation. The appearance of this line, also reported in earlier studies~\cite{HgTeQWs19,HgTeQWs20,HgTeQWs21}, was initially attributed to the coupling between the $n=0$ and $n=2$ Landau levels caused by bulk inversion asymmetry (BIA) inherent to zinc-blende semiconductors, an effect neglected in our calculations. Including BIA induces an anticrossing in the vicinity of the critical field $\mathcal{B}_c$, resulting in a partial mixing of the zero-mode Landau levels and activating the $\alpha'$ transition in the electric-dipole approximation. The observation of this transition thus served as clear experimental evidence of the inverted band structure in sample~B at low temperatures. However, subsequent studies carried out in the THz team in Montpellier~\cite{HgTeQWs22,HgTeQWs23} revealed that the original hypothesis linking the $\alpha'$ transition to BIA was not correct. The actual mechanism responsible for its activation was found to be the hybridization of the Landau-level transitions caused by electron-electron interactions. To avoid side discussion in this section, I have decided to leave a review of the results outside the scope of this m\'{e}moire but refer the interested reader to our original publications~\cite{HgTeQWs22,HgTeQWs23}

\subsection{\label{sec:PhotoHgTeQWs}Interband photoconductivity}
Apart from Landau-level spectroscopy, interband photoconductivity measurements at different temperatures can also serve as an efficient tool to probe the temperature-induced evolution of the band structure in HgTe/CdHgTe QWs. This subsection briefly summarizes the results of interband photoconductivity studies in HgTe/Cd$_{x}$Hg$_{1-x}$Te QWs, performed by Anton Ikonnikov in the group of Vladimir Gavrilenko in Nizhny Novgorod~\cite{HgTeQWs24}. Two QW samples were investigated in that work: sample~1 with a direct band ordering ($d=5$~nm, $x=0.7$) and sample~2 with an inverted band structure ($d=8$~nm, $x=0.72$) at low temperature. The details of the measurement technique can be found in the original publication~\cite{HgTeQWs24}; here, I will restrict myself to a concise overview of the experimental results and their interpretation.

\begin{figure}
\centering
\includegraphics [width=1.0\columnwidth, keepaspectratio] {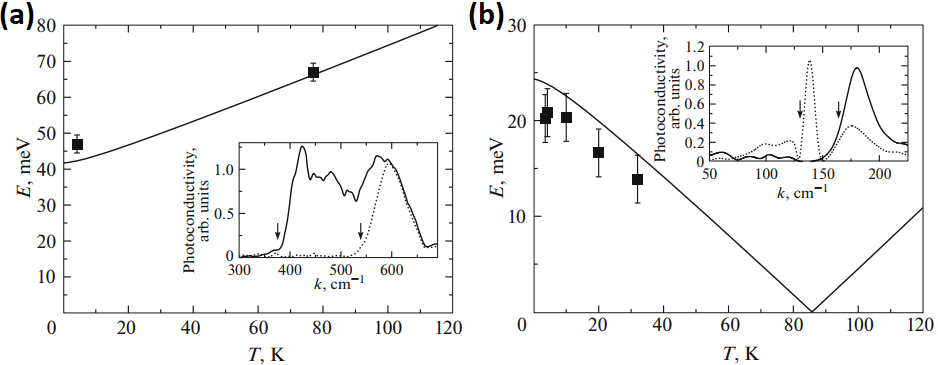}
\caption{\label{Chapter3Fig8} \textbf{(a,b)} Temperature dependences of the long-wavelength photoconductivity threshold (symbols) and the calculated band-gap (solid curve) for sample 1 in panel \textbf{(a)} and sample 2 in panel \textbf{(b)}. The insets show typical photoconductivity spectra at different temperatures. The solid curve corresponds to a temperature of $T=4.2$~K in both panels, while the dashed curve represents data at 77 K in panel (a) and 20 K in panel (b). The vertical arrows in the insets indicate the long-wavelength absorption thresholds.}
\end{figure}

The inset in Fig.~\ref{Chapter3Fig8}(a) shows the interband photoconductivity spectra measured in sample~1 at $T=4.2$~K and $T=77$~K. A clear long-wavelength absorption threshold is observed in both spectra. Increasing the temperature leads to a noticeable shift of this threshold toward higher photon energies, reflecting the band-gap increasing. The dip in the signal near 700~cm$^{-1}$ originates from absorption in the beam splitter (Mylar ML) and the black polyethylene filter used in the optical setup. Typical photoconductivity spectra for sample~2 with an inverted band structure are presented in the inset of Fig.~\ref{Chapter3Fig8}(b). In contrast to the behavior observed in sample~1, in this case the long-wavelength absorption edge shifts toward lower photon energies as the temperature increases. The positions of the absorption thresholds are indicated by arrows. The dips at approximately 145 and 120~cm$^{-1}$ correspond to absorption by transverse optical (TO) phonons in the CdHgTe barriers.

The temperature dependencies of the photoconductivity threshold for both samples are summarized in Fig.~\ref{Chapter3Fig8}. It is clearly seen that the evolution of the long-wavelength absorption threshold with temperature is fundamentally different for the two samples, reflecting distinct temperature behavior of the band-gap. In the QW with a normal band ordering (sample~1), the increase in temperature simply results in an increase of the band-gap. In contrast, in the QW with an inverted band structure (sample~2), the temperature increase first leads to a decrease of the energy separation between the E1 and H1 subbands, followed by a change in the band ordering when the critical temperature $T_c$ is reached. The solid curves in Fig.~\ref{Chapter3Fig8} show the results of my calculations of the band-gap energy as a function of temperature for both samples. A very good agreement is found for sample~1 and a reasonable agreement for sample~2, where the experimental values are slightly lower than the theoretical ones. This deviation most likely originates from the deviation of the real QW profile in sample~2 from an ideal rectangular shape. Based on my experience, such deviations are frequently observed in (013) HgTe QWs grown in Novosibirsk, probably due to an enhanced growth rate along the chosen crystallographic direction.
Nevertheless, the overall correspondence between experiment and theory confirms that the temperature evolution of the band-gap in HgTe QWs is well captured by the model calculations. Thus, interband photoconductivity measurements provide a complementary and direct experimental confirmation of the temperature-driven modification of the electronic band structure in HgTe/CdHgTe QWs, supporting the conclusions made on the basis of Landau-level spectroscopy.

\begin{figure}
\centering
\includegraphics [width=0.7\columnwidth, keepaspectratio] {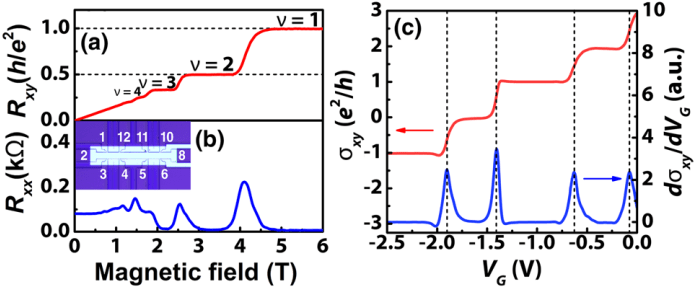}
\caption{\label{Chapter3Fig9} \textbf{(a,b)} Hall resistance $R_{xy}$ (a) and Shubnikov-de Haas (SdH) oscillations, $R_{xx}=R_{28,34}$, (b) at 1.7 K at zero gate voltage. The insert schematically shows a picture of the Hall bar. \textbf{(c)} Hall conductivity $\sigma_{xy}$ and its derivative $\partial\sigma_{xy}/\partial V_G$ as a function of gate voltage at 1.7~K and magnetic field of 2.4~T.}
\end{figure}

\subsection{\label{sec:TransportHgTeQWs}Temperature-dependent quantum magnetotransport}
In contrast to Landau-level spectroscopy and interband photoconductivity (both of which probe the system’s optical response to the electric field of the incident radiation), quantum magnetotransport measurements using gated Hall bars allow one to directly reconstruct the Landau-level spectrum of a 2D electron (or hole) system. To illustrate this concept, let us refer to Fig.~\ref{Chapter3Fig9} from our original publication~\cite{HgTeQWs25}. The figure shows the Hall resistance and the Shubnikovde Haas (SdH) oscillations measured at 1.7~K. The Hall resistance $R_{xy}$ exhibits well-defined plateaus at both even and odd multiples of $h/e^2$, reflecting the quantization of the Hall conductivity. The panel~(c) presents the Hall conductivity $\sigma_{xy}$ as a function of the gate voltage $V_G$ at a magnetic field of 2.4~T, together with its derivative $\partial\sigma_{xy}/\partial V_G$. Each peak in $\partial\sigma_{xy}/\partial V_G$ corresponds to the crossing of the Fermi level with a particular Landau level. The same information can also be extracted from the peaks of the Shubnikov-de Haas oscillations $\sigma_{xx}$. Thus, plotting the experimental data of $\partial\sigma_{xy}/\partial V_G$ and $\sigma_{xx}$ in the $(V_G,\mathcal{B})$ plane provides a powerful visualization of the Landau levels arising from the E1 and H1 subbands. An especially important advantage of this method is the possibility to determine, with high precision, the value of the critical magnetic field $\mathcal{B}_c$, corresponding to the crossing of the zero-mode Landau levels. Since $\mathcal{B}_c$ marks the inversion of band ordering, temperature-dependent mapping of these traces of zero-mode Landau levels becomes an efficient tool to probe a temperature-induced phase transition between trivial band insulator and QSHI states.

\begin{figure}
\centering
\includegraphics [width=0.75\columnwidth, keepaspectratio] {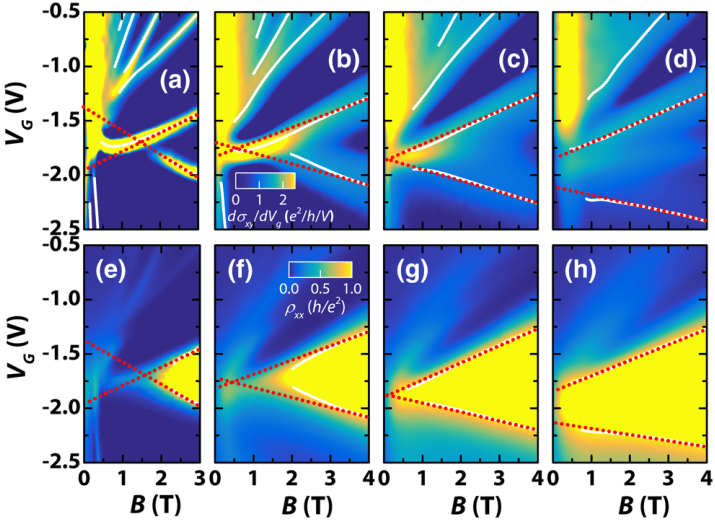}
\caption{\label{Chapter3Fig10} \textbf{(a-d)} Colormap of $\partial\sigma_{xy}/\partial V_g$ as a function of both magnetic field and gate voltage, at 1.7~K, 20~K, 27~K and 40~K respectively. The white solid curves correspond to $\sigma_{xy}= (n+1/2) e^2/h$. In the NI state ($\mathcal{B}>2.2$ T), the isolines of $\sigma_{xy}= \pm e^2/2h$ are linearly fitted by the red dotted lines giving the values of $\mathcal{B}_c$. \textbf{(e-h)} The false color maps show the longitudinal resistivity at the same temperatures as in the top panels. The white curves correspond to the isolines of $\rho_{xx}=h/e^2$. The red dotted lines are the linear fits of these curves in the field range of $\mathcal{B}=$ 2.2--6~T. The crossing point of the two red lines gives another estimate for $\mathcal{B}_c$.}
\end{figure}

To experimentally demonstrate this phase transition, we performed  the measurements of quantum Hall effect and SdH oscillations on a gated Hall bar device. The device was fabricated from a (013)-oriented HgTe/Cd$_{0.65}$Hg$_{0.35}$Te QW of 6.5~nm width, grown on a CdTe buffer~\cite{HgTeQWs25}, and it had a total length of 650~$\mu$m and a width of 50~$\mu$m. The colormaps presented in Fig.~\ref{Chapter3Fig10}(a--d) were obtained by plotting the derivative $\partial\sigma_{xy}/\partial V_G$ as a function of $\mathcal{B}$ for different temperatures. Two zero-mode Landau levels are clearly resolved above $\mathcal{B}\simeq1.5$~T. They seem to emerge from the resistance maximum at $V_G= -1.7$~V, $\mathcal{B}=0$ T. The trace of the zero-mode Landau level originating from the H1 subband broadens and gradually fades with increasing temperature, most likely due to the strong temperature dependence of hole mobility associated with their large effective mass. It becomes almost indistinguishable above 20~K. Nevertheless, the zero-mode Landau levels coincide with the characteristic condition $\sigma_{xy}=\pm e^2/2h$ and $\rho_{xx}\simeq h/e^2$, in agreement with the semicircle relation $\sigma_{xx}^2+(\sigma_{xy}\mp1/2)^2=1/4$~\cite{HgTeQWs26}.

\begin{figure}
\centering
\includegraphics [width=0.5\columnwidth, keepaspectratio] {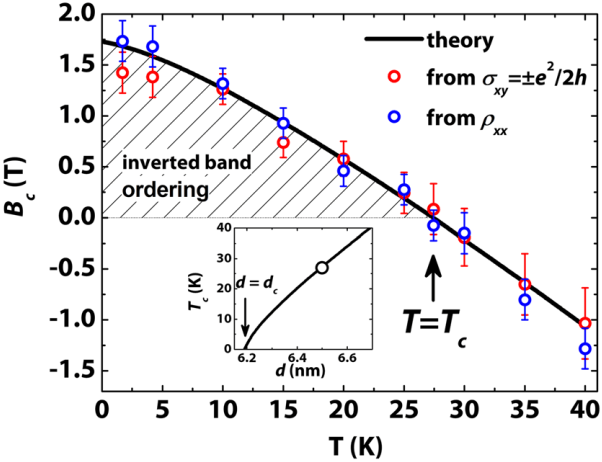}
\caption{\label{Chapter3Fig11} Theoretical (solid curve) and experimental (open symbols) values of critical magnetic field $\mathcal{B}_c$ as a function of temperature. The values marked by the red symbols are obtained from the data set in the top panels of Fig.~\ref{Chapter3Fig10}. The blue symbols are marked the values of $\mathcal{B}_c$ extracted from $\rho_{xx}$. The sparse and white regions correspond to the inverted and trivial band ordering, respectively. The inset shows $T_c$ as a function of $d$ in (013) HgTe/Cd$_{0.65}$Hg$_{0.35}$Te QWs grown on CdTe buffer. The open symbol represents experimental value measured in our sample.}
\end{figure}

In Fig.~\ref{Chapter3Fig10}(a--d), the white curves correspond to $\sigma_{xy}=(n+1/2)e^2/h$, where $n$ is an integer. These curves are well defined up to 40~K and accurately trace the Landau-level positions, in agreement with the maxima of $\partial\sigma_{xy}/\partial V_G$. The two zero-mode Landau levels deviate from lines of constant filling factor, indicating their partial overlap. As we have shown in our original paper~\cite{HgTeQWs25}, the critical magnetic field $\mathcal{B}_c$ can be estimated by linear extrapolation of the Landau-level positions toward lower magnetic fields. The red dashed lines in Fig.~\ref{Chapter3Fig10} indicate these extrapolations. At $T=1.7$~K, they intersect at $\mathcal{B}_c\simeq1.5$~T. As the temperature increases, the crossing point moves toward lower magnetic fields and eventually vanishes at $T\simeq27$~K. The crossing of the zero-mode Landau levels at zero magnetic field thus directly signals the emergence of a gapless critical state -- the hallmark of a temperature-driven topological phase transition. At higher temperatures, for instance at $T=40$~K (see Fig.~\ref{Chapter3Fig10}(d)), the extrapolated crossing occurs at negative magnetic fields, clearly indicating the formation of a trivial band insulator state. To further verify this conclusion, we also analyzed the longitudinal resistivity $\rho_{xx}$, represented in the lower panels of Fig.~\ref{Chapter3Fig10} as color-coded maps versus $\mathcal{B}$ and $V_G$ for temperatures up to 40~K. Although the complete Landau level fan chart is not as clearly visible as in the upper panels, the traces of the zero-mode Landau levels can still be identified at the edges of the main $\rho_{xx}$ peak. The red dotted lines mark the linear fittings of the isolines $\rho_{xx}=h/e^2$ within the range $\mathcal{B}=2.2$--6~T. Their crossing at lower fields provides another estimate for $\mathcal{B}_c$, consistent with the values extracted from $\sigma{xy}$.

The comparison between experimental and theoretical values of the critical magnetic field is summarized in Fig.~\ref{Chapter3Fig11}. Two independent sets of $\mathcal{B}c$ values, obtained from $\rho{xx}$ and $\sigma_{xy}$ analyses, show excellent quantitative agreement and reproduce the same temperature dependence. The critical field vanishes at $T_c\simeq27$~K, corresponding to the formation of gapless states, where the system effectively mimics massless Dirac fermions. Extending the comparison formally to negative $\mathcal{B}_c$ reveals the transition into the trivial band insulator regime. It should be noted that the experimental uncertainty in $\mathcal{B}_c$ increases with temperature, due to the broadening of the zero-mode Landau level trace from the H1 subband, as discussed above. In summary, these results provide direct and compelling experimental evidence of a temperature-induced topological phase transition in HgTe/CdHgTe QWs, which is in excellent quantitative agreement with our theoretical predictions made earlier~\cite{HgTeQWs10} (see also Section~\ref{sec:PandTinQWs}).

\section{\label{sec:HgTeDQWs}Topological states in double HgTe QWs: Bilayer graphene analogy and beyond}
As I briefly mentioned in Chapter~\ref{Chap:Career}, my interest in topological states began with conceptual studies of HgTe double QWs, consisting of two HgTe layers separated by a thin CdHgTe barrier. My attention to this system was initially drawn by two remarkable works~\cite{doubleHgTeQWs1,doubleHgTeQWs2} published by Paolo Michetti and co-authors from the Institute of Theoretical Physics and Astrophysics at the University of W\"{u}rzburg. In these papers, the authors developed the following approach to describe the electronic states in such a system in the vicinity of the $\Gamma$~point of the Brillouin zone -- each HgTe QW (front ``f'' and back ``b'') is described by the Bernevig--Hughes--Zhang (BHZ) Hamiltonian~\cite{HgTeQWs2}, while the coupling between the two layers is introduced through an additional tunneling Hamiltonian. In the basis states \{$|Ef,+\rangle$, $|Hf,+\rangle$, $|Eb,+\rangle$, $|Hb,+\rangle$, $|Ef,-\rangle$, $|Hf,-\rangle$, $|Eb,-\rangle$, $|Hb,-\rangle$\}, the effective $8\times8$ Hamiltonian for the DQW proposed by Michetti \emph{et~al.} can be written as~\cite{doubleHgTeQWs1,doubleHgTeQWs2}:
\begin{equation}
\label{eq:DQW1}
H_{\mathrm{Michetti}}(\mathbf{k})=\begin{pmatrix}
H_{\mathrm{DQW},\uparrow}(k_x,k_y) & 0 \\ 0 & H_{\mathrm{DQW},\downarrow}(k_x,k_y)\end{pmatrix},
\end{equation}
where $H_{\mathrm{DQW},\downarrow}(k_x,k_y)=H_{\mathrm{DQW},\uparrow}^{*}(-k_x,-k_y)$ with the asterisk denoting the complex conjugation due to the presence of time-reversal symmetry. The Hamiltonian for the spin up block has then the following structure:
\begin{equation}
\label{eq:DQW2}
H_{\mathrm{DQW},\uparrow}(\mathbf{k})=\begin{pmatrix}
h_{0}^{(f)}(\mathbf{k}) & h_{\mathrm{T}}(\mathbf{k}) \\ h_{\mathrm{T}}^{\dag}(\mathbf{k}) & h_{0}^{(b)}(\mathbf{k}) \end{pmatrix},
\end{equation}
where ``$\dag$'' represents the Hermitian conjugation, the tunneling Hamiltonian $h_{\mathrm{T}}(\mathbf{k})$ is given by
\begin{equation}
\label{eq:DQW3}
h_{\mathrm{T}}(\mathbf{k})=\begin{pmatrix}
\Delta_{E1} & \alpha(k_x+ik_y) \\ \alpha(k_x-ik_y) & \Delta_{H1} \end{pmatrix},
\end{equation}
and $h_{0}^{(f,b)}(\mathbf{k})$ are written as
\begin{equation}
\label{eq:DQW4}
h_{0}^{(f,b)}(\mathbf{k})=\left[C^{(f,b)}-D^{(f,b)}\mathbf{k}^2\right]\cdot{1_{2\times2}}+
\begin{pmatrix}
M^{(f,b)}-B^{(f,b)}\mathbf{k}^2 & A^{(f,b)}(k_x+ik_y) \\ A^{(f,b)}(k_x-ik_y) & -M^{(f,b)}+B^{(f,b)}\mathbf{k}^2 \end{pmatrix}.
\end{equation}
In the equations above, $k_x$ and $k_y$ are the quasimomentum components in the QW plane, while $C^{(f,b)}$, $M^{(f,b)}$, $A^{(f,b)}$, $B^{(f,b)}$, $D^{(f,b)}$, $\Delta_{E1}$, $\Delta_{H1}$ and $\alpha$ are the structure parameters being defined by the geometry of double HgTe QW~\cite{doubleHgTeQWs1,doubleHgTeQWs2}. For the reasons that will become clear below, I intentionally reproduce here the full form of the Hamiltonian proposed by Michetti \emph{et~al.} for describing the low-energy physics of double HgTe QWs. It is also important to recall that the states $Ef$ and $Eb$, as well as $Hf$ and $Hb$, correspond to the $E1$ and $H1$ subbands localized in the front ``f'' and back ``b'' HgTe layers, respectively.

Let me now emphasize two points that puzzled me, when I first encountered the Hamiltonian $H_{\mathrm{Michetti}}(\mathbf{k})$. The first issue concerns the fact that, in the model proposed by Michetti \emph{et~al.}, the tunneling coupling between the states in the two QWs is assumed to be spin-independent. This can be clearly seen from the block-diagonal structure of $H_{\mathrm{Michetti}}(\mathbf{k})$ in Eq.~(\ref{eq:DQW1}). At that time, this seemed rather strange to me, since spin-orbit interaction in materials such as HgTe is known to be extremely strong. In the presence of spin-orbit coupling, all tunneling processes are expected to be inherently spin-dependent~\cite{doubleHgTeQWs3,doubleHgTeQWs4,doubleHgTeQWs5}. The absence of spin-dependent tunneling in $H_{\mathrm{Michetti}}(\mathbf{k})$ thus immediately attracted my attention and motivated me to explore this point in more depth. The second aspect that raised my doubts in the Michetti model is the assumption that the $Ef$ and $Eb$ (as well as $Hf$ and $Hb$) states possess the same parity. As one can verify, in $h_{\mathrm{T}}(\mathbf{k})$ from Eq.~(\ref{eq:DQW3}) and $h_{\mathrm{T}}(\mathbf{k})$ from Eq.~(\ref{eq:DQW4}), the basis states of different parity are coupled through matrix elements proportional to $k_x$ and $k_y$, while those of the same parity are connected by a linear combination of constants and $\mathbf{k}^2$-dependent terms. The identical parity of the $Ef$ and $Eb$ states (as well as $Hf$ and $Hb$) thus served as a clear indication to me that this model might encounter difficulties when applied to realistic double HgTe QWs. Indeed, when tunneling between coupled QWs is properly taken into account, the functions
$|Ef,+\rangle$, $|Hf,+\rangle$, $|Eb,+\rangle$, $|Hb,+\rangle$, $|Ef,-\rangle$, $|Hf,-\rangle$, $|Eb,-\rangle$, and $|Hb,-\rangle$,
localized in the ``f'' and ``b'' layers, do not form an orthogonal basis. The correct orthogonal basis should instead be constructed from symmetric and antisymmetric linear combinations of the envelope functions localized in each QW. This point is discussed in more detail in the Supplementary Materials of our original paper~\cite{doubleHgTeQWs6}.

These two conceptual inconsistencies in the Michetti model became my main motivation to undertake a more systematic study of the double HgTe QW system. The results of this research, published in our two subsequent papers~\cite{doubleHgTeQWs6,doubleHgTeQWs7}, will be presented and discussed in the following subsections.

\subsection{\label{sec:1storderHgTe}Topological phase diagram published in 2016}
In contrast to the perturbative approach proposed by Michetti \emph{et~al.}, the starting point of my own study was a direct and non-perturbative calculation of the electronic band structure, performed within the framework of the eight-band $\mathbf{k}\cdot\mathbf{p}$ Kane Hamiltonian~\cite{HgTeQWs1}. In this approach, the double HgTe DQW, consisting of two HgTe layers of equal thickness $d$, separated by a Cd$_{0.7}$Hg$_{0.3}$Te barrier of width $t$, was treated as a single coherent system. The corresponding energy dispersions of electronic subbands were obtained by solving the eigenvalue problem for the full structure. In the following discussion, we will consider the double QW grown along the (001) crystallographic direction on a CdTe buffer, with the entire stack enclosed between top and bottom Cd$_{0.7}$Hg$_{0.3}$Te barriers.

The calculated positions of the electron-like and hole-like subbands at $\mathbf{k}=0$ as a function of the middle barrier thickness $t$ are shown in Fig.~\ref{Chapter3Fig12}(a,b). From a qualitative point of view, one can easily understand the evolution of the subband structure with changing $t$. At infinitely large barrier thickness, the states in the system correspond to those of two independent HgTe QWs, fully decoupled from each other. At finite $t$, however, each pair of subbands hybridizes, giving rise to an even--odd splitting characteristic of tunnel-coupled wells. As $t$ decreases further and the two HgTe layers start to merge, the energy levels gradually approach those of a single HgTe QW of effective width $2d$. It is clearly seen from Fig.~\ref{Chapter3Fig12}(a,b) that the splitting between the electron-like levels is substantially larger than that for the hole-like ones. This difference originates from the strong asymmetry in the effective masses of the corresponding states at $\mathbf{k}=0$, which determines both the tunneling strength and the magnitude of the even--odd splitting at a given barrier thickness $t$. Depending on the width $d$ of the HgTe layers, two qualitatively distinct types of band ordering emerge, reflecting the transition between normal and inverted regimes in the double QW system.

\begin{figure}
\centering
\includegraphics [width=0.8\columnwidth, keepaspectratio] {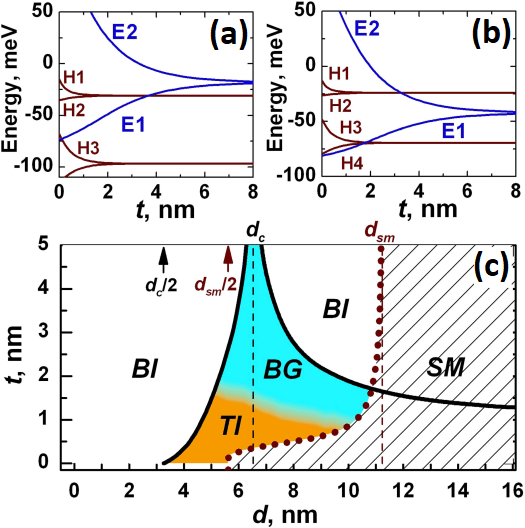}
\caption{\label{Chapter3Fig12} \textbf{(a,b)} Energy of E1, E2 (both in blue) and H1, H2 (both in red) bands at $\mathbf{k}=0$ versus barrier thickness $t$ at different QW width $d$: (a) $d_c/2<d<d_c$ and (b) $d>d_c$. \textbf{(c)} Phase diagram of double HgTe QW.
The values $d_c$ and $d_{sm}$ correspond to thickness of the single QW, at which Dirac cone and semimetal state arise, respectively (see Fig.~\ref{Chapter3Fig1}). The white-open regions are the band insulator state, while the white-striped region corresponds
to the semimetal state, when the side maxima of valence subband exceed the bottom of conduction subband. The orange and blue regions conform to 2D topological insulator (or equivalently QSHI) and BG phase, respectively. The bold black curves correspond to the arising of the Dirac cone at the $\Gamma$ point.}
\end{figure}

There are two types of the band structure ordering, which arises depending on the HgTe layers thickness $d$. The first case shown in Fig.~\ref{Chapter3Fig12}(a) takes place if $d$ varies in the range from $d_c/2$ to $d_c$, where $d_c$ corresponds to the thickness of single QW with Dirac cone at the $\Gamma$ point. In this case E2 subband always lies above H2 subband and inversion between E1 and H1 subband can take place. At higher values of $t$, the E1 subband gains a higher energy than the H1 subband, and the system has the normal band structure. When the thickness $t$ is decreased, the energy of the E1 subband reduces, whereas the H1 subband energy practically does not change. The different dependence of E1 and H1 subbands on $t$ implies that at some barrier thicknesses the band gap closes. In fact, the crossing point between E1 and H1 subbands yields a low-energy band structure with Dirac cone in the vicinity of $\mathbf{k}=0$. If the values of $t$ are small enough to induce a gap between H1 and H2 subbands, it implies a quantum phase transition between trivial band insulator (BI) and 2D TI states, as in single HgTe QW~\cite{HgTeQWs2}. However, if H1 and H2 subbands coincide at $\mathbf{k}=0$, it gives rise to an additional ``massive'' branch of valence subband at the crossing point, similar to the one of 3D Kane fermions in CdHgTe crystals shown in Fig.~\ref{Fig:Chapter2Fig1}, as well as a quantum phase transition into a specific metal phase with the following subband ordering: E2-H1-H2-E1. The reasons to assign this metal phase as a bilayer graphene (BG) phase are discussed later. The second type of band structure ordering, shown in Fig.~\ref{Chapter3Fig12}(b), is realized when $d >d_c$. In this case E1 subband always lies below H1 subband, i.e. they are inverted. At lower barrier thickness, E2 subband lies above H2 subband, while at high values of $t$, they swap their positions. At specific values of $t$, E2 and H2 subbands cross, which also leads to the appearance of a Dirac cone at the $\Gamma$ point. As for the case shown in Fig.~\ref{Chapter3Fig12}(a), additional ``massive'' branch arises as well. However, such branch corresponds to conduction H1 subband. A further increase in both the HgTe layer thickness $d$ and the barrier width $t$ leads to the band inversion between the E2 and H2 subbands.

~\\~
\fbox{
\begin{minipage}{0.975\textwidth}
    Importantly, in our earlier work published in 2016~\cite{doubleHgTeQWs6}, such a configuration was referred to as a trivial band insulator with a ``double inversion'' of the band structure. At that time, it appeared natural to classify it as trivial, since the total topological $\mathbb{Z}_2$ invariant vanishes when the parity of inverted states changes twice. However, later studies -- in particular, the 2021 paper~\cite{doubleHgTeQWs7} -- revealed that this regime actually exhibits properties of higher-order topology, rooted in the underlying cubic symmetry of the crystal lattice. A detailed discussion of these results, along with their theoretical interpretation, will be presented in Section~\ref{sec:HOTIHgTe}. For the sake of consistency with our earlier terminology, and to maintain continuity in the narrative of this section, we will nevertheless continue to refer to the state with a double inversion between the E2 and H2 subbands as a trivial band insulator, in accordance with our original 2016 classification.
\end{minipage}%
}~\\~

Figure~\ref{Chapter3Fig12}(c) shows the phase diagram, in which two bold lines correspond to the Dirac cones at the $\Gamma$ point. The left one results from the crossing between E1 and H1 subbands, while the right-side curve is connected with the crossing of E2 and H2 levels. If the middle barrier is thin enough, a gap between H1 and H2 subbands opens, and the inversion between E1 and H1 levels induces a quantum phase transition between BI and TI phases. The latter is shown by orange region. For relatively wide QWs, the so-called semimetal (SM) phase, corresponding to the white-striped region, is implemented. It is characterized by a vanishing indirect band gap, when the side maximum of the valence subband exceeds in energy the conduction subband bottom. This phase arises in single
HgTe QWs as well when the thickness goes beyond some critical value (see Fig.~\ref{Chapter3Fig1}), denoted in Fig.~\ref{Chapter3Fig12}(c) by $d_{sm}$. Bilayer graphene metal phase mentioned above corresponds to the blue region in Fig.~\ref{Chapter3Fig12}(c).

Let me now explain the reasons to call this specific metal phase with the ordering of electronic subbands E2-H1-H2-E1 a BG phase. First, we consider in details the case when both QWs have a critical thickness $d_c\simeq6.5$~nm, which at infinitely large barrier corresponds to two Dirac cones. The presence of the transparent barrier, for example of $t=3$~nm, turns from two Dirac cones in the vicinity of the $\Gamma$ point into a band structure very similar to the one of natural BG~\cite{NaturalBG1} (c.f. Fig.~\ref{Chapter3Fig13}(a)). In particular, it consists of two gapless isotropic parabolas, formed by H1 and H2 subbands. Moreover, as in BG in which non-zero band gap can be induced by breaking the inversion symmetry of two monolayers, in double HgTe QWs it could be obtained by using QWs of different thickness or by adding of one-side chemically doping. Moreover, the potential of a continuously tunable band gap through an electric field applied perpendicularly to the sample plane is of particular importance.

\begin{figure}
\centering
\includegraphics [width=0.51\columnwidth, keepaspectratio] {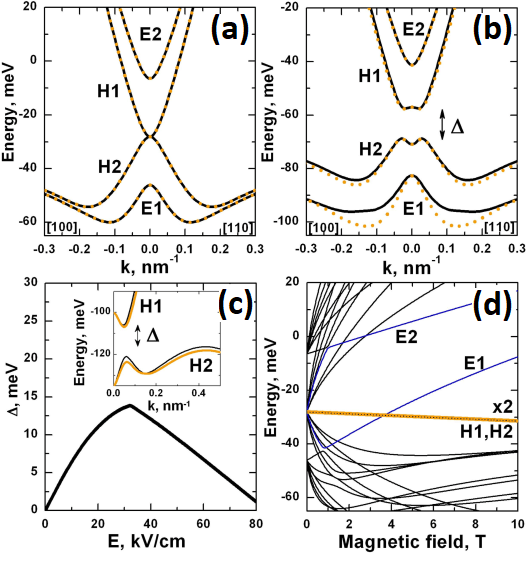}
\caption{\label{Chapter3Fig13} \textbf{(a,b)} Energy dispersions for BG phase, implemented at $d=d_c\simeq6.5$~nm and $t=3$ nm, in zero electric field (a) and in electric field of 20 kV/cm (b), oriented perpendicular the QW plane. Solid black and dotted orange curves correspond to different spin states. The presence of electric field not only opens the band gap $\Delta$ but also leads to the Rashba spin splitting. \textbf{(c)} Bulk band gap as a function of applied electric field. As it is in natural BG~\cite{NaturalBG2}, double HgTe QW in this phase has also electrically-tunable band gap. However, the presence of additional side maximum in the valence subband closes the gap in high electric field. The inset shows energy dispersions in electric field of 50 kV/cm. Solid black and orange curves correspond to different spin states. \textbf{(d)} The Landau-level fan chart for BG phase. The zero-mode Landau level, which has doubled degeneracy order as compared with other levels, is marked by bold orange line. This LL is formed by states of both H1 and H2 subbands. Landau levels, containing only the states from E1 and E2 subbands in high magnetic fields, are given in blue. The crossing between the E1 and zero-mode Landau levels, arising at critical magnetic field $B_c\simeq3.5$~T, leads to the phase
transition into normal (non-inverted) band structure, as it is in single HgTe QW.}
\end{figure}

The double HgTe QW in BG phase also holds this property. Figure~\ref{Chapter3Fig13}(b) displays energy dispersion for BG phase in perpendicular electric field of 20 kV/cm. Even in this case the dispersion curves are very similar to the band structure of natural BG in external electric field~\cite{NaturalBG1,NaturalBG2}. However, strong spin-orbit interaction in HgTe layers removes the spin degeneracy away from the $\Gamma$ point due to Rashba effect. Figure~\ref{Chapter3Fig13}(c) shows the band-gap values in the double QWs in BG phase as a function of applied electric field. Indeed, the band gap in BG phase is electrically-tunable, as it is in natural BG. However, its dependence on electric field has non-monotonic behaviour in double HgTe QWs. The reason for this is related with the additional side maximum of valence subband, which is also responsible for the formation of SM phase in zero electric field (see the diagram, shown in Fig.~\ref{Chapter3Fig13}(c)). The top of the side maximum is increasing with the strength of electric field, and the band gap reduces (see the inset, shown in Fig.~\ref{Chapter3Fig13}(c)). As a result, the band-gap is closed in high enough electric field, giving rise to the formation of the SM phase. A critical electric field, corresponding to arising of indirect-band gap, coincides with the maximum value of $\Delta$.

Another characteristic of natural BG is the unconventional quantum Hall effect, related with the absence of zero-Landau-level plateaus in Hall conductivity~\cite{NaturalBG3,NaturalBG4}. For natural BG, plateaus in the Hall conductivity $\sigma_{xy}$, occur at integer multiples of $4e^2/h$. This is similar to a conventional semiconductor with level degeneracy $g=4$ arising from the spin and valley degrees of freedom. Deviation from the conventional case occurs at low density, where there is a step in $\sigma_{xy}$ of height $8e^2/h$ across zero density, arising from the eightfold degeneracy of the zero-energy Landau level. This specific Landau level is formed by atomic orbitals of different sublattice sites from both layers~\cite{NaturalBG5}. Figure~\ref{Chapter3Fig13}(d) shows Landau level fan chart in double HgTe QW with BG phase. In the BG phase, in perpendicular magnetic field a specific zero-mode Landau level with degeneracy two times higher than other Landau levels also arises. This zero-mode Landau level is shown in Fig.~\ref{Chapter3Fig13}(d), by bold orange line. For double HgTe QW, plateaus in the Hall conductivity are expected to occur at integer numbers of $e^2/h$. However, the doubled degeneracy of zero-mode Landau level requires twice as many carriers to fill them, so the transition between the corresponding plateaus should be twice as wide in density, and the step in $\sigma_{xy}$ between the plateaus are expected to be twice as high, $2e^2/h$ instead of $e^2/h$. Since the zero-mode Landau level in double HgTe QW is formed by the states of both H1 and H2 subbands, the presence of inversion asymmetry of two HgTe layers not only opens the band gap between H1 and H2 subbands (see Fig.~\ref{Chapter3Fig13}(b)) but also splits the zero-mode Landau levels, removing the double degeneracy order. Therefore, one can also expect recovering of sequence of equidistant plateaus in the Hall conductivity as it is for gate-biased natural BG~\cite{NaturalBG4}. Besides the zero-mode Landau level, there are two additional specific Landau levels, which in high magnetic fields are formed only by the states from E1 and E2 subbands. These Landau levels are shown in Fig. 3D by blue curves. In moderate magnetic fields, these E1 and E2 Landau levels are mixed with the states from H2 and H1 subbands respectively. It results in anticrossing between E1 (E2) Landau level and Landau levels from H2 (H1) subband, which is clearly seen in magnetic fields below 1~T. The crossing between the E1 and zero-mode Landau level, arising at critical magnetic field $B_c\simeq3.5$ T, corresponds to the transition from inverted into normal band structure, similar to this observed in single HgTe QW in Section~\ref{sec:TransportHgTeQWs}.

We have considered in details the case of double HgTe QW, when both HgTe layers have a critical thickness $d_c$. However, all the mentioned properties of BG phase hold for any double QW with the values of $d$ and $t$, corresponding to the blue region in Fig.~\ref{Chapter3Fig12}(c). The only difference is the ratio $M_1$ over $M_2$, where $M_1$ parameter describes the energy gap between E1 and H1 subbands at the $\Gamma$ point, while $M_2$ corresponds to the half of the gap between E2 and H2 levels. If $d<d_c$, $2|M_2|$ exceeds $2|M_1|$, while in the opposite case of $d>d_c$, the gap between E2 and H2 subbands is lower than $2|M_1|$. An amazing property of double HgTe QW is that it shares some characteristics of natural BG even in the BI phase at $d>d_c$. In particular, the possibility of tuning the band gap by electric field and of observing the unconventional double step in plateaus in the Hall conductivity still persists.

To gain a deeper understanding of the quantum phase transitions in double HgTe QWs, in our work~\cite{doubleHgTeQWs6} we derived, in a fully explicit manner, an effective two-dimensional Hamiltonian $H_{2D}(\mathbf{k})$ directly from the realistic eight-band $\mathbf{k}\cdot\mathbf{p}$ Kane model. This Hamiltonian captures the low-energy physics in the vicinity of the $\Gamma$ point for the different topological states presented in Fig.~\ref{Chapter3Fig12}(c). In the basis $|E1{+}\rangle$, $|H1{+}\rangle$, $|H2{-}\rangle$, $|E2{-}\rangle$, $|E1{-}\rangle$, $|H1{-}\rangle$, $|H2{+}\rangle$, $|E2{+}\rangle$, the effective 2D Hamiltonian can be presented in the form:
\begin{equation}
\label{eq:DQW5}
H_{2D}(k_x,k_y)=\begin{pmatrix}
H_{4\times4}(k_x,k_y) & 0 \\ 0 & H_{4\times4}^{*}(-k_x,-k_y)\end{pmatrix},
\end{equation}
where the asterisk denotes complex conjugation, and $H_{4\times4}(k_x,k_y)$ is written as:
\begin{equation}
\label{eq:DQW6}
H_{4\times4}(k_x,k_y)=\begin{pmatrix}
\epsilon_{E1} & -A_{1}k_{+} & R_{1}k_{-}^2 & S_{0}k_{-}\\
-A_{1}k_{-} & \epsilon_{H1} & 0 & R_{2}k_{-}^2\\
R_{1}k_{+}^2 & 0 & \epsilon_{H2}  & A_{2}k_{+}\\
S_{0}k_{+} & R_{2}k_{+}^2 & A_{2}k_{-} & \epsilon_{E2} \end{pmatrix},
\end{equation}
where
\begin{eqnarray}
\label{eq:DQW7}
\epsilon_{E1}(k_x,k_y)=C_1+M_1-(D_{1}+B_{1})(k_x^2+k_y^2),~\nonumber\\
\epsilon_{H1}(k_x,k_y)=C_1-M_1-(D_{1}-B_{1})(k_x^2+k_y^2),~\nonumber\\
\epsilon_{E2}(k_x,k_y)=C_2+M_2-(D_{2}+B_{2})(k_x^2+k_y^2),~\nonumber\\
\epsilon_{H2}(k_x,k_y)=C_2-M_2-(D_{2}-B_{2})(k_x^2+k_y^2).~
\end{eqnarray}
Here, $C_{1,2}$, $M_{1,2}$, $A_{1,2}$, $B_{1,2}$, $D_{1,2}$, $S_0$ and $R_{1,2}$ are structure parameters being defined by the QW geometry and materials. The Hamiltonian $H_{2D}(k_x,k_y)$ has block-diagonal form because we kept the inversion symmetry and axial symmetry around the growth direction. Note that $H_{2D}(k_x,k_y)$ is valid for any values of $d$ and $t$. In particular for $ t=0$, it essentially describes energy dispersion in the vicinity of the $\Gamma$ point in single HgTe QW beyond the BHZ model~\cite{HgTeQWs2}.

The parameters $R_1$, $R_2$, and $S_0$ exhibit a strong dependence on the barrier thickness~$t$, gradually tending to zero as the middle CdHgTe barrier becomes sufficiently wide. In this limiting case, the double QW system effectively decouples into two independent subsystems, each described by its own BHZ Hamiltonian written for the pairs of subbands E1, H1 and E2, H2. It is important to emphasize that the structure of the effective Hamiltonian $H_{2D}(\mathbf{k})$, which we have derived directly from the realistic eight-band $\mathbf{k}\cdot\mathbf{p}$ Kane model, differs substantially from $H_{\mathrm{Michetti}}(\mathbf{k})$ introduced by Michetti~\emph{et~al.} in Eq.~(\ref{eq:DQW1}).
As the reader can readily observe, each of the Kramers blocks \lbrack$|E1{+}\rangle$, $|H1{+}\rangle$, $|H2{-}\rangle$, $|E2{-}\rangle$\rbrack~and \lbrack$|E1{-}\rangle$, $|H1{-}\rangle$, $|H2{+}\rangle$, $|E2{+}\rangle$\rbrack~contains basis states with different spin orientations. This feature leads to the appearance of a spin-dependent tunneling coupling between the two BHZ-like subsystems, in sharp contrast to the spin-independent tunneling assumed in the model of Michetti~\emph{et~al.}~\cite{doubleHgTeQWs1,doubleHgTeQWs2}. From the physical point of view, such spin-dependent coupling naturally arises from the strong spin-orbit interaction inherent to HgTe-based heterostructures, and it plays a crucial role in determining the fine structure of the low-energy band spectrum in various phases in double QWs.

The most important quantities in $H_{2D}(k_x,k_y)$ are two mass parameters $M_1$ and $M_2$.  Almost all phases in Fig.~\ref{Chapter3Fig12}(c) can be grouped into three types according to the sign of $M_1$ and $M_2$. The trivial BI corresponds to positive values of $M_1$ and $M_2$. The 2D TI (or equivalently, QSHI) and BG states arise if $M_1<0$ and $M_2>0$, and the difference between these states is defined by the gap between H1 and H2 subbands, which is zero in the case of BG state. The insulator state with double band inversion is defined by the negative values of $M_1$ and $M_2$. The SM phase, arising at all values of $M_2$ and $M_1>0$, cannot be described within the simplified model because it is only valid near the $\Gamma$ point, while the SM phase is formed by non-local overlapping of the valence band and conduction band bottom. Figure~\ref{Chapter3Fig14} presents energy dispersion for the bulk and edge states in various phases in double HgTe QWs, obtained within the simplified model. Dispersion of the bulk states are shown in black, while the orange and blue curves are the edge states, described by different blocks of $H_{2D}(\mathbf{k})$ in Eq.~(\ref{eq:DQW5}). The electrons in the edge states, marked by different colors, move in opposite directions~\cite{doubleHgTeQWs6}. The values of structure parameters used in the calculations can be found in our original work~\cite{doubleHgTeQWs6}. Figure~\ref{Chapter3Fig14} demonstrates that each change of sign of mass parameter $M_1$ or $M_2$ yields a pair of helical edge states, providing a quantum phase transition at $M_1=0$ or $M_2=0$. In particular, BG phase arising at $M_1<0$ and $M_2>0$ has a single pair of the edge states, coexisting with the gapless bulk states. If $M_1<0$ and $M_2=0$, the double HgTe QW mimics a 2D system with the presence of both bulk and edge massless fermions, whose energy dispersions are shown in Fig.~\ref{Chapter3Fig14}(c).

\begin{figure}
\centering
\includegraphics [width=0.85\columnwidth, keepaspectratio] {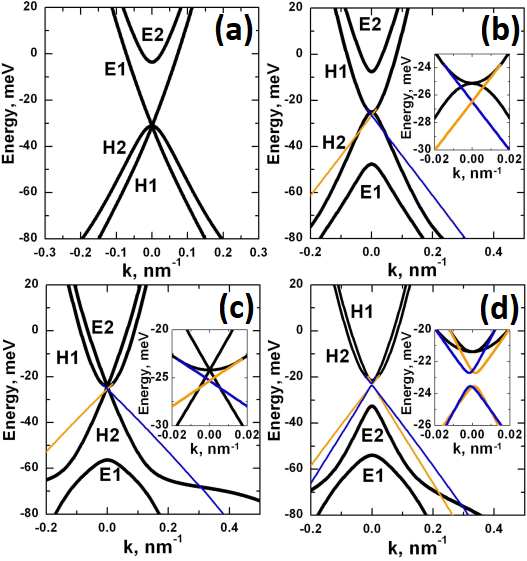}
\caption{\label{Chapter3Fig14} \textbf{(a-d)} Energy dispersions, calculated by using effective 2D Hamiltonian for (a) $M_1=0$ and $M_2>0$, (b) $M_1<0$ and $M_2>0$ (BG phase), (c) $M_1<0$ and $M_2=0$, (d) $M_1<0$ and $M_2<0$. Other band parameters are provided in Ref.~\cite{doubleHgTeQWs6}. Bulk states are shown in black. Orange and blue curves correspond to the dispersion of the edge states, obtained with open boundary conditions. Kramer's partners of the edge states, moving in opposite directions, are shown in different colors. The insets show the behavior of the dispersion curves in the vicinity of $\mathbf{k}=0$. It is seen from (a) and (c) that $M_1=0$ or $M_2=0$ corresponds to topological phase transition with arising of Dirac cone at the $\Gamma$ point of the Brillouin zone, while each negative values of $M_1$ and $M_2$ results in appearance of additional pair of the helical edge states. The BG phase (b) with $M_1<0$ and $M_2>0$ is nontrivial and characterized by coexistence of gapless bulk states with single pair of the helical edge states. The case of $M_1<0$ and $M_2<0$ shown in (d) is characterized by the presence of two pairs of the edge states. The values of structure parameters used in the calculations can be found in our original work~\cite{doubleHgTeQWs6}.}
\end{figure}

Figure~\ref{Chapter3Fig14}(d) shows a special case of the double band inversion. It perfectly illustrates that each crossing between electron-like and hole-like levels of higher indexes, also results in the appearance of a pair of helical edge states. In accordance with ``conventional'' $\texttt{Z}_2$ classification, the topological properties of corresponding insulator phase are determined by amount of inverted levels, which are connected with the number of pairs of the edge states. In the particular case of inversion of both E1 with H1 subband and E2 with H2 subband, the terms proportional to $R_1$, $R_2$ and $S_0$ induce the coupling between counter-propagating states with the same spin orientation, which could be interpreted as a spin-dependent tunneling between two layers of 2D TI . It is clear from the inset that such tunnel-coupling opens the gap in the energy spectrum of the edge states even without external one-particle scattering processes.

We will cover this case in more detail in the next section, but for now let me remark the main difference between BG phase in double HgTe QW and natural BG. The electrons in natural BG are chiral particles with $S=1/2$ pseudospin of freedom and pseudospin winding number of 2~\cite{NaturalBG3,NaturalBG4,NaturalBG5}. The later results from the commutation of operator $\mathbf{k}\cdot\mathbf{S}$ with the low-energy Hamiltonian of natural BG. As it is shown above, the simplest Hamiltonian, required for the description of bulk and edge states in the BG phase, has $J=3/2$ pseudospin of freedom, however the operator $\mathbf{k}\cdot\mathbf{J}$ does not commute with $H_{4\times4}(k_x,k_y)$ in Eq.~(\ref{eq:DQW6}), proving the non-chiral character of electron in double HgTe QWs. Thus, even though BG phase mimics some characteristics of natural BG, it is a different state of matter, exhibiting the coexistence of gapless bulk states and spin-polarized edge channels.

In conclusion, I would like to emphasize that the theoretical results presented in this section have served as a strong motivation for subsequent experimental studies of double HgTe QWs. In particular, based on the phase diagram shown in Fig.~\ref{Chapter3Fig12}, a series of double QW samples, corresponding to different topological regimes, were grown in Novosibirsk. These samples provided a unique platform to experimentally explore the rich variety of electronic phases predicted theoretically. The grown structures were later investigated by several research teams worldwide, including the THz team in Montpellier, using both magnetotransport measurements~\cite{Tr1,Tr2,Tr3,Tr4,Tr5,Tr6,Tr7} and mid-infrared Landau-level spectroscopy~\cite{LL1,LL1b,LL2,LL3,LL4}.

\subsection{\label{sec:HOTIHgTe}Higher-order topological insulator predicted in 2021}
As discussed in the previous section, the phase diagram of double HgTe QWs reveals a remarkable variety of electronic states, arising from the mutual inversion of subbands originating from two tunnel-coupled HgTe layers. Among them, there is a specific configuration, when both E1 and H1, as well as E2 and H2, subbands are inverted. Initially, such a state was regarded as a trivial band insulator, since no gapless edge states were expected to appear within the conventional $\mathbb{Z}_2$ classification. However, my subsequent theoretical studies~\cite{doubleHgTeQWs7} have demonstrated the incompleteness of this assumption. By taking into account the intrinsic cubic symmetry of HgTe-based heterostructures, I have shown that this ``trivial'' insulating phase actually realizes a 2D higher-order topological insulator (HOTI) state, characterized by the zero-dimensional (0D) corner states, induced by the cubic symmetry. The present section briefly overviews these results for the case of (001) double HgTe QWs, while a more general discussion of (0mn)-grown double QWs can be found in the original paper~\cite{doubleHgTeQWs7}.

The presence of cubic symmetry means that the diagonal blocks of $H_{2D}(\mathbf{k})$ in Eq.~(\ref{eq:DQW5}) are now split into isotropic and anisotropic parts:
\begin{equation}
\label{eq:HOTI2}
H_{4\times4}(k_x,k_y)=H_{4\times4}^{(i)}(k_x,k_y)+H_{4\times4}^{(a)}(k_x,k_y).
\end{equation}
The isotropic part $H_{4\times4}^{(i)}(k_x,k_y)$ is written as
\begin{equation}
\label{eq:HOTI3}
H_{4\times4}^{(i)}(k_x,k_y)=\begin{pmatrix}
\epsilon_{E1} & -A_{1}k_{+} & R_{1}^{(i)}k_{-}^2 & S_{0}k_{-}\\
-A_{1}k_{-} & \epsilon_{H1} & 0 & R_{2}^{(i)}k_{-}^2\\
R_{1}^{(i)}k_{+}^2 & 0 & \epsilon_{H2}  & A_{2}k_{+}\\
S_{0}k_{+} & R_{2}^{(i)}k_{+}^2 & A_{2}k_{-} & \epsilon_{E2} \end{pmatrix}
\end{equation}
with $\epsilon_{E1}(\mathbf{k})$, $\epsilon_{H1}(\mathbf{k})$, $\epsilon_{E2}(\mathbf{k})$ and $\epsilon_{H2}(\mathbf{k})$ are the same as in Eq.~(\ref{eq:DQW7}).
Here, $A_{1,2}$, $S_0$ and $R_{1,2}^{(i)}$ are now \emph{isotropic} structure parameters being defined by the QW geometry, the growth QW orientation and the materials. The isotropic term $H_{4\times 4}^{(i)}(k_x,k_y)$ preserves the rotational symmetry in the QW plane, therefore it is independent of the orientation of $x$ and $y$ axis.

In contrast, the form of the anisotropic term $H_{4\times4}^{(a)}(k_x,k_y)$ in Eq.~(\ref{eq:HOTI2}), resulting from the cubic symmetry of diamond and zinc-blende semiconductors, depend not only on the QW growth orientation but also on the orientation of $x$ and $y$ axis. For ($001$)-oriented QWs, $H_{4\times4}^{(a)}(k_x,k_y)$ has the form
\begin{equation}
\label{eq:HOTI5}
{H}_{4\times4}^{(a)}({k}_x,{k}_y)=
-\begin{pmatrix}
0 & 0 & R_{1}^{(a)}e^{i4\varphi}{k}_{+}^2 & 0\\
0 & 0 & 0 & R_{2}^{(a)}e^{i4\varphi}{k}_{+}^2\\
R_{1}^{(a)}e^{-i4\varphi}{k}_{-}^2 & 0 & 0  & 0\\
0 & R_{2}^{(a)}e^{-i4\varphi}{k}_{-}^2 & 0 & 0 \end{pmatrix}
\end{equation}
where $R_{1,2}^{(a)}$ are \emph{cubic} structure parameters, which depend on the QW geometry and materials, while the angle $\varphi$ is the angle between the $x$ axis and the ($001$) crystallographic direction.

The cubic symmetry, represented by the term ${H}_{4\times4}^{(a)}({k}_x,{k}_y)$ in Eq.~(\ref{eq:HOTI5}), leads to the appearance of 0D corner states at the junctions of edges oriented along specific crystallographic directions. To explicitly demonstrate it, one should construct an effective 1D edge Hamiltonian, in which the parameters explicitly depend on the crystallographic orientation of the edge. To this end, we decompose the full Hamiltonian $H_{2D}(\mathbf{k})$ into two parts. The first part represents two independent BHZ-like models with $M_1<0$ and $M_2<0$, corresponding respectively to the pairs of $E1$--$H1$ and $E2$--$H2$ subbands. These two decoupled blocks describe inverted-band systems that would each individually host 1D helical edge channels. The second part contains the remaining isotropic and cubic terms, which introduce the coupling between the two BHZ blocks and thereby opens a gap for the edge states. Assuming open-boundary conditions in a semi-infinite plane $y>0$, one first should solve the eigenvalue problem for the independent BHZ-like blocks to find the edge wave-functions at $k_x=0$. In this case, the edge orientation represented by the $x$ axis is defined by the angle $\varphi$ measured from ($100$) crystallographic direction. Finally, a low-energy 1D edge Hamiltonian is constructed by projecting $H_{2D}(\mathbf{k})(k_x,k_y)$ with non-zero $k_x$ onto the obtained set of the basis edge functions.

\begin{figure}
\centering
\includegraphics [width=1.0\columnwidth, keepaspectratio] {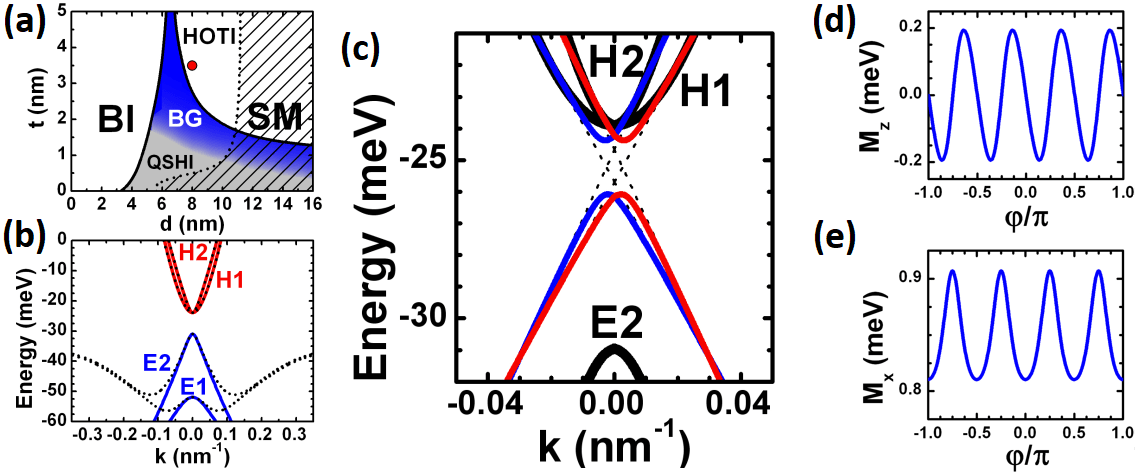}
\caption{\label{Chapter3Fig15} \textbf{(a)} The same phase diagram as in Fig.~\ref{Chapter3Fig12}c, but with the correct classification of states. The red symbol represents the parameters of the double QW, for which all subsequent calculations will be performed. \textbf{(b)}~Band structure calculated on the basis of effective 2D low-energy Hamiltonian for the QW with the layer thicknesses marked by the red symbols (a) panel. The blue and red curves represent band dispersion of electron-like and hole-like subbands, respectively. The wave vector is oriented along (100) crystallographic direction. The dotted curves represent the calculations based on realistic eight-band $\mathbf{k}\cdot\mathbf{p}$ Kane model. \textbf{(c)} Energy dispersion of the bulk and edge states based on low-energy 1D and 2D Hamiltonians. Bulk states being described by $H_{2D}(\mathbf{k})$ in Eq.~(\ref{eq:DQW5}) are shown in black. Red and blue curves correspond to different Kramer's partners of the edge states. The dashed curves are the eigenvalues of the 1D edge Hamiltonian. The wave vector and the edge are oriented along (100) crystallographic direction. \textbf{(d,e)} The angle dependent mass parameters in 1D edge Hamiltonian of ``tilted'' gapped Dirac fermions -- see Eq.~(\ref{eq:HOTI6}).}
\end{figure}

Carrying out all mathematics, as detailed in the original work~\cite{doubleHgTeQWs7}, the low-energy 1D edge Hamiltonian block, corresponding to the upper block $H_{4\times4}(k_x,k_y)$ of $H_{2D}(\mathbf{k})$ in Eq.~(\ref{eq:DQW5}), can be expressed in the following form:
\begin{equation}
\label{eq:HOTI6}
H_{\mathrm{1D}}(\delta{k})
=\varepsilon_0+v_{0}\delta{k}\mathbf{I}_2-v_{y}\delta{k}\sigma_y+M_z(\varphi)\sigma_z-M_x(\varphi)\sigma_x,
\end{equation}
where $\delta{k}=k_x-k_c$, $\mathbf{I}_2$ is a $2{\times}2$ identity matrix,  $\sigma_z$ is one of the Pauli matrices; $\varepsilon_0$ is a constant corresponding to the energy of the crossing point at $k_x=k_c$, defined by the edge states resulting from inversion of the subband pairs $|E1\rangle$--$|H1\rangle$ and $|E2\rangle$--$|H2\rangle$ in the absence of the tunneling between them:
\begin{equation}
\label{eq:HOTI7}
k_c=\dfrac{C_1-C_2+\dfrac{M_2D_2}{B_2}-\dfrac{M_1D_1}{B_1}}{\dfrac{2A_1\eta_1}{1+\eta_1^2}+\dfrac{2A_2\eta_2}{1+\eta_2^2}},
\end{equation}
where $\eta_{n}^2=(B_{n}+D_{n})/(B_{n}-D_{n})$ with $n=1$ and $2$ corresponding to the pairs of E1--H1 and E2--H2 subbands, respectively. Note that the crossing of other Kramer's partner occurs at $k_x=-k_c$.

Figure~\ref{Chapter3Fig15} shows the edge state picture of the prototype QWs with double band inversion based on low-energy 1D edge Hamiltonian for the edge oriented along ($100$) crystallographic direction, i.e. $\varphi=0$. One can see that the edge dispersion of each Kramer's partner mimics the dispersion of ``tilted'' 1D massive Dirac fermions with the velocities $v_0$ and $v_y$ that are written as
\begin{eqnarray}
\label{eq:HOTI8}
v_0=\dfrac{A_1\eta_1}{1+\eta_1^2}-\dfrac{A_2\eta_2}{1+\eta_2^2},\nonumber\\
v_y=\dfrac{A_1\eta_1}{1+\eta_1^2}+\dfrac{A_2\eta_2}{1+\eta_2^2}.
\end{eqnarray}
However, an important distinctive property of the edge states shown in Fig.~\ref{Chapter3Fig15} is that their gap is described simultaneously by two mass $M_x$ and $M_z$ parameters that prevent the gap vanishing at specific edge orientation, as clear from Fig.~~\ref{Chapter3Fig15}(d,e). Nevertheless, even such complex structure of the edge states yields the corner states in prototype QWs with double band inversion.

Let us now introduce the coordinate $x$ along the curved edge so that $x=0$ corresponds to the meeting corner, which means that $M_x$ and $M_z$ in Eq.~(\ref{eq:HOTI6}) are the function of $x$, and $\delta{k}=-i\partial/\partial{x}$. To analytically find the energy of the eigenstate localized in the vicinity of $x=0$, we consider the case of sharp corner with the single meeting point of two edges. In this case, $M_z(x)$ and $M_x(x)$ are all proportional to each other
\begin{equation}
\label{eq:HOTI9}
M_x(x)=\alpha{M_z(x)}+m,
\end{equation}
where
\begin{equation*}
\alpha=\dfrac{M_x(-\infty)-M_x(+\infty)}{M_z(-\infty)-M_z(+\infty)},
\end{equation*}
\begin{equation}
\label{eq:HOTI10}
m=\dfrac{M_z(-\infty)M_x(+\infty)-M_z(+\infty)M_x(-\infty)}{M_z(-\infty)-M_z(+\infty)}.
\end{equation}

After some mathematical calculations, one can show that under the condition expressed by Eq.~(\ref{eq:HOTI9}), the Schr\"{o}dinger equation with $H_{\mathrm{1D}}(\delta{k})$ in Eq.~(\ref{eq:HOTI6}) can be presented in the form of supersymmetric quantum mechanics, which allows an analytical solution under the following condition~\cite{doubleHgTeQWs7}:
\begin{equation}
\label{eq:HOTI11}
\left(M_z(+\infty)+{\alpha}M_x(+\infty)\right)\left(M_z(-\infty)+{\alpha}M_x(-\infty)\right)<0,
\end{equation}
the state localized in the vicinity of $x=0$ indeed exists, and its energy and wave-function $\Psi_{\mathrm{0D}}(x)$ are written as:
\begin{equation*}
E=\varepsilon_0+\dfrac{{\sigma}m}{\sqrt{1+\alpha^2}}
\dfrac{\sqrt{v_{y}^2-v_{0}^2}}{v_y},
\end{equation*}
\begin{multline}
\label{eq:HOTI12}
\Psi_{\mathrm{0D}}(x)=C
\begin{pmatrix}
v_{y}\alpha-\sigma\sqrt{1+\alpha^2}\sqrt{v_{y}^2-v_{0}^2}-iv_{0} \\
v_{y}+iv_{0}\alpha
\end{pmatrix}\cdot~\\
{\cdot}e^{\displaystyle{-ix\frac{Ev_{0}}{v_{y}^2-v_{0}^2}+\frac{\sigma}{\sqrt{1+\alpha^2}\sqrt{v_{y}^2-v_{0}^2}}\int\limits_0^{x}\left\{(1+\alpha^2)M_z(z)+m\alpha\right\}dz}},
\end{multline}
where $C$ is the normalization constant. Here, the sign of $\sigma$ should be chosen in accordance with normalized condition of $\Psi_{\mathrm{0D}}(x)$.
If $\left(M_z(+\infty)+{\alpha}M_x(+\infty)\right)>0$, $\sigma=-1$, while for $\left(M_z(+\infty)+{\alpha}M_x(+\infty)\right)<0$, $\sigma=1$.

\begin{figure}
\centering
\includegraphics [width=0.66\columnwidth, keepaspectratio] {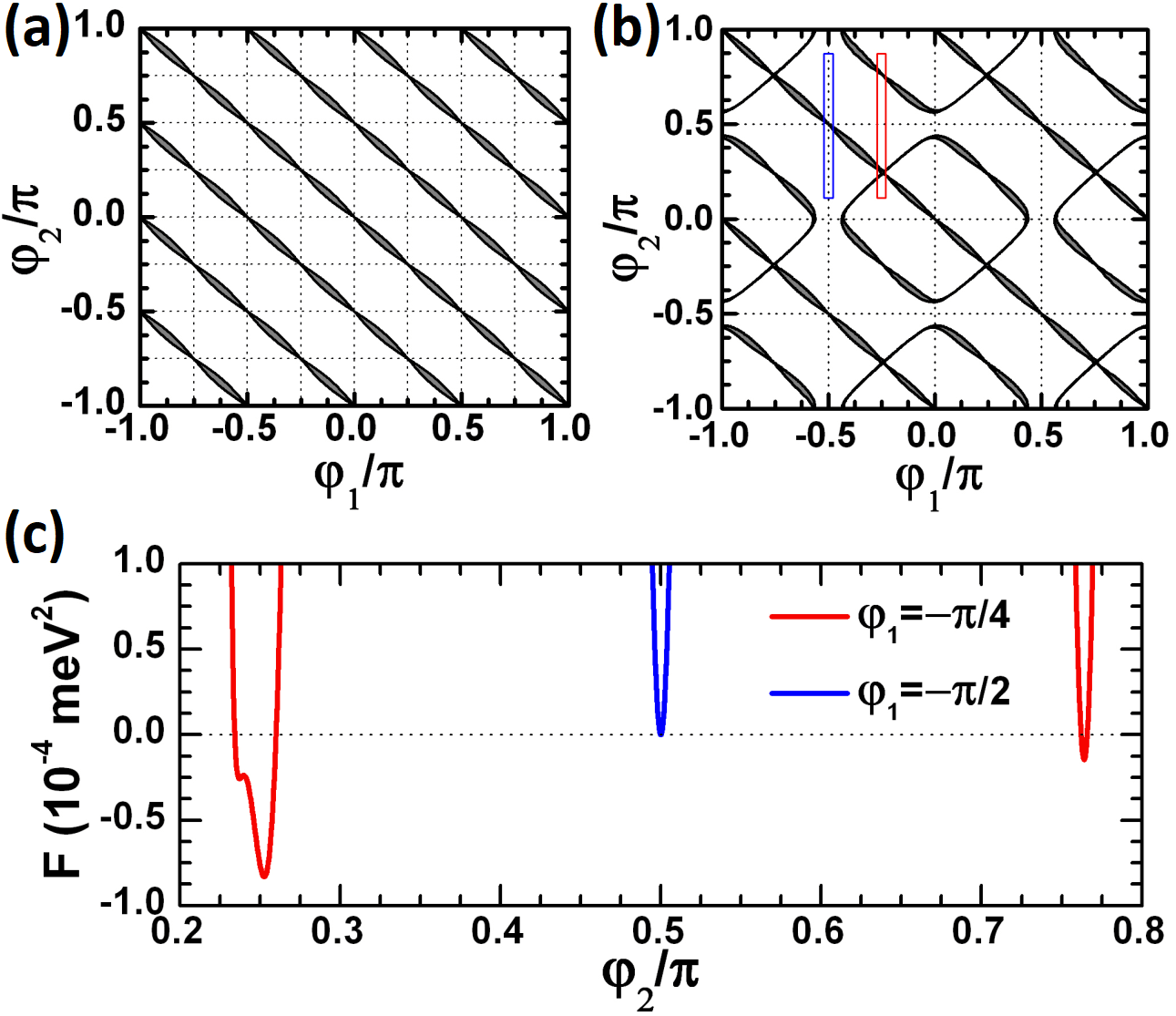}
\caption{\label{Chapter3Fig16} \textbf{(a,b)} The phase diagram for the presence (grey region) and absence (white region) of 0D topological corner states in double HgTe/CdHgTe QWs as a function of the two edge orientations. Panel (a) represents the (001)-oriented double HgTe QW, while panel (b) corresponds to the (013)-oriented QW (see Ref.~\cite{doubleHgTeQWs7} for details). \textbf{(c)} The function $F(\varphi_1,\varphi_2)=\left(M_z(\varphi_1)+{\alpha}M_x(\varphi_1)\right)\left(M_z(\varphi_2)+{\alpha}M_x(\varphi_2)\right)$
representing the existence condition of Eq.~(\ref{eq:HOTI11}) in the ranges marked with colored rectangles in the panel (b).}
\end{figure}
As it follows directly from Eq.~(\ref{eq:HOTI11}), if one of the mass parameters $M_z$ or $M_x$ equals zero, the expression reduces to the well-known Jackiw-Rebbi condition for the existence of a domain-wall bound state in a 1D Dirac system~\cite{SS3DKane1}. It is important to emphasize that if the anisotropic term $H_{4\times4}^{(a)}(k_x,k_y)$ in Eq.~(\ref{eq:HOTI2}), which accounts for the cubic symmetry of the crystal, is neglected, then $M_z$ vanishes, while $M_x$ remains positive and does not depend on the crystallographic orientation of the edge. Consequently, as one can clearly see from Eq.~(\ref{eq:HOTI11}), under these circumstances the condition for the existence of a 0D corner state is not fulfilled.

Figure~\ref{Chapter3Fig16}(a) represents the angle diagram showing the edge orientations in the (001)-oriented prototype double HgTe QW, for which the existence condition of Eq.~(\ref{eq:HOTI11}) is valid. It is seen that the values of $\varphi_1$ and $\varphi_2$ yielding the corner states are qualitatively represented as the grey regions elongated along the lines defined by
\begin{equation}
\label{eq:HOTI12}
\varphi_1+\varphi_2=n_0\left(2\pi/4\right),
\end{equation}
excluding the points lying at verticals and horizontals corresponding to $\varphi_1=m_0\pi/4$ and $\varphi_2=l_0\pi/4$, where $n_0$, $m_0$ and $l_0$ are integers. The edges at these angles coincide with $[110]$, $[1\bar{1}0]$, $[100]$ or $[010]$ crystallographic planes, being the faces of (001)-oriented QW. Since we have neglected the terms resulting from possible breaking of inversion symmetry in the system, the (001)-oriented prototype double QW preserve the mirror symmetry about the mentioned planes. Therefore, one concludes that if one of the QW edge coincides with the mirror symmetry planes, the corner states are absent in the system. Importantly, Eq.~(\ref{eq:HOTI12}) describes the corners, whose bisector corresponds to the one of the mirror symmetry planes. The latter results in reflection symmetry of the angle diagram with respect to the lines described by Eq.~(\ref{eq:HOTI12}). As expected from the four-fold rotational symmetry of (001)-oriented QWs with respect to the growth direction, the angle pattern in Fig~\ref{Chapter3Fig16}(a) also possess a $\pi/2$-periodicity in $\varphi_1$ and $\varphi_2$.

For comparison, Fig.~\ref{Chapter3Fig16}(b) provides the angle diagram of topological corner states in the ($013$)-oriented double HgTe/CdHgTe QWs with the same layer thicknesses. It is seen that the values of $\varphi_1$ and $\varphi_2$, for which the existence condition is fulfilled, represent significantly different picture of the corner states than the one for (001)-oriented QWs. We first note $\pi$-periodicity in $\varphi_1$ and $\varphi_2$ of the angle pattern due to the two-fold rotational symmetry of ($0mn$)-oriented QWs in contrast to the $\pi/2$-periodicity of (001)-oriented 2D systems. Second, due to the symmetry lowering, (013)-oriented QWs have only two $[100]$ and $[03\bar{1}]$ mirror planes perpendicular to the QW plane. These planes are represented by dotted verticals and horizontals at $\varphi_1=m_0\pi/2$ and $\varphi_2=l_0\pi/2$. It is seen that the corner states are absent if one of the QW edge coincides with the mirror symmetry planes. Finally, in contrast to (001)-oriented QWs, ($0mn$)-oriented QWs possess the corner states if one of the edges is at $\pi/4$ degrees with respect to the ($100$) crystallographic axis. The latter is clearly seen in Fig.~\ref{Chapter3Fig16}(c), which shows the negative values of the function illustrating the existence condition of Eq.~(\ref{eq:HOTI11}) for the corner states.

In conclusion, I would like to emphasize that the results presented in this section have, for the first time, demonstrated the possibility of realizing a 2D HOTI state in QW heterostructures based on zinc-blende semiconductors. In particular, a 2D system composed of two tunnel-coupled HgTe QWs with double band inversion can host 0D corner states, whose emergence is directly related to the cubic symmetry of the crystal lattice. I am convinced that similar conclusions should remain valid for any 2D system consisting of two tunnel-coupled QSHIs, provided that the crystalline symmetry is properly taken into account. Finally, it should be noted that in the original work~\cite{doubleHgTeQWs7}, alongside double HgTe QWs, I also considered QWs based on III-V semiconductors, which will be discussed in detail in Chapter~\ref{Chap:III-V}. Thus, the results presented in this section, in essence, open a new perspective for exploring higher-order topology in conventional epitaxially grown semiconductor heterostructures.

\section{\label{sec:disHgTeDQWs}Disorder-induced topological phase transitions}
The critical role of disorder in driving topological phase transitions has already been discussed in Section~\ref{sec:DisorderKane} in the context of Kane fermions in bulk HgCdTe layers. However, the very first observation of a disorder-induced topological transition was made by Li~\emph{et~al.}~\cite{TAI1} precisely in HgTe/CdHgTe QWs. By employing a tight-binding version of the BHZ Hamiltonian with Anderson-type on-site disorder~\cite{HgTeQWs2}, they demonstrated numerically that disorder can induce a novel quantum phase characterized by quantized conductance, which they called the ``topological Anderson insulator'' (TAI). Subsequently, Groth~\emph{et~al.}~\cite{TAI2} clarified the physical origin of this disorder-induced topological phase transition in HgTe/CdHgTe QWs, showing that it arises from the quadratic terms $\propto k^2\sigma_z$ in the BHZ Hamiltonian. These terms are crucial for the correct topological classification of the system~\cite{HgTeQWs3}. As demonstrated in their work, the quadratic corrections lead, in the presence of disorder, to renormalization of both the chemical potential and the mass term in the BHZ model. The TAI phase emerges when the renormalized mass term becomes negative, while the renormalized chemical potential lies inside the renormalized band-gap~\emph{et~al.}~\cite{TAI2}. This makes TAI phenomenologically indistinguishable from QSHI. Moreover, it was convincingly shown that, despite its name, the ``topological Anderson insulator'' does not correspond to an Anderson-type transition in the conventional sense. Instead, such weak-disorder-induced topological transition can be fully described within the self-consistent Born approximation (SCBA)~\cite{TAI2,TAI11,TAI12,TAI14}.

A distinctive feature of most works on disorder-induced topological phase transitions in 2D systems~\cite{TAI1,TAI2,TAI11,TAI12,TAI14} is that the emergence of a non-trivial phase was inferred from the quantized conductance values appeared in the numerical simulations, while a direct demonstration of a change of the $\mathbb{Z}_2$ invariant was missing. Indeed, the correct characterization of topological systems in the presence of various single-particle and many-body interactions (including disorder) is far from obvious. One of the ways to take into account the interaction effect in the calculations of various topological invariants is the Green's function formalism~\cite{TAI15,TAI16,TAI17,TAI18,TAI19,TAI20,TAI21,TAI22,TAI23,TAI24,TAI25,TAI26,TAI27}. Within this formalism, a certain quantity characterizing the topology of the system, is expressed in terms of the single-particle retarded Green's function
\begin{equation}
\label{eq:QWTAI1}
\hat{G}(\mathbf{k},\varepsilon)=
\left[\varepsilon-H_{0}(\mathbf{k})-\hat{\Sigma}(\mathbf{k},\varepsilon)\right]^{-1},
\end{equation}
where $H_{0}(\mathbf{k})$ is the non-interacting Hamiltonian (here, $\mathbf{k}$ is the quasimomentum, and $\varepsilon$ is considered as an external parameter, e.g. the Fermi energy), while the self-energy $\hat{\Sigma}(\mathbf{k},\varepsilon)$ incorporates the interaction effect. The latter includes all types of single-particle (electron-impurity, electron-phonon, etc.) and many-particle interactions that can be treated within the framework of single-particle Green's function approach. However, direct calculation of some topological invariants, such as the $\mathbb{Z}_2$ invariant, can be quite cumbersome, since it requires not only knowledge of the Green's function in the entire energy domain, but also the calculation of multi-fold integrals~\cite{TAI16,TAI17,TAI18,TAI19,TAI20,TAI21,TAI22,TAI23,TAI24}.

Z.~Wang and collaborators~\cite{TAI25,TAI26,TAI27} proposed a recipe for the evaluation of a topological invariants in the presence of inversion symmetry based on the zero-energy Green's function. Their main idea is based on the introduction of ``topological Hamiltonian'' as
\begin{equation}
\label{eq:QWTAI2}
H_{t}(\mathbf{k})\equiv-\hat{G}^{-1}(\mathbf{k},0)=\mathcal{H}_{0}(\mathbf{k})+\hat{\Sigma}(\mathbf{k},0),
\end{equation}
which can be used for the calculations of topological invariants in the presence of interactions. Note that the Hamiltonian introduced in this way stays Hermitian~\cite{TAI25,TAI26,TAI27,TAI28}, therefore topological characterization of the system with interactions is equivalent to the characterization of a ``non-interacting system'' with free Hamiltonian $H_{t}(\mathbf{k})$. Although the formalism based on the zero-energy Green's functions indeed often leads to correct topological characterization of different systems in the presence of single-particle and many-particle interactions~\cite{TAI27,TAI28,TAI29,TAI30,TAI31,TAI32,TAI33,TAI34}, there are still many cases of its breakdown~\cite{TAI35}. The latter is apparently due to the fact that, unlike $\hat{G}(\mathbf{k},\varepsilon)$, $H_{t}(\mathbf{k})$ does not describe the system in a full manner.

\subsection{\label{sec:qpQW}Quasiparticle picture of the phase transition}
This section will present our alternative approach based on the quasiparticle Hamiltonian, which provides a consistent describtion of the topological properties in the presence of interactions. By applying this quasiparticle approach to a 2D system described by the BHZ model, I  will demonstrate how the $\mathbb{Z}_2$ topological invariant can be calculated in the presence of disorder. In what follows, I will focus only on the main physical results, while the detailed mathematical derivations are provided in our original publication~\cite{TAI13}. As will become clear further, a disorder-induced topological phase transition represents another example of a breakdown of the topological classification based on $H_{t}(\mathbf{k})$.

Alternatively, in our original work~\cite{TAI13}, we introduce the \emph{quasiparticle} Hamiltonian
\begin{equation}
\label{eq:QWTAI3}
\mathcal{H}_{qp}(\mathbf{k},\varepsilon)=
\mathcal{H}_{0}(\mathbf{k})+\hat{\Sigma}(\mathbf{k},\varepsilon),
\end{equation}
which, like the Green's function, contains all the information about the system. In the most general case, $\mathcal{H}_{qp}(\mathbf{k},\varepsilon)$ is \emph{non-Hermitian} and therefore has complex eigenvalues. Physically, the real part of its eigenvalues characterizes the ``renormalized'' band structure, while the imaginary part describes the quasiparticle decay resulting in the finite quasiparticle lifetime. The latter however is not crucial, since the methods developed in recent years~\cite{TAI36,TAI37,TAI38,TAI39,TAI40,TAI41,TAI42,TAI43,TAI44,TAI45} allow topological characterization of non-Hermitian Hamiltonians as well. Since non-Hermitian systems have a richer phase diagram compared to Hermitian ones~\cite{TAI36,TAI37,TAI38,TAI39,TAI40,TAI41}, the topological characterization of systems with interactions in the most general way should be performed by means of the quasiparticle Hamiltonian $\mathcal{H}_{qp}(\mathbf{k},\varepsilon)$ rather than $H_{t}(\mathbf{k})$.

If the eigenvalues $E_s(\mathbf{k},\varepsilon)$ of quasiparticle Hamiltonian $\mathcal{H}_{qp}(\mathbf{k},\varepsilon)$ satisfy the condition of band separability -- when $E_s(\mathbf{k},\varepsilon){\neq}E_l(\mathbf{k},\varepsilon)$ for all $s{\neq}l$ and all $\mathbf{k}$ -- the quasiparticle description within $\mathcal{H}_{qp}(\mathbf{k},\varepsilon)$ provides a convenient and consistent way to evaluate the $\mathbb{Z}_2$ invariant in 2D system that~\cite{TAI13}. Importantly, the condition of band separability can also be satisfied even in the absence of the band-gap in the spectral function $A(\mathbf{k},\varepsilon)$ and density-of-states $D(\varepsilon)$ of 2D system that are written as
\begin{eqnarray}
\label{eq:QWTAI3b}
A(\mathbf{k},\varepsilon)=-\dfrac{1}{\pi}\sum_{s}\dfrac{\textrm{Im}E_s(\mathbf{k},\varepsilon)}{\left(\varepsilon-\textrm{Re}E_s(\mathbf{k},\varepsilon)\right)^2+\left(\textrm{Im}E_s(\mathbf{k},\varepsilon)\right)^2},\notag\\
D(\varepsilon)={\int}A(\mathbf{k},\varepsilon){d^2\mathbf{k}},~~~~~~~~~~~~~~~~~~~~~~
\end{eqnarray}
where $s$ labels the eigenvalues.

Let us assume for simplicity that, without specifying the exact type of interaction that maintains the time reversal symmetry, the quasiparticle Hamiltonian has a block-diagonal form:
\begin{equation}
\label{eq:QWTAI4}
\mathcal{H}_{qp}(\mathbf{k},\varepsilon)=\begin{pmatrix}
\mathcal{H}_{qp}^{\uparrow}(\mathbf{k},\varepsilon) & 0 \\ 0 & \mathcal{H}_{qp}^{\downarrow}(\mathbf{k},\varepsilon) \end{pmatrix}
=\mathcal{H}_{0}(\mathbf{k})+\begin{pmatrix}
\Sigma^{\uparrow}(\mathbf{k},\varepsilon) & 0 \\ 0 & \Sigma^{\downarrow}(\mathbf{k},\varepsilon)\end{pmatrix}.
\end{equation}
A block-diagonal form of $\mathcal{H}_{qp}(\mathbf{k},\varepsilon)$ reduces the calculation of the $\mathbb{Z}_2$ invariant to the calculation of
\begin{equation}
\label{eq:QWTAI5}
\mathbb{Z}_2(\varepsilon)=\mathrm{mod}\left(\dfrac{C_{\uparrow}(\varepsilon)-C_{\downarrow}(\varepsilon)}{2},2\right),
\end{equation}
where $C_{\uparrow}(\varepsilon)$ and $C_{\downarrow}(\varepsilon)$ are the Chern numbers for the upper $\mathcal{H}_{qp}^{\uparrow}(\mathbf{k},\varepsilon)$ and lower $\mathcal{H}_{qp}^{\downarrow}(\mathbf{k},\varepsilon)$ blocks, respectively. We remind that $\varepsilon$ in Eq.~(\ref{eq:QWTAI5}) should be considered as an external parameter, e.g. the Fermi energy. Since $C_{\uparrow}(\varepsilon)+C_{\downarrow}(\varepsilon)=0$ also holds for non-Hermitian systems~\cite{TAI46,TAI47} in the presence of time-reversal symmetry, on can focus only on the calculation for the upper block $\mathcal{H}_{qp}^{\uparrow}(\mathbf{k},\varepsilon)$.

To proceed further, we will focus on the 2D system described by the simplest version of BHZ Hamiltonian:
\begin{equation}
\label{eq:QWTAI6}
\mathcal{H}_{0}(k_x, k_y)=\begin{pmatrix}
H_{\mathrm{BHZ}}(\mathbf{k}) & 0 \\ 0 & H_{\mathrm{BHZ}}^{*}(-\mathbf{k})\end{pmatrix},
\end{equation}
where the asterisk stands for complex conjugation, $\textbf{k}=(k_x,k_y)$ is the momentum in the plane, and
\begin{equation}
\label{eq:QWTAI7}
H_{\mathrm{BHZ}}(\mathbf{k})=d_{0}(k)+\mathbf{d}(\mathbf{k})\cdot\vec{\sigma}.
\end{equation}
Here, $\vec{\sigma}=(\sigma_x,\sigma_y,\sigma_z)$ consists of the Pauli matrices; $d_{0}(k)=C-Dk^2$ and $\mathbf{d}=(d_x,d_y,d_z)$, where $d_x(\mathbf{k})=Ak_x$, $d_y(\mathbf{k})=Ak_y$, $d_z(k)=M-Bk^2$ and $k^2=k_x^2+k_y^2$. Here, the mass parameter $M$ describes inversion between the electron-like E1 and hole-like H1 subbands, while other parameters $A$, $B$, $C$, $D$ depend on the QW growth direction and QW geometry.

In this case, $\mathcal{H}_{qp}^{\uparrow}(\mathbf{k},\varepsilon)$ in Eq.~(\ref{eq:QWTAI4}) is represented in the form similar to Eq.~(\ref{eq:QWTAI7}):
\begin{equation}
\label{eq:QWTAI8}
\mathcal{H}_{qp}^{\uparrow}(\mathbf{k},\varepsilon)=h_{0}(\mathbf{k},\varepsilon)+\mathbf{h}(\mathbf{k},\varepsilon)\cdot\vec{\sigma},
\end{equation}
where $h_{0}$ and $\mathbf{h}=(h_x,h_y,h_z)$ are \emph{complex} functions of $\mathbf{k}$ and $\varepsilon$. Such a representation of $\mathcal{H}_{qp}^{\uparrow}(\mathbf{k},\varepsilon)$ allows one to calculate the Chern number as follows~\cite{TAI13}:
\begin{equation}
\label{eq:QWTAI9}
C^{(s)}_{\uparrow}(\varepsilon)=-\dfrac{s}{4\pi}{\int}
\dfrac{\mathbf{h}(\mathbf{k},\varepsilon)}{\lambda(\mathbf{k},\varepsilon)^3}\cdot\left[\nabla_{k_x}\mathbf{h}(\mathbf{k},\varepsilon)\times
\nabla_{k_y}\mathbf{h}(\mathbf{k},\varepsilon)\right]
{d^2\mathbf{k}},
\end{equation}
where $s={\pm}1$ and $\lambda(\mathbf{k},\varepsilon)$ is defined as
\begin{equation}
\label{eq:QWTAI10}
\lambda(\mathbf{k},\varepsilon)=\sqrt{h_x^2(\mathbf{k},\varepsilon)+h_y^2(\mathbf{k},\varepsilon)+h_z^2(\mathbf{k},\varepsilon)}.
\end{equation}
I would like to draw the reader's attention to the fact that despite the complex functions in the integrand, the Chern number always remains real and quantized.

To introduce disorder into the system, let us add a diagonal random impurity potential to the Hamiltonian $\mathcal{H}_{0}(\mathbf{k})$ in Eq.~(\ref{eq:QWTAI6}):
\begin{equation}
\label{eq:QWTAI11}
V_{imp}(\mathbf{r})=\sum_{j}v(\mathbf{r}-\mathbf{R}_j),~~~~ v(\mathbf{r})=\int\dfrac{d^2\mathbf{q}}{(2\pi)^2}\tilde{v}(\mathbf{q})e^{i\mathbf{q}\cdot\mathbf{r}},
\end{equation}
where $R_j$ denotes position of impurities and $v(\textbf{r})$ is the potential of an individual impurity. The latter is assumed to be isotropic, i.e. $\tilde{v}(\mathbf{q}) = \tilde{v}({q})$ with $|\mathbf{q}| = q$. Then, the self-energy matrix within the SCBA is written as~\cite{TAI13}:
\begin{equation}
\label{eq:QWTAI12}
\hat{\Sigma}(\mathbf{k},\varepsilon)=n_{i}\int\dfrac{d^2\mathbf{k}^\prime}{(2\pi)^2}
\tilde{v}(\mathbf{k}-\mathbf{k}^\prime)\hat{G}(\mathbf{k}^\prime,\varepsilon)\tilde{v}(\mathbf{k}^\prime-\mathbf{k}),
\end{equation}
where $n_{i}$ is the concentration of impurities.

The axial rotation symmetry of $\mathcal{H}_{0}(k_x,k_y)$ allows one to reduce Eq.~(\ref{eq:QWTAI6}) to the form~\cite{TAI13}:
\begin{equation}
\label{eq:QWTAI13}
\hat{\Sigma}(k,\varepsilon)=\begin{pmatrix}
\hat{\Sigma}^{\uparrow}(k,\varepsilon) & 0\\
0 & \hat{\Sigma}^{\downarrow}(k,\varepsilon) \end{pmatrix}
\end{equation}
with
\begin{eqnarray}
\label{eq:QWTAI13b}
\hat{\Sigma}^{\uparrow}(k,\varepsilon)=n_{i}\int\limits_0^{K_c}\dfrac{k^\prime dk^\prime}{2\pi}\begin{pmatrix}
V_0(k,k^\prime)^2G_{11}^\prime & V_{+1}(k,k^\prime)^2G_{12}^\prime\\
V_{-1}(k,k^\prime)^2G_{21}^\prime & V_0(k,k^\prime)^2G_{22}^\prime  \end{pmatrix},\nonumber\\
\hat{\Sigma}^{\downarrow}(k,\varepsilon)=n_{i}\int\limits_0^{K_c}\dfrac{k^\prime dk^\prime}{2\pi}\begin{pmatrix}
V_0(k,k^\prime)^2G_{33}^\prime & V_{-1}(k,k^\prime)^2G_{34}^\prime\\
V_{+1}(k,k^\prime)^2G_{43}^\prime & V_0(k,k^\prime)^2G_{44}^\prime  \end{pmatrix},
\end{eqnarray}
where $k=|\mathbf{k}|$, $k^\prime=|\mathbf{k}^\prime|$,
\begin{equation}
\label{eq:QWTAI14}
V_n(k,k^\prime)^2=\int\limits_0^{2\pi}\dfrac{d\theta}{2\pi}|\tilde{v}(\mathbf{k}-\mathbf{k}^\prime)|^2
\cos n\theta,
\end{equation}
and $G_{ij}^\prime\equiv G_{ij}(k^\prime,\varepsilon)$ are the component of the Green's function:
\begin{equation}
\label{eq:QWTAI15}
\hat{G}(k,\varepsilon)=\left[\varepsilon-\tilde{\mathcal{H}}_{0}(k)-\hat{\Sigma}(k,\varepsilon)\right]^{-1},
\end{equation}
with $\tilde{\mathcal{H}}_{0}(k)\equiv\mathcal{H}_{0}(k,0)$ in Eq.~(\ref{eq:QWTAI6}). In the equations above, we introduce the angle $\theta$ between $\mathbf{k}$ and $\mathbf{k}^\prime$, as well as the cut-off wave-vector $K_c=\pi/a_{\mathrm{0}}$, where $a_{\mathrm{0}}$ is the lattice constant in the plane of 2D system, which corresponds to the size of the first Brillouin zone. As clear, Eqs.~(\ref{eq:QWTAI13})--(\ref{eq:QWTAI15}) form a system of integral equations that determines both the Green's function $\hat{G}(k,\varepsilon)$ and the self-energy $\hat{\Sigma}({k},\varepsilon)$ matrices. The self-consistent solution of such integral systems in the general case is a laborious task. However, for the case of the disorder formed by the short-range impurities, for which $\tilde{v}(q)=u_0$, the solution of the problem can be significantly simplified.

Indeed, in the case of the short-range disorder, $V_n(k,k^\prime)^2=u_0^2\delta_{n,0}$ in Eq.~(\ref{eq:QWTAI14}). The latter results in a diagonal form of the self-energy matrix $\hat{\Sigma}(\varepsilon)$ in Eq.~(\ref{eq:QWTAI15}) being independent of the momentum. By means of direct calculations, one can verify that $G_{11}\left(k,\varepsilon\right)=G_{33}\left(k,\varepsilon\right)$ and $G_{22}\left(k,\varepsilon\right)=G_{44}\left(k,\varepsilon\right)$, which leads to
\begin{equation}
\label{eq:QWTAI16}
\hat{\Sigma}^{\uparrow}(\varepsilon)=\hat{\Sigma}^{\downarrow}(\varepsilon)=
\dfrac{W^2}{4\pi}\int\limits_0^{K_c^2}dx
\begin{pmatrix}
G_{11}\left(\sqrt{x},\varepsilon\right) & 0 \\[4pt]
0 & G_{22}\left(\sqrt{x},\varepsilon\right)
\end{pmatrix},
\end{equation}
where $W^2=n_{i}u_0^2$ is the disorder strength. Note that integration in Eq.~(\ref{eq:QWTAI16}) can be performed analytically~\cite{TAI13}, which transforms Eqs.~(\ref{eq:QWTAI13})--(\ref{eq:QWTAI16}) into the set of algebraic equations numerically solved by simple iterations.

Once the self-energy is known, the spectral function $A(k,\varepsilon)$ and the density-of-states (DOS) $D(\varepsilon)$ for the case of the short-range disorder can be found as
\begin{equation*}
A(k,\varepsilon)=-\dfrac{2}{\pi}\textrm{Im}\left\{G_{11}\left(k,\varepsilon\right)+G_{22}\left(k,\varepsilon\right)\right\},
\end{equation*}
\begin{equation}
\label{eq:QWTAI17}
D(\varepsilon)=\int\limits_0^{K_c}\dfrac{k dk}{2\pi}A(k,\varepsilon).
\end{equation}

\begin{figure}
\centering
\includegraphics [width=1.0\columnwidth, keepaspectratio] {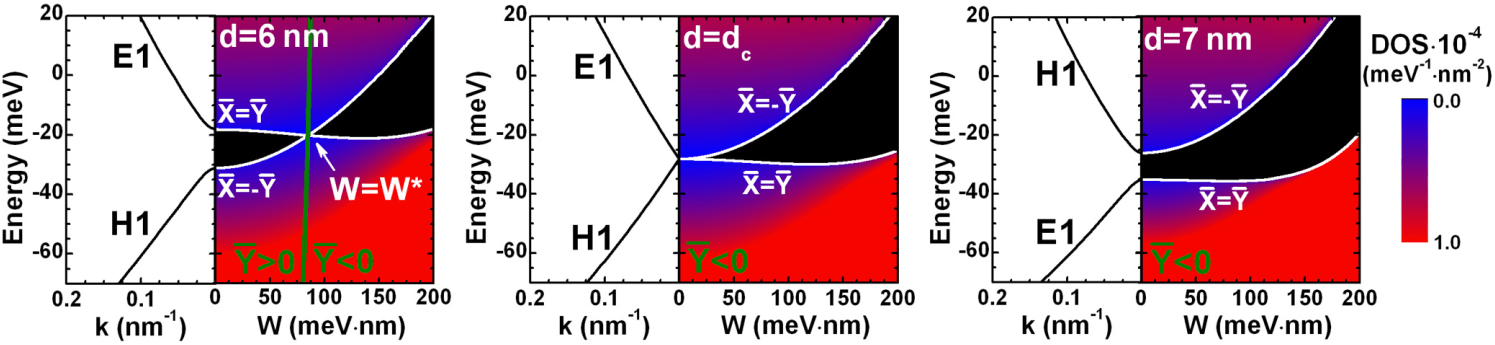}
\caption{\label{Chapter3Fig17} Band structure and color map of the DOS as a function of the short-range disorder strength~$W=\sqrt{n_{i}u_0^2}$ (see Eq.~(\ref{eq:QWTAI16})) calculated in the two-band BHZ model for HgTe/Cd$_{0.7}$Hg$_{0.3}$T QWs at different QW width: $d=6$~nm ($M>0$), $d=d_c$ ($M=0$) and $d=7$~nm ($M<0$). The white curves represent the edges of the area with the vanishing DOS. The green curve represents $\overline{Y}(\varepsilon)=\textrm{Re}\left(M+\Sigma_z(\varepsilon)\right)$.The parameters used in the calculations and meaning of $\overline{X}(\varepsilon)$ are provided in Ref.~\cite{TAI48}.}
\end{figure}

Figure~\ref{Chapter3Fig17} shows the evolution of DOS with the disorder strength $W$ for the HgTe QWs of different QW width. All details about the calculation of structural parameters involved in $\mathcal{H}_{0}(\mathbf{k})$ can be found elsewhere~\cite{TAI48}. As it is seen, the band-gap evolution strongly depends of the sign of $M$. If $M>0$, the band-gap decreases with $W$ and vanishes at a critical value $W_c=W^{*}$ and, than, it is re-opened again at $W>W_c=W^{*}$. Such behavior represents the disorder-induced topological phase transition previously discovered in the tight-binding calculations~\cite{TAI1,TAI2}. Importantly, the band-gap reopening region at $W>W_c$ was characterized by quantized conductance values inherent in the QSHI state. It is on the basis of this fact that the conclusion is made that $W=W_c=W^{*}$ corresponds to a topological phase transition. Let us now show that the band-gap behavior illustrated by the evolution of $D(\varepsilon)$ is indeed due to a change in the $\mathbb{Z}_2$ invariant. For this purpose, it is convenient to represent the self-energy blocks in Eq.~(\ref{eq:QWTAI16}) in the form
\begin{equation}
\label{eqRnD:1}
\hat{\Sigma}^{\uparrow}(\varepsilon)=\hat{\Sigma}^{\downarrow}(\varepsilon)
=\Sigma_0(\varepsilon)+\Sigma_z(\varepsilon)\sigma_z.
\end{equation}
Then, using the polar coordinate system, the Chern number for the valence band of $\mathcal{H}_{qp}^{\uparrow}(\mathbf{k},\varepsilon)$ can be calculated analytically
\begin{equation*}
C^{(-)}_{\uparrow}(\varepsilon)=\dfrac{A^2}{4}\int\limits_{0}^{\infty}
\dfrac{Bx+M+\Sigma_z(\varepsilon)}{\left[A^2x+\left(M+\Sigma_z(\varepsilon)-Bx\right)^2\right]^{3/2}}dx=\dfrac{B}{2\sqrt{B^2}}+\dfrac{M+\Sigma_z(\varepsilon)}{2\sqrt{\left(M+\Sigma_z(\varepsilon)\right)^2}}=
\end{equation*}
\begin{equation}
\label{eqRnD:2}
=\dfrac{\mathrm{sgn}\left[B\right]+\mathrm{sgn}\left[\textrm{Re}\left\{M+\Sigma_z(\varepsilon)\right\}\right]}{2}.
\end{equation}
Since parameter $B$ is always negative in HgTe QWs~\cite{TAI48,TAI49}, then Eq.~(\ref{eq:QWTAI5}) for the $\mathbb{Z}_2$ invariant becomes
\begin{equation}
\label{eqRnD:3}
\mathbb{Z}_2(\varepsilon)=\dfrac{1-\mathrm{sgn}\left[M+\textrm{Re}\Sigma_z(\varepsilon)\right]}{2}.
\end{equation}
Note that Eq.~(\ref{eqRnD:3}) is also valid at the energies $\varepsilon$ outside the band-gap region, where the imaginary part of both $\Sigma_0(\varepsilon)$ and $\Sigma_z(\varepsilon)$ is non-zero.

\begin{figure}
\centering
\includegraphics [width=0.8\columnwidth, keepaspectratio] {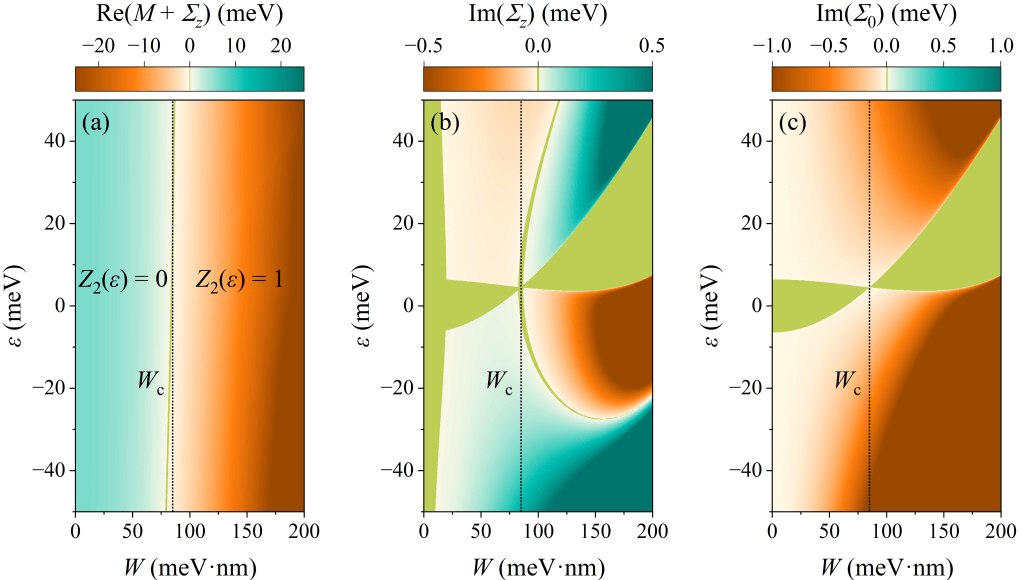}
\caption{\label{Chapter3Fig18} \textbf{(a-c)} Color map of the renormalized mass parameter $M+\textrm{Re}\Sigma_z(\varepsilon)$ in panel (a) and imaginary parts of $\Sigma_z(\varepsilon)$ in panel (b) and $\Sigma_0(\varepsilon)$ in panel (c) in the 6~nm wide HgTe/Cd$_{0.7}$Hg$_{0.3}$Te QW as a function of the short-range disorder strength $W$. The zero-energy values are marked separately in green. The dotted vertical curves represent a critical value $W=W_c=W^{*}$.}
\end{figure}

A parameter $M+\textrm{Re}\Sigma_z(\varepsilon)$ in Eq.~(\ref{eqRnD:3}) can be naturally treated as renormalized mass parameter -- a topological phase transition occurs when it changes its sign. Indeed, straightforward calculations presented in Fig.~\ref{Chapter3Fig18}(a) show that the band-gap closing at $W=W_c=W^{*}$ revealed by $D(\varepsilon)$ clearly corresponds to zero values of $M+\textrm{Re}\Sigma_z(\varepsilon)$. Therefore, the band-gap region in Fig.~\ref{Chapter3Fig17} for the 6~nm wide HgTe/Cd$_{0.7}$Hg$_{0.3}$Te QW at $W<W_c$ is characterized by the positive values of $M+\textrm{Re}\Sigma_z(\varepsilon)$ and, hence, by $\mathbb{Z}_2=0$ in accordance with Eq.~(\ref{eqRnD:3}). On the contrary, the band-gap region at $W>W_c$ corresponds to $M+\textrm{Re}\Sigma_z(\varepsilon)<0$, resulting in $\mathbb{Z}_2=1$. In this way, the above calculation of $\mathbb{Z}_2$ topological invariant
fully confirms the intuitive conclusions about the role of $M+\textrm{Re}\Sigma_z(\varepsilon)$ in the topological phase transition made by Groth \emph{et al.}~\cite{TAI2}.

As already mentioned above, in all previously published works, the disorder-induced topological phase transition was identified through the appearance of quantized conductance values under an applied bias in the numerical simulations. These quantized conductance values naturally imply the presence of disorder-induced helical edge states. Importantly, the quasiparticle approach we have developed, makes it possible to directly visualize such disorder-driven edge states. This allows us to trace the emergence of topological edge states as a direct consequence of the disorder, providing a clear physical interpretation of the topological phase transition beyond the conductance quantization. Indeed, let us consider a half-plane $y \geq 0$ of our 2D system. By applying open boundary conditions to $\mathcal{H}_{qp}(\mathbf{k},\varepsilon)$ at $y = 0$, one can analytically obtain the eigenvalues and eigenstates of the localized edge modes in complete analogy with the Hermitian case~\cite{TAI13}:
\begin{equation}
\label{eqRnD:11}
E_{\sigma}^{\mathrm{edge}}(k_x,\varepsilon)=\tilde{C}(\varepsilon)-\dfrac{D}{B}\tilde{M}(\varepsilon)
+\sigma\sqrt{1-\dfrac{D^2}{B^2}}Ak_x,
\end{equation}
where $\tilde{C}(\varepsilon)=C+\Sigma_0(\varepsilon)$ and $\tilde{M}(\varepsilon)=M+\Sigma_z(\varepsilon)$; $\sigma=+1$ and $\sigma=-1$ corresponds to $\mathcal{H}_{qp}^{\uparrow}(\mathbf{k},\varepsilon)$ and $\mathcal{H}_{qp}^{\downarrow}(\mathbf{k},\varepsilon)$, respectively.

\begin{figure}
\centering
\includegraphics [width=1.0\columnwidth, keepaspectratio] {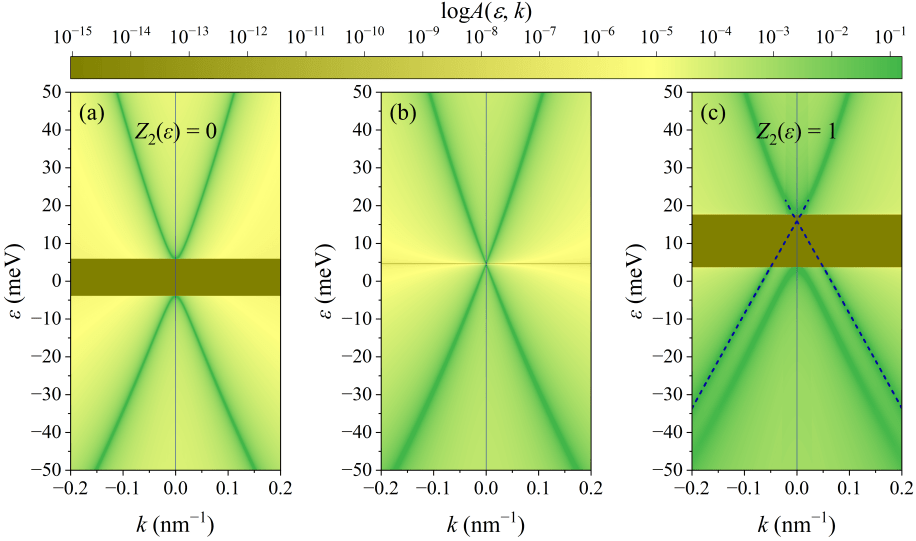}
\caption{\label{Chapter3Fig19} \textbf{(a-c)} Total spectral function $A(k,\varepsilon)+A^{\mathrm{edge}}(k,\varepsilon)$ determined by Eqs~(\ref{eq:QWTAI17}) and (\ref{eqRnD:15}) in the 6~nm wide HgTe/Cd$_{0.7}$Hg$_{0.3}$Te QW at several values of the short-range disorder strength: \textbf{(a)} $W = 40$~meV$\cdot$nm, \textbf{(b)} $W = W_c$, \textbf{(c)} $W = 130$~meV$\cdot$nm. The dashed blue lines in panel (c) correspond to the quasiparticle edge dispersion, determined by the maximum at $A^{\mathrm{edge}}(k,\varepsilon)$. Note that $A^{\mathrm{edge}}(k,\varepsilon)$ has the form of a delta-function in the band-gap region, since the damping factor $\Gamma(\varepsilon)$ in Eq.~(\ref{eqRnD:16}) vanishes therein.}
\end{figure}

Once $E_{\sigma}^{\mathrm{edge}}(k_x,\varepsilon)$ is known, in accordance with Eq.~(\ref{eq:QWTAI3b}), the spectral function for the quasiparticle edge states is written as
\begin{equation}
\label{eqRnD:15}
A^{\mathrm{edge}}(k_x,\varepsilon)=-\dfrac{1}{\pi}\sum_{\sigma}\dfrac{\Gamma(\varepsilon)}
{\left[\varepsilon-\textrm{Re}E_{\sigma}^{\mathrm{edge}}(k_x,\varepsilon)\right]^2+\Gamma(\varepsilon)^2},
\end{equation}
where $\Gamma(\varepsilon)$ is a damping factor for quasiparticle edge states:
\begin{equation}
\label{eqRnD:16}
\Gamma(\varepsilon)=\textrm{Im}\Sigma_0(\varepsilon)-\dfrac{D}{B}\textrm{Im}\Sigma_z(\varepsilon).
\end{equation}

Figure~\ref{Chapter3Fig19} shows the evolution of total spectral function determined as the sum $A(k,\varepsilon)$ in Eq.~(\ref{eq:QWTAI17}) and $A^{\mathrm{edge}}(k,\varepsilon)$ in Eq.~(\ref{eqRnD:15}) at several values of the short-range disorder strength $W$ in the 6~nm wide HgTe/Cd$_{0.7}$Hg$_{0.3}$Te QW. It is seen that although the spectral function broadens with increasing disorder strength, it still clearly represent the quasiparticle bulk dispersion. Interestingly, the bulk quasiparticles at the gapless case at $W=W_c=W^{*}$ mimics massless Dirac fermions as it is in the ``clean'' limit}. If the strength of disorder exceeds a critical value $W_c=W^{*}$, which leads to $\mathbb{Z}_2(\varepsilon)=1$, the bulk states coexist with a pair of quasiparticle helical edge states -- see Fig.~\ref{Chapter3Fig19}(c). Importantly, since $\textrm{Im}\Sigma_z(\varepsilon)$ and $\textrm{Im}\Sigma_0(\varepsilon)$ both vanish in the band-gap region (see Fig.~\ref{Chapter3Fig18}(b,c)), the damping factor $\Gamma(\varepsilon)$ turns to zero. As a result, the spectral function of the edge states in the band-gap takes the form of a delta function
\begin{equation}
\label{eqRnD:17}
A^{\mathrm{edge}}(k_x,\varepsilon)=\sum_{\sigma}\delta\left\{\varepsilon-\textrm{Re}E_{\sigma}^{\mathrm{edge}}(k_x,\varepsilon)\right\}.
\end{equation}
The latter means that the edge quasiparticles \emph{do not decay}. The absence of energy dissipation for the quasiparticle helical edge states in the band-gap explains the quantized values of the longitudinal conductance in the presence of an external bias, found previously in numerical calculations in the strip geometry~\cite{TAI1,TAI2}.

In conclusion to this section, I would like to emphasize the importance of accounting for the energy dependence of the self-energy matrix $\Sigma(\varepsilon)$ when describing topological phase transitions induced by any form of interaction. The particular case considered here -- a topological phase transition driven by short-range electrostatic disorder -- not only confirms the validity of the intuitive interpretation proposed by Groth \emph{et al.}~\cite{TAI2} in terms of the renormalization of both the chemical potential and the mass term, but also demonstrates the broader applicability of the quasiparticle concept for the topological characterization of 2D systems. This quasiparticle approach goes beyond the previously employed ``topological Hamiltonian'' $H_t(\mathbf{k})$ formalism introduced by Z.~Wang and collaborators~\cite{TAI25,TAI26,TAI27}, providing a more general and physically transparent framework for understanding interaction-driven topological phase transitions.

\subsection{\label{sec:disQW}Single HgTe QWs beyond the BHZ model}
Despite the conceptual simplicity and elegance of the BHZ model, its applicability to realistic HgTe/CdHgTe QWs~\cite{HgTeQWs2} remains restricted to the small vicinity of the $\Gamma$ point of the Brouillon zone. For a long time, the only reliable way to describe the electronic states over a broader range of quasimomentum was through numerical calculations based on the eight-band $\mathbf{k}\cdot\mathbf{p}$ Kane Hamiltonian~\cite{HgTeQWs1}. However, following our studies of double HgTe QWs, one can see an interesting alternative to the Kane-model calculations. As discussed in Section~\ref{sec:HgTeDQWs}, the low-energy effective model derived for double HgTe QWs remains valid for all values of the tunneling barrier thickness, including the limiting case of zero barrier width, where the double QW naturally transforms into a single HgTe QW. In this limiting case, the model effectively extends the accessible range of wave vectors for describing the band structure, allowing one to capture features such as the side maxima of the valence band~\cite{TAI49} that lie far beyond the range of the BHZ Hamiltonian. Figure~\ref{Chapter3Fig20} provides a comparison of the band structure for the HgTe/Cd$_{0.7}$Hg$_{0.3}$Te QWs grown on (001) CdTe buffer calculated within the Kane Hamiltonian and $H_{2D}(\mathbf{k})$ in Eq.~(\ref{eq:DQW5}) with the only isotropic part $H_{4\times4}^{(i)}(k_x,k_y)$ in Eq.~(\ref{eq:HOTI3}) for different values of $d$. It is seen that although the difference in the energy dispersion of E1 subband calculated within two different approach increases with the QW width, a reasonable agreement takes place for $d\leq9$~nm. Thus, the low-energy Hamiltonian $H_{2D}(\mathbf{k})$ introduced in Eq.~(\ref{eq:DQW5}) can be effectively used to explore a wide range of problems that go beyond the conventional BHZ model -- for instance, the study of topological edge states in wide HgTe QWs~\cite{TAI49}. In this section, however, we will briefly focus on another aspect -- the disorder-induced topological phase transition in single HgTe QWs beyond the BHZ model~\cite{TAI48}.

\begin{figure}
\centering
\includegraphics [width=1.0\columnwidth, keepaspectratio] {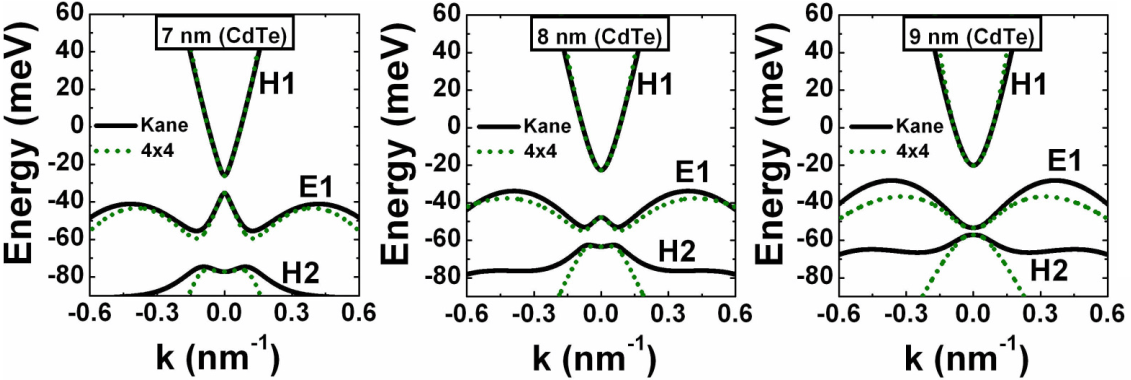}
\caption{\label{Chapter3Fig20}  Comparison of band structure for the 7, 8 and 9 nm wide HgTe/Cd$_{0.7}$Hg$_{0.3}$Te QWs grown on (001) CdTe buffer, calculated on the basis of the Kane Hamiltonian and $H_{2D}(\mathbf{k})$ in Eq.~(\ref{eq:DQW5}) with the only isotropic part $H_{4\times4}^{(i)}(k_x,k_y)$ in Eq.~(\ref{eq:HOTI3}). Positive values of $\mathbf{k}$ correspond to [100] crystallographic orientation. Parameters for $H_{4\times4}^{(i)}(\mathbf{k})$ used in the calculations are provided in Ref.~\cite{TAI49}.}
\end{figure}

For simplicity, we will neglect the anisotropic part ${H}_{4\times4}^{(a)}({k}_x,{k}y)$ in Eq.~(\ref{eq:HOTI5}). Its inclusion in $H_{2D}(\mathbf{k})$ from Eq.~(\ref{eq:DQW5}) would not qualitatively change the obtained results but would only complicate the analytical derivations. Indeed, in the absence of $H_{4\times4}^{(a)}(\mathbf{k})$, $H_{4\times4}(\mathbf{k})$ in Eq.~(\ref{eq:HOTI2}) possesses full rotational symmetry, which results in the full rotational symmetry of the Green's function $\hat{G}(\mathbf{k},\varepsilon)$:
\begin{equation}
\label{eq:BeyondBHZ1}
\hat{G}(k,\varepsilon)=\left[\varepsilon-\tilde{H}_{4\times4}(k)-\hat{\Sigma}(k,\varepsilon)\right]^{-1},
\end{equation}
which depends only on $k=|\mathbf{k}|$. In Eq.~(\ref{eq:BeyondBHZ1}) $\tilde{H}_{4\times4}(k)$ is defined by $H_{4\times4}^{(i)}(k_x,k_y)$ in Eq.~(\ref{eq:HOTI3}) as $\tilde{H}_{4\times4}(k)=H_{4\times4}^{(i)}(k,0)$, and the self-energy matrix $\hat{\Sigma}(k,\varepsilon)$ represents the presence of the random impurity potential described by Eq.~(\ref{eq:QWTAI11}). After some mathematics provided in Ref.~\cite{TAI48,TAI50}, the calculation of the self-energy matrix $\hat{\Sigma}(k,\varepsilon)$ within the SCBA is reduced to the calculation o the following integral:
\begin{equation}
\label{eq:BeyondBHZ2}
\hat{\Sigma}(k,\varepsilon)=n_{i}\int\limits_0^{K_c}\dfrac{k^\prime dk^\prime}{2\pi}\begin{pmatrix}
V_0^2G_{11}^\prime & V_{-1}^2G_{12}^\prime & V_{+2}^2G_{13}^\prime & V_{+1}^2G_{14}^\prime \\[2pt]
V_{+1}^2G_{21}^\prime & V_{0}^2G_{22}^\prime & V_{+3}^2G_{23}^\prime & V_{+2}^2G_{24}^\prime \\[2pt]
V_{-2}^2G_{31}^\prime & V_{-3}^2G_{32}^\prime & V_{0}^2G_{33}^\prime & V_{-1}^2G_{34}^\prime \\[2pt]
V_{-1}^2G_{41}^\prime & V_{-2}^2G_{42}^\prime & V_{+1}^2G_{43}^\prime & V_{0}^2G_{44}^\prime
\end{pmatrix},
\end{equation}
where $n_{i}$ is the concentration of impurities, $G_{ij}^\prime{\equiv}G_{ij}(k^\prime,\varepsilon)$ are the component of the Green's function in Eq.~(\ref{eq:BeyondBHZ1}), and $V_n^{2}{\equiv}V_n(k,k^\prime)^2$ is defined as
\begin{equation}
\label{eq:BeyondBHZ3}
V_n(k,k^\prime)^2=\int\limits_0^{2\pi}\dfrac{d\theta}{2\pi}|\tilde{v}(\textbf{k}-\textbf{k}^\prime)|^2
\cos n\theta.
\end{equation}
Here, similar to Eq.~(\ref{eq:QWTAI13b}), we introduce a cut-off wave-vector $K_c=\pi/a_0$ (where $a_0$ is the lattice constant in the QW plane, which is actually the lattice constant of the CdTe buffer), corresponding to the edge of 2D Brillouin zone. This naturally limits the integration over quasimomentum~$k$.

\begin{figure}
\centering
\includegraphics [width=1.0\columnwidth, keepaspectratio] {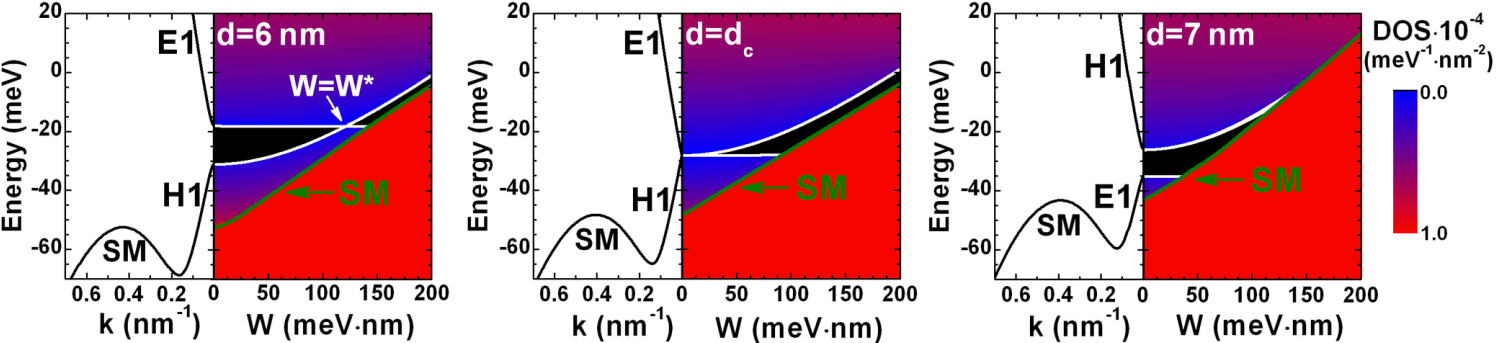}
\caption{\label{Chapter3Fig21} Band structure and the DOS as a function of the disorder strength~$W$ calculated on the basis of $H_{2D}(\mathbf{k})$ in Eq.~(\ref{eq:DQW5}) for the single HgTe QW at different QW width: $d=6$~nm, $d=d_c$ and $d=7$~nm -- the similar calculations within the BHZ model for the same QWs have been shown in Fig.~\ref{Chapter3Fig17}. The white curves represent the edges of the area with the vanishing DOS found from the numerical calculations. The green curve shows the energy edge of the side maxima identified in the DOS. Parameters for $H_{4\times4}^{(i)}(\mathbf{k})$ used in the calculations are provided in Ref.~\cite{TAI48}.}
\end{figure}

To proceed further, we assume $\tilde{v}(q)=u_0$, which corresponds to the disorder formed by the short-range impurities:
\begin{equation}
\label{eq:BeyondBHZ4}
v(\textbf{r})=\int\limits_0^{\infty}\dfrac{qdq}{2\pi}\tilde{v}(q)\int\limits_0^{2\pi}\dfrac{d\theta}{2\pi}e^{i{q}{r}\cos\theta}=\dfrac{u_0}{2\pi}\dfrac{\delta\left(r\right)}{r}.
\end{equation}
In this case, $V_n(k,k^\prime)^2=u_0^2\delta_{n,0}$ in Eq.~(\ref{eq:BeyondBHZ3}), resulting in a diagonal form of the self-energy matrix in Eq.~(\ref{eq:BeyondBHZ2}):
\begin{multline}
\label{eq:BeyondBHZ5}
\hat{\Sigma}(\varepsilon)=\begin{pmatrix}
\Sigma_{E1}(\varepsilon) & 0 & 0 & 0 \\[4pt]
0 & \Sigma_{H1}(\varepsilon) & 0 & 0 \\[4pt]
0 & 0 & \Sigma_{H2}(\varepsilon) & 0 \\[4pt]
0 & 0 & 0 & \Sigma_{E2}(\varepsilon)
\end{pmatrix}=~\\
=\dfrac{W^2}{4\pi}\int\limits_0^{K_c^2}dx
\begin{pmatrix}
G_{11}\left(\sqrt{x},\varepsilon\right) & 0 & 0 & 0 \\[4pt]
0 & G_{22}\left(\sqrt{x},\varepsilon\right) & 0 & 0 \\[4pt]
0 & 0 & G_{33}\left(\sqrt{x},\varepsilon\right) & 0 \\[4pt]
0 & 0 & 0 & G_{44}\left(\sqrt{x},\varepsilon\right)
\end{pmatrix},
\end{multline}
where $W=\sqrt{n_{i}u_0^2}$ is the disorder strength. Finally, the spectral function $A(k,\varepsilon)$ and density-of-states $D(\varepsilon)$ for the case of the short-range disorder can be written as
\begin{eqnarray}
\label{eq:BeyondBHZ6}
A(k,\varepsilon)=-\dfrac{1}{\pi}\textrm{Im}\left\{ \sum\limits_{i=1}^{4}
G_{ii}\left(k,\varepsilon\right)\right\},\notag\\
D(\varepsilon)=-\dfrac{g_S}{W^2\pi}\textrm{Im}\left\{\textrm{Tr}\left(\hat{\Sigma}(\varepsilon)\right)\right\}.~
\end{eqnarray}
We remind that in order to take into account the contribution from the lower block in Eq.~(\ref{eq:DQW5}), one should set the degeneracy factor of DOS as $g_S=2$.

Figure~\ref{Chapter3Fig21} shows the DOS evolution with the disorder for the same QW widths as in Fig.~\ref{Chapter3Fig17}, which presents the calculations within the BHZ model. As it is seen for the 6~nm HgTe QW, the disorder-induced phase transition at $W=W_{c}=W^{*}$ is still identified. However, the critical values of $W_{c}=W^{*}$ and the areas with the vanishing DOS differ significantly in two models. Another feature, which can not be addressed in the BHZ model, is the evolution of the DOS associated with the side maxima (SM) of the top valence subband. In the clean limit, the side maxima result in the large step-like increasing of the DOS. At non-zero $W$, such step-like behavior can be also used for qualitative determination of the SM position. The white and green curves in the top panels of Fig.~\ref{Chapter3Fig21} represent the evolution of the band edges in the $\Gamma$ point and the side maxima, respectively. In contrast to the BHZ model, these curves can be identified only in the numerical calculations. As the SM position primarily depends on the distance between E2 and H2 subbands~\cite{TAI49}, its evolution with the disorder remains qualitatively the same for any values of the pass parameter $M_1$, describing the band inversion between E1 and H1 subbands.
\begin{figure}
\centering
\includegraphics [width=0.55\columnwidth, keepaspectratio] {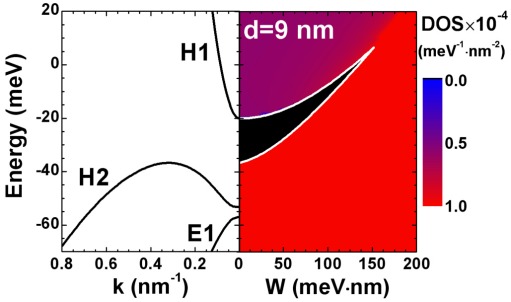}
\caption{\label{Chapter3Fig22} Band structure and the DOS as a function of the disorder strength~$W$ for the 9~nm wide HgTe QW. The white curves represent the edges of the black area with the vanishing DOS identified in the numerical calculations. Parameters for $H_{4\times4}(\mathbf{k})$ used in the calculations are given in Ref.~\cite{TAI48}.}
\end{figure}

As it is seen, the SM contribution increases with the disorder and strongly affects the area with the vanishing DOS. For the inverted HgTe QWs, the SM may result in the band-gap closing and transition into the semimetal state. The latter is clearly seen for the 9~nm wide HgTe QW representing \emph{indirect-gap} QSHI in HgTe QWs (see Fig~\ref{Chapter3Fig22}). Indeed, the upper boundary of the area with the vanishing DOS represents the evolution of the conduction band edge, while the lower boundary corresponds to the SM evolution. The semimetal state arises when the side maxima exceed the conduction band bottom. Note that such state also exists in the wide HgTe QWs in the clean limit -- see Fig.~\ref{Chapter3Fig1}(e). Thus, the disorder may not only yield to the band-inversion between E1 and H1 subbands in HgTe QWs within the quasiparticle pictures discussed in Section~\ref{sec:qpQW} but induce the semimetal state as well.

\subsection{\label{sec:disDQW}Double HgTe QWs}
Let us now apply the formalism developed in Section~\ref{sec:disQW} to the case of double HgTe QWs, for which the low-energy Hamiltonian $H_{2D}(k_x,k_y)$ in Eq.~(\ref{eq:DQW5}) -- taking into account the coupling between the E1, H1, E2, and H2 subbands -- was originally derived (see Section~\ref{sec:HgTeDQWs}). Similarly to the case of single HgTe QWs, we neglect the anisotropic terms due to the cubic crystal symmetry when describing the inversion between the E1-H1 and E2-H2 subband pairs, induced by the short-range disorder potential. Since the self-energy matrix $\hat{\Sigma}(\varepsilon)$ in Eq.~(\ref{eq:BeyondBHZ5}) for such a short-range random potential remains diagonal and the anisotropic structural parameters $R_{1}^{(a)}$ and $R_{2}^{(a)}$ in Eq.~(\ref{eq:HOTI5}) are much smaller than their isotropic counterparts $R_{1}^{(i)}$ and $R_{2}^{(i)}$ in Eq.~(\ref{eq:HOTI3}), their contribution can indeed be neglected when calculating $\hat{\Sigma}(\varepsilon)$. However, it should be emphasized that these anisotropic terms become crucial in $H_{4\times4}(k_x,k_y)$ when characterizing topological states induced by disorder within the quasiparticle approach, discussed in Section~\ref{sec:qpQW}:
\begin{equation}
\label{eq:BeyondBHZ7}
\mathcal{H}_{qp}(\mathbf{k},\varepsilon)=\begin{pmatrix}
H_{4\times4}(\mathbf{k}) & 0 \\ 0 & H_{4\times4}^{*}(-\mathbf{k})\end{pmatrix}
+\begin{pmatrix}
\hat{\Sigma}(\varepsilon) & 0 \\ 0 & \hat{\Sigma}(\varepsilon)\end{pmatrix},
\end{equation}
where the blocks $H_{4\times4}(\mathbf{k})$ and $\hat{\Sigma}(\varepsilon)$ are given by Eqs.~(\ref{eq:HOTI2}) and (\ref{eq:BeyondBHZ5}), respectively.

\begin{figure}
\centering
\includegraphics [width=1.0\columnwidth, keepaspectratio] {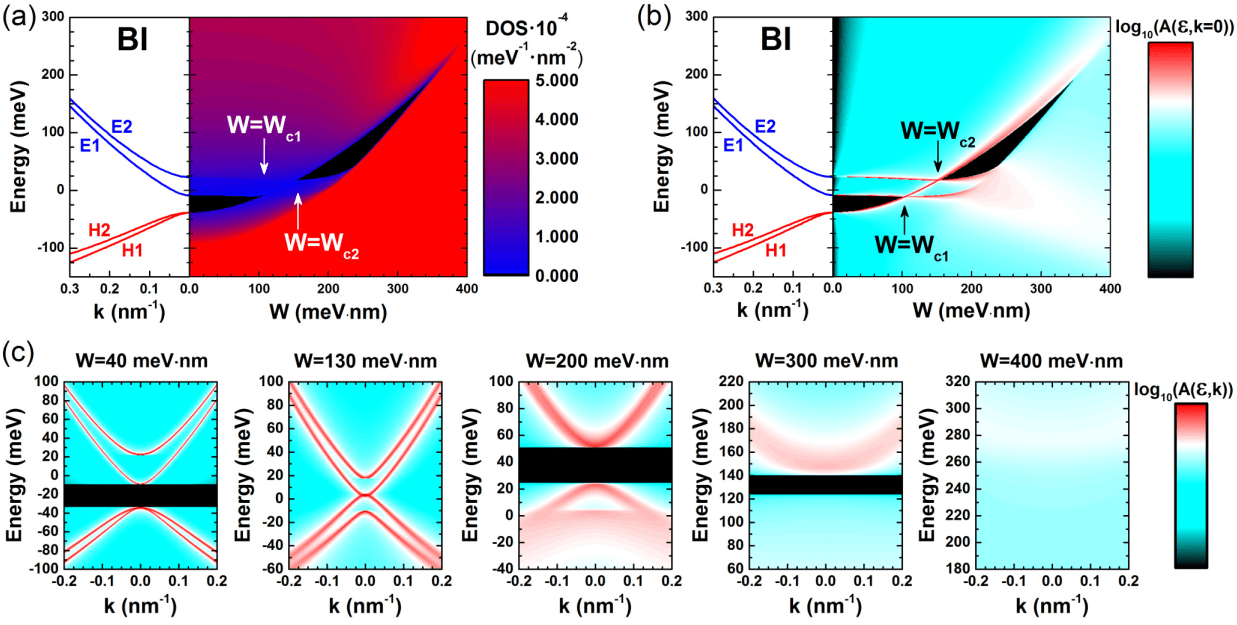}
\caption{\label{Chapter3Fig23}  \textbf{(a)} Band structure and color map of the DOS $D(\varepsilon)$ as a function of the disorder strength $W=\sqrt{{n}_{i}}u_0$ calculated for the double HgTe~QW, which has a band insulator state in the clean limit ($M_1>0$ and $M_2>0$ in $H_{2D}(k_x,k_y)$). \textbf{(b)}~Band structure and color map of the spectral function $A(k=0,\varepsilon)$ at the $\Gamma$ point for the same double HgTe QW. \textbf{(c)}~Colormap of spectral function $A(k,\varepsilon)$ at different disorder strength for the same double QW. Parameters for $H_{4\times4}(\mathbf{k})$ used in the calculations are provided in Ref.~\cite{TAI50}.}
\end{figure}

Figure~\ref{Chapter3Fig23}(a) shows the evolution of DOS with the disorder strength $W=\sqrt{{n}_{i}}u_0$ in double HgTe~QW with BI state in the clean limit represented by the positive values mass parameters $M_1$ and $M_2$ in in $H_{2D}(k_x,k_y)$. As clear, the band-gap in the DOS decreases with $W$ until it vanishes above a critical value $W_{c1}$, and then it is reopened again at $W>W_{c2}$ (see black regions in Fig.~\ref{Chapter3Fig23}(a) where the DOS vanishes). In the range $W_{c1}{\leq}W{\leq}W_{c2}$, the double HgTe QW remains gapless, which is well illustrated by the non-zero $D(\varepsilon)$. With a further increase in disorder strength, the newly opened gap closes again at $W\simeq350$~meV$\cdot$nm. Figure~\ref{Chapter3Fig23}(b) provides a disorder-induced evolution of the spectral function at the $\Gamma$ point, which perfectly reproduces two band-gap regions arising in the DOS. As clear, the first band-gap closing at $W=W_{c1}$ is attributed to the mutual inversion of \emph{E}1 and \emph{H}1 quasiparticle subbands, while the band-gap opening at $W=W_{c2}$ is related to the band inversion but in the second pair of \emph{E}2 and \emph{H}2 quasiparticle subbands. Figure~\ref{Chapter3Fig23}(c) shows the evolution of the spectral function $A(\mathbf{k},\varepsilon)$ with increasing the disorder strength. As clear, a gapless state at $W_{c1}{<}W{<}W_{c2}$ with the band ordering \emph{E}2-\emph{H}1-\emph{H}2-\emph{E}1 can be directly attributed to the BG state in the clean limit, discussed previously in Section~\ref{sec:1storderHgTe}. Moreover, the transformation of the quasiparticle spectrum with $W$ indeed resembles the changes in the band structure of double HgTe QW with the HgTe layer thicknesses (see Fig.~\ref{Chapter3Fig14} ). Thus, starting from a band insulator in the clean limit, the 2D system, under influence of disorder, successively passes into the BG state, and then into the higher-order topological insulator with double-band inversion.

In our original work~\cite{TAI50}, using the quasiparticle Hamiltonian $\mathcal{H}_{qp}(\mathbf{k},\varepsilon)$ introduced in Eq.~(\ref{eq:BeyondBHZ7}), we have mathematically demonstrated the change of the $\mathbb{Z}_2$ invariant at the critical disorder strengths $W=W_{c1}$ and $W=W_{c2}$. Moreover, we have shown that the disorder-induced double-inverted insulating state is, in fact, a higher-order topological insulator (HOTI), differing from its clean-limit counterpart only by the renormalization of the structure parameters:
\begin{eqnarray}
\label{eq:BeyondBHZ8}
C_1{\rightarrow}C_1+\dfrac{1}{2}\left[\textrm{Re}\Sigma_{E1}(\varepsilon)+\textrm{Re}\Sigma_{H1}(\varepsilon)\right],~\notag\\
M_1{\rightarrow}M_1+\dfrac{1}{2}\left[\textrm{Re}\Sigma_{E1}(\varepsilon)-\textrm{Re}\Sigma_{H1}(\varepsilon)\right],~\notag\\
C_2{\rightarrow}C_2+\dfrac{1}{2}\left[\textrm{Re}\Sigma_{E2}(\varepsilon)+\textrm{Re}\Sigma_{H2}(\varepsilon)\right],~\notag\\
M_2{\rightarrow}M_2+\dfrac{1}{2}\left[\textrm{Re}\Sigma_{E2}(\varepsilon)-\textrm{Re}\Sigma_{H2}(\varepsilon)\right],~
\end{eqnarray}
where the component of the self-energy matrix is given by Eq.~(\ref{eq:BeyondBHZ5}). Note that since $\textrm{Im}\hat{\Sigma}(\varepsilon)=0$ inside the band-gap, the self-energy matrix components in the quasiparticle edge Hamiltonian are pure real. Thus, with the replacement expressed by Eq.~(\ref{eq:BeyondBHZ8}), the problem of the emergence of  corner states in the presence of the disorder is reduced to the problem already solved in the clean limit -- see Section~\ref{sec:HOTIHgTe}. I will not dwell on this point here, since the detailed discussion can be found in the original publication~\cite{TAI50}. However, I would like to draw the reader's attention to an important fact: the band-gap closing that occurs at $W>W_{c2}$ is not related to any change in the $\mathbb{Z}_2$ invariant. Instead, this gap closing originates from the overlap of the spectral functions corresponding to the E1 and H1 quasiparticle subbands. This behavior indicates that the imaginary parts of the quasiparticle eigenvalues ${E}_{qp}^{(n)}$ become much larger than their real parts, which define the quasiparticle dispersions. Consequently, such a type of band-gap closing is apparently associated with a modification of the vorticity -- a topological invariant defined for any pair of bands as the winding number of their complex energies ${E}_{qp}^{(n)}$ in the complex-energy plane~\cite{TAI51}. A detailed clarification of this point is planned to be the subject of future research.

\begin{figure}
\centering
\includegraphics [width=0.9\columnwidth, keepaspectratio] {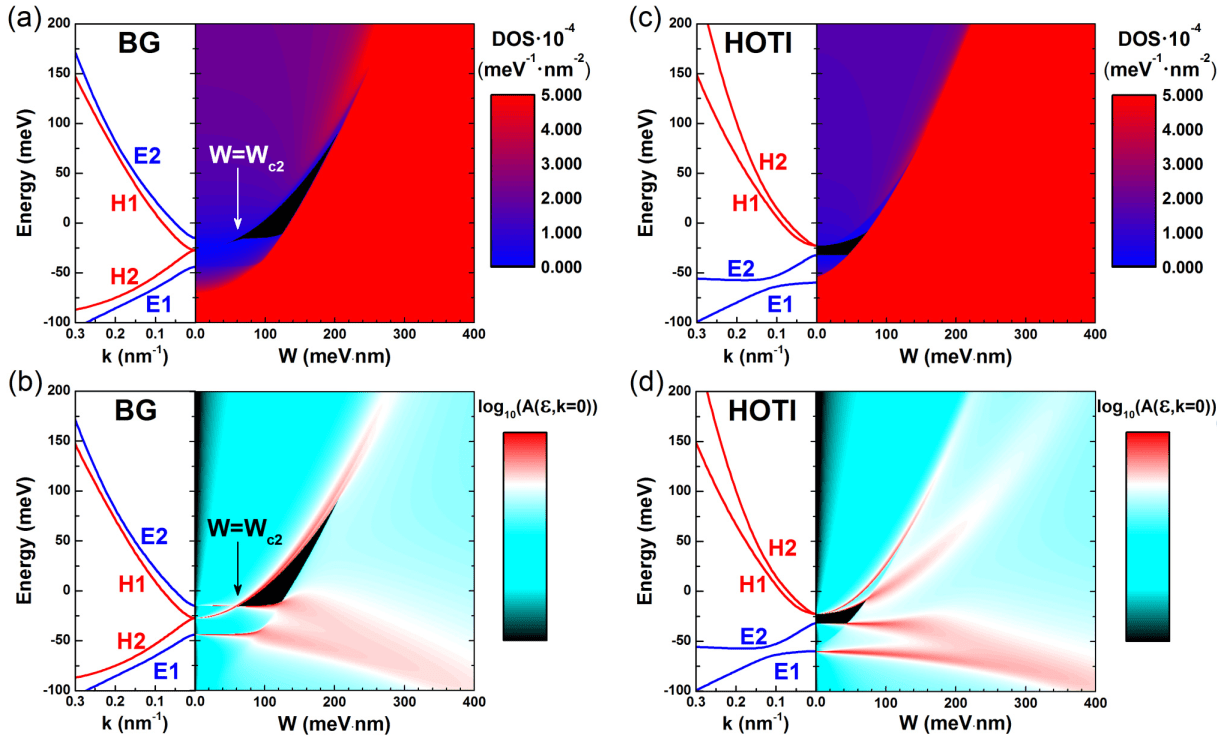}
\caption{\label{Chapter3Fig24} \textbf{(a,b)} Band structure, DOS $D(\varepsilon)$ and spectral function  at the $\Gamma$ point $A(k=0,\varepsilon)$ as a function of the disorder strength $W=\sqrt{{n}_{i}}u_0$ for double HgTe QW hosting BG state at $W=0$ ($M_1<0$ and $M_2>0$). \textbf{(c,d)}~The same as in (a) and (b) panels but for double HgTe QW hosting HOTI state in the absence of disorder ($M_1<0$ and $M_2<0$). Parameters for $H_{4\times4}(\mathbf{k})$ used in the calculations are provided in Ref.~\cite{TAI50}.}
\end{figure}

Let us now briefly consider disorder-induced phase transitions in double HgTe QWs with single- and double-band-inversion at $W=0$. Figure~\ref{Chapter3Fig24} provides the evolution of the DOS and the spectral function at the $\Gamma$ point with the disorder strength for double HgTe QWs, hosting BG and HOTI states in the clean limit, respectively. As seen from Fig.~\ref{Chapter3Fig24}(a), the double HgTe QW with BG state maintains zero band-gap even in the presence of weak disorder. When the disorder strength reaches its critical value $W_{c2}$, a band-gap opens in the DOS. As clear from Fig.~\ref{Chapter3Fig24}(b), this moment corresponds to the band inversion between \emph{E}2 and \emph{H}2 subbands. Such behavior of the spectral function unambiguously indicates that the 2D system at $W<W_{c2}$ is in the same topological state as at $W=0$, which is uniquely identified as the BG state. When $W$ exceeds its critical value $W_{c2}$, the system becomes a HOTI until its bulk band-gap vanishes at higher disorder strength (cf. Fig.~\ref{Chapter3Fig23}). A similar band-gap evolution is also seen for the double HgTe QW with a HOTI at $W=0$ as shown in Fig.~\ref{Chapter3Fig24}(c). As clear from Fig.~\ref{Chapter3Fig24}(d), such band-gap behavior is caused primarily by smearing of the edges of the E1 and H1 bands at the $\Gamma$ point that leads to the DOS overlapping between conduction and valence subbands.

\section{\label{sec:SummaryHgTe}Perspectives}
The experimental and theoretical studies of HgTe-based QW heterostructures performed in the THz team in Montpellier, with my direct participation, are by no means limited to the papers whose results were briefly presented in this chapter. In particular, several important topics have remained beyond the scope of this m\'{e}moire, such as the studies of electron-electron interaction effects and the investigations of spontaneous cyclotron emission from Dirac fermions in HgTe QWs. At present, a part of my ongoing theoretical research is devoted precisely to the interpretation of the most recent, yet unpublished, experimental results on cyclotron emission obtained by the THz team in HgTe QWs under strong electric fields. At this stage, it is still unclear which system -- HgTe/CdHgTe QWs or bulk HgCdTe films discussed in Chapter~\ref{Chap:Kane} -- is more favorable for the realization of the long-sought THz Landau level laser. As for the topological states, I am convinced that both theoretical and experimental investigations of various topological phenomena in HgTe-based QWs heterostructures, which has already spanned almost two decades, are still far from its completion. For understandable reasons, I will refrain from disclosing here all my ideas for future research, which, in one way or another, will continue to concern HgTe-based QW heterostructures. I will only hint that some of these studies will be devoted to exploring the influence of impurity potentials and cavity modes on both bulk and edge transport in these heterostructures.

\chapter{\label{Chap:III-V}Topological states in III-V semiconductor QW heterostructures}
As clear from the previous chapter, HgTe-based QW heterostructures provide a platform for realizing a wide variety of topological states. However, both the experimental exploration of these systems and their potential implementation in future topological electronics face certain intrinsic limitations. The first of these is the very limited availability of high-quality HgTe/CdHgTe QW structures, which are grown by molecular beam epitaxy only in a few specialized laboratories worldwide. Because of the weak bonding energy of mercury atoms, all processing steps must be carried out at relatively low temperatures, which severely complicates lithography and metallization procedures. As a result, even the realization of micron-sized gated Hall bars necessary for the observation of edge-state transport has been successfully achieved only by at the University of W\"{u}rzburg. This situation naturally motivated me to search for alternative material platforms that could reproduce the key features of HgTe-based topological systems while providing greater technological accessibility. From this perspective, QW heterostructures based on III-V semiconductors appear particularly promising. The technology of III-V epitaxy is among the most mature in modern semiconductor science, enabling the growth of high-quality structures by molecular-beam epitaxy on a variety of substrates, including silicon~\cite{III-VQWs1}. This compatibility with Si-based platforms opens unique perspectives for large-scale integration of topological materials into conventional electronic technologies. Furthermore, the device fabrication processes for III-V heterostructures are considerably simpler and more robust compared to those involving HgTe, allowing one to explore the physics of topological states using well-established technological tools.

\begin{figure}
\centering
\includegraphics [width=0.5\columnwidth, keepaspectratio] {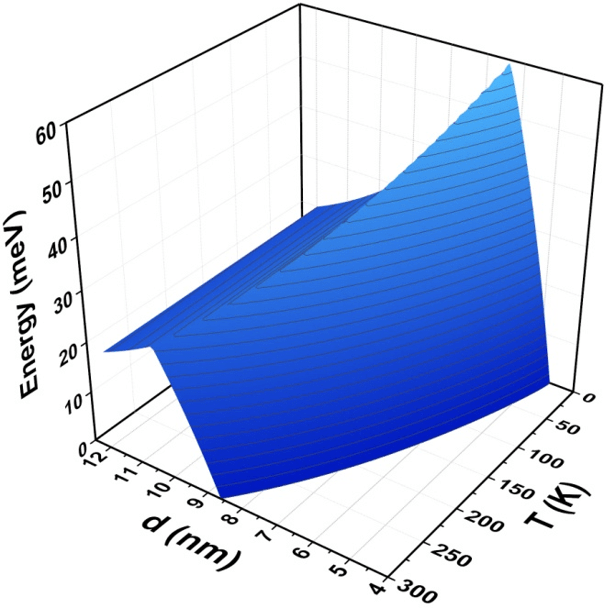}
\caption{\label{Chapter4Fig1} 3D plot of the band-gap in QSHI state as a function of QW width $d$ and temperature $T$ for (001)-oriented compressively-strained HgTe/Cd$_{0.7}$Hg$_{0.3}$Te QWs with maximum possible strain in HgTe epilayer (+1.4\%)~\cite{HgTeQWs12}.}
\end{figure}

Another strong motivation for my search for alternatives to HgTe/CdHgTe QWs was the extremely low temperature required for observing the quantum spin Hall effect (QSHE) associated with transport via edge states. In the first experimental demonstrations of this phenomenon~\cite{HgTeQWs5}, quantized conductance was observed only at temperatures of several tens of millikelvin. The appearance in 2016 of samples with compressively biaxially strained HgTe QWs~\cite{HgTeQWs12}, characterized by an inverted band-gap as large as 55 meV, initially raised hopes that the quantized conductance could be measured at much higher temperatures -- ideally above 77~K (the liquid nitrogen temperature). These expectations seemed reasonable, especially in light of the QSHE observation up to 100 K in monolayer WTe$_2$, which possesses an inverted band gap of approximately 45 meV~\cite{III-VQWs3}. It is worth emphasizing, however, that even though the inverted band-gap in HgTeQWs decreases with increasing temperature (as discussed in the previous Chapter), in compressively strained HgTe QWs it remains larger than that of monolayer WTe$_2$ up to about 150 K.

Figure~\ref{Chapter4Fig1} provides a 3D plot of the inverted band-gap as a function of the QW width $d$ and temperature $T$ for a (001)-oriented compressively strained HgTe/Cd$_{0.7}$Hg$_{0.3}$Te QW, assuming the maximum possible strain in the HgTe epilayer of +1.4\%~\cite{HgTeQWs12}. These theoretical calculations were performed using the same bulk material parameters and the same computational approach as those discussed previously in Section \ref{sec:PandTinQWs}. As one can see, a 6.5 nm-wide QW, which exhibits a band-gap of nearly 60 meV at low temperatures, undergoes a band-gap closing and a phase transition into the trivial regime at approximately 180 K. For other QW widths, the maximum achievable non-trivial band-gap becomes comparable to 45 meV only at 150 K. Thus, at least theoretically, this figure suggests that the observation of the high-temperature QSHE in HgTe QWs at temperatures above 77 K should be possible. Nevertheless, to date, the highest temperature of the QSHE observation does not exceed 15 K~\cite{III-VQWs2}. Most likely, this limitation is not of fundamental origin but rather arises from the severe technological challenges associated with fabricating micron-sized gated Hall bars from compressively strained HgTe~QWs. The mechanical fragility of strained HgTe epilayers and the delicate low-temperature processing steps make the device fabrication process extremely demanding. This, in turn, also motivates the search for more reliable material systems with a technologically stable fabrication process of short-channel quantum devices.

In this Chapter, I will present the main results of my research devoted to III-V semiconductor QW heterostructures carried out in Montpellier -- systems that, in many respects, can be regarded as III-V analogues of single and double HgTe/CdHgTe QWs. In addition, I will review the key experimental results of the THz team in Montpellier and the Technische Physik group in W\"{u}rzburg, which provided strong experimental confirmation of the theoretical predictions made in our original work~\cite{III-VQWs4}. The results presented in this Chapter not only demonstrate that III-V QW heterostructures constitute a new and reliable experimental platform for exploring the same fundamental topological phenomena as those observed in HgTe-based QWs, but they also open new perspectives for engineering and controlling topological states in technologically mature III-V semiconductor systems.

\section{\label{sec:3LIII-V}Quantum spin Hall insulator with a large bandgap, Dirac fermions, and bilayer graphene analog}
The first 2D system based on III-V semiconductors, in which the QSHI state was both theoretically predicted~\cite{III-VQWs5} and subsequently observed experimentally~\cite{III-VQWs6}, is the inverted InAs/GaSb QW bilayer. In contrast to HgTe/CdHgTe QWs, where the inversion between the E1 and H1 subbands originates from the inverted band structure of bulk HgTe, the subband inversion in InAs/GaSb QW bilayers embedded between AlSb barriers arises from the specific broken-gap band alignment of InAs and GaSb semiconductors at their interface~\cite{III-VQWs7,III-VQWs8,III-VQWs9}. For sufficiently thin InAs and GaSb layers, the electron-like E1 level localized within the InAs layer lies above the hole-like H1 level from the GaSb layer, resulting in a trivial band ordering in the QW. As the layer thicknesses increase, an inverted band ordering may occur, where the E1 subband becomes lower in energy than the H1 subband. The inherent property of bilayer InAs/GaSb QWs is the lack of inversion symmetry in the growth direction, which leads to the crossing of E1 and H1 subbands away from the $\Gamma$ point of the Brillouin zone. This, in turn, leads to the opening of a hybridization band-gap $\Delta$ due to coupling between the electron and hole states at finite quasimomentum~$\mathbf{k}$, as schematically illustrated in Fig.~\ref{Chapter4Fig2}. However, because of the relatively small inverted band-gap values -- typically lower than 3--4 meV -- the observation of quantized conductance values due to the transport via topological edge states in InAs/GaSb QW bilayers is far more challenging than in HgTe-based systems. The small band-gap results in substantial residual bulk conductivity and additional parasitic edge transport, which persists even in the trivial regime~\cite{III-VQWs10}, thereby  limiting a clear observation of the transport via topological edge states to temperatures of a few tens of millikelvin. Since the first experimental demonstration of the QSHI state in the InAs/GaSb QW bilayers, some attempts have been made to overcome the problem of the small inverted band-gap~\cite{III-VQWs11}. Lately, its value has been enhanced by replacing the GaSb layer of the composite QW with a GaInSb alloy. However, the maximum value reached in this way was only on the order of 20 meV~\cite{III-VQWs12}.

\begin{figure}
\centering
\includegraphics [width=0.65\columnwidth, keepaspectratio] {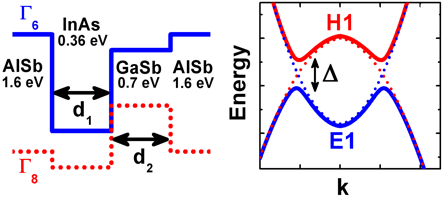}
\caption{\label{Chapter4Fig2} Schematic representation of the broken-gap alignment of the band-edges leading to in the opening of the hybridization band-gap $\Delta$ in inverted InAs/GaSb QW bilayers.
The numbers show the band-gap values in materials of the layers. Here, $d_1$ and $d_2$ are the thicknesses of InAs and GaSb layers, respectively.}
\end{figure}

As I mentioned in Chapter~\ref{Chap:Career}, while preparing the publication of my paper devoted to topological states in double HgTe QWs, an idea occurred to me that could not only overcome the limitations associated with the small band-gap in InAs/GaSb QW bilayers, but also allow the full reproduction of the physics of both single and double HgTe QWs. The concept was deceptively simple: to eliminate the structural inversion asymmetry inherent to InAs/GaSb bilayers by adding an additional layer of either InAs or GaSb. In the original publication~\cite{III-VQWs4}, such three-layer QWs were referred to as QWs with an ``InAs geometry’’ or a ``GaSb geometry'', depending on whether they contained two symmetric InAs or GaSb layers, respectively. As I will demonstrate below, the topological phase diagram of the QWs with InAs geometry closely resemble those occurring in single HgTe/CdTe QWs. In particular, depending on the layer thicknesses, these structures allow the realization of massless Dirac fermions, a semimetal (SM) state, and a QSHI state with a maximum inverted band-gap reaching approximately 16~meV. In contrast, three-layer InAs/GaSb QWs with GaSb geometry exhibit properties similar to those of two tunnel-coupled HgTe QWs discussed previously in Section~\ref{sec:HgTeDQWs}. Specifically, they display a semimetallic  phase characterized by a quadratic band touching at the $\Gamma$ point -- closely analogous to that observed in bilayer graphene -- with the possibility of realizing an unconventional quantum Hall effect~\cite{NaturalBG3} and an electrically tunable band-gap~\cite{NaturalBG4}. Finally, I will demonstrated that three-layer InAs/GaInSb QWs with InAs geometry allow for the realisation a QSHI state with a band-gap approaching 60~meV -- comparable to that of compressively biaxially strained HgTe~QWs~\cite{HgTeQWs12} -- yet, remarkably, almost insensitive to temperature.

\subsection{\label{sec:3LInAs}Three-layer InAs/Ga(In)Sb QWs with two InAs layers}
Let us first consider symmetrical InAs/GaSb/InAs QWs confined by outer AlSb barriers. The QW is supposed to be grown on (001) GaSb buffer. Because the electron-like subbands are localized in the InAs layers, whereas the hole-like subbands are localized in the GaSb layers, the InAs/GaSb/InAs QWs at $\mathbf{k}=0$ can be contingently considered as ``double QW for electrons'' with a GaSb middle barrier, which also plays a role of ``QW for holes'' (Fig.~\ref{Chapter4Fig3}A). Figure~\ref{Chapter4Fig3}B shows positions of electron-like
(E1 and E2) and hole-like (H1) subbands at zero quasi-momentum $\mathbf{k}$ as a function of InAs-layer thickness $d_1$ for GaSb-layer thickness of $d_2=4$~nm. In this case, it is intuitively clear that the electron-like levels E1 and E2 are connected with even-odd state splitting, arising from the tunnel-coupled ``QWs for electrons''. If InAs layers are thin ($d_1<d_{1c}$, where $d_{1c}$ depends on $d_2$), the E1 and E2 subbands lie above the hole-like subbands, and the three-layer QW has a trivial band ordering. For $d_1>d_{1c}$, the E1 subband lies below the H1 level, and the QW is characterized by inverted band structure.
\begin{figure}
\centering
\includegraphics [width=0.8\columnwidth, keepaspectratio] {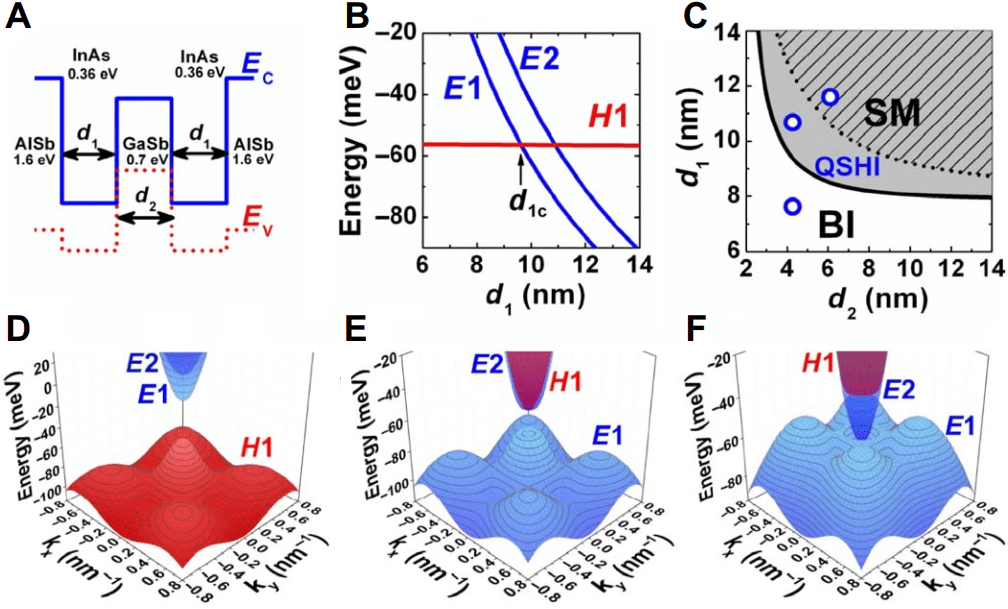}
\caption{\label{Chapter4Fig3} \textbf{(A)} Schematic representation of symmetrical three-layer InAs/GaSb QWs with InAs geometry. The numbers show the bandgap values in materials of the layers. Here, $d_1$ and $d_2$ are
the thicknesses of InAs and GaSb layers, respectively. The QW is supposed to be grown on (001) GaSb buffer. \textbf{(B)} Energy of electron-like (blue curves) and heavy hole-like (red curve) subbands at $\mathbf{k}=0$, as a function of $d_1$ at $d_2=4$~nm. Zero energy corresponds to the top of the valence band in bulk GaSb. \textbf{(C)} Phase diagram for different $d_1$ and $d_2$. The white open region is a BI state, whereas the gray-striped region defines an SM state. The gray open region corresponds to the QSHI state. \textbf{(D-F)} 3D plot of the band structure at BI (D), QSHI (E), and SM (F) states. The $x$ and $y$ axes are oriented along (100) and (010) crystallographic directions, respectively. The thicknesses of the layers for each states, used in the calculations, are marked in panel (C) by blue open symbols.}
\end{figure}

In Fig.~\ref{Chapter4Fig3}C, we provide a phase diagram for the QWs of InAs geometry with different layer thicknesses. Typical band dispersions for each phase are shown in the bottom panels. In the diagram, the solid curve describing the crossing between E1 and H1 subbands divides the $d_1$-$d_2$ plane into a white region, corresponding to trivial band insulator (BI) state with direct band ordering, and a gray region with inverted band structure. The gray region, in its turn, is split into open and striped areas, corresponding to QSHI state and  semimetallic (SM) states, respectively. As shown in Fig.~\ref{Chapter4Fig3}F, the SM phase is characterized by a vanishing indirect band-gap when the side maxima of the valence subband exceed in energy the conduction subband bottom. We note that the QWs grown along (001) direction positions of the side maxima in the valence band are independent of $d_1$ and $d_2$ and always lie along ($110$), ($\bar{1}10$), ($1\bar{1}0$), and ($\bar{1}\bar{1}0$) crystallographic directions. Thus, by varying the layer thicknesses in the QW with InAs geometry, one can realize BI, QSHI, or SM states, similar to those in single HgTe QWs shown in Fig.~\ref{Chapter3Fig1}.

To describe a quantum phase transition between BI and QSHI states, we derive effective a 2D Hamiltonian valid for the states in the vicinity of the $\Gamma$ point. Distinct from the BHZ model of HgTe/CdHgTe QWs~\cite{HgTeQWs2}, we take into account additional E2 subband, lying close in energy to the E1 and H1 states. In the basis set $|$\emph{E}1,+$\rangle$, $|$\emph{H}1,+$\rangle$, $|$\emph{E}2,-$\rangle$, $|$\emph{E}1,-$\rangle$, $|$\emph{H}1,-$\rangle$, $|$\emph{E}2,+$\rangle$, the 2D Hamiltonian for the QW of InAs geometry has the form:
\begin{equation}
\label{eq:IIIVQW1}
H_{\mathrm{InAs}}(k_x,k_y)=\begin{pmatrix}
H_1(k_x,k_y) & 0 \\ 0 & H_1^{*}(-k_x,-k_y)\end{pmatrix},
\end{equation}
where asterisk stands for complex conjugation, $k_x$ and $k_y$ are momentum $\mathbf{k}$ components in the plane, $k_{\pm}=k_x\pm ik_y$ and $H_1(k_x,k_y)$ is written as
\begin{equation}
\label{eq:IIIVQW2}
H_1(k_x,k_y)=\begin{pmatrix}
\varepsilon_{E1}(k) & -Ak_{+} & Sk_{-} \\ -Ak_{-} & \varepsilon_{H1}(k) & Rk_{-}^2 \\ Sk_{+} & Rk_{+}^2 & \varepsilon_{E2}(k)\end{pmatrix},
\end{equation}
with $\varepsilon_{E1}(k)=C+M+B_{E1}k^2$, $\varepsilon_{H1}(k)=C-M+B_{H1}k^2$ and $\varepsilon_{E2}(k)=C+M+\Delta_{E1E2}+B_{E2}k^2$. Here, $C$, $M$, $A$, $B_{E1}$, $B_{H1}$, $B_{E2}$, $R$ and $S$ are structure parameters, which depend on $d_1$ and $d_2$. Parameter $\Delta_{E1E2}$ defines the gap between the E1 and E2 subbands at $\mathbf{k}=0$. Because we have kept the inversion symmetry and axial symmetry around the growth direction, $H_{\mathrm{InAs}}(k_x,k_y)$ has a block-diagonal form. The mass parameter $M$ describes inversion between the E1 and H1 subbands: $M>0$ corresponds to a trivial BI state, whereas for a QSHI state, $M<0$. By applying the open boundary condition for $H_{\mathrm{InAs}}(k_x,k_y)$ along the $y$ axis, one can numerically calculate dispersion of helical states, arising at the sample edge at $M<0$. For instance, the edge state dispersions for the QSHI state at the layer thicknesses marked in Fig.~\ref{Chapter4Fig3}C by the open blue symbol are provided in Fig.~\ref{Chapter4Fig4}. The structure parameters for $H_{\mathrm{InAs}}(k_x,k_y)$ can be found in the Supplementary Materials of our original paper~\cite{III-VQWs4}.

\begin{figure}
\centering
\includegraphics [width=1.0\columnwidth, keepaspectratio] {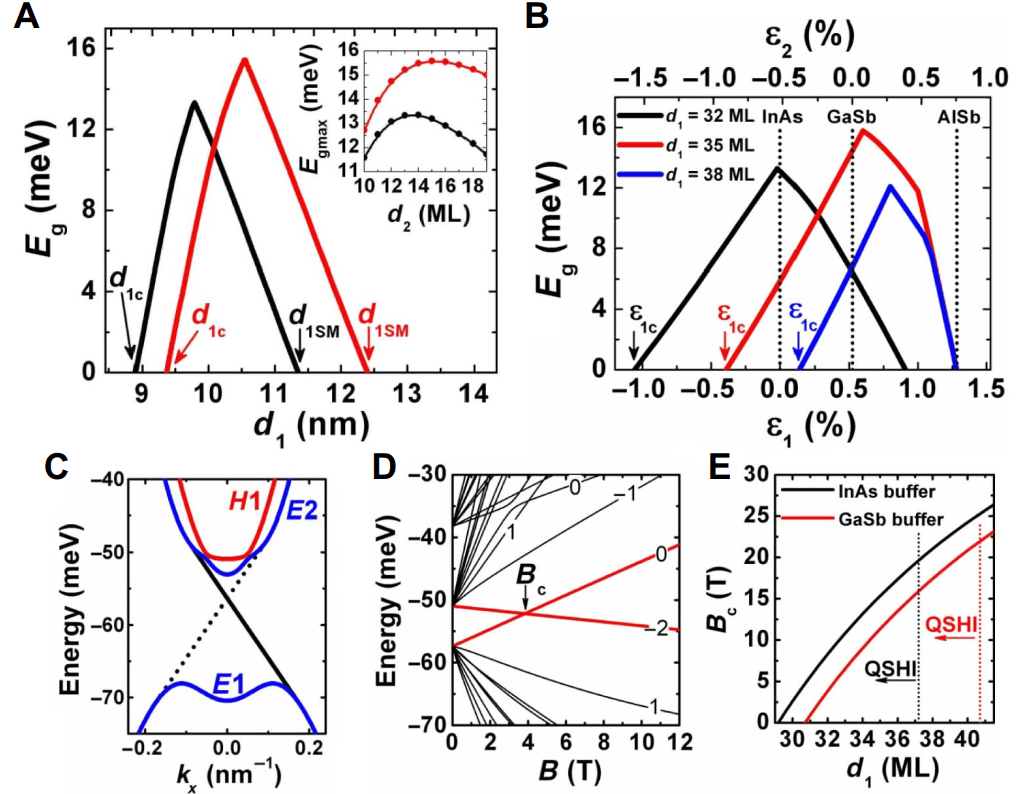}
\caption{\label{Chapter4Fig4} \textbf{(A)} Band-gap in the QSHI state as a function of $d_1$ in the InAs/GaSb QWs with InAs geometry at $d_2=14$ MLs, where 1~ML$=0.5a_0$ ($a_0$ is a lattice constant in the bulk material). The black and red curves correspond to the QWs grown on InAs and GaSb buffers, respectively. The inset shows the maximum gap, which can be achieved at a given value of $d_2$. \textbf{(B)} Band-gap in the QSHI state at different values of $d_1$ and $d_2=14$~MLs as a function of strain InAs layers $\varepsilon_1$ and GaSb layer $\varepsilon_2$. Three specific cases for the values of $\varepsilon_1$ and $\varepsilon_2$, corresponding to the InAs, GaSb, and AlSb buffers, are marked by vertical dotted lines. \textbf{(C)} Band dispersion in the QSHI state calculated on the basis of $H_{\mathrm{InAs}}(k_x,k_y)$, with the layer thicknesses marked in Fig.~\ref{Chapter4Fig3}C by blue open symbols. Electron- and heavy hole-like subbands are shown in blue and red, respectively. The black curves correspond to the dispersion of the edge states, obtained by numerical diagonalization of $H_{\mathrm{InAs}}(k_x,k_y)$ with open boundary conditions along the $y$ axis. Different Kramer's partners are shown by solid and dotted curves. \textbf{(D)}~Landau levels for the QW grown on GaSb buffer at $d_1=2$~MLs and $d_2=14$~MLs. The numbers over the curves correspond to the LL indices~\cite{HgTeQWs10}. Two red curves are the zero-mode Landau levels, which are identified within the $4\times4$ BHZ Hamiltonian~\cite{HgTeQWs2} for E1 and H1 subbands. \textbf{(E)}~Critical magnetic field $B_c$ as a function of $d_1$ with $d_2=14$ MLs. The black and red curves correspond to the structures grown on InAs and GaSb buffers, respectively. The dotted vertical lines mark the values of $d_{1SM}$.}
\end{figure}

The essential characteristic of a QSHI is the band-gap values for the bulk states. In the conventional broken-gap QW bilayers, the inverted band-gap opens away from the $\Gamma$ point of the Brillouin zone because of the hybridization between E1 andH1 subbands at nonzero quasimomentum (see Fig.~\ref{Chapter4Fig2}). In contrast, in three-layer InAs/GaSb QWs, the inverted band-gap arises in the $\Gamma$ point due to quantum confinement, whereas the E1--H1 coupling vanishes at $\mathbf{k}=0$. Since the confinement effect at $\mathbf{k}=0$ is much stronger than the E1--H1 coupling at nonzero $\mathbf{k}$, one can expect that the band-gap in the QSHI state is higher in the three-layer QWs than in InAs/GaSb QW bilayers. Figure~\ref{Chapter4Fig4}A shows the band-gap in the QSHI state as a function of $d_1$ for $d_2=14$ monolayers (MLs) of GaSb. Here, one ML in the given QW layer corresponds to a half of a lattice constant $a_0$ in the bulk material. The black and red curves correspond to the structures grown on InAs and GaSb buffers, respectively. It is seen that, starting from $d_1>d_{1c}$, increasing of $d_1$ enhances the band-gap, which occurs between the H1 and E1 levels. When the band-gap approaches to $\Delta_{E1E2}$ defined by the distance between the E1 and E2 subbands at $\mathbf{k}=0$, the system transforms into indirect-bandgap QSHI, in which the H1 level lies above E2 subband (see Fig.~\ref{Chapter4Fig4}C). Further increasing of $d_1$ decreases the indirect band-gap between the E1 and E2 subbands until it vanishes at $d_1=d_{1SM}$. The latter causes a transition into SM state with nonlocal overlapping between conduction and valence subbands. Note that $d_{1c}$ and $d_{1SM}$ depend not only on $d_2$ but also on the strain in InAs and GaSb layers, which can be varied by changing the buffer material.

Figure~\ref{Chapter4Fig4}B provides the band-gap in the QSHI state for the QWs with different values of $d_1$ and $d_2=14$ MLs as a function of strain in InAs layers $\varepsilon_1$ and GaSb layer $\varepsilon_2$ . The values of $\varepsilon_1$ and $\varepsilon_2$  are connected via the lattice constant of the buffer material. It is seen that by tuning the strain, one can successively induce transitions between the BI, QSHI, and SM states. The value of $\varepsilon_1$, corresponding to the crossing between E1 and H1 levels, for each of the QWs is marked by vertical arrow and defined as $\varepsilon_{1c}$. It is seen that by using the different values of $d_1$ and $d_2$ and by adjusting the strain, one can obtain QSHI with a band-gap of up to 16 meV. The latter is several times larger than the typical band-gap in inverted InAs/GaSb QW bilayers. Realization of the given values of $\varepsilon_1$ and $\varepsilon_2$ can be performed by using alloys of various III-V semiconductors~\cite{III-VQWs13}. The vertical dotted lines in Fig.~\ref{Chapter4Fig4}B mark the values of $\varepsilon_1$ and $\varepsilon_2$, which correspond to InAs, GaSb, and AlSb buffers. For instance, the QW with $d_1=35$ MLs and $d_2=14$ MLs grown on GaSb buffer (see Fig.~\ref{Chapter4Fig3}E) has a QSHI state with a gap close to the maximum value. Interestingly, the QWs grown on AlSb buffer do not feature a QSHI state. The band inversion defined by the positions of the subband at $\mathbf{k}=0$ is accompanied by nonlocal overlapping between conduction and valence bands. Thus, the variation of InAs and GaSb layer thicknesses drives the system from BI into the SM, avoiding the QSHI state.

As discussed in Section~\ref{sec:PandTinQWs}, a reliable fingerprint of QSHI under applied magnetic field $\mathcal{B}$ is the characteristic behavior of a particular pair of Landau levels, the so-called zero-mode Landau levels~\cite{HgTeQWs5}. The origin of this pair of Landau levels becomes apparent when using an effective $4\times4$ BHZ Hamiltonian for E1 and H1 subbands~\cite{HgTeQWs2}. Below a critical field value $\mathcal{B}_c$, the lowest zero-mode Landau level has an electron-like character and arises from the valence band, whereas the highest zero-mode Landau level has a heavy hole-like character and splits from the conduction band (see Fig.~\ref{Chapter4Fig4}D). In this case, the topological edge states are still present, whereas for larger magnetic fields $\mathcal{B}>\mathcal{B}_c$, the band ordering becomes normal, and only a trivial state can be found. At the critical magnetic field $\mathcal{B}_c$, corresponding to the crossing of zero-mode Landau levels, the quantum phase transition between the QSHI and BI states arises. Figure~\ref{Chapter4Fig4}E provides dependence of $\mathcal{B}_c$ on the thickness of InAs layers for the QWs grown on InAs and GaSb buffers. One can see that the critical magnetic field in three-layer InAs/GaSb QWs can reach 20~T, whereas the maximum values of $\mathcal{B}_c$ in HgTe/CdHgTe QWs grown on CdTe buffer do not exceed 7~T -- see Fig.~\ref{Chapter3Fig4}.

\begin{figure}
\centering
\includegraphics [width=0.8\columnwidth, keepaspectratio] {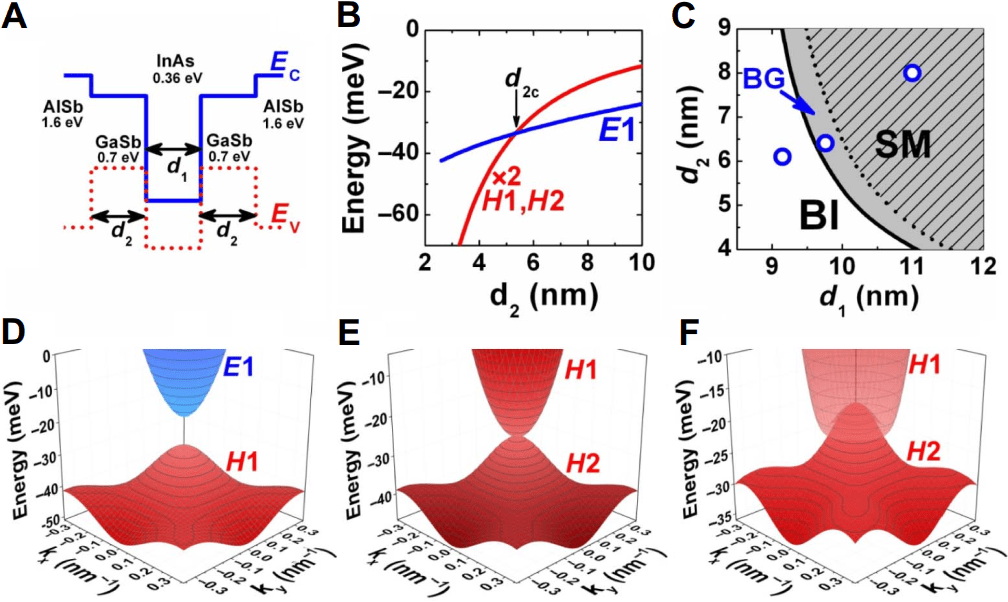}
\caption{\label{Chapter4Fig5} \textbf{(A)} Schematic representation of symmetrical three-layer InAs/GaSb QWs with GaSb geometry. The numbers show the band--gap values in materials of the layers. Here, $d_1$ and $d_2$ are the thicknesses of InAs and GaSb layers, respectively. The QW is supposed to be grown on (001) GaSb buffer. \textbf{(B)} Energy of electron-like (blue curves) and heavy hole-like (red curves) subbands at $\mathbf{k}=0$, as a function of $d_2$ at $d_1=10$ nm. Zero energy corresponds to the top of valence band in bulk GaSb. \textbf{(C)} Phase diagram for different $d_1$ and $d_2$. The white open region is a BI state, whereas the gray-striped region defines an SM state. The gray open region corresponds to a BG state. \textbf{(D-F)} 3D plot of the band structure at BI (D), BG (E), and SM (F) states. The $x$ and $y$ axes are oriented along (100) and (010) crystallographic directions, respectively. The thicknesses of the layers for each state used in the calculations are marked in panel (C) by blue open symbols.}
\end{figure}

\subsection{\label{sec:3LGaSb}Three-layer InAs/GaSb QWs with two GaSb layers}
Let us now consider symmetrical GaSb/InAs/GaSb QWs sandwiched between AlSb barriers and grown on (001) GaSb buffer. In contrast to the QWs with InAs geometry, GaSb/InAs/GaSb QWs contain two tunnel-coupled ``QWs for holes'' with InAs layer as a ``middle barrier''
(Fig.~\ref{Chapter4Fig5}A). Note that the InAs layer of reasonable thicknesses ($d_1>2$ nm) is not transparent for the particles from the hole-like subbands because of large effective mass. Therefore, the hole-like subbands localized in different GaSb layers are decoupled at $\mathbf{k}=0$. However, the mixing between heavy-hole-like and electron-like states at nonzero $\mathbf{k}$ leads to the splitting between H1 and H2 subbands, which can also be contingently considered as odd and even combinations of the hole states in each GaSb layer. Figure~\ref{Chapter4Fig5}B shows positions of E1, H1, and H2 subbands at $\mathbf{k}=0$ and the QW with $d_1=10$ nm as a function of $d_2$. If GaSb layers are thin enough so that $d_2<d_{2c}$ (where $d_{2c}$ depends on $d_1$), the E1 level lies above the H1 and H2 subbands, and the QW has a trivial band ordering. For $d_2>d_{2c}$, the QW has inverted band structure. A phase diagram for the QWs of GaSb geometry is shown in Fig.~\ref{Chapter4Fig5}C. The solid curve corresponds to the crossing between E1 and H1 subbands. It splits the $d_1$--$d_2$ plane into the white region of BI state and the gray region with inverted band structure.

The band inversion in the QWs of GaSb geometry drives the BI state into a metal phase, characterized by a band structure composed of two isotropic parabolas formed by the conduction subband H1 and valence H2 subband touching at $\mathbf{k}=0$ (see Fig.~\ref{Chapter4Fig5}E). This particular configuration closely resembles the band structure of natural bilayer graphene (BG)~\cite{NaturalBG3,NaturalBG4}. As the reader recalls, we have already encountered a similar BG phase in Chapter~\ref{Chap:HgTe}, when discussing the phase diagram of double HgTe/CdHgTe QWs -- see Fig.~\ref{Chapter3Fig12}. A further increase in the layer thicknesses $d_1$ and $d_2$ transforms the BG phase into an SM state, which, in addition to the quadratic band touching at $\mathbf{k}=0$, is also characterized by nonlocal overlapping of the valence subband with the loop of side minima in the conduction subband. This unconventional SM phase is shown by the gray-striped area. The typical band dispersions for BI, BG, and SM states are provided in the bottom panels.

A detailed proof that the BG phase in the three-layer InAs/GaSb QWs indeed possesses the same physical properties~\cite{III-VQWs4} as the analogous state in double HgTe/CdHgTe QWs (see Fig.~\ref{Chapter3Fig13}) will be omitted here. Including such a discussion would add little to what has already been extensively discussed in Chapter~\ref{Chap:HgTe} regarding the BG state in double HgTe QWs. Nevertheless, it is worth emphasizing that the key properties of this phase -- such as the unconventional quantum Hall effect~\cite{NaturalBG3} and the electric-field-induced band-gap opening ~\cite{NaturalBG2,NaturalBG4} (or more generally, any type of band-gap opening caused by non-equivalence between the GaSb layers) -- are also preserved or the SM state with quadratic band touching, represented by the gray-striped region in Fig.~\ref{Chapter4Fig5}C. These aspects will be discussed later in Section~\ref{sec:expQHE}, where I will provide the experimental observation of this unconventional quantum Hall effect in such systems.

To remain consistent in the logic of presentation, let us now introduce an effective 2D Hamiltonian describing the quantum phase transition between the BI and BG states in the QWs of GaSb geometry. By explicitly considering the coupling between the E1, H1, and H2 subbands, and using the basis set $|$\emph{E}1,+$\rangle$, $|$\emph{H}1,+$\rangle$, $|$\emph{H}2,-$\rangle$, $|$\emph{E}1,-$\rangle$, $|$\emph{H}1,-$\rangle$, $|$\emph{H}2,+$\rangle$, the effective Hamiltonian is written as~\cite{III-VQWs4}
\begin{equation}
\label{eq:IIIVQW3}
H_{\mathrm{GaSb}}(k_x,k_y)=\begin{pmatrix}
H_2(k_x,k_y) & 0 \\ 0 & H_2^{*}(-k_x,-k_y)\end{pmatrix},
\end{equation}
where
\begin{equation}
\label{eq:IIIVQW4}
H_2(k_x,k_y)=\begin{pmatrix}
\varepsilon_{E1}(k) & -Ak_{+} & Rk_{-}^2 \\ -Ak_{-} & \varepsilon_{H1}(k) & 0 \\ Rk_{+}^2 & 0 & \varepsilon_{H2}(k)\end{pmatrix},
\end{equation}
with $\varepsilon_{H2}(k)=C-M+B_{H2}k^2$. Here, the terms $\varepsilon_{E1}(k)$ and $\varepsilon_{H1}(k)$ have the same form as in Eq.~(\ref{eq:IIIVQW2}), and $B_{H2}$ depends only on thicknesses of InAs and GaSb layers. As in Eq.~(\ref{eq:IIIVQW2}), the sign of mass parameter $M$ in Eq.~(\ref{eq:IIIVQW4}) defines inversion between the E1 and H1 subbands. The block-diagonal form of $H_{\mathrm{GaSb}}(k_x,k_y)$ results from keeping the inversion and axial symmetry around the growth direction. By applying open boundary conditions to $H_{\mathrm{GaSb}}(k_x,k_y)$, one can numerically calculate the dispersion of the helical states and show their coexistence with gapless bulk states in the BG state~\cite{III-VQWs4}, as is the case in HgTe double QWs.

\subsection{\label{sec:3LDirac}Massless Dirac fermions in three-layer InAs/GaSb QWs}
As discussed above, inversion symmetry in the growth direction of the three-layer InAs/GaSb QWs leads to the crossing of the E1 and H1 subbands in the $\Gamma$ point of the Brillouin zone. Let us show that this crossing yields massless Dirac fermions in a 2D system. In the vicinity of crossing of E1 and H1 subbands, the terms quadratic in quasi-momentum in both effective Hamiltonians can be neglected. For the QWs with InAs geometry, the latter corresponds to the limit of large $\Delta_{E1E2}$. Therefore, $H_1(k_x,k_y)$ can be projected on the subspace, containing the E1 and H1 states. Since we are not interested in terms quadratic in $k$, the projection is performed by simply eliminating the third row and column of the matrix in Eq.~(\ref{eq:IIIVQW2}) and by setting $B_{E1}$ and $B_{H1}$ to zero. In the case of the QWs with GaSb geometry, keeping linear in $\mathbf{k}$ terms in $H_2(k_x,k_y)$ results to decoupling of the H2 subband from the E1 and H1 states. Therefore, the block describing the E1 and H1 subbands can be considered separately.

\begin{figure}
\centering
\includegraphics [width=0.62\columnwidth, keepaspectratio] {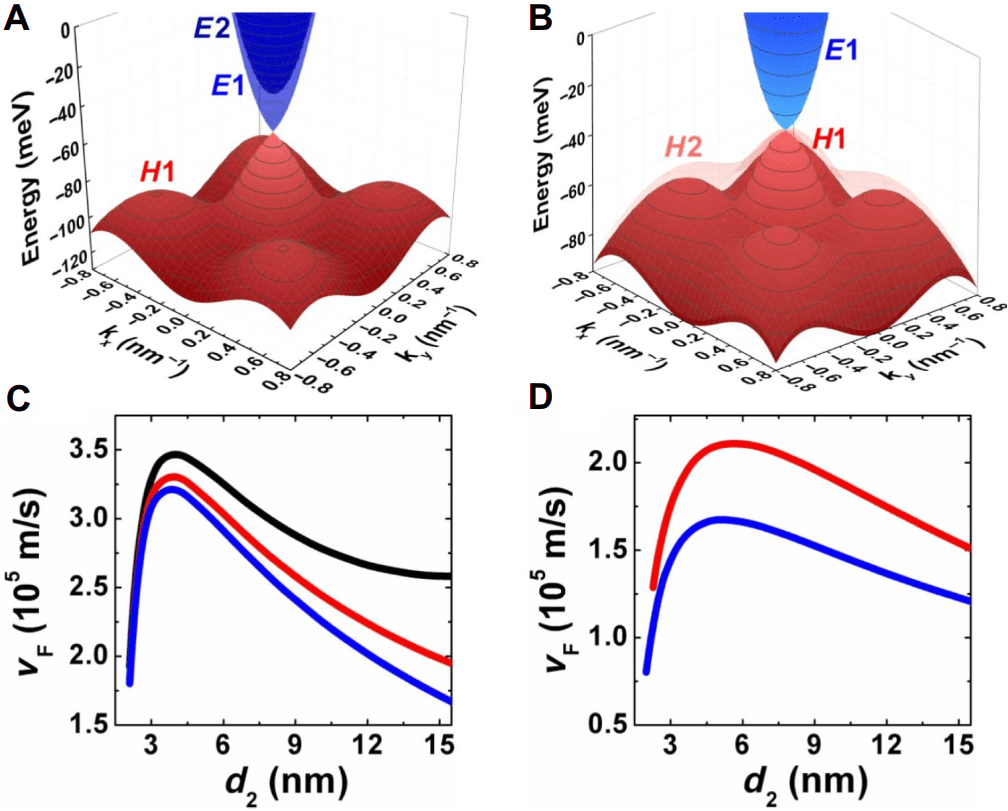}
\caption{\label{Chapter4Fig6} \textbf{Massless Dirac fermions in three-layer InAs/GaSb QWs}. \textbf{(A,B)}~3D plot of the band dispersions of three-layer InAs/GaSb QWs of InAs geometry (A) and GaSb geometry (B) both grown on GaSb buffer. The values of $d_1$ and $d_2$ correspond to the crossing of E1 and H1 subbands, shown in Figs.~\ref{Chapter4Fig3}B and \ref{Chapter4Fig5}B. \textbf{(C,D)} Fermi velocity $v_F$ of massless Dirac fermions as a function of GaSb layer thickness for the QWs of InAs geometry (C) and GaSb geometry (D) grown on different buffers: AlSb (black curves), GaSb (red curves) and InAs (blue curves).}
\end{figure}

Without loss of generality, we can set $C=0$ and the 2D Hamiltonian for the states in the vicinity of $\mathbf{k}=0$ in the crossed E1 and H1 subbands ($M=0$), for both types of the QWs takes the form
\begin{equation}
\label{eq:IIIVQW5}
H_{1,2}^{(E1+,H1+)}(k_x,k_y)\simeq\begin{pmatrix}
0 & {\hbar}v_{F}k_{+} \\ {\hbar}v_{F}k_{-} &0 \end{pmatrix}.
\end{equation}
The similar Hamiltonian can be also derived for $H_1^{*}(-k_x,-k_y)$ and $H_2^{*}(-k_x,-k_y)$. It is seen that Eq.~(\ref{eq:IIIVQW5}) corresponds to a massless Dirac Hamiltonian with Fermi velocity $v_F$, defined by parameter $A$ in Eq.~(\ref{eq:IIIVQW2})
and (\ref{eq:IIIVQW4}).

Figure~\ref{Chapter4Fig6} shows the band dispersion for the QWs of both geometries, when the E1 and H1 subbands are crossed. The layer thicknesses correspond to the crossing of the E1 and H1 subbands shown in Figs.~\ref{Chapter4Fig3}B and \ref{Chapter4Fig5}B. Because the hybridization between E1 and H1 subbands vanishes at $k=0$, it results in the gapless state with 2D massless Dirac fermions, just like in HgTe/CdTe QWs with critical thicknesses~\cite{Dirac2}. The linear band dispersion at small $k$ is seen. However, in the QWs with GaSb geometry, the Dirac cone is accompanied by an additional decoupled ``massive'' branch of H2 subband arising at the crossing point that is similar to the case of two tunnel-coupled HgTe QWs considered in Chapter~\ref{Chap:HgTe}.
The bottom panels in Fig.~\ref{Chapter4Fig6} provide $v_F$ for massless Dirac fermions as a function of $d_2$ in the QWs with InAs and GaSb geometries. It is seen that Fermi velocity is varied from $1\cdot10^5$ to $3\cdot10^5$~m/s, depending on the GaSb layer thicknesses. These values are several times lower than the ones in graphene ($\sim10^6$~m/s) and HgTe/CdHgTe QWs~\cite{Dirac2}.

Finally, we stress the effect of strain on $v_F$ of the massless particle, arising because of lattice-mismatch constant in the QW layers and buffers. It is seen that increasing tensile biaxial strain of InAs layers ($\varepsilon_1>0$) in the QWs with both geometries enhances $v_F$. As we have discussed previously, at high values of $\varepsilon_1$, the band inversion drives the QWs with InAs geometry from BI into SM phase, avoiding the QSHI state. In this case, the Dirac cone in the $\Gamma$ point of the Brillouin zone arising
at critical layer thicknesses is accompanied by the side maxima of the valence band overlapping with the conduction band bottom.

\subsection{\label{sec:3LInAsQSHI}Large-gap quantum spin Hall insulator state}
We have demonstrated that the three-layer InAs/GaSb QWs with InAs geometry and single HgTe QWs share the same possible phases and feature the comparable band-gaps in the QSHI state. As was experimentally demonstrated for HgTe QWs~\cite{HgTeQWs12}, the compressive biaxial strain extends the inverted band-gap in this 2D systems up to 60 meV at reasonable values of the strain -- see Fig.~\ref{Chapter4Fig1}. Now, we address the question of whether the band-gap of the QSHI state in three-layer InAs/GaSb the QWs can be enhanced up to the same values of compressively strained HgTe QWs. However, if either compressive or tensile strain is large enough, it destroys the QSHI state in three-layer InAs/GaSb QWs and induces a phase transition into the BI or SM phase, respectively -- see Fig.~\ref{Chapter4Fig4}B. To increase the inverted band-gap in the InAs-designed QWs, we therefore replace the middle GaSb layer in the QW by a Ga$_{1-x}$In$_{x}$Sb alloy as it was done for the InAs/GaInSb QW bilayers to increase the inverted band-gap up to 20~meV~\cite{III-VQWs12}.

\begin{figure}
\centering
\includegraphics [width=1.0\columnwidth, keepaspectratio] {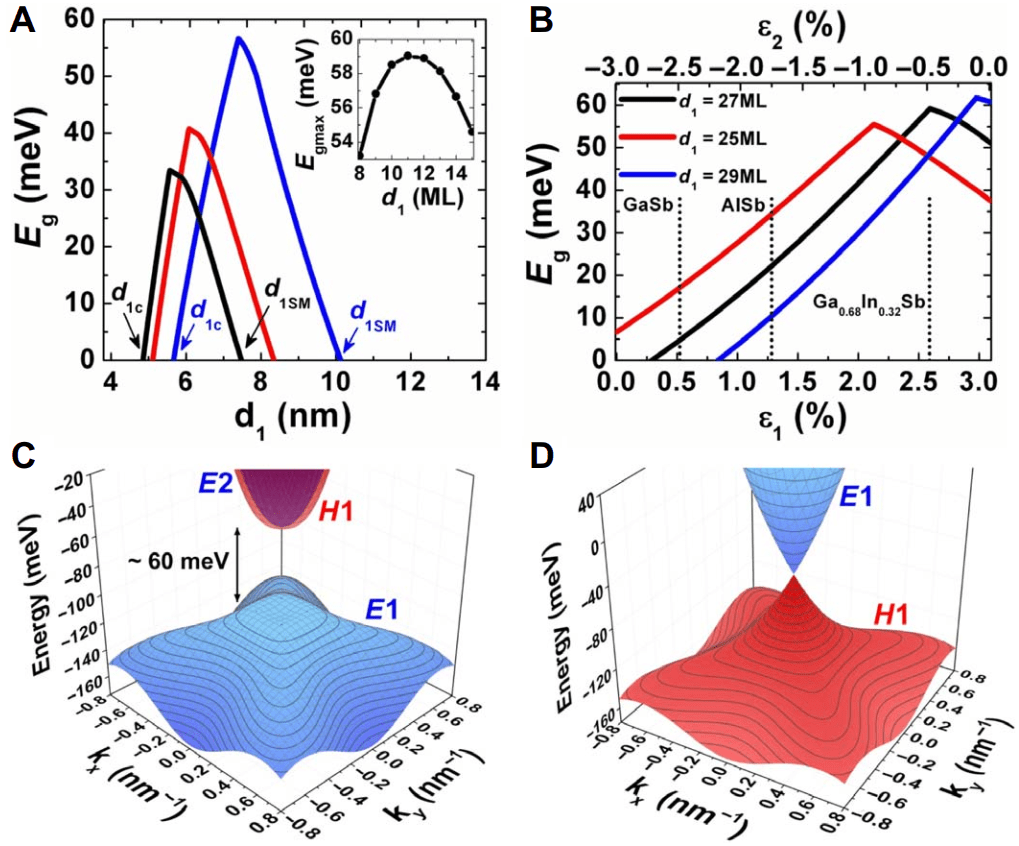}
\caption{\label{Chapter4Fig7} \textbf{(A)} Band-gap in the QSHI state for three-layer InAs/Ga$_{0.6}$In$_{0.4}$Sb QWs of InAs geometry as a function of $d_1$ at $d_2=14$~MLs (where one ML equals to a half of lattice constant in the bulk material). The black, red, and blue curves correspond to the structures grown on GaSb, AlSb, and Ga$_{0.68}$In$_{0.32}$Sb buffers, respectively. The inset shows the maximum gap, which can be achieved at a given value of $d_2$, in the QW grown on Ga$_{0.68}$In$_{0.32}$Sb buffer. \textbf{(B)}~Band-gap in the QSHI state for the InAs/Ga$_{0.6}$In$_{0.4}$Sb QWs with different values of $d_1$ and $d_2=11$~MLs as a function of strain in InAs layers $\varepsilon_1$ and Ga$_{0.6}$In$_{0.4}$Sb layer $\varepsilon_2$. Three specific cases for the values of $\varepsilon_1$ and $\varepsilon_2$, corresponding to the GaSb, AlSb, and Ga$_{0.68}$In$_{0.32}$Sb buffers, are marked by vertical dotted lines. \textbf{(C,D)} 3D plot of the band dispersion for the InAs/Ga$_{0.6}$In$_{0.4}$Sb QW grown on Ga$_{0.68}$In$_{0.32}$Sb buffer at $d_2=11$~MLs, $d_1=27$~MLs in panel (C) and $d_1=d_{1c}$ ($d_{1c}\simeq6$~nm) in panel (D). The $x$ and $y$ axes are oriented along (100) and (010) crystallographic directions, respectively.}
\end{figure}

\begin{figure}
\centering
\includegraphics [width=1.0\columnwidth, keepaspectratio] {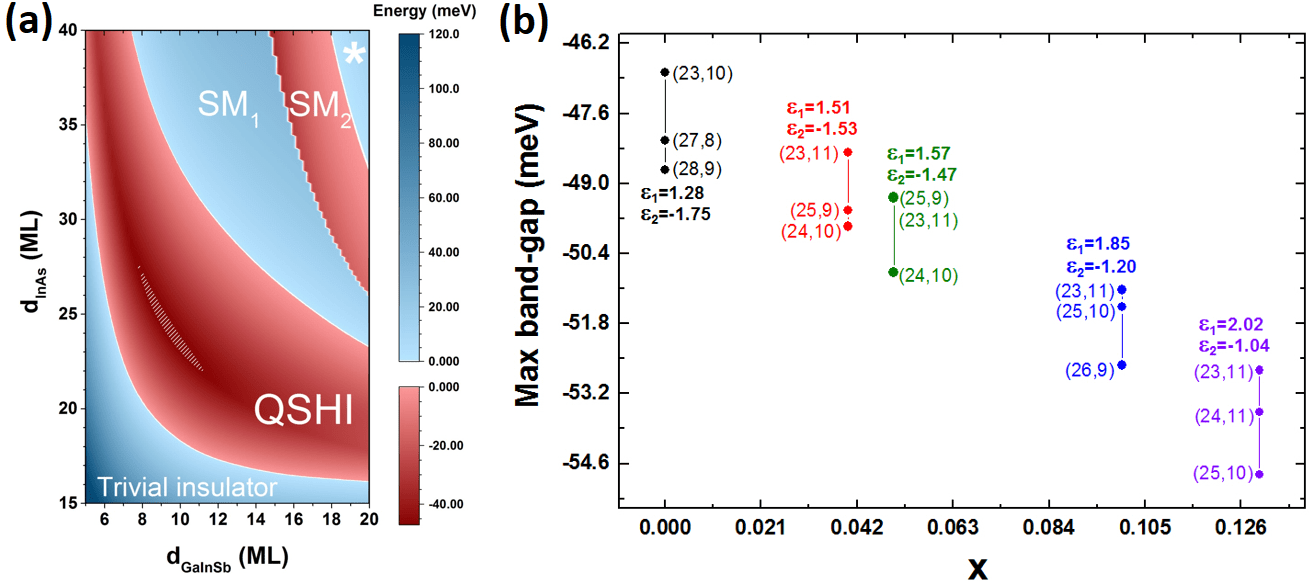}
\caption{\label{Chapter4Fig8} \textbf{(a)} Colormap diagram of symmetric three-layer InAs/Ga$_{0.60}$In$_{0.40}$Sb QWs grown on (001) AlSb buffer. The energy scale at trivial and QSHI states represents the band-gap for the bulk states  -- the positive values correspond to the trivial band ordering, whereas the negative ones are for the inverted band-gap. The regions SM$_1$ and SM$_2$ confirm the semimetals with single and double band inversions at the $\Gamma$ point of the Brillouin zone, respectively. The energy scale at these SM states represents the overlapping between conduction and valence bands. The blue region marked by asterisk corresponds to a HOTI state~\cite{doubleHgTeQWs7}, discussed previously in Section~\ref{sec:HOTIHgTe}. Here, 1 ML equals to half of the lattice constant of the bulk material. The white-sparced region represents the layer thicknesses with a band-gap higher than the inverted gap of 45 meV in the WTe$_2$ monolayer. \textbf{(b)}~The maximum values of the inverted band-gap in symmetric three-layer InAs/Ga$_{0.60}$In$_{0.40}$Sb QWs sandwiched between Al$_{1-x}$In$_{x}$Sb barriers and grown on (001) Al$_{1-x}$In$_{x}$Sb buffer. The numbers in brackets correspond to the widths of the InAs ($d_1$) and Ga$_{0.6}$In$_{0.4}$Sb ($d_2$) layers. The values of $\varepsilon_1$ and $\varepsilon_2$ conform to the strain of InAs layers $\varepsilon_1$ and Ga$_{0.6}$In$_{0.4}$Sb layer $\varepsilon_2$.}
\end{figure}

Figure~\ref{Chapter4Fig7}A shows the band-gap in the QSHI state in three-layer InAs/Ga$_{0.6}$In$_{0.4}$Sb QWs of InAs geometry as a function of $d_1$ at $d_2=14$~MLs. The black, red, and blue curves correspond to the structures grown on GaSb, AlSb, and Ga$_{0.68}$In$_{0.32}$Sb buffer, respectively. As seen, the inverted band-gap in the strained InAs/Ga$_{0.6}$In$_{0.4}$Sb QW is several times larger than the values obtained for the three-layer InAs/GaSb QWs. The inset shows the maximum gap in the strained QW grown on Ga$_{0.68}$In$_{0.32}$Sb buffer, which can be achieved at given value of $d_2$. Figure~\ref{Chapter4Fig7}B provides the QSHI band-gap in the InAs/Ga$_{0.6}$In$_{0.4}$Sb QWs with different values of $d_1$ and $d_2=11$~MLs as a function of strain InAs layers $\varepsilon_1$ and Ga$_{0.6}$In$_{0.4}$Sb layer $\varepsilon_2$. Note that the values of $\varepsilon_1$ and $\varepsilon_2$ are connected via the lattice constant of the buffer material. Figure~\ref{Chapter4Fig7}B proves theoretically that, by adjusting the strain for different values of $d_1$ and $d_2$, one can obtain a QSHI with a large band-gap. It is seen that the InAs/Ga$_{0.6}$In$_{0.4}$Sb QW with $d_1=27$ MLs and $d_2=11$ MLs grown on Ga$_{0.68}$In$_{0.32}$Sb buffer features a band-gap in the QSHI state of about 60 meV. The band structure for this QW is shown in Fig.~\ref{Chapter4Fig7}C. We have considered the case of symmetrical InAs/GaInSb/InAs QWs. Importantly, the band-gap in these QWs can be tuned by varying the strength of an applied electric field or by growing the InAs layers of different thicknesses. However, as was shown in our original paper~\cite{III-VQWs4} the maximum QSHI band-gap in the three-layer InAs/Ga$_{0.6}$In$_{0.4}$Sb QWs can be achieved only in symmetrical QW geometry. In addition to large-gap QSHI, the InAs/Ga$_{0.6}$In$_{0.4}$Sb QWs at critical thicknesses feature massless Dirac fermions with Fermi velocity (up to $7\cdot10^5$~m/s) higher than in the InAs/GaSb/InAs QWs. Moreover, a linearity of their energy dispersion persists in much higher energy ranges, as can be seen by comparing Figs.~\ref{Chapter4Fig7}D and \ref{Chapter4Fig6}A.

Looking at Fig.~\ref{Chapter4Fig7}B, a reader familiar with the MBE growth of semiconductor heterostructures may immediately notice that fabricating high-quality QWs with strain values exceeding 2\% in absolute magnitude is practically impossible, even despite the significant progress achieved in recent years in the MBE growth technology of InAs/GaInSb systems~\cite{III-VQWs14}. Particularly, an experimental attempt to grow an InAs/Ga$_{0.60}$In$_{0.40}$Sb QW on a Ga$_{0.68}$In$_{0.32}$Sb buffer -- whose band structure is shown in Fig.~\ref{Chapter4Fig7}C -- revealed a substantial number of mismatch-related crystalline defects and pronounced layer undulations in the grown sample~\cite{III-VQWs15}. For this reason, we performed additional calculations of the band-gap width in the QSHI state of three-layer InAs/Ga$_{0.60}$In$_{0.40}$Sb QWs for more realistic strain values in the layers. The results of these calculations are summarized in Fig.~\ref{Chapter4Fig8}. As seen, by growing InAs/Ga$_{0.60}$In$_{0.40}$Sb QWs on an Al$_{1-x}$In$_{x}$Sb buffer and sandwiched between the Al$_{1-x}$In$_{x}$Sb barriers, one can indeed achieve an inverted band-gap close to 60~meV at realistic strain magnitudes of $\varepsilon_1$ in the InAs layers and $\varepsilon_2$ in the Ga$_{0.60}$In$_{0.40}$Sb layer. These calculations confirm the practical feasibility of realizing large-gap QSHI states in  three-layer InAs/Ga$_{0.6}$In$_{0.4}$Sb QWs under experimentally accessible conditions.

\section{\label{sec:ExpIIIVMont}Key experimental results of THz team in Montpellier}
The theoretical results presented in the previous section served as a starting point and a strong motivation for a series of experimental investigations of three-layer InAs/Ga(In)Sb QWs conducted by the THz team in Montpellier~\cite{Dirac4,MontIIIVexp1,MontIIIVexp2,MontIIIVexp3,MontIIIVexp4,MontIIIVexp5,MontIIIVexp6}. The QW samples for these studies were grown in the NanoMIR group at the Institut d'Electronique et des Syst\`{e}mes of the University of Montpellier. In this section, I will review the most significant experimental findings, which, in my view, represent true milestones in the history of research on these QW systems. The first of these results, discussed in the next subsection, concerns the demonstration of the remarkable temperature independence of the inverted band-gap in these structures, which was an important step toward realizing higher-temperature QSHI in III-V semiconductor QWs. In what follows, I will focus only on the key experimental results, while all technical details of the measurements and data analysis can be found in our original publication~\cite{MontIIIVexp3}.

\subsection{\label{sec:TempGap}Temperature-independent inverted band-gap in three-layer \\ InAs/GaSb QWs}
To demonstrate that the band structure and the band-gap in three-layer InAs/GaSb QWs remain almost unaffected by temperature, we performed temperature-dependent studies of the Landau-level transitions in a sample with an inverted GaSb/InAs/GaSb QW, as shown in Fig.~\ref{Chapter4Fig9}a. The sample was grown on a (001)-oriented GaSb buffer and sandwiched between AlSb barriers. Magnetotransport measurements performed in the Van der Pauw geometry allowed us to determine the electron concentration and mobility at 2~K, which were found to be $n=6.2\times10^{11}$~cm$^{-2}$ and $\mu=2\times10^{4}$~cm$^{2}$/V$\cdot$s, respectively. The Fermi level position corresponding to this experimentally determined concentration is indicated in Fig.~\ref{Chapter4Fig9}a by the brown line.

\begin{figure}
\centering
\includegraphics [width=1.0\columnwidth, keepaspectratio] {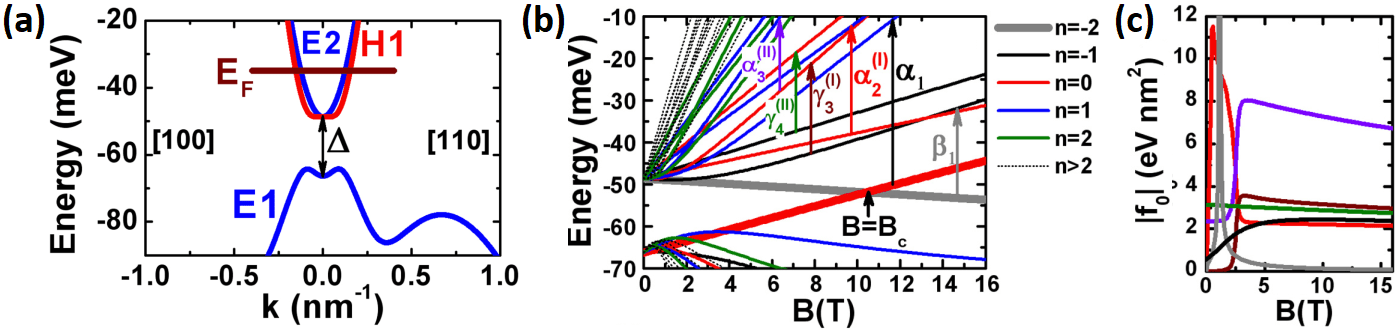}
\caption{\label{Chapter4Fig9} \textbf{(a,b)} Band structure and Landau-level fan chart of the sample under study with $d_{\mathrm{InAs}}=34$~ML and $d_{\mathrm{GaSb}}=14$~ML. Numbers in brackets in panel (a) denote crystallographic orientations of the quasimomentum. The brown line in panel (a) indicates the position of the Fermi level $E_F$ found from magnetotransport measurements. The numbers in the right side of the (b) panel correspond to the Landau level indices~\cite{HgTeQWs10}. The expected Landau-level transitions, which should be dominant in the absorption, are denoted by arrows and Greek letters. A pair of zero-mode Landau levels with indices $-2$ and $0$ is shown by bold curves. \textbf{(c)}~Oscillator strength $f_0=\hbar^2 \overline{f_\mathrm{osc}}/(2m_{0})$ for Landau level transitions shown in the (b) panel. The overline represents an averaging over orientation of the light polarization lying in the QW plane.}
\end{figure}

Figure~\ref{Chapter4Fig9}b shows the calculated Landau-level fan chart for the studied sample. The numbers on the right-hand side of the panel correspond to the Landau-level indices used within the eight-band $\mathbf{k}\cdot\mathbf{p}$ Kane Hamiltonian~\cite{HgTeQWs10}. The Landau-level transitions expected to dominate the optical absorption are indicated by arrows and Greek letters. To identify the most intense transitions, we analyzed the oscillator strength $f_\mathrm{osc}$ for all possible transitions with $\Delta n=\pm1$ involving the Landau levels with indices $n=-2$, $-1$, $0$, and $1$, which are relevant at the experimental electron concentration. The oscillator strength for a transition between the levels with multi-indices $l$ and $m$ is defined as~\cite{III-VQWs16}:
\begin{equation}
\label{eq:IIIVQWexp1}
f_\mathrm{osc}(m,l)=\dfrac{2m_{0}|\langle\mathbf{\Psi}_{m}|\mathbf{e}\cdot\widehat{\mathbf{v}}|\mathbf{\Psi}_{l}\rangle|^2}{|E_{m}-E_{l}|},
\end{equation}
where $m_0$ is the free electron mass, $\mathbf{\Psi}_{i}$ is the eight-component envelope function of the eight-band $\mathbf{k}\cdot\mathbf{p}$ Kane Hamiltonian~\cite{HgTeQWs10}, $\mathbf{e}$ is the unit vector of light polarization and $\widehat{\mathbf{v}}$ is the velocity operator, defined by
\begin{equation}
\label{eq:IIIVQWexp2}
\hbar\widehat{\mathbf{v}}=\dfrac{\partial\hat{H}_{\textbf{k}}^{8\times 8}}{\partial\widehat{\mathbf{k}}}.
\end{equation}
In Eq.~(\ref{eq:IIIVQWexp1}), $l$ and $m$ are assumed to be the complex symbols, including subband, ``spin'' indices and Landau-level numbers $n$ (or quasimomentum $\mathbf{k}_{\|}=(k_x, k_y)$ in the QW plane in the absence of magnetic field).
Note that if the orientation of the light polarization projection onto QW plane is not controlled during performing experiments, one has to perform an averaging of $f_\mathrm{osc}(m,l)$ over the orientation of $\mathbf{e}$ component lying in the QW plane.

Fig.~\ref{Chapter4Fig9}c shows the oscillator strength for the most intense Landau-level transitions, which are attributed to the cyclotron-resonance- (CR-) like transitions in E2 ($\alpha_{2}^{(\mathrm{I})}$ and $\gamma_{4}^{(\mathrm{II})}$ transitions) and H1 ($\alpha_{3}^{(\mathrm{II})}$ and $\gamma_{3}^{(\mathrm{I})}$ transitions) subbands and to the $\alpha_{1}$ and $\beta_{1}$ transitions from the zero-mode Landau levels. Other transitions, have either vanishing oscillator strength or a strength, which is several times lower than the given ones. The non-monotonic behaviour of the oscillator strengths in low magnetic fields is caused by numerous anticrossings between Landau levels with the same indices from \emph{E}2 and \emph{E}1 subbands. The latter results in non-monotonic dependence of the transition energy and the oscillator strength on magnetic field for the given Landau-level transition in the vicinity of the anticrossings.

Figures~\ref{Chapter4Fig10}a--\ref{Chapter4Fig10}c show the experimental transmission spectra at 2~K and 100~K, as well as a comparison of the experimental and theoretical transition energies. The theoretical dependencies, obtained by using band parameters at 2~K~\cite{III-VQWs13}, are given by the colored curves. The solid part of each curve represents the range of magnetic fields, in which the Landau-level transition should be observed at electron concentration measured at 2~K. The dotted line for $\beta_{1}$ transition stress the smallest oscillator strength for this transition. In low magnetic fields, the spectra show a single line attributed to the CR-like transitions in the E2 and H1 subbands. At the energies above the Reststrahlen band, the second line, corresponding to $\alpha_{1}$ transition, arises for both temperatures. It has a linear dependence at high magnetic fields. When the energy tends toward LO phonon energy in AlSb barriers ($\omega_{\mathrm{LO}}=40.9$~meV), the energies of the $\alpha_{1}$ and CR-like transitions bend, deviating from the linear behavior. The latter is clearly seen for the measurements at 2~K. We note that this cannot be attributed to electron-phonon interaction because the effect is absent in the vicinity of the phonon energies in InAs ($\omega_{\mathrm{LO}}=30$~meV and $\omega_{\mathrm{TO}}=27$~meV) and GaSb ($\omega_{\mathrm{LO}}=29.8$~meV and $\omega_{\mathrm{TO}}=28.6$~meV) layers. The observed energy bending has a pure optical origin. In our case, the optical response of the system is governed by two strongly resonant processes: the resonant dielectric function of the material at phonon's (both TO and LO) frequencies (even if the polaron effect is small) and magnetoabsorption of 2D electron system. Thus, the profile of absorption lines can be strongly modified by the singularities of the dielectric function at phonon's frequencies of the QW and barrier materials. As the symbols in Fig.~\ref{Chapter4Fig10}c define the maxima of the resulting absorption line, the latter produces the bending effect of the resonant energy in the vicinity of $\omega_{\mathrm{TO}}$ or $\omega_{\mathrm{LO}}$~\cite{ III-VQWs17}.

\begin{figure}
\centering
\includegraphics [width=1.0\columnwidth, keepaspectratio] {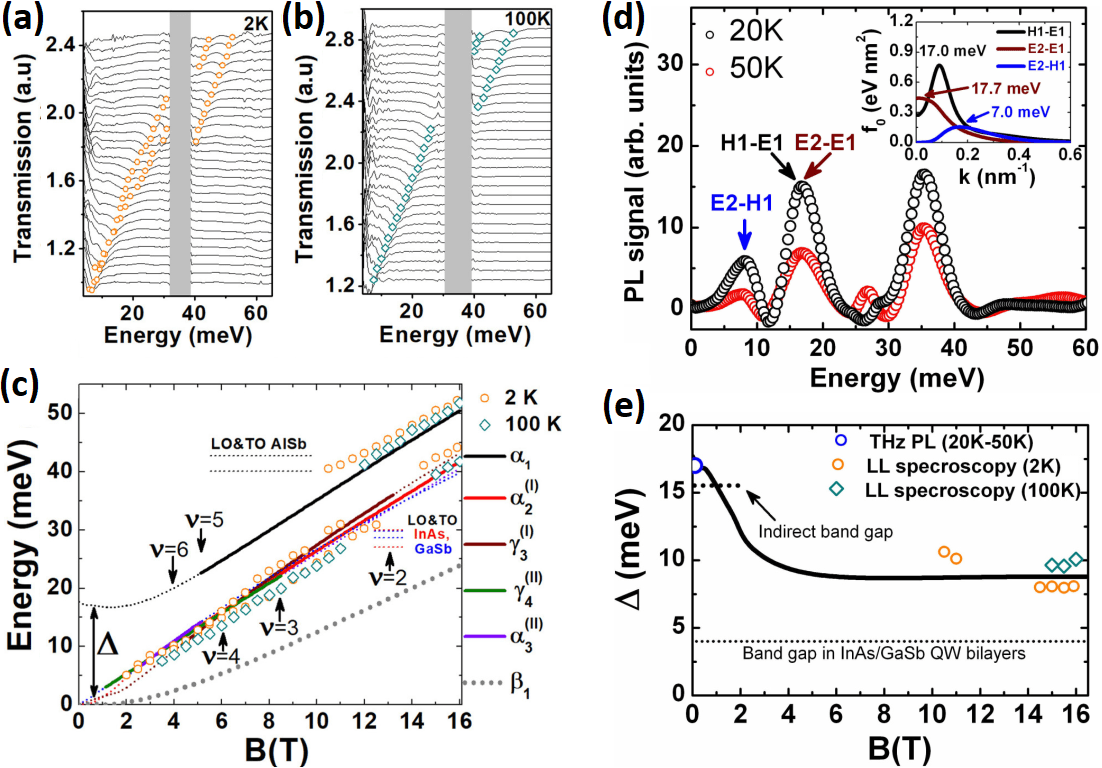}
\caption{\label{Chapter4Fig10} \textbf{(a,b)} Transmission spectra at 2~K and 100 K from 1~T (the first plot at the bottom) to 16~T (the last plot on the top) with a step for the magnetic field of 0.5~T. Shaded areas indicate the Reststrahlen band of GaAs substrate. The symbols mark positions of absorption lines. \textbf{(c)}~The most intense Landau-level transitions, marked in Fig.~\ref{Chapter4Fig9}b by arrows. Experimental data at different temperatures are represented by symbols. The arrows indicate the magnetic fields, corresponding to integer LL filling factors $\nu$ at 2~K. Horizontal dashed lines correspond to the energies of TO and LO phonons in AlSb (in black, $\omega_{\mathrm{LO}}=42.7$~meV and $\omega_{\mathrm{TO}}=40.1$~meV), InAs (in red, $\omega_{\mathrm{LO}}=30$~meV and $\omega_{\mathrm{TO}}=27$~meV) and GaSb (in blue, $\omega_{\mathrm{LO}}=29.8$~meV and $\omega_{\mathrm{TO}}=28.6$~meV). \textbf{(d)}~THz photoluminescence spectra at different temperatures measured from 45$^\circ$ direction relative to growth axis. The insets show oscillator strength $f_0=\hbar^2 \overline{f_\mathrm{osc}}/(2m_{0})$ for optical transitions between E2, E1 and H1 subbands calculated for the experimental conditions. The overline denotes an averaging over orientation of the polarization component lying in the QW plane. The numbers correspond to the transition energies at maxima of $f_0$. \textbf{(e)}~Magnetic-field dependence of $\Delta$. The solid line represents theoretical dependence extracted from Landau level fan chart. The symbols show the experimental values, obtained from Landau-level magnetospectroscopy and photoluminescence. The dotted lines mark the band-gap in our sample and the maximum inverted gap of InAs/GaSb QW bilayers.}
\end{figure}

As seen from Fig.~\ref{Chapter4Fig10}c, experimental data at both temperatures are qualitatively in good agreement with theoretical prediction. The splitting of absorption line arising at 2~K in the spectra at $B= 7-11$~T can be interpreted as a difference in the energies of CR-like transitions in the E2 and H1 subbands, respectively. Interestingly, increasing the temperature does not change the positions of the absorption lines at $B= 7-11$~T corresponding to the low-energy peaks, while the high-energy component of the splitted lines disappears completely at 100~K. The latter is directly caused by a change of the electron concentration with temperature, which affects the filling factor of the Landau levels. A larger experimental splitting of the absorption lines may result from the effects of interface inversion asymmetry (IIA) ignored in our calculations. Indeed, during the sample growth, such shutter sequences at all InAs/GaSb interfaces were organized in order to promote the formation of In--Sb bonds, which ensure higher electron mobility.
However, these In--Sb bonds give rise to non-common-atom interfaces, whose presence is taken into account by additional spin-dependent corrections in the Hamiltonian~\cite{III-VQWs18}. To simplify the calculations of Landau levels, these terms were not included in our calculations.

Importantly, as clearly seen at the energies above the Reststrahlen band, the Landau-level-transition energies are independent of temperature in high magnetic fields. Although the position of the low-energy line, which at various concentrations can be attributed to different CR-like transitions, is slightly changing with temperature, the position of the high-energy line, identified as the $\alpha_{1}$ inter-band LL transition, remains the same at 2~K and 100~K. It also indicates that the resonant energy caused by the given Landau-level transition is temperature-independent. Unfortunately, relatevely large electron concentration in our sample makes impossible to follow the energy of the $\alpha_{1}$ transition down to low magnetic fields. Therefore, we could not directly extract the gap at $k=0$ from Landau-level spectroscopy. However, we can estimate the difference in energies for CR-like and $\alpha_{1}$ transitions, marked as $\Delta$. The value of $\Delta$ at $B=0$ defines the gap. Although it decreases with magnetic field, $\Delta$ tends to a constant value at the high fields. From our experimental data, we have found $\Delta\approx8$~meV for $B>10$~T.

To determine the energy difference between the E2, E1, and H1 subbands in the vicinity of the $\Gamma$ point of the Brillouin zone for the studied sample, our colleagues from Nizhny Novgorod performed complementary investigations of the long-wavelength photoluminescence~\cite{III-VQWs19,III-VQWs20}. The results of these studies are provided in Fig.~\ref{Chapter4Fig10}d, which presents the photoluminescence spectra measured at temperatures of 20~K and 50~K under an excitation power of 300~mW. In these experiments, the sample was excited from the top surface of the structure by incident pumping radiation tilted by $45^{\circ}$ relative to the growth axis, while the emitted photoluminescence signal was collected along a direction forming the same $45^{\circ}$ angle. This geometry allowed for efficient excitation and detection of optical transitions within the QW. The insets in Fig.~\ref{Chapter4Fig10}(d) show the calculated oscillator strength of optical transitions between the E2, E1, and H1 subbands as a function of the quasimomentum oriented along the (110) crystallographic direction. As seen, the photoluminescence spectra show three well-resolved emission lines. The high-energy line observed at about 35~meV corresponds to the luminescence of the Ga antisite defects from the GaSb buffer~\cite{III-VQWs21,III-VQWs22,III-VQWs23}. The low-energy line at 7.5~meV may be related to the E2--H1 transition. Its position is in good agreement with the energy corresponding to the maximum value of the oscillator strength. The middle line at 16.5~meV is a superposition of H1-E1 and E2--E1 transitions, which have close energies. Indeed, the energy of maximum of the oscillator strength for these transitions differs by less than 1~meV. Therefore, we cannot resolve them in the photoluminescence spectra. We note that the observed E2--H1 and E2--E1 transitions are active in $z$-polarization, while the E1--H1 transition is mostly active in the polarization parallel to the QW plane. Fig.~\ref{Chapter4Fig10}d also shows that increasing temperature up to 50~K does not change the position of the emission lines for the transitions between E2, E1 and H1 subbands. This means that the gaps between E2, E1 and H1 subbands are independent of temperature, which is consistent with the results obtained by temperature-dependent Landau-level spectroscopy.

Figure~\ref{Chapter4Fig10}e summarizes our experimental results. It provides a comparison between the experimental values and the theoretical calculations of $\Delta$ defined in Fig.~\ref{Chapter4Fig10}c as a difference in energies of the $\alpha_{1}$ and $\gamma_{3}^{(\mathrm{I})}$ transitions. As clear from Fig.~\ref{Chapter4Fig9}a, the studied three-layer InAs/GaSb QW has a loop of local maxima in the valence subband in the vicinity of $\mathbf{k}=0$, and the band-gap in our sample is indirect with a value of about 15.5~meV. The experimental value of $\Delta\simeq17$~meV, extracted from the photoluminescence spectra, corresponds to the optical gap in the vicinity of the $\Gamma$ point, which is very close to the indirect band-gap. A good agreement between the experimental values of $\Delta$ and the theoretical calculations, performed with the low-temperature band parameters, is a direct evidence that the band structure and the band-gap in three-layer InAs/GaSb QWs are independent of temperature et least in the range up to 100~K.

\subsection{\label{sec:DiracIIIV}Massless Dirac fermions in three-layer InAs/GaSb QWs with two InAs layers}
Another key experimental result obtained by the THz team in Montpellier was the first experimental confirmation of massless Dirac fermions in three-layer InAs/GaSb/InAs QWs embedded between AlSb barriers~\cite{Dirac4}. These Dirac fermions arise at the point of crossing between the E1 and H1 subbands, as illustrated in Fig.~\ref{Chapter4Fig11}a. The main experimental technique used to verify the gapless state was magnetotransport measurements on a gated Hall bar, which allow one to visualize the evolution of the zero-mode Landau levels as a function of the magnetic field -- see Fig.~\ref{Chapter4Fig11}b. A similar approach has already been discussed in Section~\ref{sec:TransportHgTeQWs} in the context of temperature-driven topological phase transition in HgTe QWs~\cite{HgTeQWs25}. The sample investigated in our original work~\cite{Dirac4} was grown by molecular beam epitaxy on a semi-insulating (001) GaAs substrate with a relaxed GaSb buffer. To achieve the gapless state corresponding to the crossing of the E1 and H1 subbands, the thicknesses of the InAs and GaSb layers were carefully chosen to be 33 monolayers (MLs) and 14~MLs, respectively (see Fig.~\ref{Chapter4Fig11}). After the growth, a 50~$\mu$m-wide gated Hall bar was fabricated using a single-mesa process.

Figure~\ref{Chapter4Fig11}d provides typical gate voltage dependencies of the longitudinal $R_{xx}$ and Hall $R_{xy}$ resistances as a function of gate voltage $V_g$ at $T=1.7$~K and magnetic field $\mathcal{B}=6.5$~T. The longitudinal resistance shows clear oscillations on each side as a function of $V_g$ of the central peak occurring at $V_{DP}\simeq-4$~V when the Hall resistance presents $\pm0.5\,h/e^2$ plateaus and a sign reversal. These features demonstrate a changing of carrier concentration at different $V_g$ and the inversion of the type of carriers from electrons to holes at large negative gate voltage. Furthermore, an insulating behavior is observed at around $V_g=V_{DP}$ with $R_{xx}>h/e^2$ and small values of $R_{xy}$. The evolution of this insulating state is plotted in a 2D color map of $R_{xx}$ as a function of $V_g$ and $\mathcal{B}$ -- see Fig.~\ref{Chapter4Fig11}d. It is evident that the size of the insulating region (red area) increases with $\mathcal{B}$. The linear extrapolation down to zero field of the $R_{xx}=h/e^2$ points (black stars), corresponding to the position of the zero-mode Landau levels~\cite{Dirac2,HgTeQWs25} and delimiting the high resistance region for $B>2$~T, demonstrates that the insulating state vanishes at $B=0$~T. A similar behaviour was also observed in $d\sigma_{xy}/dV_g$ -- see our original paper for any details~\cite{Dirac4}. Thus, the crossing of the zero-mode Landau levels at $\mathcal{B}=0$~T indeed confirms the gapless state in our sample. The traces of other Landau levels of electrons and holes are also seen for $V_g-V_{DP}$ higher than $0.5$~V and lower than $-0.5$~V, respectively.

\begin{figure}
\centering
\includegraphics [width=1.0\columnwidth, keepaspectratio] {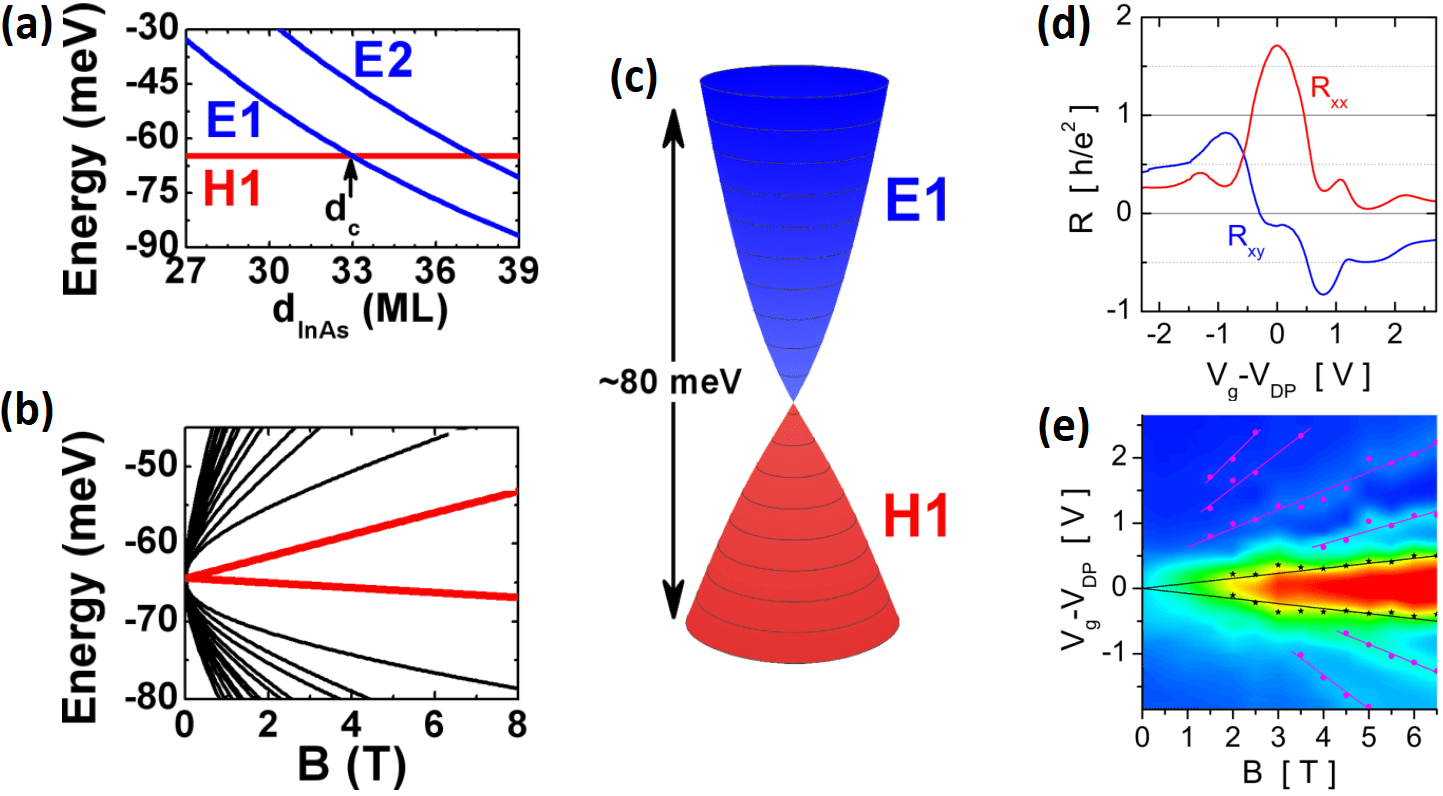}
\caption{\label{Chapter4Fig11} \textbf{(a)}~Subband energies at zero quasimomentum $\mathbf{k}$ in symmetric InAs/GaSb/InAs QWs as a function of InAs layer thickness $d_{\mathrm{InAs}}$. Blue and red curves correspond to the electron-like and hole-like states, respectively. The thickness of the GaSb layer $d_{\mathrm{GaSb}}$ equals to 12 monolayers (ML). Here, 1 ML corresponds to half of the lattice constant of the bulk material. The zero energy is referenced to the valence-band edge of bulk GaSb. \textbf{(b)}~Landau level fan chart of the studied gapless QW with $d_{\mathrm{InAs}}=33$~ML and $d_{\mathrm{GaSb}}=12$~ML. A pair of zero-mode Landau levels is presented by the red curves. \textbf{(c)}~A~3D~plot of the Dirac cone describing the energy spectrum at $\mathbf{k}<0.2$~nm$^{-1}$ for the studied gapless QW. \textbf{(d)}~Longitudinal $R_{xx}$ and Hall $R_{xy}$ resistances as a function of gate voltage $V_g$ at $T=1.7$~K and magnetic field $\mathcal{B}=6.5$~T. \textbf{(e)}~Color map of the longitudinal resistance $R_{xx}$ as a function of $V_g$ and $\mathcal{B}$. Black stars indicate $h/e^2$ values and red dots the peak values of $R_{xx}$.}
\end{figure}

The hallmark of 2D massless Dirac fermions is a specific sequence of quantum Hall plateaux observed at odd multiples of $g_ve^2/h$, where is $g_v$ valley degeneracy factor~\cite{III-VQWs24}. In HgTe QWs ($g_v=1$), the odd-integer quantum Hall sequence is much less pronounced~\cite{Dirac2} than in graphene~\cite{Dirac1} ($g_v=2$) and observed at small (less than 1~T) magnetic fields only. The latter is explained by the relatively large contribution of parabolic terms in the BHZ Hamiltonian~\cite{HgTeQWs2}, which effectively describes the gapless system caused by the crossing of the E1 and H1 subbands. The presence of these quadratic terms removes the spin degeneracy of all Landau levels, which hides the odd sequence of quantum Hall plateaux already at moderate fields, as discussed above. Since an inherent property of gapless three-layer InAs/GaSb QWs is rather low Fermi velocities $v_F$  (see Fig.~\ref{Chapter4Fig6}), the presence of quadratic terms does not allow observing the main hallmark of 2D massless Dirac fermions in the quantum Hall effect. However, an unambiguous conclusion about the presence of massless Dirac fermions in our sample can be made on the basis of other experimental studies.

\begin{figure}
\centering
\includegraphics [width=1.0\columnwidth, keepaspectratio] {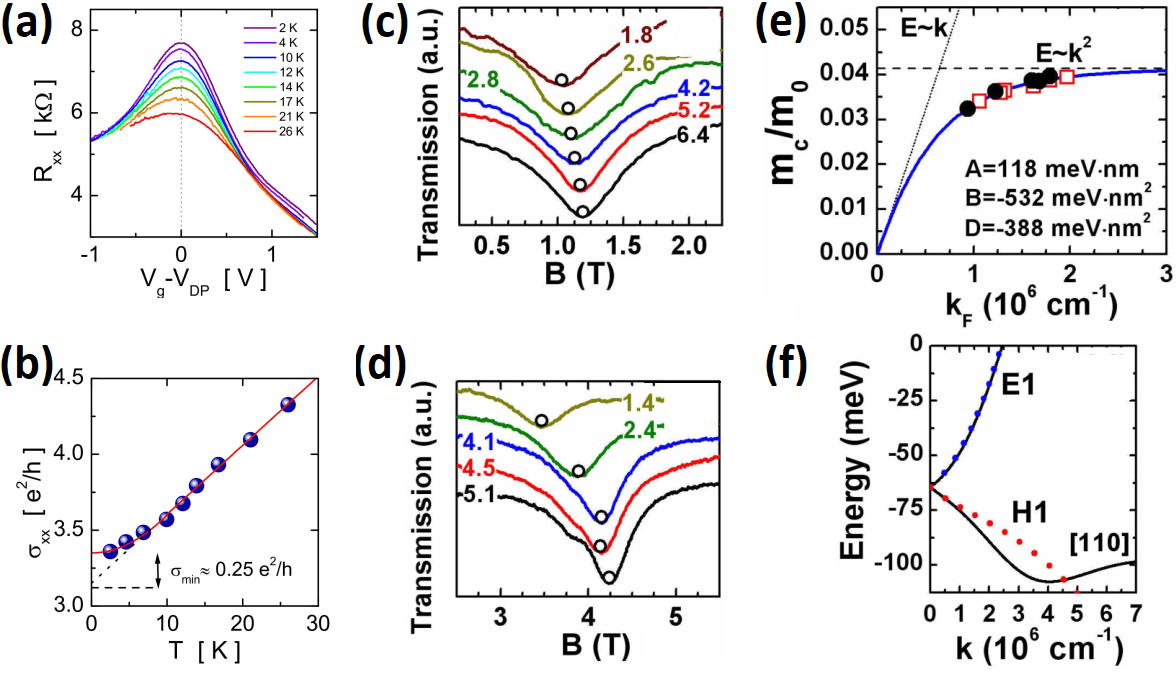}
\caption{\label{Chapter4Fig12} \textbf{(a)}~Zero-field resistance $R_{xx}$ as a function of gate voltage $V_g-V_{DP}$ at different temperatures. \textbf{(b)}~Conductivity at the Dirac point as a function of temperature. Solid circles are experimental data. The red line is a fit to exact expression of the conductivity within the Dirac-like model based on the Kubo formula derived in Ref.~\cite{Dirac2}. The black dashed line represents the high temperature behavior described by Eq.~(\ref{eq:IIIVQWexp4}).
\textbf{(c,d)} Cyclotron resonance spectra at $T=4.2$~K of electrons measured with backward wave oscillator at 845 GHz (a) and quantum cascade laser at the 3 THz (b) for different values of electron concentration. The symbols mark positions of cyclotron resonance lines. The numbers over the curves show $n_S$ in $10^{11}$~cm$^{-2}$. \textbf{(e)}~Cyclotron mass $m_c$ as a function of Fermi momentum, $k_F=\sqrt{2\pi n_S}$. The open and solid symbols correspond to the values obtained by backward wave oscillator and quantum cascade laser, respectively. The bold blue curve is the fitting to Eq.~(\ref{eq:IIIVQWexp4}). \textbf{(f)}~Comparison of band dispersion calculated within the eight-band $\mathbf{k}\cdot\mathbf{p}$ Kane Hamiltonian~\cite{HgTeQWs10} with the one described by Eq.~(\ref{eq:IIIVQWexp4b}).}
\end{figure}

Figure~\ref{Chapter4Fig12}a plots the resistance $R_{xx}$ as a function of gate voltage $V_g-V_{DP}$ at different temperatures. The graph clearly demonstrates the peaked resistivity behavior as in gapless HgTe QWs with massless Dirac fermions~\cite{Dirac2,III-VQWs24}. As seen, the maximum of $R_{xx}$ decreases with $T$. Figure~\ref{Chapter4Fig12}b shows the conductivity at the Dirac point as a function of temperature. Qualitatively, the temperature dependence of $\sigma_{xx}$ can be represented written as~\cite{Dirac2}:
\begin{equation}
\label{eq:IIIVQWexp3}
\sigma_{xx}\approx \dfrac{2}{\pi}\dfrac{e^2}{h}\dfrac{1}{1+\langle M^2\rangle/\Gamma^2}+\mathcal{O}\left(T^2\right), \text{~~~~$k_BT\ll\Gamma$,}
\end{equation}
\begin{equation}
\label{eq:IIIVQWexp4}
\sigma_{xx}\propto \dfrac{e^2}{h}\dfrac{k_BT}{\Gamma}, \text{~~~~$k_BT\geq\Gamma$,}
\end{equation}
where $\Gamma$ is the spectral broadening induced by spin-independent potential disorder and $\langle M^2\rangle$ is the variance of the gap due to spatial deviations of the layer thicknesses from their critical values. As seen from Fig.~\ref{Chapter4Fig12}b, the conductivity shows a quadratic dependence on $T$ at low temperature with the minimal value $\sigma_{xx}>3\,e^2/h$. Such a large obtained value is due to partial parallel conductivity through the buffer GaSb layer of our sample. In order to estimate the residual conductivity, the linear dependence of $\sigma_{xx}$ at high temperature is extrapolated down to $T=0$~K. By subtracting the obtained value to $\sigma_{xx}(0)$ we find $\sigma_{\text{min}}\approx0.25\,e^2/h$. As clear from Eq.~(\ref{eq:IIIVQWexp4}), the conductivity slope at high $T$ is inversely proportional to the broadening parameter $\Gamma$ induced by the disorder. By fitting the high-temperature values, we obtain $\Gamma\approx 1.9$~meV, which is in a good agreement with the LL broadening of $\Gamma=\hbar/2\tau_{tr}\approx1.8$~meV extracted from magnetotransport. From the values of $\sigma_\text{min}$ and $\Gamma$, we also evaluate the gap fluctuation in our sample $\sqrt{\langle\mathcal{M}^2\rangle}\approx 1.5 \,\Gamma \approx 3$~meV. Note that the linear dependence of $\sigma_{xx}(T)$ at high temperatures is another manifestation of the Dirac fermions in our sample~\cite{Dirac2}.

Another effective way to prove the presence of massless Dirac fermions in the gapless state is to measure the semiclassical cyclotron resonance at different positions of the Fermi level. Applying a quasiclassical quantization rule to the conduction band dispersion, calculated within the BHZ Hamiltonian~\cite{HgTeQWs2}
\begin{equation}
\label{eq:IIIVQWexp4b}
E_{\mathrm{BHZ}}(k)=C-D{k^2}+\sqrt{\left(M-B{k^2}\right)^2+A^2k^2},
\end{equation}
the cyclotron mass $m_c$ in the conduction band as a function of Fermi momentum $k_F=\sqrt{2\pi n_S}$ has the forms
\begin{equation}
\label{eq:IIIVQWexp5}
m_{c}(k_F)=\dfrac{\hbar^2k_F\sqrt{A^2+B^2k_F^2}}{A^2+2B^2k_F^2-2Dk_F\sqrt{A^2+B^2k_F^2}}.
\end{equation}
Figures~\ref{Chapter4Fig12}c and~\ref{Chapter4Fig12}d show cyclotron resonance spectra measured at $T=4.2$~K with both backward wave oscillator at 845~GHz~\cite{III-VQWs30,III-VQWs31} and quantum cascade laser at 3~THz~\cite{III-VQWs32} (pulse duration of 3~$\mu$s; repetition period of 100--200 $\mu$s) at different electron concentrations varied by using bipolar persistent photoconductivity effect~\cite{III-VQWs26,III-VQWs27,III-VQWs28,III-VQWs29}. The measurements were performed on the unprocessed sample with a changing of the electron concentration by varying the time illumination from red and blue light emitting diodes placed close to the sample. The GHz/THz radiation passing through the sample was detected either by a silicon bolometer (for backward wave oscillator) or Ge:Ga photoresistor (for quantum cascade laser). The electron concentration was determined along with the cyclotron resonance measurements via magnetotransport measurements in the van der Pauw geometry.

All the spectra contain a single CR line, defining the cyclotron mass $m_c$ at the Fermi level. As it is seen from Fig.~\ref{Chapter4Fig12}c and~\ref{Chapter4Fig12}d, cyclotron resonance lines shift toward high magnetic field with increasing of electron concentration. This indicates that cyclotron mass is strictly increasing function of $n_S$. As discussed above, this mass dependence corresponds to the gapless state with the linear band crossing. In order to extract parameters of massless Dirac fermions, we fitted our experimental data by Eq.~(\ref{eq:IIIVQWexp5}). A good agreement with experimental values is achieved for the following values of structure parameters in the BHZ model: $A=118$~meV$\cdot$nm, $B=-532$~meV$\cdot$nm$^2$ and $D=-388$~meV$\cdot$nm$^2$. Figure~\ref{Chapter4Fig12}e shows that although there are indeed massless Dirac fermions with Fermi velocity $v_F=A/\hbar=1.8\cdot10^5$~m/s, they only exist in the immediate vicinity of the $\Gamma$ point at $n_S\leq10^{10}$~cm$^{-2}$, while the terms proportional to $B$ and $D$ are relevant at higher concentrations. Additionally, we also plot the dependencies for $m_c=\hbar^2k_F/A$ (dotted curve) and $m_c=\hbar^2/(2|B|-2|D|)$ (dashed curve) for pure linear and quadratic band dispersion, respectively. Figure~\ref{Chapter4Fig12}f compares band structure of the sample numerically calculated within the eight-band $\mathbf{k}\cdot\mathbf{p}$ Kane Hamiltonian~\cite{HgTeQWs10} with the analytical expression~(\ref{eq:IIIVQWexp4b}) with the parameters extracted from the fitting $m_c(k_F)$. Parameter $C$ is chosen to have a coincidence of the band crossing point in two models. There is indeed a good agreement for the conduction band, while the valence band is well described at small quasimomentum only. This is also typical for HgTe/CdTe QWs, in which the discrepancy for the valence band is explained by the effect of the remote subbands beyond the BHZ model~\cite{TAI49}.

Thus, we have clearly demonstrated the existence of massless Dirac fermions in a three-layer InAs/GaSb/InAs QW. Magnetotransport experiments performed on the gated Hall bar unambiguously revealed the presence of gapless state in the studied sample. Furthermore, cyclotron resonance measurements performed at different electron concentrations made it possible not only to determine the characteristic Fermi velocity of the massless particles, $v_F = 1.8\times10^{5}$~m/s, but also to reveal a pronounced nonlinearity of the energy dispersion at higher energies. This latter observation provides direct evidence that, while the low-energy excitations in the system behave as Dirac fermions, the spectrum gradually deviates from the ideal linear form as one moves away from the band crossing point, reflecting the realistic band-structure effects of III-V semiconductor QWs.

\section{\label{sec:ExpIIIVWurz}Key experimental results of Technische Physik group in W\"{u}rzburg}
My theoretical predictions presented in Section~\ref{sec:3LIII-V} have also stimulated experimental investigations of three-layer InAs/Ga(In)Sb quantum wells carried out by the Technische Physik group in W\"{u}rzburg~\cite{WurzIIIVexp1,WurzIIIVexp2,WurzIIIVexp3,WurzIIIVexp4}. The long-standing expertise of this group in the molecular-beam epitaxy growth of III-V semiconductor heterostructures, as well as their advanced competence in fabricating gated Hall bars from the grown samples in their own cleanroom facilities, opened experimental possibilities that were not available to the THz team in Montpellier. In the following subsections, I will briefly review what I consider to be the most significant experimental results obtained in W\"{u}rzburg, achieved through close scientific cooperation with Montpellier.

\subsection{\label{sec:expQSHI}Observation of QSHI state at elevated temperature}
The first of these milestones is the observation of the QSHI state in three-layer InAs/GaInSb QWs at temperatures up to 60 K~\cite{WurzIIIVexp4}. The key factor that enabled the long-awaited observation of quantized conductance values in these structures was the successful fabrication by Adriana Wolf of a series of micron-sized gated Hall bars, which made it possible to directly probe the transport via topological edge states. The Hall bar devices were fabricated from a heterostructure containing a three-layer InAs/Ga$_{0.68}$In$_{0.32}$Sb QW, grown on a (001)-oriented AlSb buffer. The complete phase diagram as a function of the InAs layer thickness $d_{\mathrm{InAs}}$ and GaInSb layer thickness $d_{\mathrm{GaInSb}}$ is presented in Fig.~\ref{Chapter4Fig13}a. The white hatched region in the diagram represents the region of QSHI state with the largest band-gap values above 40 meV, which can be achieved for a given indium content of 32\% in the GaInSb alloy. The particular QW sample employed for the observation of the quantum spin Hall effect (QSHE) consisted of two InAs layers, each with a thickness of $d_{\mathrm{InAs}}=30$~ML ($\approx9.1$~nm), and a central Ga$_{0.68}$In$_{0.32}$Sb layer of $d_{\mathrm{GaInSb}}=10$~ML ($\approx3.1$~nm). These yield a moderately inverted band-gap of approximately $E_{\mathrm{gap}} = 27$~meV, as illustrated by the 3D plot in Fig.~\ref{Chapter4Fig13}b, where the $x$ and $y$ axes correspond to the [100] and [010] crystallographic directions, respectively.

Figure~\ref{Chapter4Fig13}c presents a typical scanning electron microscopy (SEM) image of the fabricated gated Hall bar devices. Each device had a total length of $L=10~\mu$m and a channel width of $W=1~\mu$m. The six Ohmic contacts were symmetrically arranged along the mesa, each with a contact length of $L_C=0.2~\mu$m, and labeled consecutively from 0 to 5. To explore the dependence of the conductance on the edge-state propagation length, several contact separation lengths $L_L$ (measured from the center of the upper contacts 1 and 2 to the lower ones 4 and 5) were implemented, with values of $L_L = 1$, $2$, and $3~\mu$m. These configurations correspond to separations between the outer contacts of $L_{NL} = 4.5$, $4.0$, and $3.5~\mu$m, thus maintaining the total channel length at $L = L_L + 2L_{NL} = 10~\mu$m. Importantly, these channel dimensions were comparable to or smaller than the typical phase coherence length $\lambda$ of the topological edge states in InAs/Ga(In)Sb-based QSHI systems~\cite{III-VQWs6,III-VQWs12,MontIIIVexp5,MontIIIVexp6}. This condition was essential for observing quantized conductance, originating from the transport through the 1D helical edge channels. To ensure the reproducibility of the experimental results, around twenty Hall bar devices were fabricated under identical conditions. Notably, no measurable top-gate leakage current was detected within the applied top-gate voltage range of $V_{TG} = +10$ to $-10$~V, confirming the excellent structural and electrical quality of the devices.

\begin{figure}
\centering
\includegraphics [width=1.0\columnwidth, keepaspectratio] {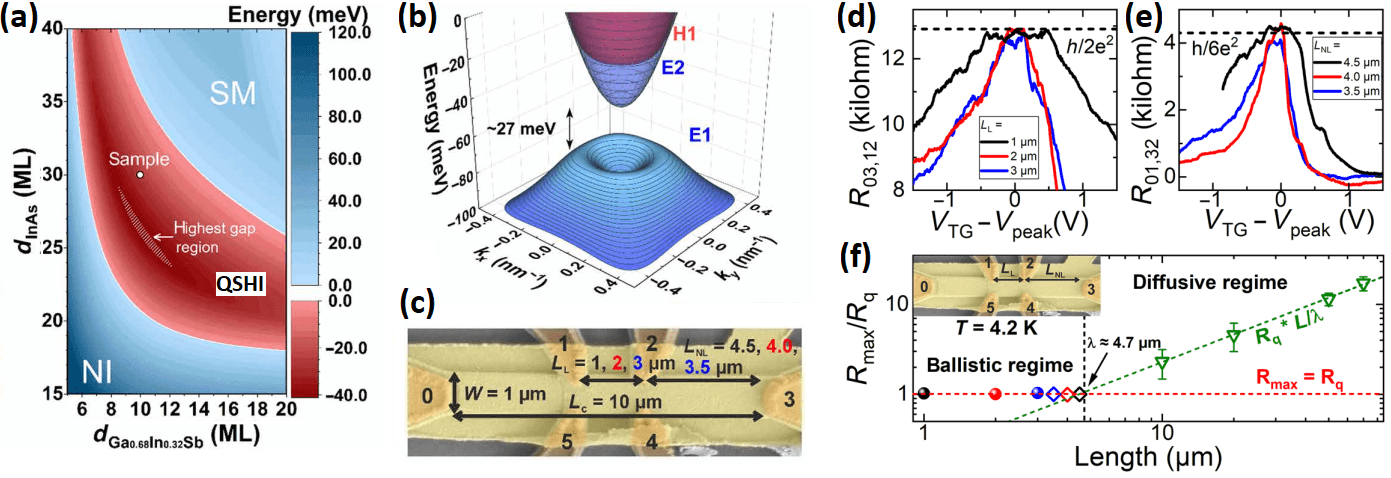}
\caption{\label{Chapter4Fig13} \textbf{(a)}  Color map diagram for symmetric three-layer InAs/Ga$_{0.68}$In$_{0.32}$Sb QWs grown on (001) AlSb buffer as a function of $d_{\mathrm{InAs}}$ and $d_{\mathrm{GaInSb}}$ with an normal insulator NI, QSHI, and SM states. The highest band-gap region in the QSHI regime and the sample studied in this manuscript are marked as a white hatched region and with an open symbol, respectively. For the SM region, the band-gap is zero, while the energy here marks the overlap between the maxima of the valence band and the minimum of conduction band. \textbf{(b)} 3D plot of the band structure of the investigated sample with the $x$ and $y$ axes being the [100] and [010] crystallographic directions, respectively. \textbf{(c)} Colored scanning electron microscopy image of a Hall bar device. The contacts are labeled from 0 to 5 with a length of $L_C = 0.2$~$\mu$m and the total length and width of the Hall bar are $L=10$~$\mu$m and $W=1$~$\mu$m, respectively. For the upper and lower contacts, different contact separation lengths $L_L=1$, $2$, and $3$~$\mu$m and $L_{NL}=4.5$, $4.0$, and $3.5$~$\mu$m were chosen to perform length-dependent measurements in local and nonlocal geometries.
\textbf{(d,e)} \textbf{Length-independent conductance quantization at 4.2 K.} Local resistance $R_{03,12}$ in (d) and nonlocal resistance $R_{01,32}$ in (e) as a function of top-gate voltage $V_{\mathrm{TG}}-V_{\mathrm{peak}}$ for gated devices with different lengths $L_L$ and $L_{NL}$. The resistance in the band-gap region is length independent and quantized to the expected value of$ h/2e^2$ and $h/6e^2$, for local and nonlocal configurations, respectively. \textbf{(f)} Maximum edge resistance values normalized to the expected quantized value $R_q$ for Hall bar devices of different lengths. Large-scaled Hall bars (green triangles) show a length dependence for the gap resistance in the diffusive regime in comparison to the Hall bars with $L_L=1$, $2$, and $3$~$\mu$m in the ballistic regime.}
\end{figure}

Figures~\ref{Chapter4Fig13}d and~\ref{Chapter4Fig13}e present the transport measurements performed in W\"{u}rzburg by Manuel Meyer and Fabian Hartmann in both the local [$R_{03,12}$, panel~(d)] and non-local [$R_{01,32}$, panel~(e)] geometries, plotted as a function of the top-gate voltage $V_{\mathrm{TG}} - V_{\mathrm{peak}}$ for devices with different channel lengths $L_L$ and $L_{NL}$. As clearly seen, the resistance peaks coincide with the quantized values of $h/2e^2$ and $h/6e^2$, expected respectively for the local and non-local configurations. This length-independent resistance within the band-gap region provides compelling evidence of the QSHE and of ballistic transport through the 1D helical edge channels. To ensure the reliability of these observations, a large number of Hall-bar devices were systematically tested. The average maximum resistance values, $R_{\mathrm{max}}$, obtained for all investigated samples were normalized to the theoretically expected quantized resistance $R_q$, and are summarized in Fig.~\ref{Chapter4Fig13}f. The condition $R_{\mathrm{max}}/R_q = 1$ corresponds to a perfect quantization. The extracted results for both local (filled circles) and non-local (diamonds) geometries demonstrate excellent agreement with theoretical expectations up to the channel lengths of $L = 4.5~\mu$m, indicating that the typical phase coherence length of the edge states is at least $\lambda\ge 4.5~\mu$m. For comparison, the data obtained from large-scale Hall bars with a width of $W = 20~\mu$m and channel lengths of $L_L = 10$, $20$, $50$, and $70~\mu$m (green triangles in Fig.~\ref{Chapter4Fig13}f) reveal a pronounced length dependence of the maximum edge resistance, as expected once $L_L>\lambda$. By extrapolating the linear fit of $R_{\mathrm{max}}/R_q$ to unity, the phase coherence length was extracted to be approximately $\lambda \simeq 4.7~\mu$m. Considering that all microscopic devices used in these experiments had contact separations shorter than $4.5~\mu$m, this result unambiguously confirms that the observed quantized resistance plateaus are not coincidental but indeed represent a clear experimental manifestation of the QSHE in three-layer InAs/GaInSb QWs.

\begin{figure}
\centering
\includegraphics [width=0.70\columnwidth, keepaspectratio] {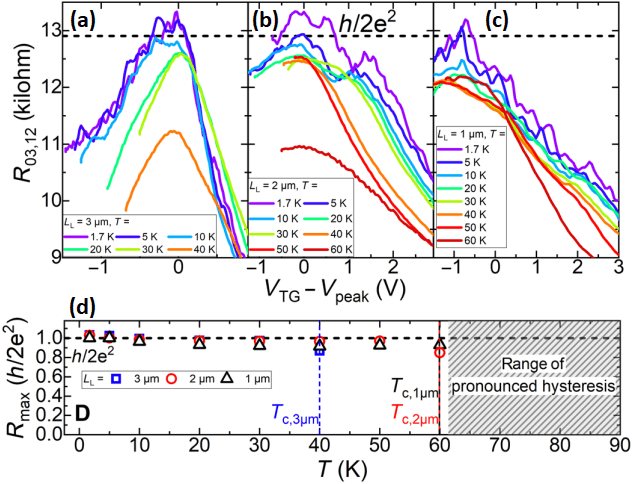}
\caption{\label{Chapter4Fig14} Robust conductance quantization up to $T=60$~K. \textbf{(a-c)}Temperature dependence of the quantized resistance value in the band-gap region for $L_L=3$, $2$ and $1$~$\mu$m in (a), (b) and (c), respectively. The resistance maximum values are summarized in panel \textbf{(d)}. For all three devices, a constant $R_{\mathrm{max}}$ at $h/2e^2$ can be observed with a maximum temperature of $T=60$~K for the $1$- and $2$-$\mu$m Hall bar devices. At larger temperatures, the band-gap region exceeds the accessible voltage range, at which the gate operates stably (grayed region).}
\end{figure}

Figure~\ref{Chapter4Fig14} presents the temperature-dependent transport measurements of the local resistance $R_{03,12}$ for Hall-bar devices with channel lengths $L_L = 3$, $2$, and $1~\mu$m. As the temperature increases, the corresponding maximum resistance values $R_{\mathrm{max}}$, are summarized in Fig.~\ref{Chapter4Fig14}d. It is clearly seen that $R_{\mathrm{max}}$ remains nearly constant at the quantized value of $h/2e^2$ up to the critical temperatures $T_c$ specific to each device: $T_{c,3\mu\mathrm{m}} = 40$~K, $T_{c,2\mu\mathrm{m}} = 60$~K, and $T_{c,1\mu\mathrm{m}} = 60$~K. Among them, the quantization appears to be most robust for the shortest channel device ($L_L = 1~\mu$m), which maintains quantization over the broadest temperature range. The onset temperature of quantized conductance was determined following the approach described in Ref.~\cite{III-VQWs3}, allowing for a maximal deviation of up to 15\%. The persistence of quantized conductance up to approximately 40~K in all investigated devices is attributed to a small residual bulk conductivity. Nonetheless, the remarkable observation of quantization up to 60~K was achieved in the devices with the shortest channel lengths ($L_L = 1$ and $2~\mu$m). This enhancement is most likely related to device-to-device variations in the residual bulk resistance, which, as confirmed by complementary measurements on large-scale Hall bars, is considerably higher in structures with short transport channels. These variations could be due to growth or processing, which requires further investigations. With further improvements in sample growth and fabrication, these variations can be minimized in the future.

It is particularly important to emphasize that the maximum temperature at which the QSHE is observed in the Hall bar device with $L_{L}=1~\mu$m is not limited by deviations from the expected quantized resistance caused by increased backscattering in the helical edge channels. Instead, as the temperature rises, the conductance plateau gradually shifts toward lower gate voltages, until the band-gap region can no longer be accessed during the up-sweep of the gate voltage. The temperature range where this effect occurs in our devices is marked by the gray-shaded area in Fig.~\ref{Chapter4Fig14}d. This shift of the entire resistance curve can be attributed to pronounced hysteresis effects between the up- and down-sweeps of the gate voltage, which ultimately limit the applicability of the gate training technique at elevated temperatures -- a procedure typically used to improve the precision of conductance quantization in the QSHE regime~\cite{III-VQWs33}. The origin of this hysteresis lies in charge accumulation at the interface between the gate dielectric and the cap layer of the QW heterostructure~\cite{III-VQWs34,III-VQWs35}.The same hysteresis effects also explain why temperature-dependent measurements for the $3~\mu$m device are shown only up to 40~K: at higher temperatures ($T=50$ and $60$~K), the gate hysteresis becomes sufficiently strong that the gap region cannot be reliably probed in the up-sweep anymore. Nevertheless, it is reasonable to expect that with further improvements in device fabrication, particularly in the stability and quality of the gate dielectric, these limitations can be overcome. This would make it possible to push the temperature for the observation of the QSHE even higher in newly fabricated devices based on the same three-layer InAs/GaInSb QWs used in the present experiments.

In addition to improving the overall device quality, it is equally important to fabricate QW heterostructures with larger inverted band-gap values in order to achieve a reliable observation of the QSHE at even higher temperatures. In particular, the three-layer InAs/Ga$_{0.68}$In$_{0.32}$Sb QW used in the present investigations possesses a moderate band-gap of approximately 27~meV. However, as illustrated in Fig.~\ref{Chapter4Fig13}a, by slightly adjusting the layer thicknesses in these QWs, the inverted band-gap can be increased up to about 42~meV. Moreover, as discussed in Section~\ref{sec:3LInAsQSHI} of this Chapter, by employing a Ga$_{0.60}$In$_{0.40}$Sb alloy and carefully selecting the buffer and barrier materials -- that remain fully compatible with high-quality pseudomorphic growth -- the band-gap in three-layer InAs/Ga$_{0.60}$In$_{0.40}$Sb QWs can exceed 50 meV at realistic strain values in the layers (see Fig.~\ref{Chapter4Fig8}).

In my opinion, although the experimental results presented in this subsection may be considered as intermediate milestones on the way toward the observation of the QSHE at temperatures above that of liquid nitrogen, they clearly demonstrate that three-layer InAs/GaInSb QWs represent an exceptionally promising material platform for realizing high-temperature QSHIs. Personally, I am convinced that the observation of quantized edge-state transport persisting well above the liquid-nitrogen temperature is merely a matter of time. When achieved, such an observation will provide complete experimental validation of my theoretical predictions~\cite{III-VQWs4} and will finally bridge the gap between fundamental research and practical topological device applications.

\subsection{\label{sec:expQHE}Unconventional quantum Hall effect in 2D semimetal with quadratic band touching}
The observation of the unconventional quantum Hall effect (QHE) in GaSb/InAs/GaSb QWs represents, in my opinion, another key experimental result of the Technische Physik group in W\"{u}rzburg. As already discussed in Sections~\ref{sec:HgTeDQWs} and~\ref{sec:3LIII-V}, both double HgTe/CdHgTe QWs and three-layer InAs/GaSb QWs with two GaSb layers host a peculiar 2D semimetallic phase characterized by quadratic band touching (QBT) at the $\Gamma$ point of the Brillouin zone.

Although such a QBT has also been theoretically predicted in other 2D systems~\cite{III-VQWs36,III-VQWs37,III-VQWs38,III-VQWs39,III-VQWs40}, the experimental confirmation of a 2D semimetal with quadratic band touching had, until recently, been achieved only in bilayer graphene. An intrinsic property of bilayer graphene is that its QBT is formed by chiral quasiparticles with a Berry phase of $2\pi$, which leads to a doubling of the ``zero-energy'' Landau level degeneracy compared to all other Landau levels~\cite{NaturalBG3}. This unique Landau level structure gives rise to the so-called unconventional QHE, characterized by the absence of a zero Hall plateau and a double step in the Hall conductivity $\sigma_{xy}$ at the charge neutrality point~\cite{NaturalBG3}. When the equivalence between the two graphene monolayers is broken -- for instance, by applying a perpendicular electric field, chemical doping, or structural asymmetry -- a band-gap opens at the QBT point~\cite{NaturalBG2}. This gap formation suppresses the unconventional behavior and restores a conventional QHE, with a zero Hall plateau $\sigma_{xy}=0$ appearing at the neutrality point~\cite{NaturalBG4}.

In contrast to bilayer graphene, the QBT in double HgTe/CdHgTe QWs and three-layer InAs/GaSb QWs is not associated with chiral quasiparticles. Nevertheless, these systems still possess a distinctive Landau level with a double degeneracy compared to other Landau levels in the presence of a magnetic field -- as illustrated in Fig.~\ref{Chapter3Fig13}d. This particular Landau level structure implies that a 2D semimetal realized in such QW heterostructures should also exhibit an unconventional QHE, provided that the QW retains perfect symmetry between the two tunnel-coupled layers -- HgTe layers in the case of double HgTe/CdHgTe QWs, and GaSb layers in the case of GaSb/InAs/GaSb QWs. It is worth noting that we made significant efforts in the THz team in Montpellier to experimentally observe this unconventional QHE in double HgTe/CdHgTe QWs~\cite{Tr3}. However, these attempts were hindered by the fundamental limitations of the existing HgTe-based MBE technology. In particular, the absence of in-situ control over layer thickness and interface sharpness using reflection high-energy electron diffraction (RHEED) makes it extremely challenging to achieve a perfectly symmetric double HgTe QW necessary for the formation of a QBT. By contrast, the far more advanced MBE technology developed for III-V semiconductor heterostructures, routinely employing RHEED for in-situ monitoring and precise control of growth parameters, opens realistic prospects for the realization of a 2D semimetal with QBT in GaSb/InAs/GaSb QWs. These structures thus provide a powerful experimental platform to verify the theoretical predictions discussed in Section~\ref{sec:3LIII-V} and to explore the emergence of the unconventional QHE beyond graphene-based materials.

\begin{figure}
\centering
\includegraphics [width=1.0\columnwidth, keepaspectratio] {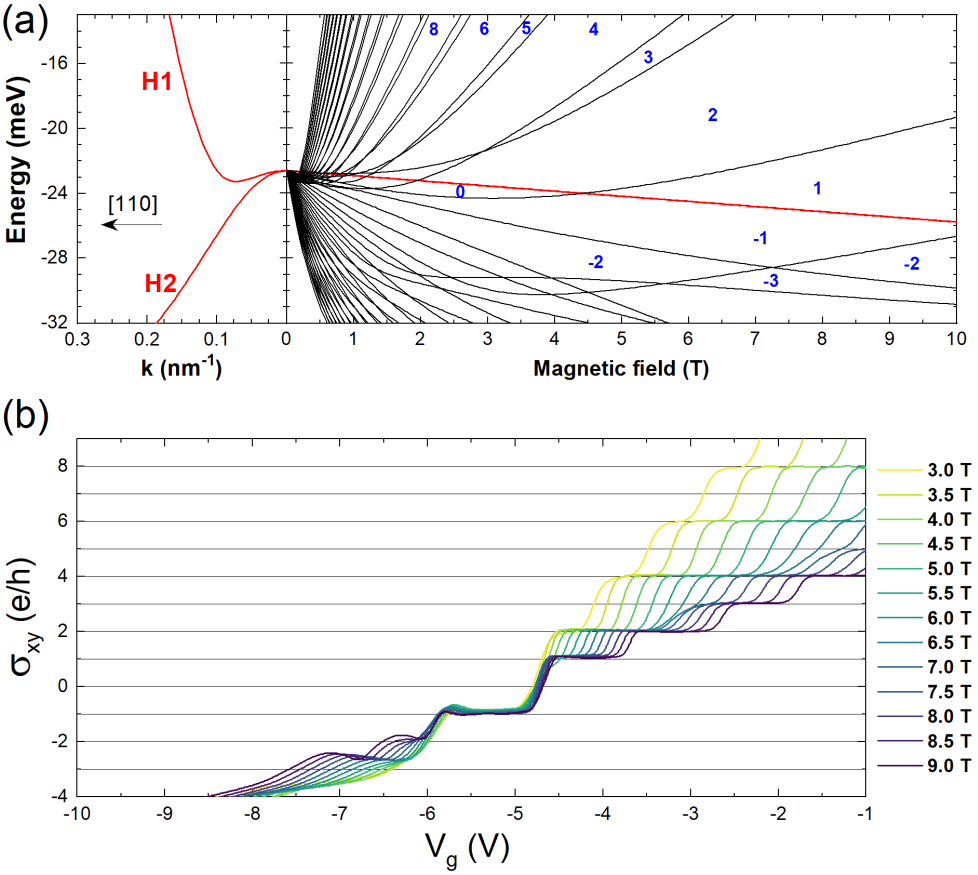}
\caption{\label{Chapter4Fig15} \textbf{(a)}~Band structure and Landau levels in sample TI126. The ``zero-mode'' Landau level, which has doubled degeneracy order as compared with other levels, is marked in red. This Landau level is formed by states at the $\Gamma$ point of both \emph{H}1 and \emph{H}2 subbands~\cite{doubleHgTeQWs6,III-VQWs4}. The blue numbers between Landau levels indicate the plateau number expected in the QHE at the given energy. \textbf{(b)}~Measured Hall conductivity  $\sigma_{xy}$ as a function of gate voltage $V_g$ in sample TI126 at 4.2~K.}
\end{figure}

To experimentally observe the unconventional QHE in a 2D semimetal with QBT, two GaSb/InAs/GaSb QW heterostructures were grown in the Technische Physik group on a (001)-oriented GaSb buffer, using AlSb as barrier layers. The heterostructures contained to a symmetric (sample TI126) or an asymmetric (sample TI104) configuration of the three-layer QW. In the symmetric QW, the InAs and GaSb layers had thicknesses of 34 and 22 MLs, respectively, whereas the asymmetric QW consisted of two GaSb layers with thicknesses of 23 and 21 MLs separated by a 34-monolayer-thick InAs layer. As before, one monolayer corresponded to half of the lattice constant of the respective bulk material. For magnetotransport experiments, gated Hall bar devices were fabricated from both heterostructures by Adriana Wolf. All devices had the width of $W=20$~$\mu$m and the distance between the lateral probes of $L=50$~$\mu$m. The electrical measurements were carried out by Sebastian Gebert and Fabian Hartmann in magnetic fields up to 9~T using the standard low-frequency lock-in technique combined with high-impedance preamplifiers to ensure optimal signal sensitivity.

Figure~\ref{Chapter4Fig15}a shows the band structure and Landau level fan chart for sample TI126. Details of calculations and notations of the LLs can be found in Ref.~\cite{HgTeQWs10}. A distinctive feature of the QBT in the symmetric three-layer GaSb/InAs/GaSb QW, arising in magnetic fields, is the presence of a particular ``zero-mode'' Landau level~\cite{doubleHgTeQWs6,III-VQWs4}, represented by the bold red curve. This ``zero-mode'' Landau level, has a pure hole-like character and is formed by the states in the vicinity of the bottom and top of the H1 and H2 subbands. Because they are touched at $\textbf{k}=0$, the ``zero-mode''  Landau level has a doubled degeneracy as compared with other levels. The latter should be the cause of the unconventional QHE, similar to that in natural bilayer graphene~\cite{NaturalBG3}. Figure~\ref{Chapter4Fig15}b shows the measured Hall conductivity $\sigma_{xy}$ of sample TI126 as a function of the applied gate voltage $V_g$ at different magnetic fields in the range from 3~T to 9~T. The observed behavior of $\sigma_{xy}(V_g)$ in the whole range of magnetic fields is in good agreement with the band structure and Landau-level calculations in Fig.~\ref{Chapter4Fig15}a. Indeed, in strong magnetic fields, all Landau levels are well separated, so when the Fermi level, varied by the gate voltage, lies inside the gap between the Landau levels, the Hall conductivity exhibits a plateau with the corresponding index, marked with blue numbers in \ref{Chapter4Fig15}a. Particularly, in the fields of more than 7~T the plateau sequence is clearly represented by $\sigma_{xy}=N{e^{2}/h}$, where $N$ is integer, with the absence of the $N=0$ plateau so that a double step of $e^{2}/h$ in height occurs at the neutrality point. This behavior of $\sigma_{xy}(V_g)$ completely replicates the unconventional QHE previously observed in bilayer graphene~\cite{NaturalBG3} with the only difference due to the valley and spin degeneracy~\cite{NaturalBG5}.

\begin{figure}
\centering
\includegraphics [width=1.0\columnwidth, keepaspectratio] {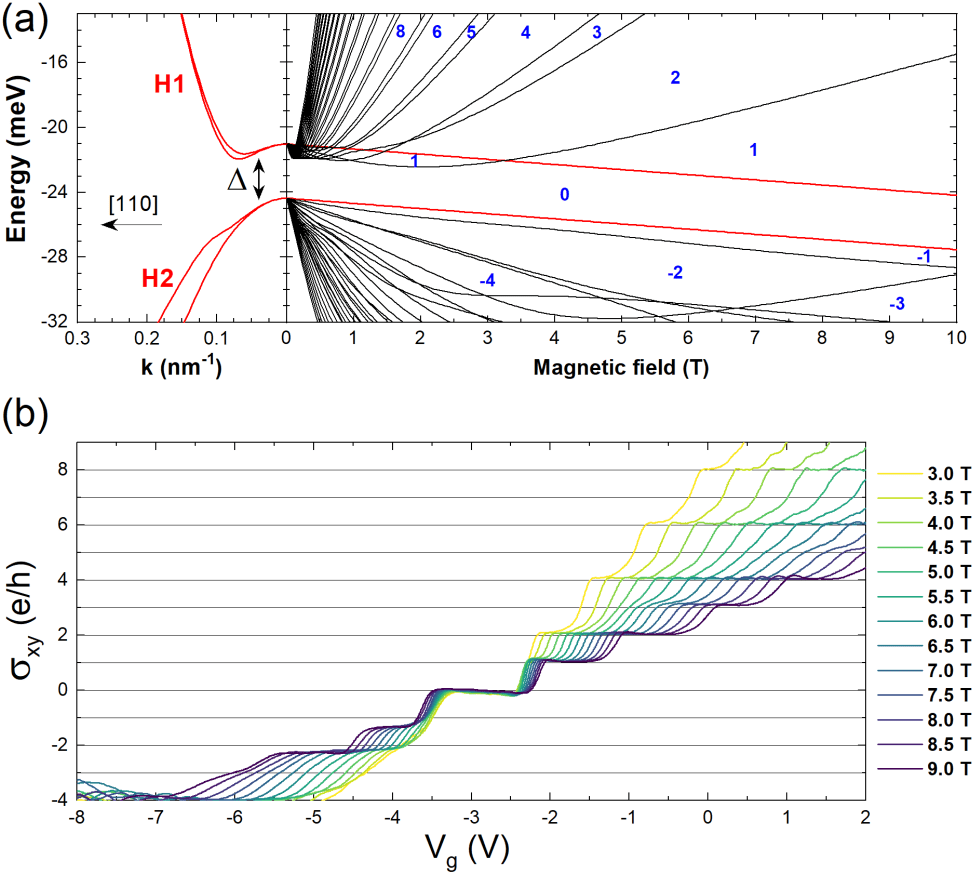}
\caption{\label{Chapter4Fig16} \textbf{(a)}~Band structure and Landau levels in sample TI104. The non-equivalence of the GaSb layers leads to the Rashba spin splitting at non-zero $\textbf{k}$ and the band-gap opening at the $\Gamma$ point. The latter also removes the double degeneracy of the ``zero-mode'' Landau levels (red curves). The blue numbers between Landau levels indicate the plateau number expected in the QHE at the given energy. \textbf{(b)}~Measured Hall conductivity  $\sigma_{xy}$ as a function of gate voltage $V_g$ in sample TI104 at 4.2~K.}
\end{figure}

A decrease in the magnetic field leads to the overlap of the densities of states (DOS) of adjacent Landau levels and, thus, to the disappearance of some plateaus at $\sigma_{xy}(V_g)$. In particular, the plateaus with odd indices $N\geq3$ disappear in the fields less than 7~T. As clear from Fig.~\ref{Chapter4Fig15}a, the plateaus with these indices correspond to the Fermi level positions between the Zeeman-split s in the conduction band. Unlike GaSb/InAs/GaSb QWs, the Zeeman splitting of Landau levels is not observed in the QHE of bilayer graphene even at magnetic fields up to 20~T due to the small g-factor values~\cite{III-VQWs41}. On the contrary, the disappearance of the plateau with $N=1$ is associated with the crossing of the lower Landau level, repeating the ``W-like'' shape of the conduction band, and the doubly degenerate ``zero-mode'' Landau level. Due to the DOS overlap of these levels in the fields less than 5.5~T, the double step of $\sigma_{xy}$ at the neutrality point turns into a triple step, corresponding to the transition between the plateaus with $N=2$ and $N=-1$. Note that this observed triple step should be absent in the QWs with the band dispersion from ``BG'' region of the diagram in (see Fig.~\ref{Chapter4Fig5}C), in which the conduction band does not have a ``W-like'' shape. Unlike its counterpart in the conduction band, the plateau with $N=-1$ remains clearly observable in smaller fields, which is also consistent with the LL fan chart in Fig.~\ref{Chapter4Fig15}a. Thus, the observed behavior of the $\sigma_{xy}$ plateaus with indices $N>-2$ is in very good agreement with the LL fan charts calculated in the axial approximation (see Ref.~\cite{HgTeQWs10} for any calculation details). To analyze the $\sigma_{xy}$ plateaus with indices $N<-2$ one has to imply more sophisticated Landau level calculations beyond the axial approximation due to the strong anisotropy of the valence band away from the QBT. But since these plateaus are not clearly visible in the QHE of sample TI126, going beyond the axial approximation does not make much sense.

As already discussed above, one of the characteristic features of bilayer graphene is the opening of a band gap upon the breaking of the equivalence between its two tunnel-coupled layers~\cite{NaturalBG2}. This non-equivalence leads to the appearance of a zero plateau in the Hall conductivity $\sigma_{xy}(V_g)$ and the disappearance of the unconventional QHE~\cite{NaturalBG4}. Let us now demonstrate the same behavior in the case of GaSb/InAs/GaSb QWs. Strictly speaking, applying a gate voltage to a nominally symmetric QW always introduces a certain non-equivalence between the two GaSb layers because of the asymmetric electric field that develops across the structure~\cite{III-VQWs42}. However, the clear observation of the unconventional QHE in the symmetric sample TI126 provides convincing evidence that, in this case, the electric field is too weak to cause a noticeable splitting of the ``zero-mode'' Landau level. To induce a more pronounced asymmetry and thereby intentionally break the equivalence of the two GaSb layers, we therefore turn our attention to the asymmetric GaSb/InAs/GaSb QW (sample TI104), in which the two GaSb layers have different thicknesses. This structural asymmetry plays the same conceptual role as an applied perpendicular electric field in bilayer graphene, allowing one to probe the transformation of the unconventional QHE into the conventional one.

Figure~\ref{Chapter4Fig16}a shows the band structure and Landau level fan chart for sample TI104 with a asymmetric three-layer GaSb/InAs/GaSb QW. Similar to bilayer graphene~\cite{NaturalBG2}, the non-equivalence of the GaSb layers indeed leads to the band-gap opening and the QBT breaking, removing the double degeneracy of the ``zero-mode'' Landau level. An additional effect of the QW asymmetry is the Rashba spin splitting, which occurs in the conduction and valence bands at non-zero $\textbf{k}$. As it is clearly seen from Fig.~\ref{Chapter4Fig16}b, the band-gap opening in an asymmetric three-layer GaSb/InAs/GaSb QW yields the~$N=0$ plateau of the Hall conductivity observed in the whole range of magnetic fields. The presence of $\sigma_{xy}=0$ recovers the sequence of equidistant plateaus in the fields of more than 7~T, representing the ``conventional'' QHE expected for 2D systems~\cite{NaturalBG3} and ``biased'' bilayer graphene~\cite{NaturalBG4}. As is the case for sample TI126, the plateaus with odd indices $N\geq3$ in sample TI104 disappear in the fields less than 7~T due to the DOS overlap between adjacent Zeeman-split LLs. Interestingly, since the band-gap opening preserves the ``W-like'' shape of the conduction band, the triple step of $\sigma_{xy}(V_g)$ in sample TI126, corresponding to the transition between the plateaus with $N=2$ and $N=-1$ in weak magnetic fields, splits into a double and a single step in sample TI103.

The experimental results presented above have not yet been published. Therefore, in this m\'{e}moire, I have chosen to focus only on the two key experimental findings that provide direct evidence for the existence of the unconventional QHE in the three-layer GaSb/InAs/GaSb QW. These findings not only confirm the main theoretical predictions discussed earlier in Section~\ref{sec:3LIII-V}, but also establish this III-V semiconductor system as a suitable experimental platform for exploring 2D semimetals with QBT. The full set of experimental and technical details, together with additional measurement data and extended analyses, will be presented in our forthcoming publication, which is currently in preparation.

\section{\label{sec:III-V}Perspectives}
Theoretical and experimental investigations of InAs/Ga(In)Sb-based QWs, carried out in Montpellier and in the Technische Physik group in W\"{u}rzburg, have already demonstrated that III-V semiconductor heterostructures constitute a robust and versatile material platform for exploring a wide spectrum of topological states initially discovered or predicted in HgTe-based QWs. However, my research on III-V-based topological systems goes far beyond the scope of the works briefly summarized in this Chapter. Particularly, a significant part of my theoretical efforts has focused on predicting the existence of 3D higher-order topological insulators and 3D higher-order Dirac semimetals in InAs/GaInSb superlattices. The preparation of a comprehensive manuscript describing these results is currently in progress.

More broadly, the studies presented in this Chapter form part of my long-term research program aimed at demonstrating that nearly all known topological phases of matter -- from conventional to higher-order and from 1D to 3D -- can be realized and tuned within the technologically mature platform of III-V semiconductor heterostructures. The key concept underlying this program is band-structure and symmetry engineering, which allows tailoring both the band inversion and the crystalline symmetries through quantum confinement, strain, specific interface design and the MBE growth of the heterostructure along specific crystallographic orientations. In this regard, InAs/Ga(In)Sb heterostructures and superlattices offer exceptional flexibility: their band alignment can be precisely controlled, their growth on various substrates is technologically established, and their integration with silicon opens perspectives for scalable topological electronics.

\begin{figure}
\centering
\includegraphics [width=1.0\columnwidth, keepaspectratio] {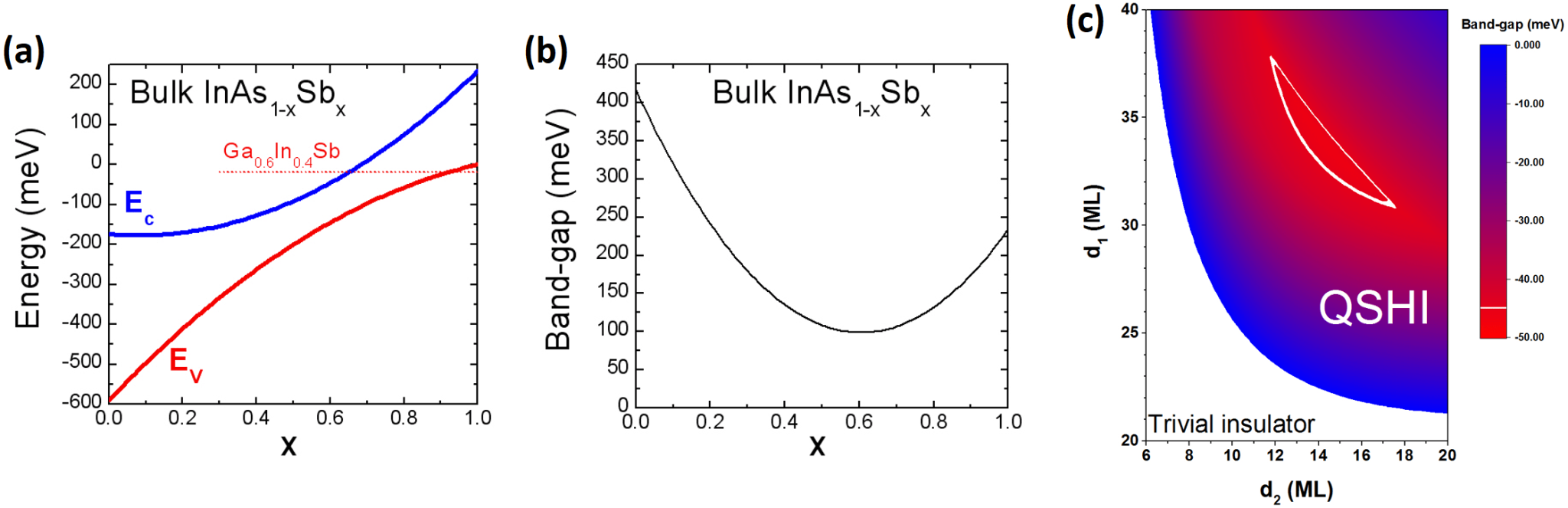}
\caption{\label{Chapter4Fig17} \textbf{(a,b)}~The position of the conduction and valence bands, as well as the band gap at the $\Gamma$ point of the Brillouin zone in the bulk InAs$_{1-x}$Sb$_{x}$ alloy as a function on the Sb content. \textbf{(c)}~Colormap diagram of symmetric three-layer InAs$_{0.70}$Sb$_{0.30}$/InAs$_{0.25}$Sb$_{0.75}$ QWs grown on (001) Al$_{0.58}$In$_{0.42}$Sb buffer with the same barrier materials. The thicknesses of InAs$_{0.70}$Sb$_{0.30}$ and InAs$_{0.25}$Sb$_{0.75}$ are represented by $d_1$ and $d_2$, respectively. The strain in the QW layers is $\varepsilon_{1}=+1.5\%$ in the InAs$_{0.70}$Sb$_{0.30}$ layer and $\varepsilon_{2}=-1.5\%$ in the InAs$_{0.25}$Sb$_{0.75}$ layer. The energy scale represents the band-gap in the QSHI state -- the negative values here emphasize the inverted band-ordering. Here, 1 ML equals to half of the lattice constant of the bulk material. The region with white boundaries marks layer thicknesses with a band gap exceeding the inverted gap of 45~meV in monolayer WTe$_2$~\cite{III-VQWs3}.}
\end{figure}

Finally, I would like to emphasize that the InAs/Ga(In)Sb heteropair is not the only III-V semiconductor system that allows for the realization of a broken-gap alignment at the heterointerface -- the key ingredient for achieving topological states in III-V semiconductor heterostructures. Another promising material system exhibiting such a band alignment is the InAs$_{1-x}$Sb$_{x}$ alloy. Figure~\ref{Chapter4Fig17} shows the evolution of the conduction- and valence-band edges in bulk InAs$_{1-x}$Sb$_{x}$ as a function of the Sb content. It is clearly seen that one can choose two compositions, $x_1$ and $x_2$, for which the conduction band minimum of one alloy lies below the valence band maximum of the other, thus forming the desired broken-gap heterointerface. Although experimental studies of the QSHI state in inverted InAsSb-based QWs have not yet been reported, such QW heterostructures were also considered in my theoretical work as potential candidates for realizing a high-temperature quantum spin Hall effect. A brief mention of this idea can be found in the third paragraph of the ``Discussion'' section of our original paper~\cite{III-VQWs4}, devoted to three-layer QWs. As illustrated by the phase diagram of symmetric three-layer InAs$_{0.70}$Sb$_{0.30}$/InAs$_{0.25}$Sb$_{0.75}$ QWs grown on a (001) Al$_{0.58}$In$_{0.42}$Sb buffer with the same barrier material, shown in Fig.~\ref{Chapter4Fig17}, the inverted band-gap in these QW structures can indeed exceed 45~meV -- the value characteristic of monolayer WTe$_2$~\cite{III-VQWs3}. Moreover, as revealed by our recent studies of non-local transport in long-channel gated Hall bars fabricated from three-layer InAs/GaInSb QWs~\cite{MontIIIVexp6}, the dominant backscattering mechanism in edge-state transport originates from spin-spin interactions between edge electrons and impurities naturally associated with intrinsic Ga-antisite defects acting as double acceptors in Ga(In)Sb. It is therefore plausible that three-layer InAsSb-based QWs, where such Ga-related defects are absent, could exhibit the quantum spin Hall effect over significantly longer channel lengths compared to their InAs/GaInSb counterparts -- see Section~\ref{sec:ExpIIIVWurz}.

Taken together, these considerations suggest that InAsSb-based heterostructures also represent an attractive and, so far, unexplored material platform for realizing robust and potentially high-temperature topological phases in III-V semiconductor heterostructures. In my view, their experimental investigation may open the next chapter in the development of III-V-based topological materials, continuing the line of research initiated by the works presented in this m\'{e}moire.

\addcontentsline{toc}{chapter}{Summary and Conclusion}
\chapter*{\label{Conclusion}Summary and Conclusion}
The scientific works given an overview in this m\'{e}moire reflect more than a decade of my research in the THz team in Montpellier, which can be collectively summarized under the title: ``\emph{Pseudo-relativistic fermions and topological states in zinc-blende semiconductors and their heterostructures}''. This choice was deliberate. It allowed me to present, in a coherent way, a body of theoretical and experimental work that not only contributed to the understanding of topological states in ``traditional'' materials, but also served as a foundation for new experimental directions, several of which are now being pursued by different scientific groups worldwide. Nevertheless, the research topics discussed here represent only part of my scientific activity. Other important contributions -- such as my investigations of collective excitations in 2D systems in the integer and fractional quantum Hall effect regimes, and the response of 2D and 3D electronic systems to electromagnetic radiation under strongly non-equilibrium carrier distributions in crossed electric and magnetic fields -- have intentionally remained beyond the scope of this m\'{e}moire. They are, however, briefly mentioned in Chapter~\ref{Chap:Career}, which retraces the main steps of my career path.

Beyond the purely scientific content, the work presented in this m\'{e}moire also demonstrates my ability to conduct independent and original research across a wide range of topics, extending far beyond a single subfield. My post-doctoral path illustrates a consistent capacity to identify emerging physical problems, develop appropriate theoretical tools, and connect these ideas to concrete experimental realizations. This scientific autonomy, developed over the years of independent work and collaboration, is a central qualification required for the \emph{Habilitation \'{a} diriger des Recherches}. A second essential component of this qualification is the ability to supervise and guide students and young researchers. Throughout my career, I have had the privilege to mentor several Bachelor, Master and PhD students both in formal and informal way. My role has extended from providing direct scientific supervision in theoretical modeling and data analysis, to helping them present their results in international conferences and prepare manuscripts for publication. These experiences have reinforced my conviction that the transmission of knowledge and the development of young researchers' critical thinking are among the most rewarding aspects of scientific life. Today, I have formulated a number of new scientific projects and research topics that can serve as the basis for future PhD theses under my supervision. The internal motivation to share my experience and knowledge with the next generation of scientists is, in fact, the principal reason, for which I am seeking the \emph{Habilitation \'{a} diriger des Recherches}, without which I cannot officially act as an independent PhD advisor in many countries around the world.

Equally important is my demonstrated capacity to lead scientific projects and manage collaborative research. In Montpellier, I often coordinated the theoretical and experimental parts of various research, which can be confirmed by the head of the THz team, Fr\'{e}d\'{e}ric Teppe. I have successfully initiated several collaborations between different groups with complementary expertise -- bringing together, for example, the THz team in Montpellier and several scientific groups from Russian to perform a number of investigations of single and double HgTe QWs. These collaborations were not accidental: they were the direct consequence of my scientific vision, my ability to formulate clear physical objectives, and my efforts to unite different communities around them. This aptitude for establishing and coordinating fruitful international collaborations is another key criterion of the \emph{Habilitation \'{a} diriger des Recherches}.

My scientific independence is also evidenced by the diversity and scope of the topics I have explored -- ranging from non-equilibrium phenomena in crossed electric and magnetic fields and collective excitations to higher-order topological states. This broad scientific vision, combined with my interdisciplinary background and hands-on understanding of both theory and experiment, ensures my ability to develop and manage new research themes in the future. From an academic perspective, my non-standard career path -- which has combined theoretical research with close engagement in experimental studies -- has provided me with a global and integrative view of physics. This multidimensional approach, I believe, constitutes a valuable asset for designing original lecture courses and seminars covering a broad range of contemporary topics: from quantum transport and semiconductor physics to topological materials and quantum technologies. Such teaching activities will form a natural complement to my research leadership.

In summary, the work presented in this m\'{e}moire demonstrates my full scientific autonomy, my ability to lead and supervise research, to manage projects and collaborations, and to develop original ideas across multiple domains of condensed-matter physics. It also reflects my commitment to education and knowledge transfer. Obtaining the academic qualification \emph{Habilitation \'{a} diriger des Recherches}  represents not only the recognition of my past achievements but also a crucial step toward my long-term goal -- to build and lead a research group dedicated to the exploration of novel quantum phenomena in semiconductor and topological materials.

\addcontentsline{toc}{chapter}{Bibliography}
\bibliography{references} 

\begin{thebibliography}{211}
\providecommand{\natexlab}[1]{#1}
\providecommand{\url}[1]{\texttt{#1}}
\expandafter\ifx\csname urlstyle\endcsname\relax
  \providecommand{\doi}[1]{doi: #1}\else
  \providecommand{\doi}{doi: \begingroup \urlstyle{rm}\Url}\fi

\bibitem[Keldysh(1964)]{Flds1}
L.~V. Keldysh.
\newblock Deep levels in semiconductors.
\newblock \emph{Sov. Phys. JETP}, 18\penalty0 (1):\penalty0 253, 1964.
\newblock ISSN 0044-4510.
\newblock URL \url{http://jetp.ras.ru/cgi-bin/e/index/e/18/1/p253?a=list}.

\bibitem[Wolff(1964)]{Flds2}
P.~A. Wolff.
\newblock Matrix elements and selection rules for the two-band model of
  bismuth.
\newblock \emph{J. Phys. Chem. Solids}, 25\penalty0 (10):\penalty0 1057--1068,
  1964.
\newblock ISSN 0022-3697.
\newblock \doi{https://doi.org/10.1016/0022-3697(64)90128-3}.
\newblock URL
  \url{https://www.sciencedirect.com/science/article/pii/0022369764901283}.

\bibitem[Novoselov et~al.(2005)Novoselov, Geim, Morozov, Jiang, Katsnelson,
  Grigorieva, Dubonos, and Firsov]{Dirac1}
K.~S. Novoselov, A.~K. Geim, S.~V. Morozov, D.~Jiang, M.~I. Katsnelson, I.~V.
  Grigorieva, S.~V. Dubonos, and A.~A. Firsov.
\newblock Two-dimensional gas of massless dirac fermions in graphene.
\newblock \emph{Nature}, 438:\penalty0 197, 2005.
\newblock \doi{10.1038/nature04233}.
\newblock URL \url{http://dx.doi.org/10.1038/nature04233}.

\bibitem[B\"{u}ttner et~al.(2011)B\"{u}ttner, Liu, Tkachov, Novik, Br\"{u}ne,
  Buhmann, Hankiewicz, Recher, Trauzettel, Zhang, and Molenkamp]{Dirac2}
B.~B\"{u}ttner, C.X. Liu, G.~Tkachov, E.G. Novik, C.~Br\"{u}ne, H.~Buhmann,
  E.M. Hankiewicz, P.~Recher, B.~Trauzettel, S.C. Zhang, and L.W. Molenkamp.
\newblock Single valley dirac fermions in zero-gap hgte quantum wells.
\newblock \emph{Nat. Phys.}, 7\penalty0 (5):\penalty0 418, 2011.
\newblock \doi{10.1038/nphys1914}.
\newblock URL \url{http://dx.doi.org/10.1038/nphys1914}.

\bibitem[Ludwig et~al.(2014)Ludwig, Vasilyev, Mikhailov, Poumirol, Jiang,
  Vafek, and Smirnov]{Dirac3}
J.~Ludwig, Yu.~B. Vasilyev, N.~N. Mikhailov, J.~M. Poumirol, Z.~Jiang,
  O.~Vafek, and D.~Smirnov.
\newblock Cyclotron resonance of single-valley dirac fermions in nearly gapless
  hgte quantum wells.
\newblock \emph{Phys. Rev. B}, 89:\penalty0 241406, Jun 2014.
\newblock \doi{10.1103/PhysRevB.89.241406}.
\newblock URL \url{http://link.aps.org/doi/10.1103/PhysRevB.89.241406}.

\bibitem[Krishtopenko et~al.(2019{\natexlab{a}})Krishtopenko, Desrat, Spirin,
  Consejo, Ruffenach, Gonzalez-Posada, Jouault, Knap, Maremyanin, Gavrilenko,
  Boissier, Torres, Zaknoune, Tourni\'e, and Teppe]{Dirac4}
S.~S. Krishtopenko, W.~Desrat, K.~E. Spirin, C.~Consejo, S.~Ruffenach,
  F.~Gonzalez-Posada, B.~Jouault, W.~Knap, K.~V. Maremyanin, V.~I. Gavrilenko,
  G.~Boissier, J.~Torres, M.~Zaknoune, E.~Tourni\'e, and F.~Teppe.
\newblock Massless dirac fermions in iii-v semiconductor quantum wells.
\newblock \emph{Phys. Rev. B}, 99:\penalty0 121405, Mar 2019{\natexlab{a}}.
\newblock \doi{10.1103/PhysRevB.99.121405}.
\newblock URL \url{https://link.aps.org/doi/10.1103/PhysRevB.99.121405}.

\bibitem[Hsieh et~al.(2008)Hsieh, Qian, Wray, Xia, Hor, Cava, and
  Hasan]{Dirac3DTI1}
D.~Hsieh, D.~Qian, L.~Wray, Y.~Xia, Y.~S. Hor, R.~J. Cava, and M.~Z. Hasan.
\newblock A topological dirac insulator in a quantum spin hall phase.
\newblock \emph{Nature}, 452:\penalty0 970, 2008.
\newblock \doi{10.1038/nature06843}.
\newblock URL \url{http://doi.org/10.1038/nature06843}.

\bibitem[Zhang et~al.(2009)Zhang, Liu, Qi, Dai, Fang, and Zhang]{Dirac3DTI2}
Haijun Zhang, Chao-Xing Liu, Xiao-Liang Qi, Xi~Dai, Zhong Fang, and Shou-Cheng
  Zhang.
\newblock Topological insulators in bi$_2$se$_3$, bi$_2$te$_3$ and sb$_2$te$_3$
  with a single dirac cone on the surface.
\newblock \emph{Nat. Phys.}, 5:\penalty0 438, 2009.
\newblock \doi{10.1038/nphys1270}.
\newblock URL \url{http://doi.org/10.1038/nphys1270}.

\bibitem[Assaf et~al.(2016)Assaf, Phuphachong, Volobuev, Inhofer, Bauer,
  Springholz, de~Vaulchier, and Guldner]{Dirac3DTI3}
B.~A. Assaf, T.~Phuphachong, V.~V. Volobuev, A.~Inhofer, G.~Bauer,
  G.~Springholz, L.~A. de~Vaulchier, and Y.~Guldner.
\newblock Massive and massless dirac fermions in pb$_{1-x}$sn$_x$te topological
  crystalline insulator probed by magneto-optical absorption.
\newblock \emph{Sci. Rep.}, 6:\penalty0 20323, 2016.
\newblock \doi{10.1038/srep20323}.
\newblock URL \url{http://doi.org/10.1038/srep20323}.

\bibitem[Liu et~al.(2014{\natexlab{a}})Liu, Zhou, Zhang, Wang, Weng,
  Prabhakaran, Mo, Shen, Fang, Dai, Hussain, and Chen]{3DDiracSM1}
Z.~K. Liu, B.~Zhou, Y.~Zhang, Z.~J. Wang, H.~M. Weng, D.~Prabhakaran, S.-K. Mo,
  Z.~X. Shen, Z.~Fang, X.~Dai, Z.~Hussain, and Y.~L. Chen.
\newblock Discovery of a three-dimensional topological dirac semimetal, na3bi.
\newblock \emph{Science}, 343\penalty0 (6173):\penalty0 864--867,
  2014{\natexlab{a}}.
\newblock ISSN 0036-8075.
\newblock \doi{10.1126/science.1245085}.
\newblock URL \url{https://science.sciencemag.org/content/343/6173/864}.

\bibitem[Liu et~al.(2014{\natexlab{b}})Liu, Jiang, Zhou, Wang, Zhang, Weng,
  Prabhakaran, Mo, Peng, Dudin, Kim, Hoesch, Fang, Dai, Shen, Feng, Hussain,
  and Chen]{3DDiracSM2}
Z.~K. Liu, J.~Jiang, B.~Zhou, Z.~J. Wang, Y.~Zhang, H.~M. Weng, D.~Prabhakaran,
  S-K. Mo, H.~Peng, P.~Dudin, T.~Kim, M.~Hoesch, Z.~Fang, X.~Dai, Z.~X. Shen,
  D.~L. Feng, Z.~Hussain, and Y.~L. Chen.
\newblock A stable three-dimensional topological dirac semimetal cd$_3$as$_2$.
\newblock \emph{Nat. Mater.}, 13:\penalty0 677, 2014{\natexlab{b}}.
\newblock \doi{10.1038/nmat3990}.
\newblock URL \url{http://doi.org/10.1038/nmat3990}.

\bibitem[Neupane et~al.(2014)Neupane, Xu, Sankar, Alidoust, Bian, Liu,
  Belopolski, Chang, Jeng, Lin, Bansil, Chou, and Hasan]{3DDiracSM3}
Madhab Neupane, Su-Yang Xu, Raman Sankar, Nasser Alidoust, Guang Bian, Chang
  Liu, Ilya Belopolski, Tay-Rong Chang, Horng-Tay Jeng, Hsin Lin, Arun Bansil,
  Fangcheng Chou, and M.~Zahid Hasan.
\newblock Observation of a three-dimensional topological dirac semimetal phase
  in high-mobility cd$_3$as$_2$.
\newblock \emph{Nat. Commun.}, 5:\penalty0 3786, 2014.
\newblock \doi{10.1038/ncomms4786}.
\newblock URL \url{http://doi.org/10.1038/ncomms4786}.

\bibitem[Lv et~al.(2015)Lv, Xu, Weng, Ma, Richard, Huang, Zhao, Chen, Matt,
  Bisti, Strocov, Mesot, Fang, Dai, Qian, Shi, and Ding]{3DWeylSM1}
B.~Q. Lv, N.~Xu, H.~M. Weng, J.~Z. Ma, P.~Richard, X.~C. Huang, L.~X. Zhao,
  G.~F. Chen, C.~E. Matt, F.~Bisti, V.~N. Strocov, J.~Mesot, Z.~Fang, X.~Dai,
  T.~Qian, M.~Shi, and H.~Ding.
\newblock Observation of weyl nodes in taas.
\newblock \emph{Nat. Phys.}, 11:\penalty0 724, 2015.
\newblock \doi{10.1038/nphys3426}.
\newblock URL \url{http://doi.org/10.1038/nphys3426}.

\bibitem[Yang et~al.(2015)Yang, Liu, Sun, Peng, Yang, Zhang, Zhou, Zhang, Guo,
  Rahn, Prabhakaran, Hussain, Mo, Felser, Yan, and Chen]{3DWeylSM2}
L.~X. Yang, Z.~K. Liu, Y.~Sun, H.~Peng, H.~F. Yang, T.~Zhang, B.~Zhou,
  Y.~Zhang, Y.~F. Guo, M.~Rahn, D.~Prabhakaran, Z.~Hussain, S.-K. Mo,
  C.~Felser, B.~Yan, and Y.~L. Chen.
\newblock Weyl semimetal phase in the non-centrosymmetric compound taas.
\newblock \emph{Nat. Phys.}, 11:\penalty0 728, 2015.
\newblock \doi{10.1038/nphys3425}.
\newblock URL \url{http://doi.org/10.1038/nphys3425}.

\bibitem[Xu et~al.(2015)Xu, Alidoust, Belopolski, Yuan, Bian, Chang, Zheng,
  Strocov, Sanchez, Chang, Zhang, Mou, Wu, Huang, Lee, Huang, Wang, Bansil,
  Jeng, Neupert, Kaminski, Lin, Jia, and Hasan]{3DWeylSM3}
Su-Yang Xu, Nasser Alidoust, Ilya Belopolski, Zhujun Yuan, Guang Bian, Tay-Rong
  Chang, Hao Zheng, Vladimir~N. Strocov, Daniel~S. Sanchez, Guoqing Chang,
  Chenglong Zhang, Daixiang Mou, Yun Wu, Lunan Huang, Chi-Cheng Lee, Shin-Ming
  Huang, BaoKai Wang, Arun Bansil, Horng-Tay Jeng, Titus Neupert, Adam
  Kaminski, Hsin Lin, Shuang Jia, and M.~Zahid Hasan.
\newblock Discovery of a weyl fermion state with fermi arcs in niobium
  arsenide.
\newblock \emph{Nat. Phys.}, 11:\penalty0 748, 2015.
\newblock \doi{10.1038/nphys3437}.
\newblock URL \url{http://doi.org/10.1038/nphys3437}.

\bibitem[Burkov et~al.(2011)Burkov, Hook, and Balents]{Exotic1}
A.~A. Burkov, M.~D. Hook, and Leon Balents.
\newblock Topological nodal semimetals.
\newblock \emph{Phys. Rev. B}, 84:\penalty0 235126, Dec 2011.
\newblock \doi{10.1103/PhysRevB.84.235126}.
\newblock URL \url{https://link.aps.org/doi/10.1103/PhysRevB.84.235126}.

\bibitem[Soluyanov et~al.(2015)Soluyanov, Gresch, Wang, Wu, Troyer, Dai, and
  Bernevig]{Exotic2}
Alexey~A. Soluyanov, Dominik Gresch, Zhijun Wang, QuanSheng Wu, Matthias
  Troyer, Xi~Dai, and B.~Andrei Bernevig.
\newblock Type-ii weyl semimetals.
\newblock \emph{Nature}, 527:\penalty0 495, 2015.
\newblock \doi{10.1038/nature15768}.
\newblock URL \url{http://doi.org/10.1038/nature15768}.

\bibitem[Wieder et~al.(2016)Wieder, Kim, Rappe, and Kane]{Exotic3}
Benjamin~J. Wieder, Youngkuk Kim, A.~M. Rappe, and C.~L. Kane.
\newblock Double dirac semimetals in three dimensions.
\newblock \emph{Phys. Rev. Lett.}, 116:\penalty0 186402, May 2016.
\newblock \doi{10.1103/PhysRevLett.116.186402}.
\newblock URL \url{https://link.aps.org/doi/10.1103/PhysRevLett.116.186402}.

\bibitem[Bradlyn et~al.(2016)Bradlyn, Cano, Wang, Vergniory, Felser, Cava, and
  Bernevig]{Exotic4}
Barry Bradlyn, Jennifer Cano, Zhijun Wang, M.~G. Vergniory, C.~Felser, R.~J.
  Cava, and B.~Andrei Bernevig.
\newblock Beyond dirac and weyl fermions: Unconventional quasiparticles in
  conventional crystals.
\newblock \emph{Science}, 353\penalty0 (6299):\penalty0 aaf5037, 2016.
\newblock ISSN 0036-8075.
\newblock \doi{10.1126/science.aaf5037}.
\newblock URL \url{https://science.sciencemag.org/content/353/6299/aaf5037}.

\bibitem[Orlita et~al.(2014)Orlita, Basko, Zholudev, Teppe, Knap, Gavrilenko,
  Mikhailov, Dvoretskii, Neugebauer, Faugeras, Barra, Martinez, and
  Potemski]{Kane1}
M.~Orlita, D.~M. Basko, M.~S. Zholudev, F.~Teppe, W.~Knap, V.~I. Gavrilenko,
  N.~N. Mikhailov, S.~A. Dvoretskii, P.~Neugebauer, C.~Faugeras, A-L. Barra,
  G.~Martinez, and M.~Potemski.
\newblock Observation of three-dimensional massless kane fermions in a
  zinc-blende crystal.
\newblock \emph{Nature Phys.}, 10:\penalty0 233--238, 2014.
\newblock \doi{10.1038/nphys2857}.
\newblock URL \url{http://dx.doi.org/10.1038/nphys2857}.

\bibitem[Laurenti et~al.(1990)Laurenti, Camassel, Bouhemadou, Toulouse, Legros,
  and Lusson]{Kane5}
J.~P. Laurenti, J.~Camassel, A.~Bouhemadou, B.~Toulouse, R.~Legros, and
  A.~Lusson.
\newblock Temperature dependence of the fundamental absorption edge of mercury
  cadmium telluride.
\newblock \emph{J. Appl. Phys.}, 67\penalty0 (10):\penalty0 6454--6460, 1990.
\newblock \doi{http://dx.doi.org/10.1063/1.345119}.
\newblock URL
  \url{http://scitation.aip.org/content/aip/journal/jap/67/10/10.1063/1.345119}.

\bibitem[Malcolm and Nicol(2015)]{Kane2}
J.~D. Malcolm and E.~J. Nicol.
\newblock Magneto-optics of massless kane fermions: Role of the flat band and
  unusual berry phase.
\newblock \emph{Phys. Rev. B}, 92:\penalty0 035118, Jul 2015.
\newblock \doi{10.1103/PhysRevB.92.035118}.
\newblock URL \url{https://link.aps.org/doi/10.1103/PhysRevB.92.035118}.

\bibitem[Raoux et~al.(2014)Raoux, Morigi, Fuchs, Pi\'echon, and
  Montambaux]{Kane3}
A.~Raoux, M.~Morigi, J.-N. Fuchs, F.~Pi\'echon, and G.~Montambaux.
\newblock From dia- to paramagnetic orbital susceptibility of massless
  fermions.
\newblock \emph{Phys. Rev. Lett.}, 112:\penalty0 026402, Jan 2014.
\newblock \doi{10.1103/PhysRevLett.112.026402}.
\newblock URL \url{https://link.aps.org/doi/10.1103/PhysRevLett.112.026402}.

\bibitem[Krishtopenko et~al.(2020{\natexlab{a}})Krishtopenko, Antezza, and
  Teppe]{Kane4}
Sergey~S. Krishtopenko, Mauro Antezza, and Fr{\'{e}}d{\'{e}}ric Teppe.
\newblock Hybridization of topological surface states with a flat band.
\newblock \emph{J. Phys.: Condens. Matter}, 32\penalty0 (16):\penalty0 165501,
  jan 2020{\natexlab{a}}.
\newblock \doi{10.1088/1361-648x/ab6741}.
\newblock URL \url{https://doi.org/10.1088/1361-648x/ab6741}.

\bibitem[Teppe et~al.(2016)Teppe, Marcinkiewicz, Krishtopenko, Ruffenach,
  Consejo, Kadykov, Desrat, But, Knap, Ludwig, Moon, Smirnov, Orlita, Jiang,
  Morozov, Gavrilenko, Mikhailov, and Dvoretskii]{Kane6}
F.~Teppe, M.~Marcinkiewicz, S.~S. Krishtopenko, S.~Ruffenach, C.~Consejo, A.~M.
  Kadykov, W.~Desrat, D.~But, W.~Knap, J.~Ludwig, S.~Moon, D.~Smirnov,
  M.~Orlita, Z.~Jiang, S.~V. Morozov, V.I. Gavrilenko, N.~N. Mikhailov, and
  S.~A. Dvoretskii.
\newblock Temperature-driven massless kane fermions in hgcdte crystals.
\newblock \emph{Nat. Commun.}, 7:\penalty0 12576, 2016.
\newblock \doi{10.1038/ncomms12576}.
\newblock URL \url{http://dx.doi.org/10.1038/ncomms12576}.

\bibitem[Szo\l{}a et~al.(2022)Szo\l{}a, Ivonyak, Przybytek, Mikhailov,
  Dvoretsky, Teppe, and Knap]{Kane7}
Maria Szo\l{}a, Yurii Ivonyak, Jacek Przybytek, Nikolai Mikhailov, Sergey
  Dvoretsky, Fr\'ed\'eric Teppe, and Wojciech Knap.
\newblock Hydrostatic deformation potentials of narrow-gap hgcdte.
\newblock \emph{Phys. Rev. B}, 106:\penalty0 195202, Nov 2022.
\newblock \doi{10.1103/PhysRevB.106.195202}.
\newblock URL \url{https://link.aps.org/doi/10.1103/PhysRevB.106.195202}.

\bibitem[Yavorskiy et~al.(2024)Yavorskiy, Ivonyak, But, Karpierz, Krajewska,
  Haras, Sai, Dub, Kazakov, Cywiński, Knap, and Łusakowski]{Kane8}
D~Yavorskiy, Y~Ivonyak, D~But, K~Karpierz, A~Krajewska, M~Haras, P~Sai, M~Dub,
  A~Kazakov, G~Cywiński, W~Knap, and J~Łusakowski.
\newblock Pressure tuning of hgcdte epitaxial layers -- the role of the highly
  disordered buffer layer.
\newblock \emph{J. Phys. D: Appl. Phys.}, 58\penalty0 (2):\penalty0 025304, oct
  2024.
\newblock \doi{10.1088/1361-6463/ad80a2}.
\newblock URL \url{https://doi.org/10.1088/1361-6463/ad80a2}.

\bibitem[Fu et~al.(2007)Fu, Kane, and Mele]{3DTI1}
Liang Fu, C.~L. Kane, and E.~J. Mele.
\newblock Topological insulators in three dimensions.
\newblock \emph{Phys. Rev. Lett.}, 98:\penalty0 106803, Mar 2007.
\newblock \doi{10.1103/PhysRevLett.98.106803}.
\newblock URL \url{https://link.aps.org/doi/10.1103/PhysRevLett.98.106803}.

\bibitem[Akrap et~al.(2016)Akrap, Hakl, Tchoumakov, Crassee, Kuba, Goerbig,
  Homes, Caha, Nov\'ak, Teppe, Desrat, Koohpayeh, Wu, Armitage, Nateprov,
  Arushanov, Gibson, Cava, van~der Marel, Piot, Faugeras, Martinez, Potemski,
  and Orlita]{Kane9}
A.~Akrap, M.~Hakl, S.~Tchoumakov, I.~Crassee, J.~Kuba, M.~O. Goerbig, C.~C.
  Homes, O.~Caha, J.~Nov\'ak, F.~Teppe, W.~Desrat, S.~Koohpayeh, L.~Wu, N.~P.
  Armitage, A.~Nateprov, E.~Arushanov, Q.~D. Gibson, R.~J. Cava, D.~van~der
  Marel, B.~A. Piot, C.~Faugeras, G.~Martinez, M.~Potemski, and M.~Orlita.
\newblock Magneto-optical signature of massless kane electrons in
  ${\mathrm{cd}}_{3}{\mathrm{as}}_{2}$.
\newblock \emph{Phys. Rev. Lett.}, 117:\penalty0 136401, Sep 2016.
\newblock \doi{10.1103/PhysRevLett.117.136401}.
\newblock URL \url{https://link.aps.org/doi/10.1103/PhysRevLett.117.136401}.

\bibitem[Desrat et~al.(2018)Desrat, Krishtopenko, Piot, Orlita, Consejo,
  Ruffenach, Knap, Nateprov, Arushanov, and Teppe]{Kane10}
W.~Desrat, S.~S. Krishtopenko, B.~A. Piot, M.~Orlita, C.~Consejo, S.~Ruffenach,
  W.~Knap, A.~Nateprov, E.~Arushanov, and F.~Teppe.
\newblock Band splitting in ${\mathrm{cd}}_{3}{\mathrm{as}}_{2}$ measured by
  magnetotransport.
\newblock \emph{Phys. Rev. B}, 97:\penalty0 245203, Jun 2018.
\newblock \doi{10.1103/PhysRevB.97.245203}.
\newblock URL \url{https://link.aps.org/doi/10.1103/PhysRevB.97.245203}.

\bibitem[Krishtopenko and Teppe(2022)]{Flds8}
Sergey~S. Krishtopenko and Fr\'ed\'eric Teppe.
\newblock Relativistic collapse of landau levels of kane fermions in crossed
  electric and magnetic fields.
\newblock \emph{Phys. Rev. B}, 105:\penalty0 125203, Mar 2022.
\newblock \doi{10.1103/PhysRevB.105.125203}.
\newblock URL \url{https://link.aps.org/doi/10.1103/PhysRevB.105.125203}.

\bibitem[Rumyantsev et~al.(2013)Rumyantsev, Morozov, Antonov, Zholudev,
  Kudryavtsev, Gavrilenko, Dvoretskii, and Mikhailov]{Kane11}
V~V Rumyantsev, S~V Morozov, A~V Antonov, M~S Zholudev, K~E Kudryavtsev, V~I
  Gavrilenko, S~A Dvoretskii, and N~N Mikhailov.
\newblock Spectra and kinetics of thz photoconductivity in narrow-gap
  hg\textsubscript{1–x}cd\textsubscript{x}te ($x<0.2$) epitaxial films.
\newblock \emph{Semicond. Sci. Technol.}, 28\penalty0 (12):\penalty0 125007,
  oct 2013.
\newblock \doi{10.1088/0268-1242/28/12/125007}.
\newblock URL \url{https://doi.org/10.1088/0268-1242/28/12/125007}.

\bibitem[Novik et~al.(2005)Novik, Pfeuffer-Jeschke, Jungwirth, Latussek,
  Becker, Landwehr, Buhmann, and Molenkamp]{HgTeQWs1}
E.~G. Novik, A.~Pfeuffer-Jeschke, T.~Jungwirth, V.~Latussek, C.~R. Becker,
  G.~Landwehr, H.~Buhmann, and L.~W. Molenkamp.
\newblock Band structure of semimagnetic
  ${\mathrm{hg}}_{1\ensuremath{-}y}{\mathrm{mn}}_{y}\mathrm{Te}$ quantum wells.
\newblock \emph{Phys. Rev. B}, 72:\penalty0 035321, Jul 2005.
\newblock \doi{10.1103/PhysRevB.72.035321}.
\newblock URL \url{https://link.aps.org/doi/10.1103/PhysRevB.72.035321}.

\bibitem[Jackiw and Rebbi(1976)]{SS3DKane1}
R.~Jackiw and C.~Rebbi.
\newblock Solitons with fermion number \textonehalf{}.
\newblock \emph{Phys. Rev. D}, 13:\penalty0 3398--3409, Jun 1976.
\newblock \doi{10.1103/PhysRevD.13.3398}.
\newblock URL \url{https://link.aps.org/doi/10.1103/PhysRevD.13.3398}.

\bibitem[Pfeffer and Zawadzki(2003)]{Kane12}
P.~Pfeffer and W.~Zawadzki.
\newblock Bychkov-rashba spin splitting and its dependence on magnetic field in
  ${\mathrm{i}\mathrm{n}\mathrm{s}\mathrm{b}/\mathrm{i}\mathrm{n}}_{0.91}{\mathrm{al}}_{0.09}\mathrm{Sb}$
  asymmetric quantum wells.
\newblock \emph{Phys. Rev. B}, 68:\penalty0 035315, Jul 2003.
\newblock \doi{10.1103/PhysRevB.68.035315}.
\newblock URL \url{https://link.aps.org/doi/10.1103/PhysRevB.68.035315}.

\bibitem[Bernardes et~al.(2007)Bernardes, Schliemann, Lee, Egues, and
  Loss]{Kane13}
Esmerindo Bernardes, John Schliemann, Minchul Lee, J.~Carlos Egues, and Daniel
  Loss.
\newblock Spin-orbit interaction in symmetric wells with two subbands.
\newblock \emph{Phys. Rev. Lett.}, 99:\penalty0 076603, Aug 2007.
\newblock \doi{10.1103/PhysRevLett.99.076603}.
\newblock URL \url{https://link.aps.org/doi/10.1103/PhysRevLett.99.076603}.

\bibitem[Pfeffer and Zawadzki(2006)]{Kane14}
P.~Pfeffer and W.~Zawadzki.
\newblock Anisotropy of spin $g$ factor in gaas/ga$_{1-x}$al$_{x}$as symmetric
  quantum wells.
\newblock \emph{Phys. Rev. B}, 74:\penalty0 233303, Dec 2006.
\newblock \doi{10.1103/PhysRevB.74.233303}.
\newblock URL \url{https://link.aps.org/doi/10.1103/PhysRevB.74.233303}.

\bibitem[Bender and Boettcher(1998)]{Kane15}
Carl~M. Bender and Stefan Boettcher.
\newblock Real spectra in non-hermitian hamiltonians having $pt$ symmetry.
\newblock \emph{Phys. Rev. Lett.}, 80:\penalty0 5243--5246, Jun 1998.
\newblock \doi{10.1103/PhysRevLett.80.5243}.
\newblock URL \url{https://link.aps.org/doi/10.1103/PhysRevLett.80.5243}.

\bibitem[Bender et~al.(2002)Bender, Brody, and Jones]{Kane16}
Carl~M. Bender, Dorje~C. Brody, and Hugh~F. Jones.
\newblock Complex extension of quantum mechanics.
\newblock \emph{Phys. Rev. Lett.}, 89:\penalty0 270401, Dec 2002.
\newblock \doi{10.1103/PhysRevLett.89.270401}.
\newblock URL \url{https://link.aps.org/doi/10.1103/PhysRevLett.89.270401}.

\bibitem[Chu et~al.(2011)Chu, Shan, Lu, and Shen]{Kane17}
Rui-Lin Chu, Wen-Yu Shan, Jie Lu, and Shun-Qing Shen.
\newblock Surface and edge states in topological semimetals.
\newblock \emph{Phys. Rev. B}, 83:\penalty0 075110, Feb 2011.
\newblock \doi{10.1103/PhysRevB.83.075110}.
\newblock URL \url{https://link.aps.org/doi/10.1103/PhysRevB.83.075110}.

\bibitem[Dyakonov and Khaetskii(1981)]{Kane18}
M.~I. Dyakonov and A.~V. Khaetskii.
\newblock Surface states in a gapless semiconductor.
\newblock \emph{JETP Lett.}, 33:\penalty0 110, Feb 1981.
\newblock URL \url{http://www.jetpletters.ac.ru/ps/1501/article_22948.shtml}.

\bibitem[Luttinger(1956)]{Kane19}
J.~M. Luttinger.
\newblock Quantum theory of cyclotron resonance in semiconductors: General
  theory.
\newblock \emph{Phys. Rev.}, 102:\penalty0 1030--1041, May 1956.
\newblock \doi{10.1103/PhysRev.102.1030}.
\newblock URL \url{https://link.aps.org/doi/10.1103/PhysRev.102.1030}.

\bibitem[Lin-Liu and Sham(1985)]{Kane20}
Y.~R. Lin-Liu and L.~J. Sham.
\newblock Interface states and subbands in hgte-cdte heterostructures.
\newblock \emph{Phys. Rev. B}, 32:\penalty0 5561--5563, Oct 1985.
\newblock \doi{10.1103/PhysRevB.32.5561}.
\newblock URL \url{https://link.aps.org/doi/10.1103/PhysRevB.32.5561}.

\bibitem[Chang et~al.(1985)Chang, Schulman, Bastard, Guldner, and Voos]{Kane21}
Yia-Chung Chang, J.~N. Schulman, G.~Bastard, Y.~Guldner, and M.~Voos.
\newblock Effects of quasi-interface states in hgte-cdte superlattices.
\newblock \emph{Phys. Rev. B}, 31:\penalty0 2557--2560, Feb 1985.
\newblock \doi{10.1103/PhysRevB.31.2557}.
\newblock URL \url{https://link.aps.org/doi/10.1103/PhysRevB.31.2557}.

\bibitem[Fu and Kane(2007)]{Kane22}
Liang Fu and C.~L. Kane.
\newblock Topological insulators with inversion symmetry.
\newblock \emph{Phys. Rev. B}, 76:\penalty0 045302, Jul 2007.
\newblock \doi{10.1103/PhysRevB.76.045302}.
\newblock URL \url{https://link.aps.org/doi/10.1103/PhysRevB.76.045302}.

\bibitem[Mahler et~al.(2019)Mahler, Mayer, Leubner, Lunczer, Di~Sante,
  Sangiovanni, Thomale, Hankiewicz, Buhmann, Gould, and Molenkamp]{Kane24}
David~M. Mahler, Julian-Benedikt Mayer, Philipp Leubner, Lukas Lunczer,
  Domenico Di~Sante, Giorgio Sangiovanni, Ronny Thomale, Ewelina~M. Hankiewicz,
  Hartmut Buhmann, Charles Gould, and Laurens~W. Molenkamp.
\newblock Interplay of dirac nodes and volkov-pankratov surface states in
  compressively strained hgte.
\newblock \emph{Phys. Rev. X}, 9:\penalty0 031034, Aug 2019.
\newblock \doi{10.1103/PhysRevX.9.031034}.
\newblock URL \url{https://link.aps.org/doi/10.1103/PhysRevX.9.031034}.

\bibitem[Kozlov et~al.(2016)Kozlov, Bauer, Ziegler, Fischer, Savchenko, Kvon,
  Mikhailov, Dvoretsky, and Weiss]{Kane25}
D.~A. Kozlov, D.~Bauer, J.~Ziegler, R.~Fischer, M.~L. Savchenko, Z.~D. Kvon,
  N.~N. Mikhailov, S.~A. Dvoretsky, and D.~Weiss.
\newblock Probing quantum capacitance in a 3d topological insulator.
\newblock \emph{Phys. Rev. Lett.}, 116:\penalty0 166802, Apr 2016.
\newblock \doi{10.1103/PhysRevLett.116.166802}.
\newblock URL \url{https://link.aps.org/doi/10.1103/PhysRevLett.116.166802}.

\bibitem[Thomas et~al.(2017)Thomas, Crauste, Haas, Jouneau, B\"auerle, L\'evy,
  Orignac, Carpentier, Ballet, and Meunier]{Kane26}
C.~Thomas, O.~Crauste, B.~Haas, P.-H. Jouneau, C.~B\"auerle, L.~P. L\'evy,
  E.~Orignac, D.~Carpentier, P.~Ballet, and T.~Meunier.
\newblock Revealing topological dirac fermions at the surface of strained hgte
  thin films via quantum hall transport spectroscopy.
\newblock \emph{Phys. Rev. B}, 96:\penalty0 245420, Dec 2017.
\newblock \doi{10.1103/PhysRevB.96.245420}.
\newblock URL \url{https://link.aps.org/doi/10.1103/PhysRevB.96.245420}.

\bibitem[Noel et~al.(2018)Noel, Thomas, Fu, Vila, Haas, Jouneau, Gambarelli,
  Meunier, Ballet, and Attan\'e]{Kane27}
P.~Noel, C.~Thomas, Y.~Fu, L.~Vila, B.~Haas, P-H. Jouneau, S.~Gambarelli,
  T.~Meunier, P.~Ballet, and J.~P. Attan\'e.
\newblock Highly efficient spin-to-charge current conversion in strained hgte
  surface states protected by a hgcdte layer.
\newblock \emph{Phys. Rev. Lett.}, 120:\penalty0 167201, Apr 2018.
\newblock \doi{10.1103/PhysRevLett.120.167201}.
\newblock URL \url{https://link.aps.org/doi/10.1103/PhysRevLett.120.167201}.

\bibitem[Aronov and Pikus(1967{\natexlab{a}})]{Flds3}
A.~G. Aronov and G.~E. Pikus.
\newblock Tunnel current in a transverse magnetic field.
\newblock \emph{Sov. Phys. JETP}, 24\penalty0 (1):\penalty0 188,
  1967{\natexlab{a}}.
\newblock ISSN 0044-4510.
\newblock URL \url{http://jetp.ras.ru/cgi-bin/e/index/e/24/1/p188?a=list}.

\bibitem[Aronov and Pikus(1967{\natexlab{b}})]{Flds4}
A.~G. Aronov and G.~E. Pikus.
\newblock Light absorption in semiconductors in crossed electric and magnetic
  fields.
\newblock \emph{Sov. Phys. JETP}, 24\penalty0 (2):\penalty0 339,
  1967{\natexlab{b}}.
\newblock ISSN 0044-4510.
\newblock URL \url{http://jetp.ras.ru/cgi-bin/e/index/e/24/2/p339?a=list}.

\bibitem[Zawadzki and Lax(1966)]{Flds5}
Wlodzimierz Zawadzki and Benjamin Lax.
\newblock Two-band model for bloch electrons in crossed electric and magnetic
  fields.
\newblock \emph{Phys. Rev. Lett.}, 16:\penalty0 1001--1003, May 1966.
\newblock \doi{10.1103/PhysRevLett.16.1001}.
\newblock URL \url{https://link.aps.org/doi/10.1103/PhysRevLett.16.1001}.

\bibitem[Weiler et~al.(1967)Weiler, Zawadzki, and Lax]{Flds6}
Margaret~H. Weiler, Wlodzimierz Zawadzki, and Benjamin Lax.
\newblock Theory of tunneling, including photon-assisted tunneling, in
  semiconductors in crossed and parallel electric and magnetic fields.
\newblock \emph{Phys. Rev.}, 163:\penalty0 733--742, Nov 1967.
\newblock \doi{10.1103/PhysRev.163.733}.
\newblock URL \url{https://link.aps.org/doi/10.1103/PhysRev.163.733}.

\bibitem[Lukose et~al.(2007)Lukose, Shankar, and Baskaran]{Flds7}
Vinu Lukose, R.~Shankar, and G.~Baskaran.
\newblock Novel electric field effects on landau levels in graphene.
\newblock \emph{Phys. Rev. Lett.}, 98:\penalty0 116802, Mar 2007.
\newblock \doi{10.1103/PhysRevLett.98.116802}.
\newblock URL \url{https://link.aps.org/doi/10.1103/PhysRevLett.98.116802}.

\bibitem[Zawadzki et~al.(1985)Zawadzki, Klahn, and Merkt]{Flds9}
W.~Zawadzki, S.~Klahn, and U.~Merkt.
\newblock Semirelativistic behavior of electrons in insb in crossed magnetic
  and electric fields.
\newblock \emph{Phys. Rev. Lett.}, 55:\penalty0 983--986, Aug 1985.
\newblock \doi{10.1103/PhysRevLett.55.983}.
\newblock URL \url{https://link.aps.org/doi/10.1103/PhysRevLett.55.983}.

\bibitem[Zawadzki et~al.(1986)Zawadzki, Klahn, and Merkt]{Flds10}
W.~Zawadzki, S.~Klahn, and U.~Merkt.
\newblock Inversion electrons on narrow-band-gap semiconductors in crossed
  electric and magnetic fields.
\newblock \emph{Phys. Rev. B}, 33:\penalty0 6916--6928, May 1986.
\newblock \doi{10.1103/PhysRevB.33.6916}.
\newblock URL \url{https://link.aps.org/doi/10.1103/PhysRevB.33.6916}.

\bibitem[Lifshitz and Kaganov(1960)]{Flds11}
I.~M. Lifshitz and M.~I. Kaganov.
\newblock Some problems of the electron theory of metals i. classical and
  quantum mechanics of electrons in metals.
\newblock \emph{Sov. Phys. Usp.}, 2:\penalty0 831--855, 1960.
\newblock \doi{10.1070/PU1960v002n06ABEH003183}.
\newblock URL \url{https://doi.org/10.1070/pu1960v002n06abeh003183}.

\bibitem[R\"{o}ssler(1984)]{Flds12}
U.~R\"{o}ssler.
\newblock Nonparabolicity and warping in the conduction band of gaas.
\newblock \emph{Solid State Commun.}, 49\penalty0 (10):\penalty0 943--947,
  1984.
\newblock ISSN 0038-1098.
\newblock \doi{https://doi.org/10.1016/0038-1098(84)90299-0}.
\newblock URL
  \url{https://www.sciencedirect.com/science/article/pii/0038109884902990}.

\bibitem[Pfeffer and Zawadzki(1990)]{Flds13}
P.~Pfeffer and W.~Zawadzki.
\newblock Conduction electrons in gaas: Five-level k\ensuremath{\cdot}p theory
  and polaron effects.
\newblock \emph{Phys. Rev. B}, 41:\penalty0 1561--1576, Jan 1990.
\newblock \doi{10.1103/PhysRevB.41.1561}.
\newblock URL \url{https://link.aps.org/doi/10.1103/PhysRevB.41.1561}.

\bibitem[Mayer and R\"{o}ssler(1991)]{Flds14}
H.~Mayer and U.~R\"{o}ssler.
\newblock Spin splitting and anisotropy of cyclotron resonance in the
  conduction band of gaas.
\newblock \emph{Phys. Rev. B}, 44:\penalty0 9048--9051, Oct 1991.
\newblock \doi{10.1103/PhysRevB.44.9048}.
\newblock URL \url{https://link.aps.org/doi/10.1103/PhysRevB.44.9048}.

\bibitem[Li et~al.(2009)Li, Chu, Jain, and Shen]{TAI1}
Jian Li, Rui-Lin Chu, J.~K. Jain, and Shun-Qing Shen.
\newblock Topological anderson insulator.
\newblock \emph{Phys. Rev. Lett.}, 102:\penalty0 136806, Apr 2009.
\newblock \doi{10.1103/PhysRevLett.102.136806}.
\newblock URL \url{https://link.aps.org/doi/10.1103/PhysRevLett.102.136806}.

\bibitem[Groth et~al.(2009)Groth, Wimmer, Akhmerov, Tworzyd\l{}o, and
  Beenakker]{TAI2}
C.~W. Groth, M.~Wimmer, A.~R. Akhmerov, J.~Tworzyd\l{}o, and C.~W.~J.
  Beenakker.
\newblock Theory of the topological anderson insulator.
\newblock \emph{Phys. Rev. Lett.}, 103:\penalty0 196805, Nov 2009.
\newblock \doi{10.1103/PhysRevLett.103.196805}.
\newblock URL \url{https://link.aps.org/doi/10.1103/PhysRevLett.103.196805}.

\bibitem[Bernevig et~al.(2006)Bernevig, Hughes, and Zhang]{HgTeQWs2}
B.~Andrei Bernevig, Taylor~L. Hughes, and Shou-Cheng Zhang.
\newblock Quantum spin hall effect and topological phase transition in hgte
  quantum wells.
\newblock \emph{Science}, 314\penalty0 (5806):\penalty0 1757--1761, 2006.
\newblock \doi{10.1126/science.1133734}.
\newblock URL \url{http://www.sciencemag.org/content/314/5806/1757.abstract}.

\bibitem[Shen et~al.(2011)Shen, Shan, and Lu]{HgTeQWs3}
Shun-Qing Shen, Wen-Yu Shan, and Hai-Zhou Lu.
\newblock Topological insulator and the dirac equation.
\newblock \emph{SPIN}, 01\penalty0 (01):\penalty0 33--44, 2011.
\newblock \doi{10.1142/S2010324711000057}.
\newblock URL \url{https://doi.org/10.1142/S2010324711000057}.

\bibitem[Guo et~al.(2010)Guo, Rosenberg, Refael, and Franz]{TAI3}
H.-M. Guo, G.~Rosenberg, G.~Refael, and M.~Franz.
\newblock Topological anderson insulator in three dimensions.
\newblock \emph{Phys. Rev. Lett.}, 105:\penalty0 216601, Nov 2010.
\newblock \doi{10.1103/PhysRevLett.105.216601}.
\newblock URL \url{https://link.aps.org/doi/10.1103/PhysRevLett.105.216601}.

\bibitem[Kobayashi et~al.(2013)Kobayashi, Ohtsuki, and Imura]{TAI4}
Koji Kobayashi, Tomi Ohtsuki, and Ken-Ichiro Imura.
\newblock Disordered weak and strong topological insulators.
\newblock \emph{Phys. Rev. Lett.}, 110:\penalty0 236803, Jun 2013.
\newblock \doi{10.1103/PhysRevLett.110.236803}.
\newblock URL \url{https://link.aps.org/doi/10.1103/PhysRevLett.110.236803}.

\bibitem[Ryu and Nomura(2012)]{TAI5}
Shinsei Ryu and Kentaro Nomura.
\newblock Disorder-induced quantum phase transitions in three-dimensional
  topological insulators and superconductors.
\newblock \emph{Phys. Rev. B}, 85:\penalty0 155138, Apr 2012.
\newblock \doi{10.1103/PhysRevB.85.155138}.
\newblock URL \url{https://link.aps.org/doi/10.1103/PhysRevB.85.155138}.

\bibitem[Chen et~al.(2015)Chen, Song, Jiang, Sun, Wang, and Xie]{TAI6}
Chui-Zhen Chen, Juntao Song, Hua Jiang, Qing-feng Sun, Ziqiang Wang, and X.~C.
  Xie.
\newblock Disorder and metal-insulator transitions in weyl semimetals.
\newblock \emph{Phys. Rev. Lett.}, 115:\penalty0 246603, Dec 2015.
\newblock \doi{10.1103/PhysRevLett.115.246603}.
\newblock URL \url{https://link.aps.org/doi/10.1103/PhysRevLett.115.246603}.

\bibitem[Shapourian and Hughes(2016)]{TAI7}
Hassan Shapourian and Taylor~L. Hughes.
\newblock Phase diagrams of disordered weyl semimetals.
\newblock \emph{Phys. Rev. B}, 93:\penalty0 075108, Feb 2016.
\newblock \doi{10.1103/PhysRevB.93.075108}.
\newblock URL \url{https://link.aps.org/doi/10.1103/PhysRevB.93.075108}.

\bibitem[Bera et~al.(2016)Bera, Sau, and Roy]{TAI8}
Soumya Bera, Jay~D. Sau, and Bitan Roy.
\newblock Dirty weyl semimetals: Stability, phase transition, and quantum
  criticality.
\newblock \emph{Phys. Rev. B}, 93:\penalty0 201302, May 2016.
\newblock \doi{10.1103/PhysRevB.93.201302}.
\newblock URL \url{https://link.aps.org/doi/10.1103/PhysRevB.93.201302}.

\bibitem[Park et~al.(2017)Park, Basa, and Gilbert]{TAI9}
Moon~Jip Park, Bora Basa, and Matthew~J. Gilbert.
\newblock Disorder-induced phase transitions of type-ii weyl semimetals.
\newblock \emph{Phys. Rev. B}, 95:\penalty0 094201, Mar 2017.
\newblock \doi{10.1103/PhysRevB.95.094201}.
\newblock URL \url{https://link.aps.org/doi/10.1103/PhysRevB.95.094201}.

\bibitem[Krishtopenko et~al.(2022)Krishtopenko, Antezza, and Teppe]{TAI10}
Sergey~S. Krishtopenko, Mauro Antezza, and Fr\'ed\'eric Teppe.
\newblock Disorder-induced topological phase transition in hgcdte crystals.
\newblock \emph{Phys. Rev. B}, 106:\penalty0 115203, Sep 2022.
\newblock \doi{10.1103/PhysRevB.106.115203}.
\newblock URL \url{https://link.aps.org/doi/10.1103/PhysRevB.106.115203}.

\bibitem[Girschik et~al.(2013)Girschik, Libisch, and Rotter]{TAI11}
Adrian Girschik, Florian Libisch, and Stefan Rotter.
\newblock Topological insulator in the presence of spatially correlated
  disorder.
\newblock \emph{Phys. Rev. B}, 88:\penalty0 014201, Jul 2013.
\newblock \doi{10.1103/PhysRevB.88.014201}.
\newblock URL \url{https://link.aps.org/doi/10.1103/PhysRevB.88.014201}.

\bibitem[Zimmermann and Schindler(2009)]{TAI12}
Roland Zimmermann and Christoph Schindler.
\newblock Coherent potential approximation for spatially correlated disorder.
\newblock \emph{Phys. Rev. B}, 80:\penalty0 144202, Oct 2009.
\newblock \doi{10.1103/PhysRevB.80.144202}.
\newblock URL \url{https://link.aps.org/doi/10.1103/PhysRevB.80.144202}.

\bibitem[Krishtopenko et~al.(2025{\natexlab{a}})Krishtopenko, Ikonnikov,
  Hartmann, H\"ofling, Jouault, and Teppe]{TAI13}
S.~S. Krishtopenko, A.~V. Ikonnikov, F.~Hartmann, S.~H\"ofling, B.~Jouault, and
  F.~Teppe.
\newblock Quasiparticle picture of topological phase transitions induced by
  interactions.
\newblock \emph{Phys. Rev. Res.}, 7:\penalty0 033116, Aug 2025{\natexlab{a}}.
\newblock \doi{10.1103/PhysRevResearch.7.033116}.
\newblock URL \url{https://link.aps.org/doi/10.1103/PhysRevResearch.7.033116}.

\bibitem[Kane and Mele(2005)]{HgTeQWs4}
C.~L. Kane and E.~J. Mele.
\newblock ${Z}_{2}$ topological order and the quantum spin hall effect.
\newblock \emph{Phys. Rev. Lett.}, 95:\penalty0 146802, Sep 2005.
\newblock \doi{10.1103/PhysRevLett.95.146802}.
\newblock URL \url{https://link.aps.org/doi/10.1103/PhysRevLett.95.146802}.

\bibitem[K\"{o}nig et~al.(2007)K\"{o}nig, Wiedmann, Br\"{u}ne, Roth, Buhmann,
  Molenkamp, Qi, and Zhang]{HgTeQWs5}
Markus K\"{o}nig, Steffen Wiedmann, Christoph Br\"{u}ne, Andreas Roth, Hartmut
  Buhmann, Laurens~W. Molenkamp, Xiao-Liang Qi, and Shou-Cheng Zhang.
\newblock Quantum spin hall insulator state in hgte quantum wells.
\newblock \emph{Science}, 318\penalty0 (5851):\penalty0 766--770, 2007.
\newblock \doi{10.1126/science.1148047}.
\newblock URL \url{http://www.sciencemag.org/content/318/5851/766.abstract}.

\bibitem[Roth et~al.(2009)Roth, Br{\"u}ne, Buhmann, Molenkamp, Maciejko, Qi,
  and Zhang]{HgTeQWs8}
Andreas Roth, Christoph Br{\"u}ne, Hartmut Buhmann, Laurens~W. Molenkamp,
  Joseph Maciejko, Xiao-Liang Qi, and Shou-Cheng Zhang.
\newblock Nonlocal transport in the quantum spin hall state.
\newblock \emph{Science}, 325\penalty0 (5938):\penalty0 294, 2009.
\newblock \doi{10.1126/science.1174736}.
\newblock URL \url{http://science.sciencemag.org/content/325/5938/294}.

\bibitem[Br\"{u}ne et~al.(2012)Br\"{u}ne, Roth, Buhmann, Hankiewicz, Molenkamp,
  Maciejko, X.-L., and Zhang]{HgTeQWs9}
C.~Br\"{u}ne, A~Roth, H.~Buhmann, E.M. Hankiewicz, L.W. Molenkamp, J.~Maciejko,
  Qi~X.-L., and S.-C. Zhang.
\newblock Spin polarization of the quantum spin hall edge states.
\newblock \emph{Nat. Phys.}, 8\penalty0 (6):\penalty0 485, 2012.
\newblock \doi{10.1038/nphys2322}.
\newblock URL \url{http://dx.doi.org/10.1038/nphys2322}.

\bibitem[Gerchikov and Subashiev(1990)]{HgTeQWs6}
L.~G. Gerchikov and A.~V. Subashiev.
\newblock Interface states in subband structure of semiconductor quantum wells.
\newblock \emph{Phys. Status Solidi B}, 160\penalty0 (2):\penalty0 443--457,
  1990.
\newblock ISSN 1521-3951.
\newblock \doi{10.1002/pssb.2221600207}.
\newblock URL \url{http://dx.doi.org/10.1002/pssb.2221600207}.

\bibitem[Kvon et~al.(2011)Kvon, Olshanetsky, Novik, Kozlov, Mikhailov, Parm,
  and Dvoretsky]{HgTeQWs7}
Z.~D. Kvon, E.~B. Olshanetsky, E.~G. Novik, D.~A. Kozlov, N.~N. Mikhailov,
  I.~O. Parm, and S.~A. Dvoretsky.
\newblock Two-dimensional electron-hole system in hgte-based quantum wells with
  surface orientation (112).
\newblock \emph{Phys. Rev. B}, 83:\penalty0 193304, May 2011.
\newblock \doi{10.1103/PhysRevB.83.193304}.
\newblock URL \url{https://link.aps.org/doi/10.1103/PhysRevB.83.193304}.

\bibitem[Krishtopenko et~al.(2016{\natexlab{a}})Krishtopenko, Yahniuk, But,
  Gavrilenko, Knap, and Teppe]{HgTeQWs10}
S.~S. Krishtopenko, I.~Yahniuk, D.~B. But, V.~I. Gavrilenko, W.~Knap, and
  F.~Teppe.
\newblock Pressure- and temperature-driven phase transitions in hgte quantum
  wells.
\newblock \emph{Phys. Rev. B}, 94:\penalty0 245402, Dec 2016{\natexlab{a}}.
\newblock \doi{10.1103/PhysRevB.94.245402}.
\newblock URL \url{https://link.aps.org/doi/10.1103/PhysRevB.94.245402}.

\bibitem[Leubner et~al.(2016)Leubner, Lunczer, Br\"une, Buhmann, and
  Molenkamp]{HgTeQWs12}
Philipp Leubner, Lukas Lunczer, Christoph Br\"une, Hartmut Buhmann, and
  Laurens~W. Molenkamp.
\newblock Strain engineering of the band gap of hgte quantum wells using
  superlattice virtual substrates.
\newblock \emph{Phys. Rev. Lett.}, 117:\penalty0 086403, Aug 2016.
\newblock \doi{10.1103/PhysRevLett.117.086403}.
\newblock URL \url{https://link.aps.org/doi/10.1103/PhysRevLett.117.086403}.

\bibitem[Olshanetsky et~al.(2015)Olshanetsky, Kvon, Gusev, Levin, Raichev,
  Mikhailov, and Dvoretsky]{HgTeQWs11}
E.~B. Olshanetsky, Z.~D. Kvon, G.~M. Gusev, A.~D. Levin, O.~E. Raichev, N.~N.
  Mikhailov, and S.~A. Dvoretsky.
\newblock Persistence of a two-dimensional topological insulator state in wide
  hgte quantum wells.
\newblock \emph{Phys. Rev. Lett.}, 114:\penalty0 126802, Mar 2015.
\newblock \doi{10.1103/PhysRevLett.114.126802}.
\newblock URL \url{https://link.aps.org/doi/10.1103/PhysRevLett.114.126802}.

\bibitem[Dvoretsky et~al.(2010)Dvoretsky, Mikhailov, Sidorov, Shvets, Danilov,
  Wittman, and Ganichev]{HgTeQWs13}
S.~Dvoretsky, N.~Mikhailov, Yu. Sidorov, V.~Shvets, S.~Danilov, B.~Wittman, and
  S.~Ganichev.
\newblock Growth of hgte quantum wells for ir to thz detectors.
\newblock \emph{J. Electron. Mater.}, 39\penalty0 (7):\penalty0 918--923, 2010.
\newblock ISSN 1543-186X.
\newblock \doi{10.1007/s11664-010-1191-7}.
\newblock URL \url{http://dx.doi.org/10.1007/s11664-010-1191-7}.

\bibitem[Tkachov and Hankiewicz(2010)]{HgTeQWs14}
G.~Tkachov and E.~M. Hankiewicz.
\newblock Ballistic quantum spin hall state and enhanced edge backscattering in
  strong magnetic fields.
\newblock \emph{Phys. Rev. Lett.}, 104:\penalty0 166803, Apr 2010.
\newblock \doi{10.1103/PhysRevLett.104.166803}.
\newblock URL \url{https://link.aps.org/doi/10.1103/PhysRevLett.104.166803}.

\bibitem[Chen et~al.(2012{\natexlab{a}})Chen, Wang, and Sun]{HgTeQWs15}
Jiang-chai Chen, Jian Wang, and Qing-feng Sun.
\newblock Effect of magnetic field on electron transport in hgte/cdte quantum
  wells: Numerical analysis.
\newblock \emph{Phys. Rev. B}, 85:\penalty0 125401, Mar 2012{\natexlab{a}}.
\newblock \doi{10.1103/PhysRevB.85.125401}.
\newblock URL \url{https://link.aps.org/doi/10.1103/PhysRevB.85.125401}.

\bibitem[Ikonnikov et~al.(2016)Ikonnikov, Krishtopenko, Drachenko, Goiran,
  Zholudev, Platonov, Kudasov, Korshunov, Maslov, Makarov, Surdin, Philippov,
  Marcinkiewicz, Ruffenach, Teppe, Knap, Mikhailov, Dvoretsky, and
  Gavrilenko]{HgTeQWs16}
A.~V. Ikonnikov, S.~S. Krishtopenko, O.~Drachenko, M.~Goiran, M.~S. Zholudev,
  V.~V. Platonov, Yu.~B. Kudasov, A.~S. Korshunov, D.~A. Maslov, I.~V. Makarov,
  O.~M. Surdin, A.~V. Philippov, M.~Marcinkiewicz, S.~Ruffenach, F.~Teppe,
  W.~Knap, N.~N. Mikhailov, S.~A. Dvoretsky, and V.~I. Gavrilenko.
\newblock Temperature-dependent magnetospectroscopy of hgte quantum wells.
\newblock \emph{Phys. Rev. B}, 94:\penalty0 155421, Oct 2016.
\newblock \doi{10.1103/PhysRevB.94.155421}.
\newblock URL \url{https://link.aps.org/doi/10.1103/PhysRevB.94.155421}.

\bibitem[Marcinkiewicz et~al.(2017)Marcinkiewicz, Ruffenach, Krishtopenko,
  Kadykov, Consejo, But, Desrat, Knap, Torres, Ikonnikov, Spirin, Morozov,
  Gavrilenko, Mikhailov, Dvoretskii, and Teppe]{HgTeQWs17}
M.~Marcinkiewicz, S.~Ruffenach, S.~S. Krishtopenko, A.~M. Kadykov, C.~Consejo,
  D.~B. But, W.~Desrat, W.~Knap, J.~Torres, A.~V. Ikonnikov, K.~E. Spirin,
  S.~V. Morozov, V.~I. Gavrilenko, N.~N. Mikhailov, S.~A. Dvoretskii, and
  F.~Teppe.
\newblock Temperature-driven single-valley dirac fermions in hgte quantum
  wells.
\newblock \emph{Phys. Rev. B}, 96:\penalty0 035405, Jul 2017.
\newblock \doi{10.1103/PhysRevB.96.035405}.
\newblock URL \url{https://link.aps.org/doi/10.1103/PhysRevB.96.035405}.

\bibitem[Schultz et~al.(1998)Schultz, Merkt, Sonntag, R\"ossler, Winkler,
  Colin, Helgesen, Skauli, and L\o{}vold]{HgTeQWs18}
M.~Schultz, U.~Merkt, A.~Sonntag, U.~R\"ossler, R.~Winkler, T.~Colin,
  P.~Helgesen, T.~Skauli, and S.~L\o{}vold.
\newblock Crossing of conduction- and valence-subband landau levels in an
  inverted hgte/cdte quantum well.
\newblock \emph{Phys. Rev. B}, 57:\penalty0 14772--14775, Jun 1998.
\newblock \doi{10.1103/PhysRevB.57.14772}.
\newblock URL \url{https://link.aps.org/doi/10.1103/PhysRevB.57.14772}.

\bibitem[Orlita et~al.(2011)Orlita, Masztalerz, Faugeras, Potemski, Novik,
  Br\"une, Buhmann, and Molenkamp]{HgTeQWs19}
M.~Orlita, K.~Masztalerz, C.~Faugeras, M.~Potemski, E.~G. Novik, C.~Br\"une,
  H.~Buhmann, and L.~W. Molenkamp.
\newblock Fine structure of zero-mode landau levels in
  hgte/hg${}_{x}$cd${}_{1\ensuremath{-}x}$te quantum wells.
\newblock \emph{Phys. Rev. B}, 83:\penalty0 115307, Mar 2011.
\newblock \doi{10.1103/PhysRevB.83.115307}.
\newblock URL \url{https://link.aps.org/doi/10.1103/PhysRevB.83.115307}.

\bibitem[Zholudev et~al.(2012)Zholudev, Teppe, Orlita, Consejo, Torres,
  Dyakonova, Czapkiewicz, Wr\'obel, Grabecki, Mikhailov, Dvoretskii, Ikonnikov,
  Spirin, Aleshkin, Gavrilenko, and Knap]{HgTeQWs20}
M.~Zholudev, F.~Teppe, M.~Orlita, C.~Consejo, J.~Torres, N.~Dyakonova,
  M.~Czapkiewicz, J.~Wr\'obel, G.~Grabecki, N.~Mikhailov, S.~Dvoretskii,
  A.~Ikonnikov, K.~Spirin, V.~Aleshkin, V.~Gavrilenko, and W.~Knap.
\newblock Magnetospectroscopy of two-dimensional hgte-based topological
  insulators around the critical thickness.
\newblock \emph{Phys. Rev. B}, 86:\penalty0 205420, Nov 2012.
\newblock \doi{10.1103/PhysRevB.86.205420}.
\newblock URL \url{https://link.aps.org/doi/10.1103/PhysRevB.86.205420}.

\bibitem[Zholudev et~al.(2015)Zholudev, Teppe, Morozov, Orlita, Consejo,
  Ruffenach, Knap, Gavrilenko, Dvoretskii, and Mikhailov]{HgTeQWs21}
M.~S. Zholudev, F.~Teppe, S.~V. Morozov, M.~Orlita, C.~Consejo, S.~Ruffenach,
  W.~Knap, V.~I. Gavrilenko, S.~A. Dvoretskii, and N.~N. Mikhailov.
\newblock Anticrossing of landau levels in hgte/cdhgte (013) quantum wells with
  an inverted band structure.
\newblock \emph{JETP Lett.}, 100\penalty0 (12):\penalty0 790--794, 2015.
\newblock \doi{10.1134/S0021364014240175}.
\newblock URL \url{http://dx.doi.org/10.1134/S0021364014240175}.

\bibitem[Krishtopenko et~al.(2020{\natexlab{b}})Krishtopenko, Kadykov, Gebert,
  Ruffenach, Consejo, Torres, Avogadri, Jouault, Knap, Mikhailov, Dvoretskii,
  and Teppe]{HgTeQWs22}
S.~S. Krishtopenko, A.~M. Kadykov, S.~Gebert, S.~Ruffenach, C.~Consejo,
  J.~Torres, C.~Avogadri, B.~Jouault, W.~Knap, N.~N. Mikhailov, S.~A.
  Dvoretskii, and F.~Teppe.
\newblock Many-particle effects in optical transitions from zero-mode landau
  levels in hgte quantum wells.
\newblock \emph{Phys. Rev. B}, 102:\penalty0 041404, Jul 2020{\natexlab{b}}.
\newblock \doi{10.1103/PhysRevB.102.041404}.
\newblock URL \url{https://link.aps.org/doi/10.1103/PhysRevB.102.041404}.

\bibitem[Ruffenach et~al.(2025)Ruffenach, Krishtopenko, Ikonnikov, Consejo,
  Torres, Baudry, Ballet, Jouault, and Teppe]{HgTeQWs23}
S.~Ruffenach, S.~S. Krishtopenko, A.~V. Ikonnikov, C.~Consejo, J.~Torres,
  X.~Baudry, P.~Ballet, B.~Jouault, and F.~Teppe.
\newblock Many-particle hybridization of optical transitions from zero-mode
  landau levels in hgte quantum wells, 2025.
\newblock URL \url{https://arxiv.org/abs/2507.00565}.

\bibitem[Ikonnikov et~al.(2017)Ikonnikov, Bovkun, Rumyantsev, Krishtopenko,
  Aleshkin, Kadykov, Orlita, Potemski, Gavrilenko, Morozov, Dvoretsky, and
  Mikhailov]{HgTeQWs24}
A.~V. Ikonnikov, L.~S. Bovkun, V.~V. Rumyantsev, S.~S. Krishtopenko, V.~Ya.
  Aleshkin, A.~M. Kadykov, M.~Orlita, M.~Potemski, V.~I. Gavrilenko, S.~V.
  Morozov, S.~A. Dvoretsky, and N.~N. Mikhailov.
\newblock On the band spectrum in p-type hgte/cdhgte heterostructures and its
  transformation under temperature variation.
\newblock \emph{Semiconductors}, 51:\penalty0 1531--1536, 2017.
\newblock \doi{10.1134/S1063782617120090}.
\newblock URL \url{http://doi.org/10.1134/S1063782617120090}.

\bibitem[Kadykov et~al.(2018)Kadykov, Krishtopenko, Jouault, Desrat, Knap,
  Ruffenach, Consejo, Torres, Morozov, Mikhailov, Dvoretskii, and
  Teppe]{HgTeQWs25}
A.~M. Kadykov, S.~S. Krishtopenko, B.~Jouault, W.~Desrat, W.~Knap,
  S.~Ruffenach, C.~Consejo, J.~Torres, S.~V. Morozov, N.~N. Mikhailov, S.~A.
  Dvoretskii, and F.~Teppe.
\newblock Temperature-induced topological phase transition in hgte quantum
  wells.
\newblock \emph{Phys. Rev. Lett.}, 120:\penalty0 086401, Feb 2018.
\newblock \doi{10.1103/PhysRevLett.120.086401}.
\newblock URL \url{https://link.aps.org/doi/10.1103/PhysRevLett.120.086401}.

\bibitem[Shahar et~al.(1997)Shahar, Tsui, Shayegan, Shimshoni, and
  Sondhi]{HgTeQWs26}
D.~Shahar, D.~C. Tsui, M.~Shayegan, E.~Shimshoni, and S.~L. Sondhi.
\newblock A different view of the quantum hall plateau-to-plateau transitions.
\newblock \emph{Phys. Rev. Lett.}, 79:\penalty0 479--482, Jul 1997.
\newblock \doi{10.1103/PhysRevLett.79.479}.
\newblock URL \url{https://link.aps.org/doi/10.1103/PhysRevLett.79.479}.

\bibitem[Michetti et~al.(2012)Michetti, Budich, Novik, and
  Recher]{doubleHgTeQWs1}
Paolo Michetti, Jan~C. Budich, Elena~G. Novik, and Patrik Recher.
\newblock Tunable quantum spin hall effect in double quantum wells.
\newblock \emph{Phys. Rev. B}, 85:\penalty0 125309, Mar 2012.
\newblock \doi{10.1103/PhysRevB.85.125309}.
\newblock URL \url{https://link.aps.org/doi/10.1103/PhysRevB.85.125309}.

\bibitem[Michetti and Trauzettel(2013)]{doubleHgTeQWs2}
Paolo Michetti and Björn Trauzettel.
\newblock Devices with electrically tunable topological insulating phases.
\newblock \emph{Appl. Phys. Lett.}, 102\penalty0 (6):\penalty0 063503, 02 2013.
\newblock ISSN 0003-6951.
\newblock \doi{10.1063/1.4792275}.
\newblock URL \url{https://doi.org/10.1063/1.4792275}.

\bibitem[Tarasenko et~al.(2004)Tarasenko, Perel', and
  Yassievich]{doubleHgTeQWs3}
S.~A. Tarasenko, V.~I. Perel', and I.~N. Yassievich.
\newblock In-plane electric current is induced by tunneling of spin-polarized
  carriers.
\newblock \emph{Phys. Rev. Lett.}, 93:\penalty0 056601, Jul 2004.
\newblock \doi{10.1103/PhysRevLett.93.056601}.
\newblock URL \url{https://link.aps.org/doi/10.1103/PhysRevLett.93.056601}.

\bibitem[Perel' et~al.(2003)Perel', Tarasenko, Yassievich, Ganichev, Bel'kov,
  and Prettl]{doubleHgTeQWs4}
V.~I. Perel', S.~A. Tarasenko, I.~N. Yassievich, S.~D. Ganichev, V.~V. Bel'kov,
  and W.~Prettl.
\newblock Spin-dependent tunneling through a symmetric semiconductor barrier.
\newblock \emph{Phys. Rev. B}, 67:\penalty0 201304, May 2003.
\newblock \doi{10.1103/PhysRevB.67.201304}.
\newblock URL \url{https://link.aps.org/doi/10.1103/PhysRevB.67.201304}.

\bibitem[Wang et~al.(2005)Wang, Yang, Chang, and Chan]{doubleHgTeQWs5}
L.~G. Wang, Wen Yang, Kai Chang, and K.~S. Chan.
\newblock Spin-dependent tunneling through a symmetric semiconductor barrier:
  The dresselhaus effect.
\newblock \emph{Phys. Rev. B}, 72:\penalty0 153314, Oct 2005.
\newblock \doi{10.1103/PhysRevB.72.153314}.
\newblock URL \url{https://link.aps.org/doi/10.1103/PhysRevB.72.153314}.

\bibitem[Krishtopenko et~al.(2016{\natexlab{b}})Krishtopenko, Knap, and
  Teppe]{doubleHgTeQWs6}
S.~S. Krishtopenko, W.~Knap, and F.~Teppe.
\newblock Phase transitions in two tunnel-coupled hgte quantum wells: Bilayer
  graphene analogy and beyond.
\newblock \emph{Sci. Rep.}, 6:\penalty0 30755, 2016{\natexlab{b}}.
\newblock \doi{10.1038/srep30755}.
\newblock URL \url{http://dx.doi.org/10.1038/srep30755}.

\bibitem[Krishtopenko(2021)]{doubleHgTeQWs7}
S.~S. Krishtopenko.
\newblock Higher-order topological insulator in cubic semiconductor quantum
  wells.
\newblock \emph{Sci. Rep.}, 11:\penalty0 21060, 2021.
\newblock \doi{10.1038/s41598-021-00577-z}.
\newblock URL \url{http://doi.org/10.1038/s41598-021-00577-z}.

\bibitem[Ohta et~al.(2006)Ohta, Bostwick, Seyller, Horn, and
  Rotenberg]{NaturalBG1}
Taisuke Ohta, Aaron Bostwick, Thomas Seyller, Karsten Horn, and Eli Rotenberg.
\newblock Controlling the electronic structure of bilayer graphene.
\newblock \emph{Science}, 313\penalty0 (5789):\penalty0 951--954, 2006.
\newblock \doi{10.1126/science.1130681}.
\newblock URL \url{https://www.science.org/doi/abs/10.1126/science.1130681}.

\bibitem[McCann(2006)]{NaturalBG2}
Edward McCann.
\newblock Asymmetry gap in the electronic band structure of bilayer graphene.
\newblock \emph{Phys. Rev. B}, 74:\penalty0 161403, Oct 2006.
\newblock \doi{10.1103/PhysRevB.74.161403}.
\newblock URL \url{https://link.aps.org/doi/10.1103/PhysRevB.74.161403}.

\bibitem[Novoselov et~al.(2006)Novoselov, McCann, Morozov, Fal'ko, Katsnelson,
  Zeitler, Jiang, Schedin, and Geim]{NaturalBG3}
K.~S. Novoselov, E.~McCann, S.~V. Morozov, V.~I. Fal'ko, M.~I. Katsnelson,
  U.~Zeitler, D.~Jiang, F.~Schedin, and A.~K. Geim.
\newblock Unconventional quantum hall effect and berry's phase of $2\pi$ in
  bilayer graphene.
\newblock \emph{Nat. Phys.}, 2\penalty0 (3):\penalty0 177, 2006.
\newblock \doi{10.1038/nphys245}.
\newblock URL \url{http://dx.doi.org/10.1038/nphys245}.

\bibitem[Castro et~al.(2007)Castro, Novoselov, Morozov, Peres, dos Santos,
  Nilsson, Guinea, Geim, and Neto]{NaturalBG4}
Eduardo~V. Castro, K.~S. Novoselov, S.~V. Morozov, N.~M.~R. Peres, J.~M.
  B.~Lopes dos Santos, Johan Nilsson, F.~Guinea, A.~K. Geim, and A.~H.~Castro
  Neto.
\newblock Biased bilayer graphene: Semiconductor with a gap tunable by the
  electric field effect.
\newblock \emph{Phys. Rev. Lett.}, 99:\penalty0 216802, Nov 2007.
\newblock \doi{10.1103/PhysRevLett.99.216802}.
\newblock URL \url{https://link.aps.org/doi/10.1103/PhysRevLett.99.216802}.

\bibitem[McCann and Fal'ko(2006)]{NaturalBG5}
Edward McCann and Vladimir~I. Fal'ko.
\newblock Landau-level degeneracy and quantum hall effect in a graphite
  bilayer.
\newblock \emph{Phys. Rev. Lett.}, 96:\penalty0 086805, Mar 2006.
\newblock \doi{10.1103/PhysRevLett.96.086805}.
\newblock URL \url{https://link.aps.org/doi/10.1103/PhysRevLett.96.086805}.

\bibitem[Yakunin et~al.(2016)Yakunin, Krishtopenko, Podgornykh, Popov, Neverov,
  Mikhailov, and Dvoretsky]{Tr1}
M.~V. Yakunin, S.~S. Krishtopenko, S.~M. Podgornykh, M.~R. Popov, V.~N.
  Neverov, N.~N. Mikhailov, and S.~A. Dvoretsky.
\newblock Hgte/cdhgte double quantum well with a spectrum of bilayer graphene
  and peculiarities of its magnetotransport.
\newblock \emph{JETP Lett.}, 104\penalty0 (6):\penalty0 403--410, 2016.
\newblock \doi{10.1134/S0021364016180132}.
\newblock URL \url{http://dx.doi.org/10.1134/S0021364016180132}.

\bibitem[Podgornykh et~al.(2019)Podgornykh, Yakunin, Krishtopenko, Popov,
  Mikhailov, and Dvoretskii]{Tr2}
S.~M. Podgornykh, M.~V. Yakunin, S.~S. Krishtopenko, M.~R. Popov, N.~N.
  Mikhailov, and S.~A. Dvoretskii.
\newblock On the thermal activation of conductivity electrons in a $p$-type
  hgte/cdhgte double quantum well with hgte layers of critical width.
\newblock \emph{Semiconductors}, 53\penalty0 (7):\penalty0 919--922, 2019.
\newblock \doi{10.1134/S1063782619070194}.
\newblock URL \url{https://doi.org/10.1134/S1063782619070194}.

\bibitem[Yakunin et~al.(2020)Yakunin, Krishtopenko, Desrat, Podgornykh, Popov,
  Neverov, Dvoretsky, Mikhailov, Teppe, and Jouault]{Tr3}
M.~V. Yakunin, S.~S. Krishtopenko, W.~Desrat, S.~M. Podgornykh, M.~R. Popov,
  V.~N. Neverov, S.~A. Dvoretsky, N.~N. Mikhailov, F.~Teppe, and B.~Jouault.
\newblock Unconventional reentrant quantum hall effect in a hgte/cdhgte double
  quantum well.
\newblock \emph{Phys. Rev. B}, 102:\penalty0 165305, Oct 2020.
\newblock \doi{10.1103/PhysRevB.102.165305}.
\newblock URL \url{https://link.aps.org/doi/10.1103/PhysRevB.102.165305}.

\bibitem[Gusev et~al.(2020)Gusev, Olshanetsky, Hernandez, Raichev, Mikhailov,
  and Dvoretsky]{Tr4}
G.~M. Gusev, E.~B. Olshanetsky, F.~G.~G. Hernandez, O.~E. Raichev, N.~N.
  Mikhailov, and S.~A. Dvoretsky.
\newblock Two-dimensional topological insulator state in double hgte quantum
  well.
\newblock \emph{Phys. Rev. B}, 101:\penalty0 241302, Jun 2020.
\newblock \doi{10.1103/PhysRevB.101.241302}.
\newblock URL \url{https://link.aps.org/doi/10.1103/PhysRevB.101.241302}.

\bibitem[Gusev et~al.(2021)Gusev, Olshanetsky, Hernandez, Raichev, Mikhailov,
  and Dvoretsky]{Tr5}
G.~M. Gusev, E.~B. Olshanetsky, F.~G.~G. Hernandez, O.~E. Raichev, N.~N.
  Mikhailov, and S.~A. Dvoretsky.
\newblock Multiple crossings of landau levels of two-dimensional fermions in
  double hgte quantum wells.
\newblock \emph{Phys. Rev. B}, 103:\penalty0 035302, Jan 2021.
\newblock \doi{10.1103/PhysRevB.103.035302}.
\newblock URL \url{https://link.aps.org/doi/10.1103/PhysRevB.103.035302}.

\bibitem[Nikolaev et~al.(2022)Nikolaev, Kazakov, Drozdov, Bannikov, Spirin,
  Menshchikov, Dvoretsky, Mikhailov, Khokhlov, and Ikonnikov]{Tr6}
I.~Nikolaev, A.~Kazakov, K.~Drozdov, M.~Bannikov, K.~Spirin, R.~Menshchikov,
  S.~Dvoretsky, N.~Mikhailov, D.~Khokhlov, and A.~Ikonnikov.
\newblock Bipolar persistent photoconductivity in hgte/cdhgte double quantum
  well heterostructures and its application for reversible change in the
  conductivity type.
\newblock \emph{J. Appl. Phys.}, 132:\penalty0 234301, Dec 2022.
\newblock \doi{10.1063/5.0121364}.
\newblock URL \url{http://dx.doi.org/10.1063/5.0121364}.

\bibitem[Sotnichuk et~al.(2023)Sotnichuk, Kazakov, Nikolaev, Drozdov,
  Menshchikov, Dvoretsky, Mikhailov, Khokhlov, and Ikonnikov]{Tr7}
Mikhail~K. Sotnichuk, Aleksei~S. Kazakov, Ilya~D. Nikolaev, Konstantin~A.
  Drozdov, Roman~V. Menshchikov, Sergey~A. Dvoretsky, Nikolay~N. Mikhailov,
  Dmitry~R. Khokhlov, and Anton~V. Ikonnikov.
\newblock Cap layer effect on key features of persistent photoconductivity
  spectra in hgte/cdhgte double quantum well heterostructures.
\newblock \emph{Photonics}, 10\penalty0 (8):\penalty0 877, 2023.
\newblock ISSN 2304-6732.
\newblock \doi{10.3390/photonics10080877}.
\newblock URL \url{https://www.mdpi.com/2304-6732/10/8/877}.

\bibitem[Bovkun et~al.(2016)Bovkun, Krishtopenko, Ikonnikov, Aleshkin, Kadykov,
  Ruffenach, Consejo, Teppe, Knap, Orlita, Piot, Potemski, Mikhailov,
  Dvoretskii, and Gavrilenko]{LL1}
L.~S. Bovkun, S.~S. Krishtopenko, A.~V. Ikonnikov, V.~Ya. Aleshkin, A.~M.
  Kadykov, S.~Ruffenach, C.~Consejo, F.~Teppe, W.~Knap, M.~Orlita, B.~Piot,
  M.~Potemski, N.~N. Mikhailov, S.~A. Dvoretskii, and V.~I. Gavrilenko.
\newblock Magnetospectroscopy of double hgte/cdhgte quantum wells.
\newblock \emph{Semiconductors}, 50:\penalty0 1532--1538, Nov 2016.
\newblock \doi{10.1134/S1063782616110063}.
\newblock URL \url{https://doi.org/10.1134/S1063782616110063}.

\bibitem[Bovkun et~al.(2018)Bovkun, Ikonnikov, Aleshkin, Krishtopenko,
  Mikhailov, Dvoretskii, Potemski, Piot, Orlita, and Gavrilenko]{LL1b}
L.~S. Bovkun, A.~V. Ikonnikov, V.~Ya. Aleshkin, S.~S. Krishtopenko, N.~N.
  Mikhailov, S.~A. Dvoretskii, M.~Potemski, B.~Piot, M.~Orlita, and V.~I.
  Gavrilenko.
\newblock Polarization-sensitive fourier-transform spectroscopy of hgte/cdhgte
  quantum wells in the far infrared range in a magnetic field.
\newblock \emph{JETP Lett.}, 108\penalty0 (5):\penalty0 329--334, 2018.
\newblock \doi{10.1134/S0021364018170058}.
\newblock URL \url{http://dx.doi.org/10.1134/S0021364018170058}.

\bibitem[Bovkun et~al.(2019)Bovkun, Ikonnikov, Aleshkin, Maremyanin, Mikhailov,
  Dvoretskii, Krishtopenko, Teppe, Piot, Potemski, Orlita, and Gavrilenko]{LL2}
L.~S. Bovkun, A.~V. Ikonnikov, V.~Ya. Aleshkin, K.~V. Maremyanin, N.~N.
  Mikhailov, S.~A. Dvoretskii, S.~S. Krishtopenko, F.~Teppe, B.~A. Piot,
  M.~Potemski, M.~Orlita, and V.~I. Gavrilenko.
\newblock Magnetospectroscopy of double hgte/cdhgte qws with inverted band
  structure in high magnetic fields up to 30 t.
\newblock \emph{Opto-electron. Rev.}, 27:\penalty0 213--218, Jun 2019.
\newblock \doi{10.1016/j.opelre.2019.06.001}.
\newblock URL \url{https://doi.org/10.1016/j.opelre.2019.06.001}.

\bibitem[Ikonnikov et~al.(2022)Ikonnikov, Krishtopenko, Bovkun, Mikhailov,
  Dvoretskii, Piot, Potemski, Orlita, Teppe, and Gavrilenko]{LL3}
A.~V. Ikonnikov, S.~S. Krishtopenko, L.~S. Bovkun, N.~N. Mikhailov, S.~A.
  Dvoretskii, B.~A. Piot, M.~Potemski, M.~Orlita, F.~Teppe, and V.~I.
  Gavrilenko.
\newblock Origin of structure inversion asymmetry in double hgte quantum wells.
\newblock \emph{JETP Lett.}, 116\penalty0 (8):\penalty0 547--555, 2022.
\newblock \doi{10.1134/S0021364022601889}.
\newblock URL \url{http://dx.doi.org/10.1134/S0021364022601889}.

\bibitem[Bovkun et~al.(2023)Bovkun, Krishtopenko, Aleshkin, Mikhailov,
  Dvoretsky, Teppe, Orlita, Gavrilenko, and Ikonnikov]{LL4}
L.~S. Bovkun, S.~S. Krishtopenko, V.~Ya. Aleshkin, N.~N. Mikhailov, S.~A.
  Dvoretsky, F.~Teppe, M.~Orlita, V.~I. Gavrilenko, and A.~V. Ikonnikov.
\newblock Simultaneous observation of the cyclotron resonances of electrons and
  holes in a hgte/cdhgte double quantum well under ``optical gate'' effect.
\newblock \emph{JETP Lett.}, 118\penalty0 (11):\penalty0 867--874, 2023.
\newblock \doi{10.1134/S0021364023603536}.
\newblock URL \url{http://dx.doi.org/10.1134/S0021364023603536}.

\bibitem[Chen et~al.(2012{\natexlab{b}})Chen, Liu, Lin, Zhang, and
  Jiang]{TAI14}
Liang Chen, Qin Liu, Xulin Lin, Xiaogang Zhang, and Xunya Jiang.
\newblock Disorder dependence of helical edge states in {HgTe}/{CdTe} quantum
  wells.
\newblock \emph{New J. of Phys.}, 14\penalty0 (4):\penalty0 043028, apr
  2012{\natexlab{b}}.
\newblock \doi{10.1088/1367-2630/14/4/043028}.
\newblock URL \url{https://doi.org/10.1088/1367-2630/14/4/043028}.

\bibitem[Niu et~al.(1985)Niu, Thouless, and Wu]{TAI15}
Qian Niu, D.~J. Thouless, and Yong-Shi Wu.
\newblock Quantized hall conductance as a topological invariant.
\newblock \emph{Phys. Rev. B}, 31:\penalty0 3372--3377, Mar 1985.
\newblock \doi{10.1103/PhysRevB.31.3372}.
\newblock URL \url{https://link.aps.org/doi/10.1103/PhysRevB.31.3372}.

\bibitem[Volovik and Yakovenko(1989)]{TAI16}
G.~E. Volovik and V.~M. Yakovenko.
\newblock Fractional charge, spin and statistics of solitons in superfluid
  $^3$he film.
\newblock \emph{J. Condens. Matter Phys.}, 1\penalty0 (31):\penalty0 5263, aug
  1989.
\newblock \doi{10.1088/0953-8984/1/31/025}.
\newblock URL \url{https://dx.doi.org/10.1088/0953-8984/1/31/025}.

\bibitem[Volovik(2003)]{TAI17}
G.~E. Volovik.
\newblock \emph{The Universe in a Helium Droplet}.
\newblock Clarendon Press, Oxford, 2003.
\newblock URL
  \url{http://www.issp.ac.ru/ebooks/books/open/The_Universe_in_a_Helium_Droplet.pdf}.

\bibitem[Wang et~al.(2010)Wang, Qi, and Zhang]{TAI18}
Zhong Wang, Xiao-Liang Qi, and Shou-Cheng Zhang.
\newblock Topological order parameters for interacting topological insulators.
\newblock \emph{Phys. Rev. Lett.}, 105:\penalty0 256803, Dec 2010.
\newblock \doi{10.1103/PhysRevLett.105.256803}.
\newblock URL \url{https://link.aps.org/doi/10.1103/PhysRevLett.105.256803}.

\bibitem[Gurarie(2011)]{TAI19}
V.~Gurarie.
\newblock Single-particle green's functions and interacting topological
  insulators.
\newblock \emph{Phys. Rev. B}, 83:\penalty0 085426, Feb 2011.
\newblock \doi{10.1103/PhysRevB.83.085426}.
\newblock URL \url{https://link.aps.org/doi/10.1103/PhysRevB.83.085426}.

\bibitem[Essin and Gurarie(2011)]{TAI20}
Andrew~M. Essin and Victor Gurarie.
\newblock Bulk-boundary correspondence of topological insulators from their
  respective green's functions.
\newblock \emph{Phys. Rev. B}, 84:\penalty0 125132, Sep 2011.
\newblock \doi{10.1103/PhysRevB.84.125132}.
\newblock URL \url{https://link.aps.org/doi/10.1103/PhysRevB.84.125132}.

\bibitem[Kawabata et~al.(2019{\natexlab{a}})Kawabata, Shiozaki, Ueda, and
  Sato]{TAI21}
Kohei Kawabata, Ken Shiozaki, Masahito Ueda, and Masatoshi Sato.
\newblock Symmetry and topology in non-hermitian physics.
\newblock \emph{Phys. Rev. X}, 9:\penalty0 041015, Oct 2019{\natexlab{a}}.
\newblock \doi{10.1103/PhysRevX.9.041015}.
\newblock URL \url{https://link.aps.org/doi/10.1103/PhysRevX.9.041015}.

\bibitem[Peralta~Gavensky et~al.(2023)Peralta~Gavensky, Sachdev, and
  Goldman]{TAI22}
Lucila Peralta~Gavensky, Subir Sachdev, and Nathan Goldman.
\newblock Connecting the many-body chern number to luttinger's theorem through
  st\ifmmode \check{r}\else \v{r}\fi{}eda's formula.
\newblock \emph{Phys. Rev. Lett.}, 131:\penalty0 236601, Dec 2023.
\newblock \doi{10.1103/PhysRevLett.131.236601}.
\newblock URL \url{https://link.aps.org/doi/10.1103/PhysRevLett.131.236601}.

\bibitem[Blason and Fabrizio(2023)]{TAI23}
Andrea Blason and Michele Fabrizio.
\newblock Unified role of green's function poles and zeros in correlated
  topological insulators.
\newblock \emph{Phys. Rev. B}, 108:\penalty0 125115, Sep 2023.
\newblock \doi{10.1103/PhysRevB.108.125115}.
\newblock URL \url{https://link.aps.org/doi/10.1103/PhysRevB.108.125115}.

\bibitem[Setty et~al.(2024)Setty, Xie, Sur, Chen, Vergniory, and Si]{TAI24}
Chandan Setty, Fang Xie, Shouvik Sur, Lei Chen, Maia~G. Vergniory, and Qimiao
  Si.
\newblock Electronic properties, correlated topology, and green's function
  zeros.
\newblock \emph{Phys. Rev. Res.}, 6:\penalty0 033235, Sep 2024.
\newblock \doi{10.1103/PhysRevResearch.6.033235}.
\newblock URL \url{https://link.aps.org/doi/10.1103/PhysRevResearch.6.033235}.

\bibitem[Wang and Zhang(2012)]{TAI25}
Zhong Wang and Shou-Cheng Zhang.
\newblock Simplified topological invariants for interacting insulators.
\newblock \emph{Phys. Rev. X}, 2:\penalty0 031008, Aug 2012.
\newblock \doi{10.1103/PhysRevX.2.031008}.
\newblock URL \url{https://link.aps.org/doi/10.1103/PhysRevX.2.031008}.

\bibitem[Wang et~al.(2012)Wang, Qi, and Zhang]{TAI26}
Zhong Wang, Xiao-Liang Qi, and Shou-Cheng Zhang.
\newblock Topological invariants for interacting topological insulators with
  inversion symmetry.
\newblock \emph{Phys. Rev. B}, 85:\penalty0 165126, Apr 2012.
\newblock \doi{10.1103/PhysRevB.85.165126}.
\newblock URL \url{https://link.aps.org/doi/10.1103/PhysRevB.85.165126}.

\bibitem[Wang and Yan(2013)]{TAI27}
Zhong Wang and Binghai Yan.
\newblock Topological hamiltonian as an exact tool for topological invariants.
\newblock \emph{J. Condens. Matter Phys.}, 25\penalty0 (15):\penalty0 155601,
  mar 2013.
\newblock \doi{10.1088/0953-8984/25/15/155601}.
\newblock URL \url{https://dx.doi.org/10.1088/0953-8984/25/15/155601}.

\bibitem[Tran et~al.(2022)Tran, Nguyen, Nguyen, and Tran]{TAI28}
Minh-Tien Tran, Duong-Bo Nguyen, Hong-Son Nguyen, and Thanh-Mai~Thi Tran.
\newblock Topological green function of interacting systems.
\newblock \emph{Phys. Rev. B}, 105:\penalty0 155112, Apr 2022.
\newblock \doi{10.1103/PhysRevB.105.155112}.
\newblock URL \url{https://link.aps.org/doi/10.1103/PhysRevB.105.155112}.

\bibitem[Budich et~al.(2012)Budich, Thomale, Li, Laubach, and Zhang]{TAI29}
Jan~Carl Budich, Ronny Thomale, Gang Li, Manuel Laubach, and Shou-Cheng Zhang.
\newblock Fluctuation-induced topological quantum phase transitions in quantum
  spin-hall and anomalous-hall insulators.
\newblock \emph{Phys. Rev. B}, 86:\penalty0 201407, Nov 2012.
\newblock \doi{10.1103/PhysRevB.86.201407}.
\newblock URL \url{https://link.aps.org/doi/10.1103/PhysRevB.86.201407}.

\bibitem[Hung et~al.(2013)Hung, Wang, Gu, and Fiete]{TAI30}
Hsiang-Hsuan Hung, Lei Wang, Zheng-Cheng Gu, and Gregory~A. Fiete.
\newblock Topological phase transition in a generalized kane-mele-hubbard
  model: A combined quantum monte carlo and green's function study.
\newblock \emph{Phys. Rev. B}, 87:\penalty0 121113, Mar 2013.
\newblock \doi{10.1103/PhysRevB.87.121113}.
\newblock URL \url{https://link.aps.org/doi/10.1103/PhysRevB.87.121113}.

\bibitem[Yoshida et~al.(2014)Yoshida, Peters, Fujimoto, and Kawakami]{TAI31}
Tsuneya Yoshida, Robert Peters, Satoshi Fujimoto, and Norio Kawakami.
\newblock Characterization of a topological mott insulator in one dimension.
\newblock \emph{Phys. Rev. Lett.}, 112:\penalty0 196404, May 2014.
\newblock \doi{10.1103/PhysRevLett.112.196404}.
\newblock URL \url{https://link.aps.org/doi/10.1103/PhysRevLett.112.196404}.

\bibitem[Budich et~al.(2013)Budich, Trauzettel, and Sangiovanni]{TAI32}
Jan~Carl Budich, Bj\"orn Trauzettel, and Giorgio Sangiovanni.
\newblock Fluctuation-driven topological hund insulators.
\newblock \emph{Phys. Rev. B}, 87:\penalty0 235104, Jun 2013.
\newblock \doi{10.1103/PhysRevB.87.235104}.
\newblock URL \url{https://link.aps.org/doi/10.1103/PhysRevB.87.235104}.

\bibitem[Amaricci et~al.(2015)Amaricci, Budich, Capone, Trauzettel, and
  Sangiovanni]{TAI33}
A.~Amaricci, J.~C. Budich, M.~Capone, B.~Trauzettel, and G.~Sangiovanni.
\newblock First-order character and observable signatures of topological
  quantum phase transitions.
\newblock \emph{Phys. Rev. Lett.}, 114:\penalty0 185701, May 2015.
\newblock \doi{10.1103/PhysRevLett.114.185701}.
\newblock URL \url{https://link.aps.org/doi/10.1103/PhysRevLett.114.185701}.

\bibitem[He et~al.(2016{\natexlab{a}})He, Wu, Meng, and Lu]{TAI34}
Yuan-Yao He, Han-Qing Wu, Zi~Yang Meng, and Zhong-Yi Lu.
\newblock Topological invariants for interacting topological insulators. i.
  efficient numerical evaluation scheme and implementations.
\newblock \emph{Phys. Rev. B}, 93:\penalty0 195163, May 2016{\natexlab{a}}.
\newblock \doi{10.1103/PhysRevB.93.195163}.
\newblock URL \url{https://link.aps.org/doi/10.1103/PhysRevB.93.195163}.

\bibitem[He et~al.(2016{\natexlab{b}})He, Wu, Meng, and Lu]{TAI35}
Yuan-Yao He, Han-Qing Wu, Zi~Yang Meng, and Zhong-Yi Lu.
\newblock Topological invariants for interacting topological insulators. ii.
  breakdown of single-particle green's function formalism.
\newblock \emph{Phys. Rev. B}, 93:\penalty0 195164, May 2016{\natexlab{b}}.
\newblock \doi{10.1103/PhysRevB.93.195164}.
\newblock URL \url{https://link.aps.org/doi/10.1103/PhysRevB.93.195164}.

\bibitem[Lee(2016)]{TAI36}
Tony~E. Lee.
\newblock Anomalous edge state in a non-hermitian lattice.
\newblock \emph{Phys. Rev. Lett.}, 116:\penalty0 133903, Apr 2016.
\newblock \doi{10.1103/PhysRevLett.116.133903}.
\newblock URL \url{https://link.aps.org/doi/10.1103/PhysRevLett.116.133903}.

\bibitem[Leykam et~al.(2017)Leykam, Bliokh, Huang, Chong, and Nori]{TAI37}
Daniel Leykam, Konstantin~Y. Bliokh, Chunli Huang, Y.~D. Chong, and Franco
  Nori.
\newblock Edge modes, degeneracies, and topological numbers in non-hermitian
  systems.
\newblock \emph{Phys. Rev. Lett.}, 118:\penalty0 040401, Jan 2017.
\newblock \doi{10.1103/PhysRevLett.118.040401}.
\newblock URL \url{https://link.aps.org/doi/10.1103/PhysRevLett.118.040401}.

\bibitem[Yao et~al.(2018)Yao, Song, and Wang]{TAI38}
Shunyu Yao, Fei Song, and Zhong Wang.
\newblock Non-hermitian chern bands.
\newblock \emph{Phys. Rev. Lett.}, 121:\penalty0 136802, Sep 2018.
\newblock \doi{10.1103/PhysRevLett.121.136802}.
\newblock URL \url{https://link.aps.org/doi/10.1103/PhysRevLett.121.136802}.

\bibitem[Yao and Wang(2018)]{TAI39}
Shunyu Yao and Zhong Wang.
\newblock Edge states and topological invariants of non-hermitian systems.
\newblock \emph{Phys. Rev. Lett.}, 121:\penalty0 086803, Aug 2018.
\newblock \doi{10.1103/PhysRevLett.121.086803}.
\newblock URL \url{https://link.aps.org/doi/10.1103/PhysRevLett.121.086803}.

\bibitem[Shen et~al.(2018{\natexlab{a}})Shen, Zhen, and Fu]{TAI40}
Huitao Shen, Bo~Zhen, and Liang Fu.
\newblock Topological band theory for non-hermitian hamiltonians.
\newblock \emph{Phys. Rev. Lett.}, 120:\penalty0 146402, Apr
  2018{\natexlab{a}}.
\newblock \doi{10.1103/PhysRevLett.120.146402}.
\newblock URL \url{https://link.aps.org/doi/10.1103/PhysRevLett.120.146402}.

\bibitem[Jiang et~al.(2018)Jiang, Yang, and Chen]{TAI41}
Hui Jiang, Chao Yang, and Shu Chen.
\newblock Topological invariants and phase diagrams for one-dimensional
  two-band non-hermitian systems without chiral symmetry.
\newblock \emph{Phys. Rev. A}, 98:\penalty0 052116, Nov 2018.
\newblock \doi{10.1103/PhysRevA.98.052116}.
\newblock URL \url{https://link.aps.org/doi/10.1103/PhysRevA.98.052116}.

\bibitem[Philip et~al.(2018)Philip, Hirsbrunner, and Gilbert]{TAI42}
Timothy~M. Philip, Mark~R. Hirsbrunner, and Matthew~J. Gilbert.
\newblock Loss of hall conductivity quantization in a non-hermitian quantum
  anomalous hall insulator.
\newblock \emph{Phys. Rev. B}, 98:\penalty0 155430, Oct 2018.
\newblock \doi{10.1103/PhysRevB.98.155430}.
\newblock URL \url{https://link.aps.org/doi/10.1103/PhysRevB.98.155430}.

\bibitem[Chen and Zhai(2018)]{TAI43}
Yu~Chen and Hui Zhai.
\newblock Hall conductance of a non-hermitian chern insulator.
\newblock \emph{Phys. Rev. B}, 98:\penalty0 245130, Dec 2018.
\newblock \doi{10.1103/PhysRevB.98.245130}.
\newblock URL \url{https://link.aps.org/doi/10.1103/PhysRevB.98.245130}.

\bibitem[Hirsbrunner et~al.(2019)Hirsbrunner, Philip, and Gilbert]{TAI44}
Mark~R. Hirsbrunner, Timothy~M. Philip, and Matthew~J. Gilbert.
\newblock Topology and observables of the non-hermitian chern insulator.
\newblock \emph{Phys. Rev. B}, 100:\penalty0 081104, Aug 2019.
\newblock \doi{10.1103/PhysRevB.100.081104}.
\newblock URL \url{https://link.aps.org/doi/10.1103/PhysRevB.100.081104}.

\bibitem[Liu et~al.(2019)Liu, Zhang, Ai, Gong, Kawabata, Ueda, and Nori]{TAI45}
Tao Liu, Yu-Ran Zhang, Qing Ai, Zongping Gong, Kohei Kawabata, Masahito Ueda,
  and Franco Nori.
\newblock Second-order topological phases in non-hermitian systems.
\newblock \emph{Phys. Rev. Lett.}, 122:\penalty0 076801, Feb 2019.
\newblock \doi{10.1103/PhysRevLett.122.076801}.
\newblock URL \url{https://link.aps.org/doi/10.1103/PhysRevLett.122.076801}.

\bibitem[Sato et~al.(2012)Sato, Hasebe, Esaki, and Kohmoto]{TAI46}
Masatoshi Sato, Kazuki Hasebe, Kenta Esaki, and Mahito Kohmoto.
\newblock {Time-Reversal Symmetry in Non-Hermitian Systems}.
\newblock \emph{Prog. Theor. Phys.}, 127\penalty0 (6):\penalty0 937--974, 06
  2012.
\newblock ISSN 0033-068X.
\newblock \doi{10.1143/PTP.127.937}.
\newblock URL \url{https://doi.org/10.1143/PTP.127.937}.

\bibitem[Kawabata et~al.(2019{\natexlab{b}})Kawabata, Higashikawa, Gong,
  Ashida, and Ueda]{TAI47}
K.~Kawabata, S.~Higashikawa, Z.~Gong, Y.~Ashida, and M.~Ueda.
\newblock Topological unification of time-reversal and particle-hole symmetries
  in non-hermitian physics.
\newblock \emph{Nat. Commun.}, 10:\penalty0 297, 2019{\natexlab{b}}.
\newblock \doi{10.1038/s41467-018-08254-y}.
\newblock URL \url{http://doi.org/10.1038/s41467-018-08254-y}.

\bibitem[Krishtopenko et~al.(2020{\natexlab{c}})Krishtopenko, Antezza, and
  Teppe]{TAI48}
Sergey~S. Krishtopenko, Mauro Antezza, and Fr\'ed\'eric Teppe.
\newblock Disorder-induced phase transition in dirac systems beyond the linear
  approximation.
\newblock \emph{Phys. Rev. B}, 101:\penalty0 205424, May 2020{\natexlab{c}}.
\newblock \doi{10.1103/PhysRevB.101.205424}.
\newblock URL \url{https://link.aps.org/doi/10.1103/PhysRevB.101.205424}.

\bibitem[Krishtopenko and Teppe(2018{\natexlab{a}})]{TAI49}
S.~S. Krishtopenko and F.~Teppe.
\newblock Realistic picture of helical edge states in hgte quantum wells.
\newblock \emph{Phys. Rev. B}, 97:\penalty0 165408, Apr 2018{\natexlab{a}}.
\newblock \doi{10.1103/PhysRevB.97.165408}.
\newblock URL \url{https://link.aps.org/doi/10.1103/PhysRevB.97.165408}.

\bibitem[Krishtopenko et~al.(2025{\natexlab{b}})Krishtopenko, Ikonnikov,
  Jouault, and Teppe]{TAI50}
S.~S. Krishtopenko, A.~V. Ikonnikov, B.~Jouault, and F.~Teppe.
\newblock Disorder-induced phase transitions in double hgte quantum wells.
\newblock \emph{Phys. Rev. B}, 111:\penalty0 045431, Jan 2025{\natexlab{b}}.
\newblock \doi{10.1103/PhysRevB.111.045431}.
\newblock URL \url{https://link.aps.org/doi/10.1103/PhysRevB.111.045431}.

\bibitem[Shen et~al.(2018{\natexlab{b}})Shen, Zhen, and Fu]{TAI51}
Huitao Shen, Bo~Zhen, and Liang Fu.
\newblock Topological band theory for non-hermitian hamiltonians.
\newblock \emph{Phys. Rev. Lett.}, 120:\penalty0 146402, Apr
  2018{\natexlab{b}}.
\newblock \doi{10.1103/PhysRevLett.120.146402}.
\newblock URL \url{https://link.aps.org/doi/10.1103/PhysRevLett.120.146402}.

\bibitem[Tourni\'{e} et~al.(2022)Tourni\'{e}, Bartolome, Calvo, Loghmari,
  D\'{\i}az-Thomas, Teissier, Baranov, Cerutti, and Rodriguez]{III-VQWs1}
Eric Tourni\'{e}, Laura~Monge Bartolome, Marta~Rio Calvo, Zeineb Loghmari,
  Daniel~A. D\'{\i}az-Thomas, Roland Teissier, Alexei~N. Baranov, Laurent
  Cerutti, and Jean-Baptiste Rodriguez.
\newblock Mid-infrared iii-v semiconductor lasers epitaxially grown on si
  substrates.
\newblock \emph{Light Sci. Appl.}, 11:\penalty0 165, 2022.
\newblock \doi{10.1038/s41377-022-00850-4}.
\newblock URL \url{http://doi.org/10.1038/s41377-022-00850-4}.

\bibitem[Wu et~al.(2018)Wu, Fatemi, Gibson, Watanabe, Taniguchi, Cava, and
  Jarillo-Herrero]{III-VQWs3}
Sanfeng Wu, Valla Fatemi, Quinn~D. Gibson, Kenji Watanabe, Takashi Taniguchi,
  Robert~J. Cava, and Pablo Jarillo-Herrero.
\newblock Observation of the quantum spin hall effect up to 100 kelvin in a
  monolayer crystal.
\newblock \emph{Science}, 359\penalty0 (6371):\penalty0 76--79, 2018.
\newblock \doi{10.1126/science.aan6003}.
\newblock URL \url{https://www.science.org/doi/abs/10.1126/science.aan6003}.

\bibitem[Bendias et~al.(2018)Bendias, Shamim, Herrmann, Budewitz, Shekhar,
  Leubner, Kleinlein, Bocquillon, Buhmann, and Molenkamp]{III-VQWs2}
Kalle Bendias, Saquib Shamim, Oliver Herrmann, Andreas Budewitz, Pragya
  Shekhar, Philipp Leubner, Johannes Kleinlein, Erwann Bocquillon, Hartmut
  Buhmann, and Laurens~W. Molenkamp.
\newblock High mobility hgte microstructures for quantum spin hall studies.
\newblock \emph{Nano Lett.}, 18\penalty0 (8):\penalty0 4831--4836, 2018.
\newblock \doi{10.1021/acs.nanolett.8b01405}.
\newblock URL \url{http://doi.org/10.1021/acs.nanolett.8b01405}.

\bibitem[Krishtopenko and Teppe(2018{\natexlab{b}})]{III-VQWs4}
Sergey~S. Krishtopenko and Fr\'{e}d\'{e}ric Teppe.
\newblock Quantum spin hall insulator with a large bandgap, dirac fermions, and
  bilayer graphene analog.
\newblock \emph{Sci. Adv.}, 4\penalty0 (4):\penalty0 eaap7529,
  2018{\natexlab{b}}.
\newblock \doi{10.1126/sciadv.aap7529}.
\newblock URL \url{https://www.science.org/doi/abs/10.1126/sciadv.aap7529}.

\bibitem[Liu et~al.(2008)Liu, Hughes, Qi, Wang, and Zhang]{III-VQWs5}
Chaoxing Liu, Taylor~L. Hughes, Xiao-Liang Qi, Kang Wang, and Shou-Cheng Zhang.
\newblock Quantum spin hall effect in inverted type-ii semiconductors.
\newblock \emph{Phys. Rev. Lett.}, 100:\penalty0 236601, Jun 2008.
\newblock \doi{10.1103/PhysRevLett.100.236601}.
\newblock URL \url{https://link.aps.org/doi/10.1103/PhysRevLett.100.236601}.

\bibitem[Knez et~al.(2011)Knez, Du, and Sullivan]{III-VQWs6}
Ivan Knez, Rui-Rui Du, and Gerard Sullivan.
\newblock Evidence for helical edge modes in inverted
  $\mathrm{InAs}/\mathrm{GaSb}$ quantum wells.
\newblock \emph{Phys. Rev. Lett.}, 107:\penalty0 136603, Sep 2011.
\newblock \doi{10.1103/PhysRevLett.107.136603}.
\newblock URL \url{https://link.aps.org/doi/10.1103/PhysRevLett.107.136603}.

\bibitem[Naveh and Laikhtman(1995)]{III-VQWs7}
Y.~Naveh and B.~Laikhtman.
\newblock Band‐structure tailoring by electric field in a weakly coupled
  electron‐hole system.
\newblock \emph{Appl. Phys. Lett.}, 66\penalty0 (15):\penalty0 1980--1982, 04
  1995.
\newblock ISSN 0003-6951.
\newblock \doi{10.1063/1.113297}.
\newblock URL \url{https://doi.org/10.1063/1.113297}.

\bibitem[Yang et~al.(1997)Yang, Yang, Bennett, and Shanabrook]{III-VQWs8}
M.~J. Yang, C.~H. Yang, B.~R. Bennett, and B.~V. Shanabrook.
\newblock Evidence of a hybridization gap in ``semimetallic'' inas/gasb
  systems.
\newblock \emph{Phys. Rev. Lett.}, 78:\penalty0 4613--4616, Jun 1997.
\newblock \doi{10.1103/PhysRevLett.78.4613}.
\newblock URL \url{https://link.aps.org/doi/10.1103/PhysRevLett.78.4613}.

\bibitem[Drndic et~al.(1997)Drndic, Grimshaw, Cooper, Ritchie, and
  Patel]{III-VQWs9}
M.~Drndic, M.~P. Grimshaw, L.~J. Cooper, D.~A. Ritchie, and N.~K. Patel.
\newblock Tunable electron-hole gases in gated inas/gasb/alsb systems.
\newblock \emph{Appl. Phys. Lett.}, 70\penalty0 (4):\penalty0 481--483, 01
  1997.
\newblock ISSN 0003-6951.
\newblock \doi{10.1063/1.118187}.
\newblock URL \url{https://doi.org/10.1063/1.118187}.

\bibitem[Nichele et~al.(2016)Nichele, Suominen, Kjaergaard, Marcus, Sajadi,
  Folk, Qu, Beukman, Vries, Veen, Nadj-Perge, Kouwenhoven, Nguyen, Kiselev, Yi,
  Sokolich, Manfra, Spanton, and Moler]{III-VQWs10}
Fabrizio Nichele, Henri~J Suominen, Morten Kjaergaard, Charles~M Marcus,
  Ebrahim Sajadi, Joshua~A Folk, Fanming Qu, Arjan J~A Beukman, Folkert K~de
  Vries, Jasper~van Veen, Stevan Nadj-Perge, Leo~P Kouwenhoven, Binh-Minh
  Nguyen, Andrey~A Kiselev, Wei Yi, Marko Sokolich, Michael~J Manfra, Eric~M
  Spanton, and Kathryn~A Moler.
\newblock Edge transport in the trivial phase of inas/gasb.
\newblock \emph{New J. Phys.}, 18\penalty0 (8):\penalty0 083005, jul 2016.
\newblock \doi{10.1088/1367-2630/18/8/083005}.
\newblock URL \url{https://doi.org/10.1088/1367-2630/18/8/083005}.

\bibitem[Du et~al.(2015)Du, Knez, Sullivan, and Du]{III-VQWs11}
Lingjie Du, Ivan Knez, Gerard Sullivan, and Rui-Rui Du.
\newblock Robust helical edge transport in gated $\mathrm{InAs}/\mathrm{GaSb}$
  bilayers.
\newblock \emph{Phys. Rev. Lett.}, 114:\penalty0 096802, Mar 2015.
\newblock \doi{10.1103/PhysRevLett.114.096802}.
\newblock URL \url{https://link.aps.org/doi/10.1103/PhysRevLett.114.096802}.

\bibitem[Du et~al.(2017)Du, Li, Lou, Wu, Liu, Han, Zhang, Sullivan, Ikhlassi,
  Chang, and Du]{III-VQWs12}
Lingjie Du, Tingxin Li, Wenkai Lou, Xingjun Wu, Xiaoxue Liu, Zhongdong Han, Chi
  Zhang, Gerard Sullivan, Amal Ikhlassi, Kai Chang, and Rui-Rui Du.
\newblock Tuning edge states in strained-layer $\mathrm{InAs}/\mathrm{GaInSb}$
  quantum spin hall insulators.
\newblock \emph{Phys. Rev. Lett.}, 119:\penalty0 056803, Aug 2017.
\newblock \doi{10.1103/PhysRevLett.119.056803}.
\newblock URL \url{https://link.aps.org/doi/10.1103/PhysRevLett.119.056803}.

\bibitem[Vurgaftman et~al.(2001)Vurgaftman, Meyer, and Ram-Mohan]{III-VQWs13}
I.~Vurgaftman, J.~R. Meyer, and L.~R. Ram-Mohan.
\newblock Band parameters for iii--v compound semiconductors and their alloys.
\newblock \emph{J. Appl. Phys.}, 89\penalty0 (11):\penalty0 5815--5875, 06
  2001.
\newblock ISSN 0021-8979.
\newblock \doi{10.1063/1.1368156}.
\newblock URL \url{https://doi.org/10.1063/1.1368156}.

\bibitem[Patrashin et~al.(2017)Patrashin, Akahane, Sekine, and
  Hosako]{III-VQWs14}
Mikhail Patrashin, Kouichi Akahane, Norihiko Sekine, and Iwao Hosako.
\newblock Molecular beam epitaxy of strained-layer inas/gainsb superlattices
  for long-wavelength photodetectors.
\newblock \emph{J. Cryst. Growth}, 477:\penalty0 86--90, 2017.
\newblock ISSN 0022-0248.
\newblock \doi{https://doi.org/10.1016/j.jcrysgro.2017.02.030}.
\newblock URL
  \url{https://www.sciencedirect.com/science/article/pii/S0022024817301033}.

\bibitem[Haugan et~al.(2022)Haugan, Das, Bharadwaj, Ram-Mohan, Corbett, Smith,
  Gupta, Mahalingam, Bedford, and Eyink]{III-VQWs15}
H.~J. Haugan, D.~Das, S.~Bharadwaj, L.~R. Ram-Mohan, J.~P. Corbett, R.~K.
  Smith, J.~A. Gupta, K.~Mahalingam, R.~G. Bedford, and K.~G. Eyink.
\newblock Optimization of hybridized inassb/ingasb semiconductor topological
  materials.
\newblock \emph{Appl. Phys. Lett.}, 121\penalty0 (6):\penalty0 062109, 08 2022.
\newblock ISSN 0003-6951.
\newblock \doi{10.1063/5.0099721}.
\newblock URL \url{https://doi.org/10.1063/5.0099721}.

\bibitem[Krishtopenko et~al.(2017)Krishtopenko, Ikonnikov, Maremyanin, Bovkun,
  Spirin, Kadykov, Marcinkiewicz, Ruffenach, Consejo, Teppe, Knap, Semyagin,
  Putyato, Emelyanov, Preobrazhenskii, and Gavrilenko]{MontIIIVexp1}
S.~S. Krishtopenko, A.~V. Ikonnikov, K.~V. Maremyanin, L.~S. Bovkun, K.~E.
  Spirin, A.~M. Kadykov, M.~Marcinkiewicz, S.~Ruffenach, C.~Consejo, F.~Teppe,
  W.~Knap, B.~R. Semyagin, M.~A. Putyato, E.~A. Emelyanov, V.~V.
  Preobrazhenskii, and V.~I. Gavrilenko.
\newblock Cyclotron resonance of {D}irac fermions in {InAs/GaSb/InAs} quantum
  wells.
\newblock \emph{Semiconductors}, 51\penalty0 (1):\penalty0 38--42, Jan 2017.
\newblock ISSN 1090-6479.
\newblock \doi{10.1134/S1063782617010109}.
\newblock URL \url{https://doi.org/10.1134/S1063782617010109}.

\bibitem[Ruffenach et~al.(2017)Ruffenach, Krishtopenko, Bovkun, Ikonnikov,
  Marcinkiewicz, Consejo, Potemski, Piot, Orlita, Semyagin, Putyato, Emelyanov,
  Preobrazhenskii, Knap, Gonzalez-Posada, Boissier, Tourni\'{e}, Teppe, and
  Gavrilenko]{MontIIIVexp2}
S.~Ruffenach, S.~S. Krishtopenko, L.~S. Bovkun, A.~V. Ikonnikov,
  M.~Marcinkiewicz, C.~Consejo, M.~Potemski, B.~Piot, M.~Orlita, B.~R.
  Semyagin, M.~A. Putyato, E.~A. Emelyanov, V.~V. Preobrazhenskii, W.~Knap,
  F.~Gonzalez-Posada, G.~Boissier, E.~Tourni\'{e}, F.~Teppe, and V.~I.
  Gavrilenko.
\newblock Magnetoabsorbtion of {D}irac fermions in gapless three-layer
  {InAs/GaSb/InAs} quantum wells.
\newblock \emph{JETP Lett.}, 106\penalty0 (11):\penalty0 727, 2017.
\newblock \doi{10.1134/S0021364017230102}.
\newblock URL \url{http://dx.doi.org/10.1134/S0021364017230102}.

\bibitem[Krishtopenko et~al.(2018)Krishtopenko, Ruffenach, Gonzalez-Posada,
  Boissier, Marcinkiewicz, Fadeev, Kadykov, Rumyantsev, Morozov, Gavrilenko,
  Consejo, Desrat, Jouault, Knap, Tourni\'e, and Teppe]{MontIIIVexp3}
S.~S. Krishtopenko, S.~Ruffenach, F.~Gonzalez-Posada, G.~Boissier,
  M.~Marcinkiewicz, M.~A. Fadeev, A.~M. Kadykov, V.~V. Rumyantsev, S.~V.
  Morozov, V.~I. Gavrilenko, C.~Consejo, W.~Desrat, B.~Jouault, W.~Knap,
  E.~Tourni\'e, and F.~Teppe.
\newblock Temperature-dependent terahertz spectroscopy of inverted-band
  three-layer inas/gasb/inas quantum well.
\newblock \emph{Phys. Rev. B}, 97:\penalty0 245419, Jun 2018.
\newblock \doi{10.1103/PhysRevB.97.245419}.
\newblock URL \url{https://link.aps.org/doi/10.1103/PhysRevB.97.245419}.

\bibitem[Krishtopenko et~al.(2019{\natexlab{b}})Krishtopenko, Ruffenach,
  Gonzalez-Posada, Consejo, Desrat, Jouault, Knap, Fadeev, Kadykov, Rumyantsev,
  Morozov, Boissier, Tourni{\'{e}}, Gavrilenko, and Teppe]{MontIIIVexp4}
S.~S. Krishtopenko, S.~Ruffenach, F.~Gonzalez-Posada, C.~Consejo, W.~Desrat,
  B.~Jouault, W.~Knap, M.~A. Fadeev, A.~M. Kadykov, V.~V. Rumyantsev, S.~V.
  Morozov, G.~Boissier, E.~Tourni{\'{e}}, V.~I. Gavrilenko, and F.~Teppe.
\newblock Terahertz spectroscopy of two-dimensional semimetal in three-layer
  {InAs}/{GaSb}/{InAs} quantum well.
\newblock \emph{{JETP} Letters}, 109\penalty0 (2):\penalty0 96--101, jan
  2019{\natexlab{b}}.
\newblock \doi{10.1134/s0021364019020085}.
\newblock URL \url{http://dx.doi.org/10.1134/s0021364019020085}.

\bibitem[Avogadri et~al.(2022)Avogadri, Gebert, Krishtopenko, Castillo,
  Consejo, Ruffenach, Roblin, Bray, Krupko, Juillaguet, Contreras, Wolf,
  Hartmann, H\"ofling, Boissier, Rodriguez, Nanot, Tourni\'e, Teppe, and
  Jouault]{MontIIIVexp5}
C.~Avogadri, S.~Gebert, S.~S. Krishtopenko, I.~Castillo, C.~Consejo,
  S.~Ruffenach, C.~Roblin, C.~Bray, Y.~Krupko, S.~Juillaguet, S.~Contreras,
  A.~Wolf, F.~Hartmann, S.~H\"ofling, G.~Boissier, J.-B. Rodriguez, S.~Nanot,
  E.~Tourni\'e, F.~Teppe, and B.~Jouault.
\newblock Large inverted band gap in strained three-layer inas/gainsb quantum
  wells.
\newblock \emph{Phys. Rev. Res.}, 4:\penalty0 L042042, Dec 2022.
\newblock \doi{10.1103/PhysRevResearch.4.L042042}.
\newblock URL \url{https://link.aps.org/doi/10.1103/PhysRevResearch.4.L042042}.

\bibitem[Benlemqwanssa et~al.(2024)Benlemqwanssa, Krishtopenko, Meyer,
  Benhamou-Bui, Bonnet, Wolf, Bray, Consejo, Ruffenach, Nanot, Rodriguez,
  Tourni\'e, Hartmann, H\"ofling, Teppe, and Jouault]{MontIIIVexp6}
S.~Benlemqwanssa, S.S. Krishtopenko, M.~Meyer, B.~Benhamou-Bui, L.~Bonnet,
  A.~Wolf, C.~Bray, C.~Consejo, S.~Ruffenach, S.~Nanot, J.-B. Rodriguez,
  E.~Tourni\'e, F.~Hartmann, S.~H\"ofling, F.~Teppe, and B.~Jouault.
\newblock Multiprobe analysis to separate edge currents from bulk currents in
  quantum spin hall insulators and to analyze their temperature dependence.
\newblock \emph{Phys. Rev. Appl.}, 22:\penalty0 064059, Dec 2024.
\newblock \doi{10.1103/PhysRevApplied.22.064059}.
\newblock URL \url{https://link.aps.org/doi/10.1103/PhysRevApplied.22.064059}.

\bibitem[Winkler et~al.(1996)Winkler, Merkler, Darnhofer, and
  R\"ossler]{III-VQWs16}
R.~Winkler, M.~Merkler, T.~Darnhofer, and U.~R\"ossler.
\newblock Theory for the cyclotron resonance of holes in strained asymmetric
  ge-sige quantum wells.
\newblock \emph{Phys. Rev. B}, 53:\penalty0 10858--10865, Apr 1996.
\newblock \doi{10.1103/PhysRevB.53.10858}.
\newblock URL \url{https://link.aps.org/doi/10.1103/PhysRevB.53.10858}.

\bibitem[Ziesmann et~al.(1987)Ziesmann, Heitmann, and Chang]{III-VQWs17}
M.~Ziesmann, D.~Heitmann, and L.~L. Chang.
\newblock Interaction of optical phonons with electrons in an inas quantum
  well.
\newblock \emph{Phys. Rev. B}, 35:\penalty0 4541--4544, Mar 1987.
\newblock \doi{10.1103/PhysRevB.35.4541}.
\newblock URL \url{https://link.aps.org/doi/10.1103/PhysRevB.35.4541}.

\bibitem[Ivchenko et~al.(1996)Ivchenko, Kaminski, and R\"ossler]{III-VQWs18}
E.~L. Ivchenko, A.~Yu. Kaminski, and U.~R\"ossler.
\newblock Heavy-light hole mixing at zinc-blende (001) interfaces under normal
  incidence.
\newblock \emph{Phys. Rev. B}, 54:\penalty0 5852--5859, Aug 1996.
\newblock \doi{10.1103/PhysRevB.54.5852}.
\newblock URL \url{https://link.aps.org/doi/10.1103/PhysRevB.54.5852}.

\bibitem[Morozov et~al.(2014{\natexlab{a}})Morozov, Rumyantsev, Antonov,
  Kadykov, Maremyanin, Kudryavtsev, Mikhailov, Dvoretskii, and
  Gavrilenko]{III-VQWs19}
S.~V. Morozov, V.~V. Rumyantsev, A.~V. Antonov, A.~M. Kadykov, K.~V.
  Maremyanin, K.~E. Kudryavtsev, N.~N. Mikhailov, S.~A. Dvoretskii, and V.~I.
  Gavrilenko.
\newblock Time resolved photoluminescence spectroscopy of narrow gap
  hg$_{1-x}$cd$_{x}$te/cd$_{y}$hg$_{1-y}$te quantum well heterostructures.
\newblock \emph{Appl. Phys. Lett.}, 105\penalty0 (2):\penalty0 022102, 07
  2014{\natexlab{a}}.
\newblock ISSN 0003-6951.
\newblock \doi{10.1063/1.4890416}.
\newblock URL \url{https://doi.org/10.1063/1.4890416}.

\bibitem[Morozov et~al.(2014{\natexlab{b}})Morozov, Rumyantsev, Antonov,
  Maremyanin, Kudryavtsev, Krasilnikova, Mikhailov, Dvoretskii, and
  Gavrilenko]{III-VQWs20}
S.~V. Morozov, V.~V. Rumyantsev, A.~V. Antonov, K.~V. Maremyanin, K.~E.
  Kudryavtsev, L.~V. Krasilnikova, N.~N. Mikhailov, S.~A. Dvoretskii, and V.~I.
  Gavrilenko.
\newblock Efficient long wavelength interband photoluminescence from hgcdte
  epitaxial films at wavelengths up to 26~$\mu$m.
\newblock \emph{Appl. Phys. Lett.}, 104\penalty0 (7):\penalty0 072102, 02
  2014{\natexlab{b}}.
\newblock ISSN 0003-6951.
\newblock \doi{10.1063/1.4866006}.
\newblock URL \url{https://doi.org/10.1063/1.4866006}.

\bibitem[Ling et~al.(2004)Ling, Lui, Ma, Chen, Fung, and Beling]{III-VQWs21}
C.~C. Ling, M.~K. Lui, S.~K. Ma, X.~D. Chen, S.~Fung, and C.~D. Beling.
\newblock Nature of the acceptor responsible for p-type conduction in liquid
  encapsulated czochralski-grown undoped gallium antimonide.
\newblock \emph{Appl. Phys. Lett.}, 85\penalty0 (3):\penalty0 384--386, 2004.
\newblock \doi{10.1063/1.1773934}.
\newblock URL \url{http://dx.doi.org/10.1063/1.1773934}.

\bibitem[Kujala et~al.(2014)Kujala, Segercrantz, Tuomisto, and
  Slotte]{III-VQWs22}
J.~Kujala, N.~Segercrantz, F.~Tuomisto, and J.~Slotte.
\newblock Native point defects in gasb.
\newblock \emph{J. Appl. Phys.}, 116\penalty0 (14):\penalty0 143508, 2014.
\newblock \doi{10.1063/1.4898082}.
\newblock URL \url{http://dx.doi.org/10.1063/1.4898082}.

\bibitem[Segercrantz et~al.(2014)Segercrantz, Slotte, Makkonen, Kujala,
  Tuomisto, Song, and Wang]{III-VQWs23}
N.~Segercrantz, J.~Slotte, I.~Makkonen, J.~Kujala, F.~Tuomisto, Y.~Song, and
  S.~Wang.
\newblock Point defect balance in epitaxial gasb.
\newblock \emph{Appl. Phys. Lett.}, 105\penalty0 (8):\penalty0 082113, 2014.
\newblock \doi{10.1063/1.4894473}.
\newblock URL \url{http://dx.doi.org/10.1063/1.4894473}.

\bibitem[Castro~Neto et~al.(2009)Castro~Neto, Guinea, Peres, Novoselov, and
  Geim]{III-VQWs24}
A.~H. Castro~Neto, F.~Guinea, N.~M.~R. Peres, K.~S. Novoselov, and A.~K. Geim.
\newblock The electronic properties of graphene.
\newblock \emph{Rev. Mod. Phys.}, 81:\penalty0 109--162, Jan 2009.
\newblock \doi{10.1103/RevModPhys.81.109}.
\newblock URL \url{https://link.aps.org/doi/10.1103/RevModPhys.81.109}.

\bibitem[Aleshkin et~al.(2005{\natexlab{a}})Aleshkin, Gavrilenko, Ikonnikov,
  Sadofyev, Bird, Johnson, and Zhang]{III-VQWs30}
V.~Ya. Aleshkin, V.~I. Gavrilenko, A.~V. Ikonnikov, Yu.~G. Sadofyev, J.~P.
  Bird, S.~R. Johnson, and Y.~H. Zhang.
\newblock Cyclotron resonance in doped and undoped inas/alsb heterostructures
  with quantum wells.
\newblock \emph{Semiconductors}, 39\penalty0 (1):\penalty0 62--66, Jan
  2005{\natexlab{a}}.
\newblock ISSN 1090-6479.
\newblock \doi{10.1134/1.1852647}.
\newblock URL \url{https://doi.org/10.1134/1.1852647}.

\bibitem[Kalinin et~al.(2013)Kalinin, Krishtopenko, Maremyanin, Spirin,
  Gavrilenko, Biryukov, Baidus, and Zvonkov]{III-VQWs31}
K.~P. Kalinin, S.~S. Krishtopenko, K.~V. Maremyanin, K.~E. Spirin, V.~I.
  Gavrilenko, A.~A. Biryukov, N.~V. Baidus, and B.~N. Zvonkov.
\newblock Rashba spin splitting and cyclotron resonance in strained ingaas/inp
  heterostructures with a two-dimensional electron gas.
\newblock \emph{Semiconductors}, 47\penalty0 (11):\penalty0 1485--1491, Nov
  2013.
\newblock ISSN 1090-6479.
\newblock \doi{10.1134/S1063782613110092}.
\newblock URL \url{https://doi.org/10.1134/S1063782613110092}.

\bibitem[Ikonnikov et~al.(2010)Ikonnikov, Antonov, Lastovkin, Gavrilenko,
  Sadof'ev, and Samal]{III-VQWs32}
A.~V. Ikonnikov, A.~V. Antonov, A.~A. Lastovkin, V.~I. Gavrilenko, Yu.~G.
  Sadof'ev, and N.~Samal.
\newblock High-resolution emission spectra of pulsed terahertz quantum-cascade
  lasers.
\newblock \emph{Semiconductors}, 44\penalty0 (11):\penalty0 1467--1471, Nov
  2010.
\newblock ISSN 1090-6479.
\newblock \doi{10.1134/S1063782610110175}.
\newblock URL \url{https://doi.org/10.1134/S1063782610110175}.

\bibitem[Gauer et~al.(1993)Gauer, Scriba, Wixforth, Kotthaus, Nguyen, Tuttle,
  English, and Kroemer]{III-VQWs26}
Ch~Gauer, J~Scriba, A~Wixforth, J~P Kotthaus, C~Nguyen, G~Tuttle, J~H English,
  and H~Kroemer.
\newblock Photoconductivity in alsb/inas quantum wells.
\newblock \emph{Semicond. Sci. Technol.}, 8\penalty0 (1S):\penalty0 S137, Jan
  1993.
\newblock \doi{10.1088/0268-1242/8/1S/031}.
\newblock URL \url{https://doi.org/10.1088/0268-1242/8/1S/031}.

\bibitem[Sadofyev et~al.(2005)Sadofyev, Ramamoorthy, Bird, Johnson, and
  Zhang]{III-VQWs27}
Yu.~G. Sadofyev, A.~Ramamoorthy, J.~P. Bird, S.~R. Johnson, and Y.-H. Zhang.
\newblock Large negative persistent photoconductivity in inas/alsb quantum
  wells.
\newblock \emph{Appl. Phys. Lett.}, 86\penalty0 (19):\penalty0 192109, 2005.
\newblock \doi{10.1063/1.1926407}.
\newblock URL \url{https://doi.org/10.1063/1.1926407}.

\bibitem[Aleshkin et~al.(2005{\natexlab{b}})Aleshkin, Gavrilenko, Gaponova,
  Ikonnikov, Maremyanin, Morozov, Sadofyev, Johnson, and Zhang]{III-VQWs28}
V.~Ya. Aleshkin, V.~I. Gavrilenko, D.~M. Gaponova, A.~V. Ikonnikov, K.~V.
  Maremyanin, S.~V. Morozov, Yu.~G. Sadofyev, S.~R. Johnson, and Y.~H. Zhang.
\newblock Spectra of persistent photoconductivity in inas/alsb quantum-well
  heterostructures.
\newblock \emph{Semiconductors}, 39\penalty0 (1):\penalty0 22--26, Jan
  2005{\natexlab{b}}.
\newblock ISSN 1090-6479.
\newblock \doi{10.1134/1.1852637}.
\newblock URL \url{https://doi.org/10.1134/1.1852637}.

\bibitem[Gavrilenko et~al.(2010)Gavrilenko, Ikonnikov, Krishtopenko, Lastovkin,
  Maremyanin, Sadofyev, and Spirin]{III-VQWs29}
V.~I. Gavrilenko, A.~V. Ikonnikov, S.~S. Krishtopenko, A.~A. Lastovkin, K.~V.
  Maremyanin, Yu.~G. Sadofyev, and K.~E. Spirin.
\newblock Persistent photoconductivity in inas/alsb heterostructures with
  double quantum wells.
\newblock \emph{Semiconductors}, 44\penalty0 (5):\penalty0 616--622, May 2010.
\newblock ISSN 1090-6479.
\newblock \doi{10.1134/S106378261005012X}.
\newblock URL \url{https://doi.org/10.1134/S106378261005012X}.

\bibitem[Meyer et~al.(2021)Meyer, Schmid, Jabeen, Bastard, Hartmann, and
  H\"ofling]{WurzIIIVexp1}
M.~Meyer, S.~Schmid, F.~Jabeen, G.~Bastard, F.~Hartmann, and S.~H\"ofling.
\newblock Topological band structure in inas/gasb/inas triple quantum wells.
\newblock \emph{Phys. Rev. B}, 104:\penalty0 085301, Aug 2021.
\newblock \doi{10.1103/PhysRevB.104.085301}.
\newblock URL \url{https://link.aps.org/doi/10.1103/PhysRevB.104.085301}.

\bibitem[Schmid et~al.(2022)Schmid, Meyer, Jabeen, Bastard, Hartmann, and
  H\"ofling]{WurzIIIVexp2}
S.~Schmid, M.~Meyer, F.~Jabeen, G.~Bastard, F.~Hartmann, and S.~H\"ofling.
\newblock Exploring the phase diagram of inas/gasb/inas trilayer quantum wells.
\newblock \emph{Phys. Rev. B}, 105:\penalty0 155304, Apr 2022.
\newblock \doi{10.1103/PhysRevB.105.155304}.
\newblock URL \url{https://link.aps.org/doi/10.1103/PhysRevB.105.155304}.

\bibitem[Meyer et~al.(2023)Meyer, Schmid, Jabeen, Bastard, Hartmann, and
  Höfling]{WurzIIIVexp3}
M~Meyer, S~Schmid, F~Jabeen, G~Bastard, F~Hartmann, and S~Höfling.
\newblock Voltage control of the quantum scattering time in inas/gasb/inas
  trilayer quantum wells.
\newblock \emph{New J. Phys.}, 25\penalty0 (2):\penalty0 023035, mar 2023.
\newblock \doi{10.1088/1367-2630/acbab7}.
\newblock URL \url{https://doi.org/10.1088/1367-2630/acbab7}.

\bibitem[Meyer et~al.(2025)Meyer, Baumbach, Krishtopenko, Wolf, Emmerling,
  Schmid, Kamp, Jouault, Rodriguez, Tournie, Müller, Thomale, Bastard, Teppe,
  Hartmann, and Höfling]{WurzIIIVexp4}
Manuel Meyer, Jonas Baumbach, Sergey Krishtopenko, Adriana Wolf, Monika
  Emmerling, Sebastian Schmid, Martin Kamp, Benoit Jouault, Jean-Baptiste
  Rodriguez, Eric Tournie, Tobias Müller, Ronny Thomale, Gerald Bastard,
  Frederic Teppe, Fabian Hartmann, and Sven Höfling.
\newblock Quantum spin hall effect in iii-v semiconductors at elevated
  temperatures: Advancing topological electronics.
\newblock \emph{Sci. Adv.}, 11\penalty0 (43):\penalty0 eadz2408, 2025.
\newblock \doi{10.1126/sciadv.adz2408}.
\newblock URL \url{https://www.science.org/doi/abs/10.1126/sciadv.adz2408}.

\bibitem[Lunczer et~al.(2019)Lunczer, Leubner, Endres, M\"uller, Br\"une,
  Buhmann, and Molenkamp]{III-VQWs33}
Lukas Lunczer, Philipp Leubner, Martin Endres, Valentin~L. M\"uller, Christoph
  Br\"une, Hartmut Buhmann, and Laurens~W. Molenkamp.
\newblock Approaching quantization in macroscopic quantum spin hall devices
  through gate training.
\newblock \emph{Phys. Rev. Lett.}, 123:\penalty0 047701, Jul 2019.
\newblock \doi{10.1103/PhysRevLett.123.047701}.
\newblock URL \url{https://link.aps.org/doi/10.1103/PhysRevLett.123.047701}.

\bibitem[Mueller et~al.(2015)Mueller, Pal, Karalic, Tschirky, Charpentier,
  Wegscheider, Ensslin, and Ihn]{III-VQWs34}
Susanne Mueller, Atindra~Nath Pal, Matija Karalic, Thomas Tschirky, Christophe
  Charpentier, Werner Wegscheider, Klaus Ensslin, and Thomas Ihn.
\newblock Nonlocal transport via edge states in inas/gasb coupled quantum
  wells.
\newblock \emph{Phys. Rev. B}, 92:\penalty0 081303, Aug 2015.
\newblock \doi{10.1103/PhysRevB.92.081303}.
\newblock URL \url{https://link.aps.org/doi/10.1103/PhysRevB.92.081303}.

\bibitem[Mueller et~al.(2017)Mueller, Mittag, Tschirky, Charpentier,
  Wegscheider, Ensslin, and Ihn]{III-VQWs35}
Susanne Mueller, Christopher Mittag, Thomas Tschirky, Christophe Charpentier,
  Werner Wegscheider, Klaus Ensslin, and Thomas Ihn.
\newblock Edge transport in inas and inas/gasb quantum wells.
\newblock \emph{Phys. Rev. B}, 96:\penalty0 075406, Aug 2017.
\newblock \doi{10.1103/PhysRevB.96.075406}.
\newblock URL \url{https://link.aps.org/doi/10.1103/PhysRevB.96.075406}.

\bibitem[Sun et~al.(2009)Sun, Yao, Fradkin, and Kivelson]{III-VQWs36}
Kai Sun, Hong Yao, Eduardo Fradkin, and Steven~A. Kivelson.
\newblock Topological insulators and nematic phases from spontaneous symmetry
  breaking in 2d fermi systems with a quadratic band crossing.
\newblock \emph{Phys. Rev. Lett.}, 103:\penalty0 046811, Jul 2009.
\newblock \doi{10.1103/PhysRevLett.103.046811}.
\newblock URL \url{https://link.aps.org/doi/10.1103/PhysRevLett.103.046811}.

\bibitem[Nishimoto et~al.(2010)Nishimoto, Nakamura, O'Brien, and
  Fulde]{III-VQWs37}
Satoshi Nishimoto, Masaaki Nakamura, Aroon O'Brien, and Peter Fulde.
\newblock Metal-insulator transition of fermions on a kagome lattice at 1/3
  filling.
\newblock \emph{Phys. Rev. Lett.}, 104:\penalty0 196401, May 2010.
\newblock \doi{10.1103/PhysRevLett.104.196401}.
\newblock URL \url{https://link.aps.org/doi/10.1103/PhysRevLett.104.196401}.

\bibitem[Wen et~al.(2010)Wen, R\"uegg, Wang, and Fiete]{III-VQWs38}
Jun Wen, Andreas R\"uegg, C.-C.~Joseph Wang, and Gregory~A. Fiete.
\newblock Interaction-driven topological insulators on the kagome and the
  decorated honeycomb lattices.
\newblock \emph{Phys. Rev. B}, 82:\penalty0 075125, Aug 2010.
\newblock \doi{10.1103/PhysRevB.82.075125}.
\newblock URL \url{https://link.aps.org/doi/10.1103/PhysRevB.82.075125}.

\bibitem[Zhu et~al.(2016)Zhu, Gong, Zeng, Fu, and Sheng]{III-VQWs39}
W.~Zhu, Shou-Shu Gong, Tian-Sheng Zeng, Liang Fu, and D.~N. Sheng.
\newblock Interaction-driven spontaneous quantum hall effect on a kagome
  lattice.
\newblock \emph{Phys. Rev. Lett.}, 117:\penalty0 096402, Aug 2016.
\newblock \doi{10.1103/PhysRevLett.117.096402}.
\newblock URL \url{https://link.aps.org/doi/10.1103/PhysRevLett.117.096402}.

\bibitem[Ji et~al.(2022)Ji, Gao, Yue, Wang, Wu, Dai, and Weng]{III-VQWs40}
Xuecong Ji, Jiacheng Gao, Changming Yue, Zhijun Wang, Hua Wu, Xi~Dai, and
  Hongming Weng.
\newblock Interaction-driven topological phase transition in monolayer
  ${\mathrm{crcl}}_{2}$(pyrazine)${}_{2}$.
\newblock \emph{Phys. Rev. B}, 106:\penalty0 235103, Dec 2022.
\newblock \doi{10.1103/PhysRevB.106.235103}.
\newblock URL \url{https://link.aps.org/doi/10.1103/PhysRevB.106.235103}.

\bibitem[Bray et~al.(2022)Bray, Maussang, Consejo, Delgado-Notario,
  Krishtopenko, Yahniuk, Gebert, Ruffenach, Dinar, Moench, Eroms, Indykiewicz,
  Jouault, Torres, Meziani, Knap, Yurgens, Ganichev, and Teppe]{III-VQWs41}
C.~Bray, K.~Maussang, C.~Consejo, J.~A. Delgado-Notario, S.~Krishtopenko,
  I.~Yahniuk, S.~Gebert, S.~Ruffenach, K.~Dinar, E.~Moench, J.~Eroms,
  K.~Indykiewicz, B.~Jouault, J.~Torres, Y.~M. Meziani, W.~Knap, A.~Yurgens,
  S.~D. Ganichev, and F.~Teppe.
\newblock Temperature-dependent zero-field splittings in graphene.
\newblock \emph{Phys. Rev. B}, 106:\penalty0 245141, Dec 2022.
\newblock \doi{10.1103/PhysRevB.106.245141}.
\newblock URL \url{https://link.aps.org/doi/10.1103/PhysRevB.106.245141}.

\bibitem[Meyer et~al.(2024)Meyer, F\"ahndrich, Schmid, Wolf, Krishtopenko,
  Jouault, Bastard, Teppe, Hartmann, and H\"ofling]{III-VQWs42}
M.~Meyer, T.~F\"ahndrich, S.~Schmid, A.~Wolf, S.~S. Krishtopenko, B.~Jouault,
  G.~Bastard, F.~Teppe, F.~Hartmann, and S.~H\"ofling.
\newblock Coexistence of topological and normal insulating phases in
  electro-optically tuned inas/gasb bilayer quantum wells.
\newblock \emph{Phys. Rev. B}, 109:\penalty0 L121303, Mar 2024.
\newblock \doi{10.1103/PhysRevB.109.L121303}.
\newblock URL \url{https://link.aps.org/doi/10.1103/PhysRevB.109.L121303}.

\end{thebibliography}
\end{document}